\documentclass[showpacs,nofootinbib,showkeys,eqsecnum,prd,aps,report]{revtex4-1}
\usepackage{amsfonts,amsmath,amssymb,bm,graphicx}

\begin{document}

\title{ Quantization of charged fields in the presence of critical potential
steps}
\author{S.~P.~Gavrilov${}^{a,c}$}
\email{gavrilovsergeyp@yahoo.com}
\author{D.~M.~Gitman${}^{a,b,d}$}
\email{gitman@if.usp.br}
\date{\today }

\begin{abstract}
QED with strong external backgrounds that can create particles from the
vacuum is well developed for the so-called $t$-electric potential steps,
which are time-dependent external electric fields that are switched on and
off at some time instants. However, there exist many physically interesting
situations where external backgrounds do not switch off at the time
infinity. E.g., these are time-independent nonuniform electric fields that
are concentrated in restricted space areas. The latter backgrounds represent
a kind of spatial $x$-electric potential steps for charged particles. They
can also create particles from the vacuum, the Klein paradox being closely
related to this process. Approaches elaborated for treating quantum effects
in the $t$-electric potential steps are not directly applicable to the $x$%
-electric potential steps and their generalization for $x$-electric
potential steps was not sufficiently developed. We believe that the present
work represents a consistent solution of the latter problem. We have
considered a canonical quantization of the Dirac and scalar fields with $x$%
-electric potential step and have found in- and out-creation and
annihilation operators that allow one to have particle interpretation of the
physical system under consideration. To identify in- and out-operators we
have performed a detailed mathematical and physical analysis of solutions of
the relativistic wave equations with an $x$-electric potential step with
subsequent QFT analysis of correctness of such an identification. We
elaborated a nonperturbative (in the external field) technique that allows
one to calculate all characteristics of zero-order processes, such, for
example, scattering, reflection, and electron-positron pair creation,
without radiation corrections, and also to calculate Feynman diagrams that
describe all characteristics of processes with interaction between the
\textrm{in}-, \textrm{out}-particles and photons. These diagrams have
formally the usual form, but contain special propagators. Expressions for
these propagators in terms of \textrm{in}- and \textrm{out}-solutions are
presented. We apply the elaborated approach to two popular exactly solvable
cases of $x$-electric potential steps, namely, to the Sauter potential and
to the Klein step.
\end{abstract}


\affiliation{${}^{a}$Department of Physics, Tomsk State University, 634050 Tomsk, Russia;\\
${}^{b}$P.N. Lebedev Physical Institute, 53 Leninskiy prospect, 119991 Moscow,
Russia;\\
${}^{c}$Department of General and Experimental Physics, Herzen State
Pedagogical University of Russia, Moyka embankment 48, 191186
St.~Petersburg, Russia\\
${}^{d}$Institute of Physics, University of S\~{a}o Paulo, CP 66318, CEP
05315-970 S\~{a}o Paulo, SP, Brazil}

\maketitle
\tableofcontents

\section{Introduction\label{S1}}

The effect of particle creation by strong electromagnetic and gravitational
fields has been attracting attention already for a long time. The effect has
a pure quantum nature and was first considered in the framework of the
relativistic quantum mechanics with understanding that all the questions can
be answered only in the framework of quantum field theory (QFT). QFT with
external background is to a certain extent an appropriate model for such
calculations. In the framework of such a model, the particle creation is
closely related to a violation of the vacuum stability with time.
Backgrounds (external fields) that violate the vacuum stability are to be
electriclike fields that are able to produce nonzero work when interacting
with charged particles. Depending on the structure of such backgrounds,
different approaches for calculating the effect were proposed and realized.
Initially, the effect of particle creation was considered for time-dependent
external electric fields that are switched on and off at the initial and the
final time instants respectively. In what follows, we call such kind of
external fields the $t$-electric potential steps. Scattering, particle
creation from the vacuum and particle annihilation by the $t$-electric
potential steps were considered in the framework of the relativistic quantum
mechanics, see Refs. \cite{Nikis70a,Nikis79,GMR85,ruffini}; a more complete
list of relevant publications can be found in \cite{ruffini}. A general
formulation of quantum electrodynamics (QED) and QFT with $t$-electric
potential steps was developed in Refs. \cite{Gitman}. However, there exist
many physically interesting situations where external backgrounds formally
are not switched off at the time infinity, the corresponding backgrounds
formally being not $t$-electric potential steps. As an example, we may point
out time-independent nonuniform electric fields that are concentrated in
restricted space areas. The latter fields represent a kind of spatial or, as
we call them conditionally, $x$-electric potential steps for charged
particles. The $x$-electric potential steps can also create particles from
the vacuum, the Klein paradox being closely related to this process \cite%
{Klein27,Sauter31a,Sauter-pot}. We recall that Klein considered the
reflection and transmission of relativistic electrons incident on a
sufficiently high rectangular potential step (the Klein step) and he had
found that there exists a range of energy where the transmission coefficient
is negative and the reflection coefficient is greater than one. There would
apparently be more reflected fermions than incoming. This is called the
Klein paradox. One can find a broader interpretation of what should be
called\ the Klein paradox, e.g. see\ \cite{Hol98,KrekSuGr04,Gerr+etal11}.
These authors propose to speak about the Klein paradox when one encounters a
special behavior of stationary solutions both for fermions and bosons,
unusual for the nonrelativistic quantum mechanics.\ For example, if the
electron kinetic energy belongs to the so-called Klein zone, which is
situated just below the range of obvious total reflection, the stationary
solutions penetrate through the step, the sign of the kinetic energy being,
however, reversed. Just after the original Klein's paper the problem was
studied by Sauter, who considered both the Klein step \cite{Sauter31a} and a
more realistic smoothed potential step, $-\alpha E\tanh \left( x/\alpha
\right) $,\ which is\ called the Sauter potential \cite{Sauter-pot}. To
avoid confusion, the Klein paradox should be distinguished from the Klein\
tunneling through the square barrier, e.g., see \cite{DomCal99} and
references therein. This tunneling without an exponential suppression occurs
when an electron is incident on a high barrier, even when it is not high
enough to create particles. Approaches elaborated for treating quantum
effects in the $t$-electric potential steps are not directly applicable to
the $x$-electric potential steps. Some heuristic calculations of the
particle creation by $x$-electric potential steps in the framework of the
relativistic quantum mechanics were presented by Nikishov in Refs. \cite%
{Nikis79,Nikis70b} and later developed by Hansen and Ravndal in Ref. \cite%
{HansRavn81}. One should also mention the Damour work\ \cite{Damour77}, that
contributed significantly in applying semiclassical methods for treating
strong field problems in astrophysics. In fact, this work presents a first
step to bridge the gap between approaches to quantum effects in potential
steps developed within relativistic quantum mechanics and QFT. Using the
Damour's approach, mean numbers of pairs created by a strong uniform
electric field confined between two capacitor plates separated by a finite
distance was calculated in Ref.~\cite{WongW88}. A detailed historical review
can be found in Refs. \cite{DomCal99,HansRavn81}. Nikishov had tested his
way of calculation using the special case of a constant and uniform electric
field, which is possible both for the $t$-electric potential steps and the $%
x $-electric potential steps, see \cite{Nikis79,Nikis70b,Nikis04}). At that
time, however, no justification for such calculations from the QFT point of
view was known.

Thus, we face a situation when a material that is commonly treated as a part
of an introductory discussion to the relativistic quantum mechanics, see,
e.g., \cite{RQM93}, is not studied completely and has not an unique
interpretation in the research literature. For example, Nikishov \cite%
{Nikis04} has pointed out an inconsistency in the interpretation given by
Hansen and Ravndal \cite{HansRavn81}. In spite of\ the recognized
achievements\textbf{\ }of this author in this area, there was no response to
his observation. For the first time numerical simulations on space-time
resolved data for the Klein paradox in a three-particle problem were
reported in \cite{KrekSuGr04}. It was shown how electron-positron pairs are
created and found that the results contradict to conclusions made in several
works where the incoming electron was noted to \textquotedblleft knock
out\textquotedblright\ electrons from the step \cite{GMR85} or to
\textquotedblleft stimulate\textquotedblright\ \cite{NKM99} pair production.
A clear reduction\ was shown instead of the suggested enhancement of the
pair-production rate at those time instances, when the incoming electron
wave packet overlaps spatially with the potential step. Recently, quantum
simulations for the evolution of the Dirac spinor in the presence of linear
potential with trapped ions were interpreted again as an electron transition
to the negative energy branch, see Ref.~\cite{Gerr+etal11}.

Although the attempts to formulate a consistent QFT with potential steps as
a background field \cite{Nikis79,Nikis70b,HansRavn81,Nikis04,Damour77} have
not been completed, a general understanding was achieved that the Klein
paradox is absent from a future consistent QFT. In the recent years the main
attention of researchers was drawn to improving calculation technics, as
well as to special calculations and applications.\ Though our present work
is devoted to the formulation of such consistent QFT and contains some
applications only in order to illustrate the general construction, \ we
overview briefly, for completeness, some recent results obtained after the
review\ \cite{DomCal99} had appeared. We restrict this brief review only to
works, where the particle creation effect by potential steps is considered,
leaving the works devoted to quantum relativistic motion in potential
barriers or well potentials, where the problem of bound states is principal,
aside. The main motivation for applications is due to a close connection
between particle creation by strong electrostatic potentials, in particular,
by the steps, and the Unruh effect, the phenomenon of particle emission from
black holes and cosmological horizons, e.g., see reviews \cite%
{ruffini,CHM08,AndMot14}. Recent progress in laser physics allows one to
hope that particle creation effect will be experimentally observed in
laboratory conditions in the near future (see Refs.~\cite{Dun09} for a
review). In achieving extreme field strengths, the inhomogeneity of the
realistic field becomes important. There appears an interest to study the
effect in realistic inhomogeneous fields, in the\ main by using
semiclassical and numerical methods. The recent experimental demonstration
of the dynamic Casimir effect, and a soon expected verification of an\
analog of\ the Hawking radiation by using superconducting circuit devices,
associates these two academic problems with possible applications to
condensed matter physics \cite{NJBN12}. In the\ recent years, in\ what
concerns\ applications to condensed matter physics, particle creation by
external fields became an observable effect in physics of graphene and
similar nanostructures, say, in topological insulators and\ Weyl semimetals,
this area being currently under intense development \cite%
{castroneto,dassarma,top-insul11,VafVish14}. This is explained by the fact
that although the physics that gives rise to the massless Dirac fermions in
each of the above-mentioned materials is different, the low-energy
properties are governed by the same Dirac kinematics. The gap between the
upper and lower branches in the corresponding Dirac particle spectra is very
small, so that the particle creation effect turns out to be dominant (under
certain conditions) as a response to the external electriclike field action
on such materials. In particular, the particle creation effect is crucial
for understanding the conductivity of graphene, especially in the so-called
nonlinear regime. Electron-hole pair creation (which is an analog of the
electron-positron pair creation from the vacuum) was recently observed in
graphene by its indirect influence on the graphene conductivity \cite%
{Vandecasteele10}. The conductivity of graphene modified by the particle
creation was calculated in the framework of QED with $t$-electric potential
steps in ~\cite{GavGitY12}. It should be noted that the proof of the
masslessness of charge carriers in graphene was obtained in studying the
Klein tunneling through\ special potential barriers there. Soon after its
theoretical prediction \cite{KatsNovG06}, the Klein tunneling in graphene,
where the role of potential steps is played by p-n junctions, was observed
by several experimental groups \cite{Klein-tunn-exp09} (For a
colloquium-style introduction to this subject see \cite{been08} and recent
review \cite{dassarma}). Numerical modeling of the Klein tunneling through
potential barriers in n-p-n and n-n' junctions are in good agreement with
theoretical predictions \cite{Log+etal15}. Note that a sharp n-n' junction
can be described by the step potential. However, electron's kinetic energies
considered in that work do not allow one to speak about particle creation
effect. Possible experimental configurations for testing the pair creation
by a linear step of finite length were proposed in \cite{allor}. Observation
of the Klein paradox in the context of so-called quantum quenches, which can
nowadays be performed in experiments with ultracold atoms, is proposed in
\cite{quench15}. In achieving extreme field strengths, the inhomogeneity of
the field becomes important. A number of works have been done to estimate
the pair production rate for electric fields, whose direction is fixed and
whose magnitude varies in one spatial dimension \cite{KimPage06,KimLeeY10}.
These approaches are semiclassical, essentially WKB. A Monte Carlo world
line loop method has been developed and applied to the vacuum pair
production problem in Ref.~\cite{GieK05}. The world line instanton method,
in which the Monte Carlo sum is effectively dominated by a single instanton
loop, was extended to one-dimensional inhomogeneous fields in Refs.~\cite%
{DunS05}. The agreement between these results is excellent. The world line
instanton technique was extended in Ref.~\cite{DunW06} to compute the vacuum
pair production rate for spatially inhomogeneous electric background fields,
with the spatial inhomogeneity being genuinely two or three dimensional. A
mathematical analysis of pair production rate by the Sauter potential was
presented in Refs. \cite{ChK09,ChK11}. Loop corrections to two-point
correlation functions in the case of the time-independent electric field
given by a linear potential step were studied in Ref.~\cite{AkhmP15}.

In the present article, we consider a canonical quantization of the Dirac
and scalar (Klein-Gordon) fields in the presence of the $x$-electric
potential step as a background [quantization of the action (\ref{2.1})] in
terms of adequate \textrm{in-} and \textrm{out}-particles and develop a
calculation technique of different quantum processes, such as scattering,
reflection, and electron-positron pair creation. At the first stage of the
quantization, a Dirac Heisenberg operator $\hat{\Psi}\left( X\right) $ that
satisfies equal-time anticommutation relations and the Dirac equation, as
well, is assigned to the Dirac field $\psi \left( X\right) $ , see Sec. \ref%
{S4}. However, the complete program of quantization includes also the second
stage. At that stage we have to construct a Hilbert space of state vectors,
where anticommutation relations are realized, and construct operators of all
physical quantities of the system under consideration. For the free Dirac
field this second stage of the quantization program is well known, see, e.g.
\cite{Schwe61}\emph{.} The result is formulated in terms of a Fock space
constructed on the base of creation and annihilation operators of free Dirac
particles. For the Dirac field interacting with any external background, the
first stage of the canonical quantization of the action (\ref{2.1}) gives
the same above-mentioned result. But the second stage of the program has no
universal solution suitable for any background. Each specific background, or
a class of backgrounds, has to be analyzed separately. Here it is desirable
to find \textrm{in}- and \textrm{out}-creation and annihilation operators
that allow one to have particle interpretation of the quantized Dirac field,
as it was done in the case of $t$-electric potential steps\ in \cite{Gitman}%
. In the case under consideration, the corresponding construction is
possible, but it is more complicated (probably, not any background allows
such interpretation of the quantized Dirac field). Another problem is
related to the time-independence of the background under consideration.
Whereas when considering $t$-electric potential steps that vanish at the
time infinity one can naturally introduce certain in- and out-creation and
annihilation operators starting first with the Schr\"{o}dinger
representation and then$\ $passing to the Heisenberg picture, this way of
action is not possible in time-independent backgrounds. Now one must
quantize in the Heisenberg picture\ from the very beginning. Then a new
problem appears: how to identify in- and out-operators. When doing this, we
believe that the\ time independent $x$-electric potential steps, as well as
any constant electromagnetic field, is an idealization. In fact, any
external field was switched on at a remote time instant $t_{in},$ then it\
was acting during a very large\ period of time $T$, and finally it was
switched off at another remote time instant $t_{out}=t_{in}+T.$ We also
believe that if $T$ is large enough one can ignore effects of switching the
external field on and off. Besides, to identify in- and out-operators it is
important to perform a detailed mathematical and physical analysis of
solutions of the Dirac and Klein-Gordon equations with an $x$-electric
potential step. Such an analysis is presented below in Secs. \ref{S2} and %
\ref{S3}, \ where we introduce special in- and out- solutions of the Dirac
and Klein-Gordon equations that are used in the quantization program. It is
here that we give their physical interpretation to be confirmed further
after the quantization has been fulfilled in Secs. \ref{S5}, \ref{S6}, and %
\ref{S7}. In the latter sections we present a calculation of different
quantum processes such as scattering, reflection, and electron-positron pair
creation. In Sec. \ref{S8} we discuss quantum processes in complete QED and
construct effective particle propagators in terms of the in- and
out-solutions introduced.\emph{\ }In Sec. \ref{S9} the developed theory is
applied to exactly solvable cases of $x$-electric potential steps, namely,
to the Sauter potential, and to the Klein step. Finally, we present a
consistent QFT treatment of processes where a naive one-particle
consideration may lead to the Klein paradox. In Appendix \ref{Abosons} we
consider some peculiarities of the quantization of the scalar field in the
Klein zone. In Appendix \ref{t-const} we have studied the orthogonality
relations on the hyperplane $t=\ $\textrm{const.} for the different ranges
of quantum numbers. In Appendix \ref{mean} we have studied one-particle mean
values of the charge, the kinetic energy, the number of particles, the
current, and the energy flux through the surfaces{\large \ }$x=x_{\mathrm{L}%
} ${\large \ }and{\large \ }$x=x_{\mathrm{R}}${\large . }In Appendix \ref%
{Aloc} we show that stable electron wave packets in the Klein zone can exist
only in the left asymptotic region, whereas stable positron wave packets can
exist only in the right asymptotic region. The main contributions to the
particle creation processes in a slowly alternating Sauter field are found
in Appendix \ref{Auniform}.

\section{Dirac equation with $x$-electric potential step as an external
background\label{S2}}

First, we describe a typical $x$-electric potential step, for which our
construction is elaborated.

Potentials of an external electromagnetic field $A^{\mu }\left( X\right) $
in $d=D+1$ dimensional Minkowski space-time parametrized by coordinates $X,$
\begin{equation}
X=\left( X^{\mu },\ \mu =0,1,\ldots ,D\right) =\left( t,\mathbf{r}\right) ,\
X^{0}=t,\ \mathbf{r}=\left( X^{1},\ldots ,X^{D}\right)  \label{2.2}
\end{equation}%
that correspond to an $x$-electric potential step are chosen to be%
\begin{equation}
A^{\mu }\left( X\right) =\left( A^{0}\left( x\right) ,A^{j}=0,\
j=1,2,...,D\right) ,\ x=X^{1},  \label{2.3}
\end{equation}%
so that the magnetic field $\mathbf{B}$ is zero and the electric field $%
\mathbf{E}$ reads%
\begin{equation}
\mathbf{E}\left( X\right) =\mathbf{E}\left( x\right) =\left( E_{x}\left(
x\right) ,0,...,0\right) ,\ \ E_{x}\left( x\right) =-A_{0}^{\prime }\left(
x\right) =E\left( x\right) .  \label{2.4}
\end{equation}

The electric field (\ref{2.4}) is directed along the axis $x^{1}=x$, it is
inhomogeneous in the $x$-direction, and does not depend on time $t$ [$%
\mathbf{E}\left( x\right) $ is a constant field]. The main property of any $%
x $-electric potential step is%
\begin{equation}
A_{0}\left( x\right) \overset{x\rightarrow \pm \infty }{\longrightarrow }%
A_{0}\left( \pm \infty \right) ,\ \ E\left( x\right) \overset{\left\vert
x\right\vert \rightarrow \infty }{\longrightarrow }0,  \label{2.5}
\end{equation}%
where $A_{0}\left( \pm \infty \right) $ are some constant quantities, which
means that the electric field under consideration is switched off at spatial
infinity. In addition, we suppose\footnote{%
That means that we do not consider external electric fields that can create
bound states for charged particles.} that the first derivative of the scalar
potential $A_{0}\left( x\right) $ does not change its sign for any $x\in
\mathbb{R}.$ For definiteness sake, we suppose that%
\begin{equation}
A_{0}^{\prime }\left( x\right) \leq 0\Longrightarrow \left\{
\begin{array}{l}
E\left( x\right) =-A_{0}^{\prime }\left( x\right) \geq 0 \\
A_{0}\left( -\infty \right) >A_{0}\left( +\infty \right)%
\end{array}%
\right. ,  \label{2.7}
\end{equation}%
and that there exist points $x_{\mathrm{L}}$ and $x_{\mathrm{R}}$ ($x_{%
\mathrm{R}}>x_{\mathrm{L}}$) such that for $x\in $ $S_{\mathrm{L}}=(-\infty
,x_{\mathrm{L}}]$ and for $x\in S_{\mathrm{R}}=[x_{\mathrm{R}},\infty )$)
the electric field is already switched off,%
\begin{eqnarray}
\left. A_{0}\left( x\right) \right\vert _{x\in S_{\mathrm{L}}}
&=&A_{0}\left( -\infty \right) ,\ \ \left. E\left( x\right) \right\vert
_{x\in S_{\mathrm{L}}}=0,  \notag \\
\left. A_{0}\left( x\right) \right\vert _{x\in S_{\mathrm{R}}}
&=&A_{0}\left( +\infty \right) ,\ \ \left. E\left( x\right) \right\vert
_{x\in S_{\mathrm{R}}}=0,  \label{2.8}
\end{eqnarray}%
whereas the electric field is not zero in the region $S_{\mathrm{int}%
}=\left( x_{\mathrm{L}},x_{\mathrm{R}}\right) $. It accelerates positrons
along the axis $x$ in positive direction, and electrons in the negative
direction.

As an example of the $x$-electric potential step, we refer to the Sauter
potential \cite{Sauter-pot} given by Eq. (\ref{10.1}).

\begin{figure}[tbp]
\centering\includegraphics[scale=.6]{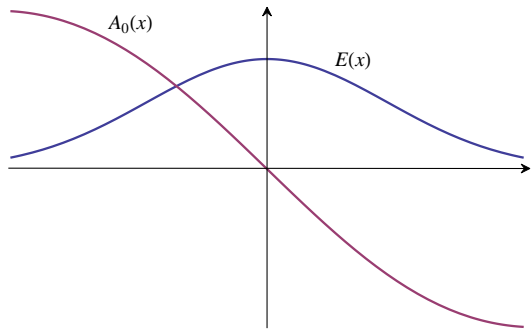}
\caption{$x$-electric potential step}
\label{1}
\end{figure}

Both the scalar potential $A_{0}(x)$ and the corresponding electric field $%
E(x)$ are shown on the same figure Fig.~\ref{1}.

Then in the $d=D+1$ dimensional (dim.) Minkowski space-time, we consider a
classical Dirac field $\psi \left( X\right) $ interacting with an external
electromagnetic field $A^{\mu }(X)$ (external background) representing an $x$%
-electric potential step. The action of this system has the form\footnote{%
Here we are using the natural system of units $\hslash =c=1$.}%
\begin{equation}
S=\int \bar{\psi}\left( X\right) \left[ \gamma ^{\mu }P_{\mu }-m\right] \psi
\left( X\right) dX,\ \ P_{\mu }=i\partial _{\mu }-qA_{\mu }(X),  \label{2.1}
\end{equation}%
where $\overline{\psi }=\psi ^{\dagger }\gamma ^{0}$, and$\ dX=dtd\mathbf{r.}
$

The classical Dirac field describes particles of a mass $m$ and with a
charge $q.$ The corresponding quantum theory is already charge symmetric,
and contains particles of the both signs of charges $\pm |q|$. We suppose
that\footnote{$e>0$ is the magnitude of the electron charge.} $q=-e$, which
means that we consider electrons as basic particles and positrons as their
antiparticles.

In $d$ dimensions, the Dirac field $\psi \left( X\right) $\ is a column with
$2^{\left[ d/2\right] }$ components (we call it just a spinor in what
follows), and $\gamma ^{\mu }$ are $2^{\left[ d/2\right] }\times 2^{\left[
d/2\right] }$ gamma-matrices, see e.g. \cite{BraWe35},
\begin{equation*}
\lbrack \gamma ^{\mu },\gamma ^{\nu }]_{+}=2\eta ^{\mu \nu },\;\eta ^{\mu
\nu }=\mathrm{diag}(\underbrace{1,-1,\ldots ,-1}_{d}),\;\mu ,\nu =0,1,\ldots
,D.
\end{equation*}

The Dirac field $\psi \left( X\right) $ satisfies the Dirac equation $\left(
\gamma ^{\mu }P_{\mu }-m\right) \psi =0,$ which, when written in the
Hamiltonian form, reads (we have taken into account that $\mathbf{A}=0$ for $%
x$-electric potential steps),
\begin{equation}
i\partial _{0}\psi \left( X\right) =\hat{H}\psi \left( X\right) ,\ \ \hat{H}%
=\gamma ^{0}\left( -i\gamma ^{j}\partial _{j}+m\right) +U\left( x\right) ,
\label{e1}
\end{equation}%
where $\hat{H}$ is the one-particle Dirac Hamiltonian and%
\begin{equation}
U\left( x\right) =qA_{0}\left( x\right) =-eA_{0}\left( x\right)  \label{2.60}
\end{equation}%
is the potential energy of the electron in the $x$-electric potential step.
It should be noted that $\hat{H}$ does not depend on the time $t$ in the
background under consideration.

\section{Stationary states with given left and right asymptotics\label{S3}}

\subsection{General\label{SS3.1}}

Let us consider stationary solutions of Dirac equation (\ref{e1}), having
the following form%
\begin{eqnarray}
&&\psi \left( X\right) =\exp \left( -ip_{0}t+i\mathbf{p}_{\bot }\mathbf{r}%
_{\bot }\right) \psi _{p_{0},\mathbf{p}_{\bot }}\left( x\right) ,\ \
X=\left( t,x,\mathbf{r}_{\bot }\right) ,  \notag \\
&&\psi _{p_{0},\mathbf{p}_{\bot }}\left( x\right) =\left\{ \gamma ^{0}\left[
p_{0}-U\left( x\right) \right] -\gamma ^{1}\hat{p}_{x}-\boldsymbol{\gamma }%
_{\bot }\mathbf{p}_{\bot }+m\right\} \phi _{p_{0},\mathbf{p}_{\bot }}(x),
\notag \\
&&\mathbf{r}_{\bot }=\left( X^{2},\ldots ,X^{D}\right) ,\ \mathbf{p}_{\bot
}=\left( p^{2},\ldots ,p^{D}\right) ,\ \boldsymbol{\gamma }_{\bot }=\left(
\gamma ^{2},...,\gamma ^{D}\right) ,\ \hat{p}_{x}=-i\partial _{x},
\label{e2}
\end{eqnarray}%
where $\psi _{p_{0},\mathbf{p}_{\bot }}\left( x\right) $ and $\phi _{p_{0},%
\mathbf{p}_{\bot }}(x)$ are spinors that depend on $x$ alone. In fact these
are stationary states with the energy $p_{0}$ and with definite momenta $%
\mathbf{p}_{\bot }$ in the perpendicular to the axis $x$ directions.
Substituting (\ref{e2}) into Dirac equation (\ref{e1}) (i.e., partially
squaring the Dirac equation), we obtain a second-order differential equation
for the spinor $\phi _{p_{0},\mathbf{p}_{\bot }}(x)$,
\begin{equation}
\left\{ \hat{p}_{x}^{2}-i\gamma ^{0}\gamma ^{1}U^{\prime }\left( x\right) -%
\left[ p_{0}-U\left( x\right) \right] ^{2}+\mathbf{p}_{\bot
}^{2}+m^{2}\right\} \phi _{p_{0},\mathbf{p}_{\bot }}(x)=0.  \label{e2.1}
\end{equation}

We can separate spinning variables by the substitution $\phi _{p_{0},\mathbf{%
p}_{\bot }}(x)=\varphi \left( x\right) v,$ where $\nu $ is a constant spinor
that is an eigenvector for the operator $\gamma ^{0}\gamma ^{1},$ and $%
\varphi \left( x\right) $ are some scalar functions. The latter operator has
two eigenvalues $\chi =\pm 1,$ such that $\gamma ^{0}\gamma ^{1}v_{\chi
}=\chi v_{\chi }.$ In addition, in $d>3,$ equation (\ref{e2.1}) allows one
to subject the constant spinors $v_{\chi }$ to some supplementary conditions
that, for example, can be chosen in the form%
\begin{eqnarray}
&&i\gamma ^{2s}\gamma ^{2s+1}v_{\chi ,\sigma }=\sigma _{s}v_{\chi ,\sigma }\
\mathrm{for}\ \mathrm{even\ }d,  \notag \\
&&i\gamma ^{2s+1}\gamma ^{2s+2}v_{\chi ,\sigma }=\sigma _{s}v_{\chi ,\sigma
}\ \mathrm{for\ odd}\ d,  \notag \\
&&\sigma =(\sigma _{s}=\pm 1,\ \ s=1,2,...,\left[ d/2\right] -1)),
\label{e2.5}
\end{eqnarray}%
where the quantum numbers $\sigma _{s}$ describe spin degrees of freedom (in
($1+1$) and $\left( 2+1\right) $ dimensions ($d=2,3$) there are no spin
degrees of freedom described by the quantum numbers $\sigma $). In what
follows we choose the spinors $v_{\chi ,\sigma }$ orthonormalized, $v_{\chi
,\sigma }^{\dag }v_{\chi ^{\prime },\sigma }=\delta _{\chi ,\chi ^{\prime
}}\delta _{\sigma ,\sigma ^{\prime }}.$ Taking all the said into account, we
use the following substitution in equation (\ref{e2.1}):
\begin{equation}
\phi _{p_{0},\mathbf{p}_{\bot }}(x)=\varphi _{n}^{\left( \chi \right)
}\left( x\right) v_{\chi ,\sigma },\;\ n=(p_{0},\mathbf{p}_{\bot },\sigma
).\   \label{2.33}
\end{equation}%
Then scalar functions $\varphi _{n}^{\left( \chi \right) }\left( x\right) $
have to obey the second order differential equation%
\begin{equation}
\left\{ \hat{p}_{x}^{2}-i\chi U^{\prime }\left( x\right) -\left[
p_{0}-U\left( x\right) \right] ^{2}+\mathbf{p}_{\bot }^{2}+m^{2}\right\}
\varphi _{n}^{\left( \chi \right) }\left( x\right) =0.  \label{e3}
\end{equation}

Thus, we are going to deal with solutions (\ref{e2}) of the form%
\begin{eqnarray}
&&\psi _{n}^{\left( \chi \right) }\left( X\right) =\exp \left( -ip_{0}t+i%
\mathbf{p}_{\bot }\mathbf{r}_{\bot }\right) \varrho _{n}^{\left( \chi
\right) }\left( x\right) ,\ \ \hat{H}\psi _{n}^{\left( \chi \right)
}=p_{0}\psi _{n}^{\left( \chi \right) },  \notag \\
&&\varrho _{n}^{\left( \chi \right) }\left( x\right) =\left\{ \gamma ^{0}
\left[ p_{0}-U\left( x\right) \right] -\gamma ^{1}\hat{p}_{x}-\boldsymbol{%
\gamma }_{\bot }\mathbf{p}_{\bot }+m\right\} \varphi _{n}^{\left( \chi
\right) }\left( x\right) v_{\chi ,\sigma }\ .  \label{e7}
\end{eqnarray}

One can easily verify (this is a well-known property related to the specific
structure of the projection operator in the brackets $\left\{ ...\right\} $)
that solutions $\psi _{n}^{\left( \chi \right) }\left( X\right) $ and $\psi
_{n}^{\left( \chi ^{\prime }\right) }\left( X\right) $ (\ref{e7}), which
only differ by values of $\chi $ are linearly dependent. Because of this, it
suffices to work with solutions corresponding to one of the possible two
values of $\chi $. That is why, we sometimes omit the subscript $\chi $ in
the solutions, in such cases it is supposed that the spin quantum number $%
\chi $ is fixed in a certain way. Due to the same reason, there exists, in
fact, only $J_{(d)}=2^{[d/2]-1}$ different spin states (labeled by the
quantum numbers $\sigma $) for a given set $p_{0},\mathbf{p}_{\bot }$.
Special examples of solutions (\ref{e7}) are given in Sec. \ref{S9}.

A formal transition to the case of scalar field can be done by setting $\chi
=0$\ in Eq.~(\ref{e3}), $\varphi _{n}\left( x\right) =\varphi _{n}^{\left(
0\right) }\left( x\right) $. Then one can find a complete set of solutions
of the Klein-Gordon equation in the following form%
\begin{equation}
\psi _{n}\left( X\right) =\exp \left( -ip_{0}t+i\mathbf{p}_{\bot }\mathbf{r}%
_{\bot }\right) \varphi _{n}\left( x\right) ,\;\ n=(p_{0},\mathbf{p}_{\bot
}).  \label{e7b}
\end{equation}

In what follows, we are going to use solutions (\ref{e7}) and (\ref{e7b})
with special left and right asymptotics. Let us describe these solutions. In
such solutions the functions $\varphi _{n}\left( x\right) $ (with the
omitted index $\chi $) are denoted as $_{\;\zeta }\varphi _{n}\left(
x\right) $ or $^{\;\zeta }\varphi _{n}\left( x\right) $, respectively, and
satisfy the following asymptotic conditions [we recall that $A_{0}^{\prime
}\left( x\right) =0$ and the scalar potential $A_{0}\left( x\right) $ takes
constant values in the regions $S_{\mathrm{L}}$ and $S_{\mathrm{R}}$, see
Eqs. (\ref{2.8}).]
\begin{eqnarray}
&&_{\;\zeta }\varphi _{n}\left( x\right) =\varphi _{n_{,}\zeta }^{\mathrm{L}%
}\left( x\right) ,\ x\in S_{\mathrm{L}}=(-\infty ,x_{\mathrm{L}}]\ ,  \notag
\\
&&\left\{ \hat{p}_{x}^{2}-\left[ \pi _{0}\left( \mathrm{L}\right) \right]
^{2}+\pi _{\bot }^{2}\right\} \varphi _{n}^{\mathrm{L}}\left( x\right) =0;
\label{2.36a} \\
&&^{\;\zeta }\varphi _{n}\left( x\right) =\varphi _{n,\zeta }^{\mathrm{R}%
}\left( x\right) ,\ x\in S_{\mathrm{R}}=[x_{\mathrm{R}},\infty ),  \notag \\
&&\left\{ \hat{p}_{x}^{2}-\left[ \pi _{0}\left( \mathrm{R}\right) \right]
^{2}+\pi _{\bot }^{2}\right\} \varphi _{n}^{\mathrm{R}}\left( x\right) =0.
\label{2.36b}
\end{eqnarray}%
Here we have introduced the quantities $\pi _{0}\left( \mathrm{L}/\mathrm{R}%
\right) ,$%
\begin{eqnarray}
&&\pi _{0}\left( \mathrm{R}\right) =p_{0}-U_{\mathrm{R}},\;\pi _{0}\left(
\mathrm{L}\right) =p_{0}-U_{\mathrm{L}},\ \ \pi _{\bot }=\sqrt{\mathbf{p}%
_{\bot }^{2}+m^{2}},  \notag \\
&&U_{\mathrm{L}}=U\left( -\infty \right) ,\ \ U_{\mathrm{R}}=U\left( +\infty
\right) ,  \label{2.59}
\end{eqnarray}%
Since $p_{0}$ is the total energy of a particle, we interpret $\pi
_{0}\left( \mathrm{R}\right) $ and $\pi _{0}\left( \mathrm{L}\right) $ as
its asymptotic kinetic energies in the regions $S_{\mathrm{R}}$ and $S_{%
\mathrm{L}}$ respectively. We call the quantity $\pi _{\bot }$ the total
transversal energy or, for simplicity, the transversal energy (in spite of
the fact that it includes the rest energy as well). The introduced kinetic
energies satisfy the relation%
\begin{equation}
\pi _{0}\left( \mathrm{L}\right) =\pi _{0}\left( \mathrm{R}\right) +\mathbb{U%
}>\pi _{0}\left( \mathrm{R}\right) ,  \label{2.62}
\end{equation}%
where $\mathbb{U}$ is the magnitude of the electric step,%
\begin{equation}
\mathbb{U}=U_{\mathrm{R}}-U_{\mathrm{L}}>0.  \label{2.61}
\end{equation}

At the same time one can see that in the asymptotic regions $S_{\mathrm{L}}$
and $S_{\mathrm{R}}$ solutions of the Dirac equation $\psi _{n}\left(
X\right) $ are eigenfunctions of the operator $\hat{H}-U\left( x\right) $
with the eigenvalues $\pi _{0}\left( \mathrm{L}\right) $ and $\pi _{0}\left(
\mathrm{R}\right) $, respectively. Thus, it is natural to call this operator
the kinetic energy operator $\hat{H}^{\mathrm{kin}},$
\begin{equation}
\hat{H}^{\mathrm{kin}}=\hat{H}-U\left( x\right) ,\ \left. \ \hat{H}^{\mathrm{%
kin}}\psi _{n}\left( X\right) \right\vert _{x\rightarrow \pm \infty }=\pi
_{0}\left( \mathrm{R}/\mathrm{L}\right) \left. \ \psi _{n}\left( X\right)
\right\vert _{x\rightarrow \pm \infty }\ .  \label{2.61a}
\end{equation}

One-particle kinetic energy operator $\hat{H}^{\mathrm{kin}}$\ of the scalar
field [this field satisfies the Klein-Gordon equation written in the
Hamiltonian form (\ref{a0})] has the same asymptotic properties.

Equation (\ref{2.36a}) has a complete set of\ solutions (right asymptotics)
in the form of plane waves$,$%
\begin{equation}
\ \varphi _{n,\zeta }^{\mathrm{R}}\left( x\right) =\ ^{\zeta }\mathcal{N}%
\exp \left( ip^{\mathrm{R}}\ x\right) ,  \label{2.62a}
\end{equation}%
with real momenta $p^{\mathrm{R}}$ along the axis $x,$%
\begin{equation}
p^{\mathrm{R}}=\zeta \sqrt{\left[ \pi _{0}\left( \mathrm{R}\right) \right]
^{2}-\pi _{\bot }^{2}},\ \ \zeta =\mathrm{sgn}\ \left( p^{\mathrm{R}}\right)
=\pm .  \label{2.62b}
\end{equation}%
The corresponding solutions (\ref{e7}) of the Dirac equation are denoted as $%
^{\zeta }\psi _{n}\left( X\right) .$ They satisfy the condition%
\begin{equation}
\hat{p}_{x}\ ^{\zeta }\psi _{n}\left( X\right) =p^{\mathrm{R}}\ ^{\zeta
}\psi _{n}\left( X\right) ,\ \ x\rightarrow +\infty ,  \label{2.62l}
\end{equation}%
i.e., these are states with definite momenta $p^{\mathrm{R}}$ as $%
x\rightarrow +\infty $.

Nontrivial solutions $^{\zeta }\psi _{n}\left( X\right) $ exist only for
quantum numbers $n$ that obey the following relation%
\begin{equation}
\left[ \pi _{0}\left( \mathrm{R}\right) \right] ^{2}>\pi _{\bot
}^{2}\Longleftrightarrow \left\{
\begin{array}{l}
\pi _{0}\left( \mathrm{R}\right) >\pi _{\bot } \\
\pi _{0}\left( \mathrm{R}\right) <-\pi _{\bot }%
\end{array}%
\right. .  \label{2.62d}
\end{equation}

Equation (\ref{2.36b}) has a complete set of\ solutions (left asymptotics) $%
\varphi _{n}^{\mathrm{L}}\left( x\right) $ in the form of plane waves,%
\begin{equation}
\varphi _{n,\zeta }^{\mathrm{L}}\left( x\right) =\ _{\zeta }\mathcal{N}\exp
\left( ip^{\mathrm{L}}\ x\right) ,  \label{2.6}
\end{equation}%
with real momenta $p^{\mathrm{L}}$ along the axis $x,$%
\begin{equation}
p^{\mathrm{L}}=\zeta \sqrt{\left[ \pi _{0}\left( \mathrm{L}\right) \right]
^{2}-\pi _{\bot }^{2}},\ \ \zeta =\mathrm{sgn}\ \left( p^{\mathrm{L}}\right)
=\pm \ .  \label{2.6a}
\end{equation}%
The corresponding solutions (\ref{e7}) and (\ref{e7b}) are denoted as $%
_{\zeta }\psi _{n}\left( X\right) $. They satisfy the condition%
\begin{equation}
\hat{p}_{x}\ _{\zeta }\psi _{n}\left( X\right) =p^{\mathrm{L}}\ _{\zeta
}\psi _{n}\left( X\right) ,\ \ x\rightarrow -\infty \ ,  \label{2.6l}
\end{equation}%
i.e., these are states with definite momenta $p^{\mathrm{L}}$ as $%
x\rightarrow -\infty $.

The normalization factors $\mathcal{N}$ are determined in the next section.

There exists a useful relation between absolute values of the momenta $p^{%
\mathrm{R}}$ and $p^{\mathrm{L}},$%
\begin{eqnarray}
\left\vert p^{\mathrm{L}}\right\vert &=&\sqrt{\left\vert p^{\mathrm{R}%
}\right\vert ^{2}+2\eta _{\mathrm{R}}\mathbb{U}\sqrt{\left\vert p^{\mathrm{R}%
}\right\vert ^{2}+\pi _{\bot }^{2}}+\mathbb{U}^{2}},  \notag \\
\left\vert p^{\mathrm{R}}\right\vert &=&\sqrt{\left\vert p^{\mathrm{L}%
}\right\vert ^{2}-2\eta _{\mathrm{L}}\mathbb{U}\sqrt{\left\vert p^{\mathrm{L}%
}\right\vert ^{2}+\pi _{\bot }^{2}}+\mathbb{U}^{2}},  \label{2.61b}
\end{eqnarray}%
where $\eta _{\mathrm{L}}=\mathrm{sgn\ }\pi _{0}\left( \mathrm{L}\right) $,\
$\ \eta _{\mathrm{R}}=\mathrm{sgn\ }\pi _{0}\left( \mathrm{R}\right) .$

Nontrivial solutions $_{\zeta }\psi _{n}\left( X\right) $ exist only for
quantum numbers $n$ that obey the following relation%
\begin{equation}
\left[ \pi _{0}\left( \mathrm{L}\right) \right] ^{2}>\pi _{\bot
}^{2}\Longleftrightarrow \left\{
\begin{array}{l}
\pi _{0}\left( \mathrm{L}\right) >\pi _{\bot } \\
\pi _{0}\left( \mathrm{L}\right) <-\pi _{\bot }%
\end{array}%
\right. .  \label{2.6c}
\end{equation}

In what follows we distinguish two types of electric steps, noncritical and
critical, by their magnitudes as follows:%
\begin{equation}
\mathbb{U}=\left\{
\begin{array}{l}
\mathbb{U}<\mathbb{U}_{c}=2m\ ,\ \mathrm{noncritical\ steps} \\
\mathbb{U}>\mathbb{U}_{c},\ \mathrm{critical\ steps}%
\end{array}%
\right. .  \label{2.63}
\end{equation}

\subsection{Ranges of quantum numbers\label{SS3.2}}

There exist some ranges of quantum numbers $n$\ where solutions $\varphi
_{n}^{\mathrm{L/R}}\left( x\right) $\ have similar forms. These ranges and
the corresponding solutions are described below.

\begin{figure}[tbp]
\centering\includegraphics[scale=.3]{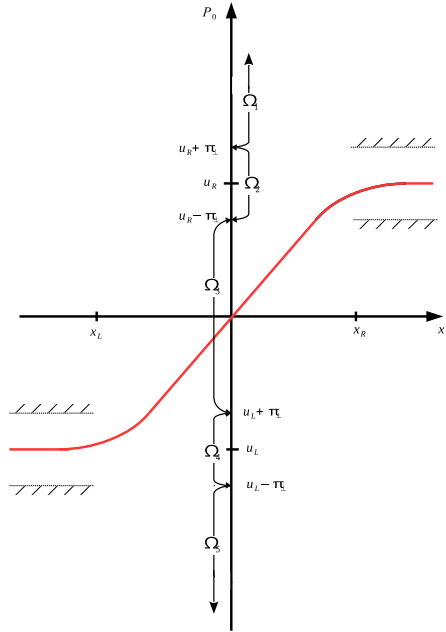}
\caption{Potential energy $U\left( x\right) $ of electrons in an $x$%
-electric step }
\label{2}
\end{figure}

In the case of the critical steps, which is of the main interest for us in
the present article, there exist five ranges\emph{\ }$\Omega _{k}$\emph{, }$%
k=1,...,5.$ We denote\emph{\ }the corresponding quantum numbers by $n_{k}$,\
so that $n_{k}\in \Omega _{k}$, see Fig.~\ref{2}.

It should be noted that the curve plotted on Fig. 2 is the potential energy $%
U\left( x\right) $ (\ref{2.60}) of an electron in the $x$-electric step,
such that electrons are accelerated to the left and positrons to the right
by the electric field (\ref{2.4}).

In the case of critical steps\emph{\ }$\mathbb{U}>\mathbb{U}_{c}$,{\large \ }%
the range{\large \ }$\Omega _{3}${\large \ }does exist for the quantum
numbers $p_{\bot }$\ restricted by the inequality,%
\begin{equation}
2\pi _{\bot }\leq \mathbb{U}\ .  \label{2.47}
\end{equation}%
{\large \ }Note that the range $\Omega _{3}$ often referred to as the Klein
zone.\emph{\ }In the case of noncritical steps\emph{\ }$\mathbb{U}<\mathbb{U}%
_{c}$, there exist only four ranges, the range $\Omega _{3}$ is absent.

The manifold of all the quantum numbers $n$ is denoted by $\Omega ,$ so that
$\Omega =\Omega _{1}\cup \cdots \cup \ \Omega _{5}.$

Below, we describe all the ranges in detail. In this connection, it should
be noted that the ranges $\Omega _{4}$ and $\Omega _{5}$ are similar to the
ranges $\Omega _{2}$ and $\Omega _{1}$ under the change of electrons by
positrons.

\subsubsection{Ranges $\Omega _{1}$ and $\Omega _{5}$\label{SSS3.2.1}}

The ranges $\Omega _{1}$ and $\Omega _{5}$ exist for any $\mathbb{U}$ and
includes the quantum numbers $n_{1}$ and $n_{5}$, respectively, that obey
the inequalities%
\begin{eqnarray}
p_{0} &\geq &U_{\mathrm{R}}+\pi _{\bot }\Longleftrightarrow \pi _{0}\left(
\mathrm{R}\right) \geq \pi _{\bot }\ \mathrm{\;if\;}n\in \Omega _{1},  \notag
\\
p_{0} &\leq &U_{\mathrm{L}}-\pi _{\bot }\Longleftrightarrow -\pi _{0}\left(
\mathrm{L}\right) \geq \pi _{\bot }\ \mathrm{\;if\;}n\in \Omega _{5}
\label{2.38}
\end{eqnarray}%
for a given $\pi _{\bot },$ see Fig. 2. Note that the inequalities (\ref%
{2.38}) imply the inequalities (\ref{2.6c}) for $n_{1}$ and (\ref{2.62d})
for $n_{5}$. Then it follows from Eqs. (\ref{2.38}) that in the ranges $%
\Omega _{1}$ and $\Omega _{5}$ there exist solutions $\ ^{\zeta }\psi
_{n}\left( X\right) $ and $\ _{\zeta }\psi _{n}\left( X\right) $. \ $^{\zeta
}\psi _{n}\left( X\right) $ can be interpreted as either a wave function of
electron for $n_{1}$ or a wave function of positron for $n_{5}$ with momenta
$p^{\mathrm{R}}$ along the axis $x$, given by Eq. (\ref{2.62b}), whereas $\
_{\zeta }\psi _{n}\left( X\right) $ can be interpreted as either a wave
function of electron for $n_{1}$ or a wave function of positron for $n_{5}$
with momenta $p^{\mathrm{L}}$ along the axis $x,$ given by Eq. (\ref{2.6a}).
The further analysis of these solutions in the framework of QED, given in
the Sec. \ref{S4} confirms this interpretation.

We believe that each pair of solutions $\ _{\zeta }\psi _{n}\left( X\right) $%
, $\zeta =\pm $ , forms a complete set for any given $n\in \Omega _{1}\cup \
\Omega _{5}$, the same is true for each pair of solutions $\ ^{\zeta }\psi
_{n}\left( X\right) $. This means that any given solution $\ ^{\zeta }\psi
_{n}\left( X\right) $ can be decomposed in terms of two solutions $\
_{-}\psi _{n}\left( X\right) $ and $\ _{+}\psi _{n}\left( X\right) ,$
whereas any given solution $\ _{\zeta }\psi _{n}\left( X\right) $ can be
decomposed in terms of two solutions $\ ^{+}\psi _{n}\left( X\right) $ and $%
\ ^{-}\psi _{n}\left( X\right) $ in the ranges $\Omega _{1}$ and $\Omega
_{5} $ .

\subsubsection{Ranges $\Omega _{2}$ and $\Omega _{4}$\label{SSS3.2.2}}

The ranges $\Omega _{2}$\ and $\Omega _{4}$ exist for any $\mathbb{U}$ and
include the quantum numbers $n_{2}$ that obey the inequalities%
\begin{eqnarray}
U_{\mathrm{R}}-\pi _{\bot } &<&p_{0}<U_{\mathrm{R}}+\pi _{\bot },\ \ \pi
_{0}\left( \mathrm{L}\right) >\pi _{\bot }\ \mathrm{if\;}2\pi _{\bot }\leq
\mathbb{U},  \notag \\
U_{\mathrm{L}}+\pi _{\bot } &<&p_{0}<U_{\mathrm{R}}+\pi _{\bot }\ \mathrm{%
if\;}2\pi _{\bot }>\mathbb{U},  \label{2.43}
\end{eqnarray}%
and the quantum numbers $n_{4}$ that obey the inequalities%
\begin{eqnarray}
U_{\mathrm{L}}-\pi _{\bot } &<&p_{0}<U_{\mathrm{L}}+\pi _{\bot },\ \ \pi
_{0}\left( \mathrm{R}\right) <-\pi _{\bot }\ \mathrm{if\;}2\pi _{\bot }\leq
U,  \notag \\
U_{\mathrm{L}}-\pi _{\bot } &<&p_{0}<U_{\mathrm{R}}-\pi _{\bot }\ \mathrm{%
if\;}2\pi _{\bot }>U.  \label{2.53}
\end{eqnarray}%
As a consequence of Eqs.~(\ref{2.43}) and (\ref{2.53}) there exist solutions
$_{\zeta }\psi _{n_{2}}\left( X\right) $ with definite left asymptotics and $%
\ ^{\zeta }\psi _{n_{4}}\left( X\right) $ with definite right asymptotics. $%
_{\zeta }\psi _{n_{2}}\left( X\right) $ and $^{\zeta }\psi _{n_{4}}\left(
X\right) $ can be interpreted as a wave function of electron and positron,
respectively. Nontrivial solutions $\ ^{\zeta }\psi _{n_{2}}\left( X\right) $
and $\ _{\zeta }\psi _{n_{4}}\left( X\right) $ do not exist, since Eq.~(\ref%
{2.43}) contradicts Eq.~(\ref{2.62d}) and Eq.~(\ref{2.53}) contradicts Eq.~(%
\ref{2.6c}).

The fact that any solution with quantum numbers $n_{2}$ has zero right
asymptotic and any solution with quantum numbers $n_{4}$ has zero left
asymptotic imposes restrictions on the form of the solutions $\ _{\zeta
}\psi _{n_{2}}\left( X\right) $ and $^{\zeta }\psi _{n_{4}}\left( X\right) $%
. In particular, they cannot be independent for different $\zeta .$ For
example, because the set $\ _{\zeta }\psi _{n_{2}}\left( X\right) $ is
complete, any solution $\psi _{n_{2}}\left( X\right) $ can be represented as%
\begin{equation}
\psi _{n_{2}}\left( X\right) =_{\ +}\psi _{n_{2}}\left( X\right) c_{+}+\
_{-}\psi _{n_{2}}\left( X\right) c_{-}\ .  \label{2.46d}
\end{equation}%
In the region $S_{\mathrm{R}}$ the superposition $\psi _{n_{2}}\left(
X\right) $ has zero asymptotics and therefore the corresponding Dirac
current in $x$-direction is zero. Using this consideration one can easily
find that $|c_{+}|\ =|c_{-}|,$ such that Eq.~(\ref{2.46d}) can be written as%
\begin{equation}
\psi _{n_{2}}\left( X\right) =_{\ +}\psi _{n_{2}}\left( X\right) e^{+i\theta
_{n_{2}}}+\ _{-}\psi _{n_{2}}\left( X\right) e^{-i\theta _{n_{2}}}\ .
\label{2.46b}
\end{equation}%
Following by the same logic, one sees that in the region $S_{\mathrm{L}}$
the only allowed superposition of $\ ^{+}\psi _{n_{4}}\left( X\right) $ and $%
\ ^{-}\psi _{n_{4}}\left( X\right) $ has zero asymptotics and therefore any
solution $\psi _{n_{2}}\left( X\right) $ can be written as
\begin{equation}
\psi _{n_{4}}\left( X\right) =\ ^{+}\psi _{n_{4}}\left( X\right) e^{+i\theta
_{n_{4}}}+\ ^{-}\psi _{n_{4}}\left( X\right) e^{-i\theta _{n_{4}}}\ .
\label{2.46a}
\end{equation}

In fact, $\psi _{n_{2}}\left( X\right) $\ are wave functions that describe
an unbounded motion in $x\rightarrow -\infty $ direction while $\psi
_{n_{4}}\left( X\right) $\ are wave functions that describe an unbounded
motion toward $x=+\infty $. Equations (\ref{2.46b}) and (\ref{2.46a})
provide asymptotic forms for these wave functions. These forms are sums of
two waves traveling in opposite directions, with equal in magnitude
currents, which means that we deal with standing waves.

\subsubsection{Range $\Omega _{3}$\label{SSS3.2.3}}

In the $\Omega _{3}$-range (in the Klein zone) the quantum numbers $\mathbf{p%
}_{\bot }$ are restricted by the inequality (\ref{2.47}) and for any of such
$\pi _{\bot }$\ quantum numbers $p_{0}$ obey the double inequality, see Fig.~%
\ref{2},%
\begin{equation}
U_{\mathrm{L}}+\pi _{\bot }\leq p_{0}\leq U_{\mathrm{R}}-\pi _{\bot }.
\label{2.48}
\end{equation}

Inequality (\ref{2.48}) implies also that $\pi _{0}\left( \mathrm{L}\right)
\geq \pi _{\bot }$ and$\ \pi _{0}\left( \mathrm{R}\right) \leq -\pi _{\bot
}. $ It follows from this inequality that there exist solutions $\ ^{\zeta
}\psi _{n_{3}}\left( X\right) $ (see condition (\ref{2.62d})). On the other
hand, since the inequality (\ref{2.48}) implies $\pi _{0}\left( \mathrm{L}%
\right) >\pi _{\bot }$ and (\ref{2.6c}), there exist solutions $\ _{\zeta
}\psi _{n_{3}}\left( X\right) $ as well. Thus, in the range $\Omega _{3}$\
there exist the following sets of solutions%
\begin{equation}
\left\{ \ _{\zeta }\psi _{n_{3}}\left( X\right) \right\} ,\ \ \left\{ \
^{\zeta }\psi _{n_{3}}\left( X\right) \right\} ,\ \ \zeta =\pm .
\label{2.48a}
\end{equation}

However, the one-particle interpretation of these solutions based on energy
spectrum in a similar way as has been done in{\large \ }the ranges $\Omega
_{1}$\ and $\Omega _{5}${\large \ }becomes inconsistent. Indeed, it is
enough to see the following contradiction: from the point of view of the
left asymptotic area, only electron states are possible in the range $\Omega
_{3}$, whereas from the point of view of the right asymptotic area, only
positron states are possible in this range. Detailed consideration in the
framework of QED shows (see Sec. \ref{S7}) that in a certain sense the
solutions $\ ^{\zeta }\psi _{n_{3}}\left( X\right) $ describe electrons,
whereas the solutions $\ _{\zeta }\psi _{n_{3}}\left( X\right) $ describe
positrons.

\subsection{Orthogonality, normalization and completeness\label{SS3.3}}

In this subsection we study orthonormalization, mutual decompositions and
completeness of the solutions introduced above. To this end it is convenient
to use two types of inner products in the Hilbert space of Dirac spinors.
One of them is defined on the hyperplane $x=\mathrm{const}$, and the\ other
on the hyperplane $t=\ $\textrm{const. }

\subsubsection{Using inner product on $x$-constant hyperplane}

Let us start with the inner product on the hyperplane $x=\mathrm{const.}$
For any two Dirac spinors $\psi \left( X\right) $ and $\psi ^{\prime }\left(
X\right) $ it has the form%
\begin{equation}
\left( \psi ,\psi ^{\prime }\right) _{x}=\int \psi ^{\dag }\left( X\right)
\gamma ^{0}\gamma ^{1}\psi ^{\prime }\left( X\right) dtd\mathbf{r}_{\bot }\ .
\label{IP}
\end{equation}

Due to the $t$-independence of the external field that corresponds to the $x$%
-electric potential steps, we can provide $x$-independence of the inner
product (\ref{IP}) for two solutions of the Dirac equation by imposing
periodic boundary conditions\ in the variables $t$ and $X^{j}$, $j=2,...,D$.
Thus, we consider our theory in a large space-time box that has a spatial
volume $V_{\bot }=\prod\limits_{j=2}^{D}K_{j}$ and the time dimension $T$,
where all $K_{j}$ and $T$ are macroscopically large. It is supposed that all
the solutions $\psi \left( X\right) $ are periodic under transitions from
one box to another. Then the integration in (\ref{IP}) over the transverse
coordinates is fulfilled from $-K_{j}/2$ to $+K_{j}/2$, and over the time $t$
from $-T/2$ to $+T/2$. Under these suppositions, the inner product (\ref{IP}%
) does not depend on $x$. The limits $K_{j}\rightarrow \infty $ and $%
T\rightarrow \infty $ are assumed in final expressions.

It should be noted that usually QFT deals with physical quantities that are
presented by volume integrals on the hyperplane\emph{\ }$t=\mathrm{const}$%
\emph{. }However, if\textrm{\ }we wish to extract results of the
one-particle scattering theory from a classical field theory, all the
constituent quantities, such as reflection and transmission coefficients
etc., have to be represented with the help of the inner product (\ref{IP})
on the hyperplane $x=\mathrm{const}$.\emph{\ }Indeed, in such a theory we
observe, for example, electric current, energy flux, or other types of
currents flowing through surfaces $x=\mathrm{const}$ situated in asymptotic
regions. In addition, we suppose that all the measurements are performed
during a macroscopic time (say, the time $T$) when the external field can be
considered as constant. Then the currents under consideration can be
represented by integral of the form\emph{\ }$T^{-1}\int_{T}dt...$ .

We note that for $\psi ^{\prime }=\psi $ the inner product (\ref{IP})
divided by $T$ represents the current of the Dirac field $\psi \left(
X\right) $ across the hyperplane $x=\mathrm{const}$. For nondecaying wave
functions this current differs from zero.

For two different solutions of the form (\ref{e7}) the integral in the
right-hand side of Eq. (\ref{IP}) can be easily calculated to be%
\begin{equation}
\left( \psi _{n},\psi _{n^{\prime }}^{\prime }\right) _{x}=V_{\bot }T\delta
_{n,n^{\prime }}\mathcal{I},\ \ \mathcal{I}=\varrho _{n}^{\dag }\left(
x\right) \gamma ^{0}\gamma ^{1}\varrho _{n}^{\prime }\left( x\right) .
\label{e5}
\end{equation}%
Using the structure of spinors $v_{\sigma }$, one can represent the current
density $\mathcal{I}$ as follows%
\begin{equation}
\mathcal{I}=\varphi _{n}^{\ast }\left( x\right) \left( i\overleftarrow{%
\partial }_{x}-i\overrightarrow{\partial }_{x}\right) \left[ p_{0}-U\left(
x\right) +\chi i\partial _{x}\right] \varphi _{n}^{\prime }\left( x\right) .
\label{e6}
\end{equation}

Let us consider solutions $\left\{ \ _{\zeta }\psi _{n}\left( X\right)
\right\} $ and $\left\{ \ ^{\zeta }\psi _{n}\left( X\right) \right\} $ with
left and right plane-wave asymptotics, respectively. They are orthogonal for
different $n$ and can be subjected to the following orthonormality conditions%
\begin{eqnarray}
&&\left( \ _{\zeta }\psi _{n},\ _{\zeta ^{\prime }}\psi _{n^{\prime
}}\right) _{x}=\zeta \eta _{\mathrm{L}}\delta _{\zeta ,\zeta ^{\prime
}}\delta _{n,n^{\prime }},\ \ \eta _{\mathrm{L}}=\mathrm{sgn\ }\pi
_{0}\left( \mathrm{L}\right) ,\ n\in \Omega _{1}\cup \Omega _{2}\cup \Omega
_{3}\cup \Omega _{5},  \notag \\
&&\left( \ ^{\zeta }\psi _{n},\ ^{\zeta ^{\prime }}\psi _{n^{\prime
}}\right) _{x}=\zeta \eta _{\mathrm{R}}\delta _{\zeta ,\zeta ^{\prime
}}\delta _{n,n^{\prime }},\ \ \eta _{\mathrm{R}}=\mathrm{sgn\ }\pi
_{0}\left( \mathrm{R}\right) ,\ n\in \Omega _{1}\cup \Omega _{3}\cup \Omega
_{4}\cup \Omega _{5}.  \label{c3}
\end{eqnarray}%
In justification of (\ref{c3}) one has to take into account that for these
states the relations%
\begin{equation*}
\left\vert \pi _{0}\left( \mathrm{L}\right) \right\vert >\left\vert p^{%
\mathrm{L}}\right\vert ,\ \left\vert \pi _{0}\left( \mathrm{R}\right)
\right\vert >\left\vert p^{\mathrm{R}}\right\vert
\end{equation*}%
hold. This fact explains why the sign of the inner products, which coincides
with the sign of $\mathcal{I}$ is due to the sign of the quantity $\pi
_{0}\left( \mathrm{L}/\mathrm{R}\right) $.

Then the normalization factors [with respect to the inner product (\ref{IP}%
)] in solutions of the Dirac equation that have plane-wave asymptotics (\ref%
{2.62a}) and (\ref{2.6}) have the form%
\begin{eqnarray}
\ _{\zeta }\mathcal{N} &\mathcal{=}&_{\zeta }CY,\;\;\ ^{\zeta }\mathcal{N}=\
^{\zeta }CY,\;\;Y=\left( V_{\bot }T\right) ^{-1/2},  \notag \\
\ _{\zeta }C &=&\left[ 2\left\vert p^{\mathrm{L}}\right\vert \left\vert \pi
_{0}\left( \mathrm{L}\right) -\chi p^{\mathrm{L}}\right\vert \right]
^{-1/2},\;\ ^{\zeta }C=\left[ 2\left\vert p^{\mathrm{R}}\right\vert
\left\vert \pi _{0}\left( R\right) -\chi p^{\mathrm{R}}\right\vert \right]
^{-1/2}.  \label{e8b}
\end{eqnarray}

In the $K_{j}\rightarrow \infty $ and $T\rightarrow \infty $ limit one has
to replace $\delta _{n,n^{\prime }}$ in normalization conditions (\ref{c3})
by $\delta _{\sigma ,\sigma ^{\prime }}\delta \left( p_{0}-p_{0}^{\prime
}\right) \delta \left( \mathbf{p}_{\bot }-\mathbf{p}_{\bot }^{\prime
}\right) $\ and to set $Y=\left( 2\pi \right) ^{-\left( d-1\right) /2}$ in
Eqs. (\ref{e8b}).

As was assumed, see subsection \ref{SS3.2}, each pair of solutions $_{\zeta
}\psi _{n}\left( X\right) $ and $^{\zeta }\psi _{n}\left( X\right) $ with
given quantum numbers $n\in \Omega _{1}\cup \Omega _{3}\cup \Omega _{5}$ is
complete in the space of solutions with each $n.$ Due to (\ref{c3}) the
corresponding mutual decompositions of such solutions have the form%
\begin{eqnarray}
\eta _{\mathrm{L}}\ ^{\zeta }\psi _{n}\left( X\right) &=&\ _{+}\psi
_{n}\left( X\right) g\left( _{+}\left\vert ^{\zeta }\right. \right) -\
_{-}\psi _{n}\left( X\right) g\left( _{-}\left\vert ^{\zeta }\right. \right)
,  \notag \\
\eta _{\mathrm{R}}\ _{\zeta }\psi _{n}\left( X\right) &=&\ ^{+}\psi
_{n}\left( X\right) g\left( ^{+}\left\vert _{\zeta }\right. \right) -\
^{-}\psi _{n}\left( X\right) g\left( ^{-}\left\vert _{\zeta }\right. \right)
,  \label{rel1}
\end{eqnarray}%
where decomposition coefficients $g$ are given by the relations:
\begin{equation}
\left( \ _{\zeta }\psi _{n},\ ^{\zeta ^{\prime }}\psi _{n^{\prime }}\right)
_{x}=\delta _{n,n^{\prime }}g\left( \ _{\zeta }\left\vert ^{\zeta ^{\prime
}}\right. \right) ,\ \ g\left( \ ^{\zeta ^{\prime }}\left\vert _{\zeta
}\right. \right) =g\left( \ _{\zeta }\left\vert ^{\zeta ^{\prime }}\right.
\right) ^{\ast },\ \ n\in \Omega _{1}\cup \Omega _{3}\cup \Omega _{5}\ .
\label{c12}
\end{equation}

Substituting (\ref{rel1}) into orthonormality conditions (\ref{c3}), we
derive the following unitary relations for the decomposition coefficients%
\footnote{%
Similar relations are known in the $t$-electric potential steps, see \cite%
{Gitman}.}:%
\begin{eqnarray}
&&g\left( \ ^{\zeta ^{\prime }}\left\vert _{+}\right. \right) g\left( \
_{+}\left\vert ^{\zeta }\right. \right) -g\left( \ ^{\zeta ^{\prime
}}\left\vert _{-}\right. \right) g\left( \ _{-}\left\vert ^{\zeta }\right.
\right) =\zeta \eta _{\mathrm{L}}\eta _{\mathrm{R}}\delta _{\zeta ,\zeta
^{\prime }}\ ,  \notag \\
&&g\left( \ _{\zeta ^{\prime }}\left\vert ^{+}\right. \right) g\left( \
^{+}\left\vert _{\zeta }\right. \right) -g\left( \ _{\zeta ^{\prime
}}\left\vert ^{-}\right. \right) g\left( \ ^{-}\left\vert _{\zeta }\right. \
\right) =\zeta \eta _{\mathrm{L}}\eta _{\mathrm{R}}\delta _{\zeta ,\zeta
^{\prime }}\ .  \label{UR}
\end{eqnarray}%
In particular, these relations imply that%
\begin{equation}
\left\vert g\left( _{-}\left\vert ^{+}\right. \right) \right\vert
^{2}=\left\vert g\left( _{+}\left\vert ^{-}\right. \right) \right\vert
^{2},\;\left\vert g\left( _{+}\left\vert ^{+}\right. \right) \right\vert
^{2}=\left\vert g\left( _{-}\left\vert ^{-}\right. \right) \right\vert
^{2},\;\frac{g\left( _{+}\left\vert ^{-}\right. \right) }{g\left(
_{-}\left\vert ^{-}\right. \right) }=\frac{g\left( ^{+}\left\vert
_{-}\right. \right) }{g\left( ^{+}\left\vert _{+}\right. \right) }.
\label{UR2}
\end{equation}%
One can see that all the coefficients $g$ can be expressed via only two of
them, e.g. via $g\left( _{+}\left\vert ^{+}\right. \right) $ and $g\left(
_{+}\left\vert ^{-}\right. \right) $. However, even the latter coefficients
are not completely independent, they are related as follows:%
\begin{equation}
\left\vert g\left( _{+}\left\vert ^{-}\right. \right) \right\vert
^{2}-\left\vert g\left( _{+}\left\vert ^{+}\right. \right) \right\vert
^{2}=-\eta _{\mathrm{L}}\eta _{\mathrm{R}}.  \label{UR1}
\end{equation}%
Nevertheless, in what follows, we will use both coefficients $g\left(
_{+}\left\vert ^{-}\right. \right) $ and $g\left( _{+}\left\vert ^{+}\right.
\right) $ in our consideration. This maintains a certain symmetry in writing
formulas and, moreover, allows one to generalize the consideration to the
cases when the inner potential may depend on several space coordinates and
to the case when solutions $\left\{ \ _{\zeta }\psi _{n}\left( X\right)
\right\} $ and $\left\{ \ ^{\zeta }\psi _{m}\left( X\right) \right\} $ are
characterized by different sets of quantum numbers, i.e., the sets $n$ and $%
m $ do not coincide.

For any two solutions $\psi \left( X\right) $ and $\psi ^{\prime }\left(
X\right) $ of the Klein-Gordon equation, the inner product on the hyperplane
$x=\mathrm{const}$ has the form%
\begin{equation}
\left( \psi ,\psi ^{\prime }\right) _{x}=\int \psi ^{\ast }\left( X\right)
\left( i\overleftarrow{\partial }_{x}-i\overrightarrow{\partial }_{x}\right)
\psi ^{\prime }\left( X\right) dtd\mathbf{r}_{\bot }.  \label{scip}
\end{equation}%
Orthonormality conditions for solutions of the Klein-Gordon equation can be
presented in form (\ref{c3}) where $\eta _{\mathrm{L}}=\eta _{\mathrm{R}}=1$%
. The normalization factors with respect to inner product (\ref{scip}) are%
\begin{equation}
\ _{\zeta }\mathcal{N}=\ _{\zeta }CY,\;\;\ ^{\zeta }\mathcal{N}=\ ^{\zeta
}CY,\;\ _{\zeta }C=\left\vert 2p^{\mathrm{L}}\right\vert ^{-1/2},\;\ ^{\zeta
}C=\left\vert 2p^{\mathrm{R}}\right\vert ^{-1/2},  \label{a3b}
\end{equation}%
where the factor $Y$ is given by (\ref{e8b}). In the Klein-Gordon case all
the relations presented above and that include coefficients $g$ hold true
with the setting{\large \ }$\eta _{\mathrm{L}}=\eta _{\mathrm{R}}=1$.

\subsubsection{Using inner product on $t$-constant hyperplane}

We recall that usually the inner product between two solutions $\psi \left(
X\right) $ and $\psi ^{\prime }\left( X\right) $ of the Dirac equation is
defined on $t$-\textrm{const.} hyperplane as follows:%
\begin{equation}
\left( \psi ,\psi ^{\prime }\right) =\int \psi ^{\dag }\left( X\right) \psi
^{\prime }\left( X\right) d\mathbf{r}.  \label{t3}
\end{equation}%
Such an inner product does not depend on the choice of the hyperplane (does
not depend on $t$) if the solutions obey certain boundary conditions that
allow one to integrate by parts in (\ref{t3}) neglecting boundary terms.
Since physical states are wave packets that vanish on the remote boundaries,
the inner product (\ref{t3}) is time-independent for such states. However,
considering solutions that are generalized states, which do not vanish at
the spatial infinity, one should take some additional steps to keep the
inner product (\ref{t3}) time independent. Sometimes to do this it is enough
to impose periodic boundary conditions (in all spatial directions) on Dirac
wave functions and on the corresponding external field. However, in the case
under consideration, the external field $A_{0}\left( x\right) $ of $x$%
-electric potential steps with different asymptotics at $x\rightarrow \pm
\infty $ cannot be subjected to any periodic boundary conditions in $x$%
-direction without changing its physical meaning. That is why, to provide
the time independence of the inner product, one has to extend the definition
of the inner product. Under the assumption that solutions with quantum
numbers $n$ form a complete set of function in the corresponding Hilbert
space at each time instant $t,$ it is enough to make such an extension for a
pair $\psi _{n}\left( X\right) $ and $\psi _{n^{\prime }}^{\prime }\left(
X\right) $ with all possible $n$ and $n^{\prime }$. Such an inner product is
described below.

Let $\psi _{n}\left( X\right) $ and $\psi _{n^{\prime }}^{\prime }\left(
X\right) $ be solutions of the Dirac equation that were described in the
previous subsections \ref{SS3.1} and \ref{SS3.2}. They allow one to impose
periodic conditions in the coordinates $X^{j},$ $j=2,...,D$ with periodicity
$K_{j}$ (of course it implies quantization of the corresponding transverse
momenta). Then the inner product for the pair $\psi _{n}\left( X\right) $
and $\psi _{n^{\prime }}^{\prime }\left( X\right) $ is defined as follows:%
\begin{equation}
\left( \psi _{n},\psi _{n^{\prime }}^{\prime }\right) =\int_{V_{\bot }}d%
\mathbf{r}_{\bot }\int\limits_{-K^{\left( \mathrm{L}\right) }}^{K^{\left(
\mathrm{R}\right) }}\psi _{n}^{\dag }\left( X\right) \psi _{n^{\prime
}}^{\prime }\left( X\right) dx,\ \ V_{\bot }=\prod\limits_{j=2}^{D}K_{j}\ ,
\label{t4}
\end{equation}%
and the limits $K^{\left( \mathrm{L}/\mathrm{R}\right) }\rightarrow \infty $
are assumed in final expressions. As it is demonstrated in Appendix {\large %
\ }\ref{t-const}, the so-defined inner product is time-independent. In what
follows the improper integral over $x$ in the right-hand side of Eq. (\ref%
{t4}) is reduced to its special principal value to provide a certain
additional property important for us .

Finally, we obtain the following orthonormality relations, see details in
Appendix \ref{t-const},
\begin{eqnarray}
&&\left( \ _{\zeta }\psi _{n},\ _{\zeta }\psi _{n^{\prime }}\right) =\left(
\ ^{\zeta }\psi _{n},\ ^{\zeta }\psi _{n^{\prime }}\right) =\delta
_{n,n^{\prime }}\mathcal{M}_{n}\ ,\ \ n\in \Omega _{1}\cup \Omega _{3}\cup
\Omega _{5}\ ,  \notag \\
&&\left( \psi _{n},\psi _{n^{\prime }}\right) =\delta _{n,n^{\prime }}%
\mathcal{M}_{n},\;\;n,n^{\prime }\in \Omega _{2}\cup \Omega _{4};  \notag \\
&&\mathcal{M}_{n}=2\frac{K^{\left( \mathrm{R}\right) }}{T}\left\vert \frac{%
\pi _{0}\left( \mathrm{R}\right) }{p^{\mathrm{R}}}\right\vert \left\vert
g\left( _{+}\left\vert ^{+}\right. \right) \right\vert ^{2}+O\left( 1\right)
,\ \ n\in \Omega _{1}\cup \Omega _{5},  \notag \\
&&\mathcal{M}_{n}=2\frac{K^{\left( \mathrm{R}\right) }}{T}\left\vert \frac{%
\pi _{0}\left( \mathrm{R}\right) }{p^{\mathrm{R}}}\right\vert \left\vert
g\left( _{+}\left\vert ^{-}\right. \right) \right\vert ^{2}+O\left( 1\right)
,\ \ n\in \Omega _{3},  \label{i12}
\end{eqnarray}%
where $\mathcal{M}_{n}\mathcal{\ }$for $n\in \Omega _{2}\cup \Omega _{4}$
are given by Eqs.~(\ref{i10a}). It follows from Eqs. (\ref{i12}) that
densities (in the $x$-direction) of the wave functions $_{\zeta }\psi
_{n_{i}},\ i=1,3,5$ are dominating in the region $S_{\mathrm{R}}$, whereas
the densities of the wave functions $^{\zeta }\psi _{n_{i}},\ i=1,3,5$ are
dominating in the region $S_{\mathrm{L}}$.

In the limit $K^{\left( \mathrm{L}/\mathrm{R}\right) }\rightarrow \infty $,
with the\ account of condition (\ref{i8}), we obtain solutions normalized to
the $\delta $-function,%
\begin{eqnarray}
&&\left( \ _{\zeta }\psi _{n},\ _{\zeta }\psi _{n^{\prime }}\right) =\left(
\ ^{\zeta }\psi _{n},\ ^{\zeta }\psi _{n^{\prime }}\right) =\delta _{\sigma
,\sigma ^{\prime }}\delta \left( p_{0}-p_{0}^{\prime }\right) \delta \left(
\mathbf{p}_{\bot }-\mathbf{p}_{\bot }^{\prime }\right) \mathcal{M}_{n},\ \
n,n^{\prime }\in \Omega _{1}\cup \Omega _{3}\cup \Omega _{5}\ ,  \notag \\
&&\left( \psi _{n},\psi _{n^{\prime }}\right) =\delta _{\sigma ,\sigma
^{\prime }}\delta \left( p_{0}-p_{0}^{\prime }\right) \delta \left( \mathbf{p%
}_{\bot }-\mathbf{p}_{\bot }^{\prime }\right) ,\;\;n,n^{\prime }\in \Omega
_{2}\cup \Omega _{4};  \notag \\
&&\mathcal{M}_{n}=\left\vert g\left( _{+}\left\vert ^{+}\right. \right)
\right\vert ^{2},\ \ n\in \Omega _{1}\cup \Omega _{5};\ \ \mathcal{M}%
_{n}=\left\vert g\left( _{+}\left\vert ^{-}\right. \right) \right\vert
^{2},\ \ n\in \Omega _{3}\mathrm{.}  \label{i13a}
\end{eqnarray}

We recall that the inner product between two solutions $\psi \left( X\right)
$\ and $\psi ^{\prime }\left( X\right) $\ of the Klein-Gordon equation is
defined on $t$-const. hyperplane as follows as a charge,%
\begin{equation}
\left( \psi ,\psi ^{\prime }\right) =\int \Phi ^{\dag }\left( X\right)
\left(
\begin{array}{cc}
0 & 1 \\
1 & 0%
\end{array}%
\right) \Phi \left( X\right) d\mathbf{r,}\ \ \Phi \left( X\right) =\left(
\begin{array}{c}
\left[ i\partial _{t}-U\left( x\right) \right] \psi \left( X\right) \\
\psi \left( X\right)%
\end{array}%
\right) .  \label{a12}
\end{equation}%
It is proportional to the field charge if $\psi =\psi ^{\prime }.$ It is
assumed that the improper integral over $x$\ in the right-hand side of Eq. (%
\ref{a12}) is treated as its special principal value, in the same spirit as
in the Dirac case.{\large \ }One can verify that with this inner product the
following orthonormality relations hold
\begin{eqnarray}
&&\left( \psi _{n},\psi _{n^{\prime }}^{\prime }\right) =0,\mathrm{\;}n\neq
n^{\prime },\ \ \forall n,n^{\prime };\ \ \left( \psi _{n_{2}},\psi
_{n_{2}^{\prime }}\right) =\delta _{n_{2},n_{2}^{\prime }}\mathcal{M}%
_{n_{2}},\;\;\left( \psi _{n_{4}},\psi _{n_{4}^{\prime }}\right) =-\delta
_{n_{4},n_{4}^{\prime }}\mathcal{M}_{n_{4}};  \notag \\
&&\left( \ _{\zeta }\psi _{n},\ ^{-\zeta }\psi _{n}\right) =0,\ \ \left( \
^{\zeta }\psi _{n},\ ^{\zeta }\psi _{n^{\prime }}\right) =\left( \ _{\zeta
}\psi _{n},\ _{\zeta }\psi _{n^{\prime }}\right) =\mathrm{sgn\ }\pi
_{0}\left( \mathrm{L}\right) \delta _{n,n^{\prime }}\mathcal{M}_{n}\ ,\ \
n\in \Omega _{1}\cup \Omega _{5}\ ;  \notag \\
&&\left( \ _{\zeta }\psi _{n},\ ^{\zeta }\psi _{n}\right) =0,\ \left( \
^{\zeta }\psi _{n},\ ^{\zeta }\psi _{n^{\prime }}\right) =-\left( \ _{\zeta
}\psi _{n},\ _{\zeta }\psi _{n^{\prime }}\right) =\delta _{n,n^{\prime }}%
\mathcal{M}_{n},\ \ n\in \Omega _{3},  \label{a14}
\end{eqnarray}%
where $\mathcal{M}_{n}$ are given by Eqs. (\ref{i12}). Unlike the Dirac
case, there are $\pm $\ signs in the right-hand side of Eq. (\ref{a14}).%
{\large \ }Then it is natural to suppose that the solutions $\ _{\zeta }\psi
_{n_{3}}$, $\psi _{n_{4}}$, $\ _{\zeta }\psi _{n_{5},}$\ and $\ ^{\zeta
}\psi _{n_{5}}$\ describe positron states.{\large \ }

Thus, both for the Dirac and the Klein-Gordon equations,{\large \ }for each
set of quantum numbers $n,$ there exist one or two complete sets of solutions

(a) For $\forall n\in \Omega _{1}\cup \Omega _{5}$ we have two $\left( \zeta
=\pm \right) $ sets: $\left\{ \ _{\zeta }\psi _{n}\left( X\right) ,\
^{-\zeta }\psi _{n}\left( X\right) \right\} $.

(b) For $\forall n\in \Omega _{3}\ $we have two $\left( \zeta =\pm \right) $
sets:$\ \left\{ \ _{\zeta }\psi _{n}\left( X\right) ,\ ^{\zeta }\psi
_{n}\left( X\right) \right\} .$

(c) For $\forall n\in \Omega _{2}\cup \Omega _{4}$ we have the set $\left\{
\psi _{n}\left( X\right) \right\} .$

We believe that all these solutions allow one to construct two complete (at
any time instant $t)$ systems both in the Hilbert space of Dirac spinors and
in the\ Hilbert space of scalar fields. In the Dirac case, this assumption
is equivalent to the existence of the propagation function $G\left(
X,X^{\prime }\right) $ in the space of solutions, which satisfies the
boundary condition \ \ \ \ \ \ \ \ \ \ \ \ \ \ \ \ \ \ \ \ \ \ \ \ \ \ \ \ \
\ \ \ \ \ \ \ \ \ \ \ \ \ \ \ \ \ \ \ \ \ \ \ \ \ \ \ \ \ \ \ \ \ \ \ \
\begin{equation}
\left. G\left( X,X^{\prime }\right) \right\vert _{t=t^{\prime }}=\delta
\left( \mathbf{r-r}^{\prime }\right) ,  \label{i14a}
\end{equation}%
and has the following form%
\begin{eqnarray}
&&G\left( X,X^{\prime }\right) =\sum_{i=1}^{5}G_{i}\left( X,X^{\prime
}\right) ,  \notag \\
&&G_{i}\left( X,X^{\prime }\right) =\sum_{n_{i}}\mathcal{M}_{n_{i}}^{-1}\psi
_{n_{i}}\left( X\right) \psi _{n_{i}}^{\dag }\left( X^{\prime }\right) ,\
i=2,4;  \notag \\
&&G_{i}\left( X,X^{\prime }\right) =\sum_{n_{i}}\mathcal{M}_{n_{i}}^{-1}%
\left[ \;_{+}\psi _{n_{i}}\left( X\right) \;_{+}\psi _{n_{i}}^{\dag }\left(
X^{\prime }\right) +\;^{-}\psi _{n_{i}}\left( X\right) \;^{-}\psi
_{n_{i}}^{\dag }\left( X^{\prime }\right) \right]  \notag \\
&&\ =\sum_{n_{i}}\mathcal{M}_{n_{i}}^{-1}\left[ \ _{-}\psi _{n_{i}}\left(
X\right) \;_{-}\psi _{n_{i}}^{\dag }\left( X^{\prime }\right) +\ ^{+}\psi
_{n_{i}}\left( X\right) \ ^{+}\psi _{n_{i}}^{\dag }\left( X^{\prime }\right) %
\right] ,\ i=1,5;  \notag \\
&&G_{3}\left( X,X^{\prime }\right) =\sum_{n_{3}}\mathcal{M}_{n_{3}}^{-1}%
\left[ \;_{+}\psi _{n_{3}}\left( X\right) \;_{+}\psi _{n_{3}}^{\dag }\left(
X^{\prime }\right) +\;^{+}\psi _{n_{3}}\left( X\right) \;^{+}\psi
_{n_{3}}^{\dag }\left( X^{\prime }\right) \right]  \notag \\
&&\ =\sum_{n_{3}}\mathcal{M}_{n_{3}}^{-1}\left[ \;_{-}\psi _{n_{3}}\left(
X\right) \;_{-}\psi _{n_{3}}^{\dag }\left( X^{\prime }\right) +\;^{-}\psi
_{n_{3}}\left( X\right) \;^{-}\psi _{n_{3}}^{\dag }\left( X^{\prime }\right) %
\right] \ .  \label{i14}
\end{eqnarray}%
The propagation function $G\left( X,X^{\prime }\right) $\ in the space of
scalar fields satisfies the boundary condition
\begin{equation}
\left. G\left( X,X^{\prime }\right) \right\vert _{t=t^{\prime
}}=0,\;\;\left. \left[ i\partial _{t}-U\left( x\right) \right] G\left(
X,X^{\prime }\right) \right\vert _{t=t^{\prime }}=\delta (\mathbf{r}-\mathbf{%
r}^{\prime }),  \label{a16}
\end{equation}%
and can be presented as
\begin{equation}
G\left( X,X^{\prime }\right) =\sum_{i=1}^{3}G_{i}\left( X,X^{\prime }\right)
-\sum_{i=4}^{5}G_{i}\left( X,X^{\prime }\right) ,  \label{a17}
\end{equation}%
where $G_{i}\left( X,X^{\prime }\right) $, $i=1,2,4,5$, have the form given
by Eq.~(\ref{i14}) with clear understanding that $\psi _{n_{i}}\left(
X\right) $ are scalar fields. The only $G_{3}\left( X,X^{\prime }\right) $
component has distinct form
\begin{eqnarray}
G_{3}\left( X,X^{\prime }\right) &=&\sum_{n_{3}}\mathcal{M}_{n_{3}}^{-1}%
\left[ -\;_{+}\psi _{n_{3}}\left( X\right) \;_{+}\psi _{n_{3}}^{\ast }\left(
X^{\prime }\right) +\;^{+}\psi _{n_{3}}\left( X\right) \;^{+}\psi
_{n_{3}}^{\ast }\left( X^{\prime }\right) \right]  \notag \\
\ &=&\sum_{n_{3}}\mathcal{M}_{n_{3}}^{-1}\left[ -\;_{-}\psi _{n_{3}}\left(
X\right) \;_{-}\psi _{n_{3}}^{\ast }\left( X^{\prime }\right) +\;^{-}\psi
_{n_{3}}\left( X\right) \;^{-}\psi _{n_{3}}^{\ast }\left( X^{\prime }\right) %
\right] \ .  \label{a17b}
\end{eqnarray}

It should be stressed that there are two equivalent representations for each
propagation function $G_{i}\left( X,X^{\prime }\right) ,\ i=1,3,5.$ In
addition, our further construction is based on the assumption that Dirac
spinors (a) and (b) are divided in the \textrm{in-}and \textrm{out-}%
solutions as follows:
\begin{eqnarray}
&&\mathrm{in-solutions:\ }_{+}\psi _{n_{1}},\ ^{-}\psi _{n_{1}};\mathrm{\ }%
_{-}\psi _{n_{5}},\ ^{+}\psi _{n_{5}};\ \ _{-}\psi _{n_{3}},\ ^{-}\psi
_{n_{3}}\ ,  \notag \\
&&\mathrm{out-solutions:\ }_{-}\psi _{n_{1}},\ ^{+}\psi _{n_{1}};\ _{+}\psi
_{n_{5}},\ ^{-}\psi _{n_{5}};\ \ _{+}\psi _{n_{3}},\ ^{+}\psi _{n_{3}},
\label{in-out}
\end{eqnarray}%
while for scalar fields, the classification in the range $\Omega _{3}$\
differs from the one given by Eq.~(\ref{in-out}) due to positions of\ the
left superscripts and subscripts $\pm $,
\begin{eqnarray}
&&\mathrm{in-solutions:\ }_{+}\psi _{n_{1}},\ ^{-}\psi _{n_{1}};\mathrm{\ }%
_{-}\psi _{n_{5}},\ ^{+}\psi _{n_{5}};\ \ _{+}\psi _{n_{3}},\ ^{+}\psi
_{n_{3}}\ ,  \notag \\
&&\mathrm{out-solutions:\ }_{-}\psi _{n_{1}},\ ^{+}\psi _{n_{1}};\ _{+}\psi
_{n_{5}},\ ^{-}\psi _{n_{5}};\ \ _{-}\psi _{n_{3}},\ ^{-}\psi _{n_{3}}.
\label{a18}
\end{eqnarray}

There exist difficulties in interpreting the states $\ _{\zeta }\psi
_{n_{3}}\left( X\right) $ and $\ ^{\zeta }\psi _{n_{3}}\left( X\right) $ in
the framework of the one-particle theory. There existed different point of
view on such an interpretation, see \cite{Nikis79,Nikis04} and \cite%
{HansRavn81}.\textrm{\ }A consistent interpretation can be obtained in the
framework of QED and is presented in Sec. \ref{S7}.

\section{Quantization in terms of particles\label{S4}}

In this and in the following sections, we treat the Dirac and the scalar
fields in terms of adequate in- and out-particles.{\large \ }On the base of
results of Sec. \ref{S3}, we decompose quantum field operators in complete
sets of the corresponding solutions introducing some annihilation and
creation operators. Then using the well-known equal time (anti)commutation
relations for the quantum fields we establish (anti)commutation relations
for the introduced operators. The problem of an identification of these
operators as in and out annihilation and creation operators is considered in
Secs. \ref{S5} - \ref{S7}.

The Dirac Heisenberg operator $\hat{\Psi}\left( X\right) $ is assigned to
the Dirac field $\psi \left( X\right) .$ This operator satisfies the Dirac
equation (\ref{e1}) and the anticommutation relations%
\begin{equation}
\left. \left[ \hat{\Psi}\left( X\right) ,\hat{\Psi}\left( X^{\prime }\right) %
\right] _{+}\right\vert _{t=t^{\prime }}=0,\ \ \left. \left[ \hat{\Psi}%
\left( X\right) ,\hat{\Psi}^{\dagger }\left( X^{\prime }\right) \right]
_{+}\right\vert _{t=t^{\prime }}=\delta \left( \mathbf{r-r}^{\prime }\right)
,  \label{3.3}
\end{equation}%
e.g., see \cite{Schwe61,GitTy90}.

The Klein-Gordon Heisenberg operator $\hat{\Psi}\left( X\right) $ is
assigned to the scalar field $\psi \left( X\right) .$ In terms of the
canonical pair

\begin{equation*}
\hat{\Phi}\left( X\right) =\left(
\begin{array}{c}
i\hat{\Pi}^{\dag }\left( X\right) \\
\hat{\Psi}\left( X\right)%
\end{array}%
\right) ,
\end{equation*}%
it satisfies the Klein-Gordon equation,%
\begin{equation}
\left[ i\partial _{t}-U\left( x\right) \right] \hat{\Phi}\left( X\right) =%
\hat{H}^{\mathrm{kin}}\hat{\Phi}\left( X\right) ,\ \ \hat{H}^{\mathrm{kin}%
}=\left(
\begin{array}{cc}
0 & -\left( \partial _{j}\right) ^{2}+m^{2} \\
1 & 0%
\end{array}%
\right) ,  \label{a0}
\end{equation}%
and the commutation relations%
\begin{equation}
\left. \left[ \hat{\Phi}\left( X\right) ,\hat{\Phi}\left( X^{\prime }\right) %
\right] _{-}\right\vert _{t=t^{\prime }}=0,\ \ \left. \left[ \hat{\Phi}%
\left( X\right) ,\hat{\Phi}^{\dag }\left( X^{\prime }\right) \right]
_{-}\right\vert _{t=t^{\prime }}=\delta \left( \mathbf{r-r}^{\prime }\right)
\left(
\begin{array}{cc}
0 & 1 \\
1 & 0%
\end{array}%
\right) ,  \label{a19}
\end{equation}%
e.g., see \cite{Schwe61,GitTy90}.{\large \ }Here $\hat{H}^{\mathrm{kin}}$ is
the one-particle kinetic energy operator $\hat{H}^{\mathrm{kin}}$. It
follows from Eq.~(\ref{a0}) that
\begin{equation*}
i\hat{\Pi}^{\dag }\left( X\right) =\left[ i\partial _{t}-U\left( x\right) %
\right] \hat{\Psi}\left( X\right) .
\end{equation*}

The further quantization of the scalar field can be done in the same manner
as the quantization for Dirac field. That is why, in what follows up to Sec. %
\ref{S8}, we consider only the quantization of Dirac field in detail, adding
some comments about the scalar field when necessary, see Appendix \ref%
{Abosons} for some peculiarities of the quantization in the range $\Omega
_{3}$.

\subsection{Introducing creation and annihilation operators\label{SS4.1}}

We consider the most interesting case $\mathbb{U}>2m$ of the critical steps.
In this case we decompose the Heisenberg operator of Dirac field $%
\hat{\Psi}\left( X\right) $ in two sets of solutions$\left\{ \ _{\zeta }\psi
_{n}\left( X\right) \right\} $ and $\left\{ \ ^{\zeta }\psi _{n}\left(
X\right) \right\} $ of the Dirac equation (\ref{e1}) complete on the
hyperplane $t=\mathrm{const}$. Operator-valued coefficients in such
decompositions do not depend on coordinates because both $\hat{\Psi}\left(
X\right) $ and the complete sets satisfy the same Dirac equation (\ref{e1}).
Our division of the quantum numbers $n$ in five ranges, implies the
representation for $\hat{\Psi}\left( X\right) $ as a sum of five operators $%
\hat{\Psi}_{i}\left( X\right) $, $i=1,2,3,4,5$,%
\begin{equation}
\hat{\Psi}\left( X\right) =\sum_{i=1}^{5}\hat{\Psi}_{i}\left( X\right) .
\label{2.20}
\end{equation}

For each of three operators $\hat{\Psi}_{i}\left( X\right) $,$\ i=1,3,5,$
there exist two possible decompositions according to the existence of two
different complete sets of solutions with the same quantum numbers $n$ in
the ranges $\Omega _{1}$, $\Omega _{3}$, and $\Omega _{5},$ see completeness
relation (\ref{i14}). Thus, we have:%
\begin{eqnarray}
&&\hat{\Psi}_{1}\left( X\right) =\sum_{n_{1}}\mathcal{M}_{n_{1}}^{-1/2}\left[
\ _{+}a_{n_{1}}(\mathrm{in})\ _{+}\psi _{n_{1}}\left( X\right) +\
^{-}a_{n_{1}}(\mathrm{in})\ ^{-}\psi _{n_{1}}\left( X\right) \right]  \notag
\\
&&\ =\sum_{n_{1}}\mathcal{M}_{n_{1}}^{-1/2}\left[ \ ^{+}a_{n_{1}}(\mathrm{out%
})\ ^{+}\psi _{n_{1}}\left( X\right) +\ _{-}a_{n_{1}}(\mathrm{out})\
_{-}\psi _{n_{1}}\left( X\right) \right] ,  \notag \\
&&\hat{\Psi}_{3}\left( X\right) =\sum_{n_{3}}\mathcal{M}_{n_{3}}^{-1/2}\left[
\ ^{-}a_{n_{3}}(\mathrm{in})\ ^{-}\psi _{n_{3}}\left( X\right) +\
_{-}b_{n_{3}}^{\dagger }(\mathrm{in})\ _{-}\psi _{n_{3}}\left( X\right) %
\right]  \notag \\
&&\ =\sum_{n_{3}}\mathcal{M}_{n_{3}}^{-1/2}\left[ \ ^{+}a_{n_{3}}(\mathrm{out%
})\ ^{+}\psi _{n_{3}}\left( X\right) +\ _{+}b_{n_{3}}^{\dagger }(\mathrm{out}%
)\ _{+}\psi _{n_{3}}\left( X\right) \right] ,  \notag \\
&&\hat{\Psi}_{5}\left( X\right) =\sum_{n_{5}}\mathcal{M}_{n_{5}}^{-1/2}\left[
\ ^{+}b_{n_{5}}^{\dag }(\mathrm{in})\ ^{+}\psi _{n_{5}}\left( X\right) +\
_{-}b_{n_{5}}^{\dag }(\mathrm{in})\ _{-}\psi _{n_{5}}\left( X\right) \right]
\notag \\
&&\ =\sum_{n_{5}}\mathcal{M}_{n_{5}}^{-1/2}\left[ \ _{+}b_{n_{5}}^{\dag }(%
\mathrm{out})\ _{+}\psi _{n_{5}}\left( X\right) +\ ^{-}b_{n_{5}}^{\dag }(%
\mathrm{out})\ ^{-}\psi _{n_{5}}\left( X\right) \right] .  \label{2.23}
\end{eqnarray}

There may\ exist only one complete set of solutions with the same quantum
numbers $n_{2}$ and $n_{4}$. Therefore, we have only one possible
decomposition for each of the\ two operators $\hat{\Psi}_{i}\left( X\right)
, $ $i=2,4$,%
\begin{equation}
\hat{\Psi}_{2}\left( X\right) =\sum_{n_{2}}\mathcal{M}%
_{n_{2}}^{-1/2}a_{n_{2}}\psi _{n_{2}}\left( X\right) ,\ \ \hat{\Psi}%
_{4}\left( X\right) =\sum_{n_{4}}\mathcal{M}_{n_{4}}^{-1/2}b_{n_{4}}^{%
\dagger }\psi _{n_{4}}\left( X\right) .  \label{2.21}
\end{equation}

We interpret all $a$ and $b\ $as annihilation and all $a^{\dag }$ and $%
b^{\dag }$ as creation operators; all $a$ and $a^{\dag }$ are interpreted\
as describing electrons and all $b$ and $b^{\dag }$ as describing positrons;
all the operators labeled by the argument \textrm{in} are interpreted\ as
\textrm{in}-operators, whereas all the operators labeled by the argument
\textrm{out} as \textrm{out}-operators.

In this connection, we reiterate that the time-independence of the external
field under consideration is an idealization. In fact, it is supposed that
the external field was switched on at a time instant $t_{in},$ then it was
acting as a constant field during a large time $T$, and finally it was
switched off at a time instant $t_{out}=t_{in}+T,$ and that\ one can ignore
effects of its switching on and off. Then we suppose that in the Schr\"{o}%
dinger picture one can introduce creation and annihilation operators of
particles at the initial and final time instants. In the Heisenberg
representation, these operators when being developed to zero time instant
are called \textrm{in}-operators and \textrm{out}-operators. It is the\
well-known procedure in QFT with $t$-electric potential steps.{\large \ }%
Technical realization of this construction was presented in Refs. \cite%
{Gitman}. In QED with constant fields, in particular, with the$\ x$-electric
potential steps, we quantize directly in the Heisenberg representation. And
here we encounter the problem of identification of \textrm{in}-operators and
\textrm{out}-operators. Its final solution is presented in Sec. \ref{S5}, %
\ref{S6}, and \ref{S7}.

Taking into account the orthonormalization relations (\ref{i12}) and the
completeness relations (\ref{i14}), we find that the anticommutation
relations (\ref{3.3}) for the Heisenberg operator (\ref{2.20}) yield the
following anticommutation rules for the introduced creation and annihilation
\textrm{in}- or \textrm{out-}operators (here we do not discuss commutation
relations between sets of \textrm{in} and \textrm{out}-operators):

All creation (annihilation) operators with different quantum numbers $n$
anticommute between themselves; all the operators from different ranges $%
\Omega _{i}$ anticommute between themselves; all anticommutators in each
range $\Omega _{i},\ i=1,2,3,4,5$ have the form:%
\begin{eqnarray}
&&\ [\ _{+}a_{n_{1}}(\mathrm{in}),\ _{+}a_{n_{1}^{\prime }}^{\dag }(\mathrm{%
in})]_{+}=[\ ^{-}a_{n_{1}}(\mathrm{in}),\ ^{-}a_{n_{1}^{\prime }}^{\dag }(%
\mathrm{in})]_{+}  \notag \\
&&\ =[\ ^{+}a_{n_{1}}(\mathrm{out}),\ ^{+}a_{n_{^{\prime }1}}^{\dag }(%
\mathrm{out})]_{+}=[\ _{-}a_{n_{1}}(\mathrm{out}),\ _{-}a_{n_{^{\prime
}1}}^{\dag }(\mathrm{out})]_{+}=\delta _{n_{1},n_{1}^{\prime }}\ ;  \notag \\
&&\ [a_{n_{2}},a_{n_{2}^{\prime }}^{\dag }]_{+}=\delta _{n_{2},n_{2}^{\prime
}}\ ;\ \ [b_{n_{4}},b_{n_{4}^{\prime }}^{\dag }]_{+}=\delta
_{n_{4},n_{4}^{\prime }}\ ;  \notag \\
&&\ [\ ^{-}a_{n_{3}}(\mathrm{in}),\ ^{-}a_{n_{3}^{\prime }}^{\dag }(\mathrm{%
in})]_{+}=[\ ^{+}a_{n_{3}}(\mathrm{out}),\ ^{+}a_{n_{3}^{\prime }}^{\dag }(%
\mathrm{out})]_{+}  \notag \\
&&\ =[\ _{-}b_{n_{3}}(\mathrm{in}),\ _{-}b_{n_{3}^{\prime }}^{\dag }(\mathrm{%
in})]_{+}=[\ _{+}b_{n_{3}}(\mathrm{out}),\ _{+}b_{n_{3}^{\prime }}^{\dag }(%
\mathrm{out})]_{+}=\delta _{n_{3},n_{3}^{\prime }}\ ;  \notag \\
&&\ [\ ^{+}b_{n_{5}}(\mathrm{in}),\ ^{+}b_{n_{5}^{\prime }}^{\dag }(\mathrm{%
in})]_{+}=[\ _{-}b_{n_{5}}(\mathrm{in}),\ _{-}b_{n_{5}^{\prime }}^{\dag }(%
\mathrm{in})]_{+}=  \notag \\
&&\ =[\ _{+}b_{n_{5}}(\mathrm{out}),\ _{+}b_{n_{5}^{\prime }}^{\dag }(%
\mathrm{out})]_{+}=[\ ^{-}b_{n_{5}}(\mathrm{out}),\ ^{-}b_{n_{5}^{\prime
}}^{\dag }(\mathrm{out})]_{+}=\delta _{n_{5},n_{5}^{\prime }}\ .
\label{2.24}
\end{eqnarray}

Some preliminary remarks about the division of creation and annihilation
operators into \textrm{in}- and \textrm{out}-type are in order.{\Huge \ }%
Usually, when\ quantizing a field theory with a time-dependent external
background, we work in the Schr\"{o}dinger picture, where we have to define
initial and final asymptotic states. Even if the external field is switched
off at the time infinity, its potentials may be different from zero there.
Thus, the Schr\"{o}dinger initial and final asymptotic states are different
as it occurs in QED with $t$-electric potential steps. Then the Schr\"{o}%
dinger initial and final asymptotic states give rise to in$-$ and out$-$
states (the corresponding operators) in the Heisenberg picture. In the case
under consideration, where we formally deal with time-independent
backgrounds, and quantize directly in the Heisenberg picture, there appears
the problem of identifying \textrm{in}- and \textrm{out}-operators. In the
case of $x$-electric potential steps, we will be guided by the following
physical considerations: All the \textrm{in}-particles (created by the
\textrm{in}-creation operators from the vacuum) are moving from the
asymptotic region $S_{\mathrm{L}}$ or $S_{\mathrm{R}}$ to the step, whereas
all the \textrm{out}-particles (created by the \textrm{out}-creation
operators from the vacuum) are moving from the step to the asymptotic region
$S_{\mathrm{L}}$ or $S_{\mathrm{R}}$.

Below, after analyzing properties of one-particle states created by the
introduced operators, we confirm the consistence of their division into
\textrm{in}- and \textrm{out}-types.

\subsection{Physical quantities\label{SS4.2}}

\subsubsection{Classical physical quantities}

Energy $\mathcal{H}$ of the classical Dirac field has the form%
\begin{equation}
\mathcal{H}=\int \psi ^{\dagger }\left( X\right) \hat{H}\psi \left( X\right)
d\mathbf{r}\ \mathbf{,}  \label{i14d}
\end{equation}%
where one-particle Hamiltonian $\hat{H}$ is given by Eq. (\ref{e1}). The
energy (\ref{i14d}) is a gauge dependent quantity.{\Huge \ }Substituting the
one-particle kinetic energy operator $\hat{H}^{\mathrm{kin}}$, given by Eq. (%
\ref{2.61a}), for $\hat{H}$ in the right-hand side (RHS) of Eq. (\ref{i14d})
we obtain a gauge invariant quantity $\mathcal{H}^{\mathrm{kin}}$, which we
call the kinetic energy of the classical Dirac field $\psi \left( X\right) ,$%
\begin{equation}
\mathcal{H}^{\mathrm{kin}}=\int \psi ^{\dagger }\left( X\right) \hat{H}^{%
\mathrm{kin}}\psi \left( X\right) d\mathbf{r}\ \mathbf{,\ }\hat{H}^{\mathrm{%
kin}}=\hat{H}-U\left( x\right) .  \label{i15}
\end{equation}

Decomposing the field $\psi \left( X\right) $ over the complete set $\psi
_{n}\left( X\right) $, and dividing integral (\ref{i15}) in three integrals
within the regions $S_{\mathrm{L}}$, $S_{\mathrm{int}}$ and $S_{\mathrm{R}}$%
, as was done for the quantity $\mathcal{R}$ in Eq. (\ref{i2}), we reduce
calculating the quantity (\ref{i15}) to calculating the following matrix
elements%
\begin{eqnarray}
&&\mathcal{H}_{nl}^{\mathrm{kin}}=\mathcal{H}_{nl}^{\mathrm{L}}+\mathcal{H}%
_{nl}^{\mathrm{int}}+\mathcal{H}_{nl}^{\mathrm{R}},  \notag \\
&&\mathcal{H}_{nl}^{\mathrm{L}}=\int_{-K^{\left( \mathrm{L}\right) }}^{x_{%
\mathrm{L}}}h_{nl}dx,\;\;\mathcal{H}_{nl}^{\mathrm{int}}=\int_{x_{\mathrm{L}%
}}^{x_{\mathrm{R}}}h_{nl}dx,\;\;\mathcal{H}_{nl}^{\mathrm{R}}=\int_{x_{%
\mathrm{R}}}^{K^{\left( \mathrm{R}\right) }}h_{nl}dx\ ,  \notag \\
&&h_{nl}=\int \psi _{n}^{\dagger }\left( X\right) \hat{H}^{\mathrm{kin}}\psi
_{l}\left( X\right) d\mathbf{r}_{\bot }\ \mathbf{.}  \label{classKinEn}
\end{eqnarray}

The matrix elements $\mathcal{H}_{nl}^{\mathrm{int}}$ are finite for any $n$
and $l$, and the terms $\mathcal{H}_{nl}^{\mathrm{L}}$ and $\mathcal{H}%
_{nl}^{\mathrm{R}}$ dominate in the limit $K^{\left( \mathrm{L}/\mathrm{R}%
\right) }\rightarrow \infty $. In the asymptotic regions $S_{\mathrm{L}}$
and $S_{\mathrm{R}}$ solutions $\psi _{n}\left( X\right) $ with any $n$ are
eigenfunctions of the operator $\hat{H}^{\mathrm{kin}}$ with the eigenvalues
$\pi _{0}\left( \mathrm{L}\right) $ and $\pi _{0}\left( \mathrm{R}\right) $,
respectively. That is why only diagonal matrix elements $\mathcal{H}_{nl}^{%
\mathrm{L}}$ and $\mathcal{H}_{nl}^{\mathrm{R}}$ differ from zero in the
limit $K^{\left( \mathrm{L}/\mathrm{R}\right) }\rightarrow \infty $. Thus,
the\ stationary states introduced in Sec. \ref{SS3.1} diagonalize the
introduced kinetic energy (\ref{i15}). This is an important necessary
condition in the quantization procedure which provides in what follows an
interpretation in terms of particles.

The kinetic energy of a stationary state $\psi _{n}\left( X\right) $ reads%
\begin{equation}
\mathcal{E}_{n}=\mathcal{M}_{n}^{-1}\int \psi _{n}^{\dag }\left( X\right) %
\left[ p_{0}-U\left( x\right) \right] \psi _{n}\left( X\right) d\mathbf{r}.
\label{i16}
\end{equation}%
The kinetic energy of a wave packet $\psi \left( X\right) $ composed of
stationary states $\psi _{n}\left( X\right) $ is a sum of the partial
energies (\ref{i16}).

One can easily see that stationary states with quantum numbers $n$ from the
regions $\Omega _{2}$ and $\Omega _{4}$ have the following kinetic energies
\begin{equation}
\mathcal{E}_{n_{2}}=\pi _{0}\left( \mathrm{L}\right) ,\;\mathcal{E}%
_{n_{4}}=\pi _{0}\left( \mathrm{R}\right) .  \label{i17a}
\end{equation}

Let us consider stationary states $\left\{ \ _{\zeta }\psi _{n}\left(
X\right) \right\} $ and $\left\{ \ ^{\zeta }\psi _{n}\left( X\right)
\right\} $ with quantum numbers $n$ from the ranges $\Omega _{1},\Omega
_{3}, $ and $\Omega _{5}.$ Their kinetic energies are denoted as $\ _{\zeta }%
\mathcal{E}_{n}$ and $\ ^{\zeta }\mathcal{E}_{n}$, respectively. In the same
manner, which was used in finding the orthonormality relations of the
corresponding solutions, and retaining only the leading terms in the limit $%
K^{\left( \mathrm{L}/\mathrm{R}\right) }\rightarrow \infty $, we obtain
\begin{eqnarray}
\ _{\zeta }\mathcal{E}_{n} &\mathcal{=}&V_{\bot }\mathcal{M}_{n}^{-1}\left(
_{\zeta }E_{n}^{\mathrm{L}}+\ _{\zeta }E_{n}^{\mathrm{R}}\right)
,\;\;^{\zeta }\mathcal{E}_{n}=V_{\bot }\mathcal{M}_{n}^{-1}\left( \ ^{\zeta
}E_{n}^{\mathrm{L}}\ \mathcal{+\ }^{\zeta }E_{n}^{\mathrm{R}}\right) ,
\notag \\
\ _{\zeta }E_{n}^{\mathrm{L}} &=&\pi _{0}\left( \mathrm{L}\right) \ _{\zeta }%
\mathcal{R}_{\mathrm{L}},\ \;_{\zeta }E_{n}^{\mathrm{R}}=\pi _{0}\left(
\mathrm{R}\right) \ _{\zeta }\mathcal{R}_{\mathrm{R}}\ ,  \notag \\
\ ^{\zeta }E_{n}^{\mathrm{L}} &=&\pi _{0}\left( \mathrm{L}\right) \ ^{\zeta }%
\mathcal{R}_{\mathrm{L}},\;\ ^{\zeta }E_{n}^{\mathrm{R}}=\pi _{0}\left(
\mathrm{R}\right) \ ^{\zeta }\mathcal{R}_{\mathrm{R}}\ ,  \label{i17b}
\end{eqnarray}%
where $n\in \Omega _{1}\cup \Omega _{3}\cup \Omega _{5}$, and the quantities
$\ ^{\zeta }\mathcal{R}_{\mathrm{L}/\mathrm{R}}$ and $\ _{\zeta }\mathcal{R}%
_{\mathrm{L}/\mathrm{R}}$ are given by Eqs. (\ref{i10}) and (\ref{i11})$;$
in fact, they depend also on the index $n$. Then, using Eqs. (\ref{i10}),(%
\ref{i11}),(\ref{UR}), and Eq. (\ref{i8}), we find%
\begin{eqnarray}
\ _{\zeta }\mathcal{E}_{n} &=&\pi _{0}\left( \mathrm{R}\right) +\frac{%
\mathbb{U}}{2}\left\vert g\left( _{+}\left\vert ^{+}\right. \right)
\right\vert ^{-2},\ \;^{\zeta }\mathcal{E}_{n}=\pi _{0}\left( \mathrm{L}%
\right) -\frac{\mathbb{U}}{2}\left\vert g\left( _{+}\left\vert ^{+}\right.
\right) \right\vert ^{-2},\ n\in \Omega _{1}\cup \Omega _{5},  \label{i18a}
\\
\ _{\zeta }\mathcal{E}_{n} &\mathcal{=}&\pi _{0}\left( \mathrm{R}\right) +%
\frac{\mathbb{U}}{2}\left\vert g\left( _{+}\left\vert ^{-}\right. \right)
\right\vert ^{-2},\;\;^{\zeta }\mathcal{E}_{n}=\pi _{0}\left( \mathrm{L}%
\right) -\frac{\mathbb{U}}{2}\left\vert g\left( _{+}\left\vert ^{-}\right.
\right) \right\vert ^{-2},\ n\in \Omega _{3}.  \label{i18b}
\end{eqnarray}

The energies $\mathcal{E}_{n_{i}},\ i=1,2$ are positive, and the
corresponding solutions in Eq. (\ref{i16}) are electron states, whereas all
the energies $\mathcal{E}_{n_{i}},\ i=4,5$ are negative, and the
corresponding solutions in Eq. (\ref{i16}) are positron states,%
\begin{equation}
\mathcal{E}_{n}>0,\ \ \forall n\in \Omega _{1}\cup \Omega _{2}\ ;\ \
\mathcal{E}_{n}<0,\ \ \forall n\in \Omega _{4}\cup \Omega _{5}\ .
\label{i19a}
\end{equation}

To estimate signs of energies (\ref{i18b}), we note that they contain two
terms with opposite signs. It follows from Eq. (\ref{UR1}) that $\left\vert
g\left( _{+}\left\vert ^{-}\right. \right) \right\vert ^{-2}\leqslant 1$.
Then Eqs. (\ref{i18b}) imply
\begin{equation}
\ ^{\zeta }\mathcal{E}_{n_{3}}-\ _{\zeta }\mathcal{E}_{n_{3}}=\mathbb{U}%
\left( 1-\left\vert g\left( _{+}\left\vert ^{-}\right. \right) \right\vert
^{-2}\right) \geqslant 0.  \label{i19}
\end{equation}%
It has to be stressed that the latter condition is enough to provide the
existing of a vacuum in the further quantum field theory.

Note that energies $\mathcal{E}_{n_{i}}$, $i=1,2,4,5$\ for the stationary
states of bosons have the same form as for fermions in (\ref{i17a}) and (\ref%
{i18a}). However, the corresponding relations for bosons
\begin{equation}
\ _{\zeta }\mathcal{E}_{n}\ \mathcal{=\ }\pi _{0}\left( \mathrm{R}\right) -%
\frac{\mathbb{U}}{2}\left\vert g\left( _{+}\left\vert ^{-}\right. \right)
\right\vert ^{-2},\;\;\ ^{\zeta }\mathcal{E}_{n}=\pi _{0}\left( \mathrm{L}%
\right) +\frac{\mathbb{U}}{2}\left\vert g\left( _{+}\left\vert ^{-}\right.
\right) \right\vert ^{-2},\ n\in \Omega _{3}.  \label{a15b}
\end{equation}%
differ from their fermionic versions~(\ref{i18b}).{\large \ }Equations (\ref%
{a15b}) imply the inequalities%
\begin{equation}
\ ^{\zeta }\mathcal{E}_{n_{3}}>0,\ \ _{\zeta }\mathcal{E}_{n_{3}}<0,
\label{a15d}
\end{equation}%
which provide the positiveness of any boson excitations over\ the vacuum.

\subsubsection{Quantum physical quantities}

After the second quantization, physical quantities of the classical Dirac
field turn out to be operators. In what follows, we are going to consider
some of them. The first one is kinetic energy operator $\widehat{\mathbb{H}}%
^{\mathrm{kin}}$, which, just from the beginning, we write in a renormalized
form,%
\begin{equation}
\widehat{\mathbb{H}}^{\mathrm{kin}}=\int \hat{\Psi}^{\dagger }\left(
X\right) \hat{H}^{\mathrm{kin}}\ \hat{\Psi}\left( X\right) d\mathbf{r}-%
\mathbb{H}_{0}\mathbf{\ ,}  \label{cq05}
\end{equation}%
where the constant (in general, infinite) term $\mathbb{H}_{0}$\ corresponds
to the energy of vacuum fluctuations. Its explicit expression will be given
below.

Inserting decompositions (\ref{2.20}), (\ref{2.23}), and (\ref{2.21}) in the
right-hand side of Eq. (\ref{cq05}), we obtain a representation of the
kinetic energy in terms of the introduced creation and annihilation
operators,
\begin{eqnarray}
&&\widehat{\mathbb{H}}^{\mathrm{kin}}=\sum_{i=1}^{5}\sum_{n_{i}}\widehat{%
\mathbb{H}}_{n_{i}}\mathbf{,\ \ }\mathbb{H}_{0}=\sum_{n_{3}}\ _{+}\mathcal{E}%
_{n_{3}}+\sum_{n_{4}}\pi _{0}\left( \mathrm{R}\right) +\sum_{n_{5}}\left( \
_{+}\mathcal{E}_{n_{5}}+\ ^{-}\mathcal{E}_{n_{5}}\right) \mathbf{,}  \notag
\\
&&\widehat{\mathbb{H}}_{n_{1}}=\ _{+}\mathcal{E}_{n_{1}}\
_{+}a_{n_{1}}^{\dag }(\mathrm{in})\ _{+}a_{n_{1}}(\mathrm{in})+\ ^{-}%
\mathcal{E}_{n_{1}}\ ^{-}a_{n_{1}}^{\dag }(\mathrm{in})\ ^{-}a_{n_{1}}(%
\mathrm{in})  \notag \\
&=&\ _{-}\mathcal{E}_{n_{1}}\ _{-}a_{n_{1}}^{\dag }(\mathrm{out})\
_{-}a_{n_{1}}(\mathrm{out})+\ ^{+}\mathcal{E}_{n_{1}}\ ^{+}a_{n_{1}}^{\dag }(%
\mathrm{out})\ ^{+}a_{n_{1}}(\mathrm{out}),  \notag \\
&&\widehat{\mathbb{H}}_{n_{2}}=\pi _{0}\left( \mathrm{L}\right)
a_{n_{2}}^{\dagger }a_{n_{2}},\;\;\widehat{\mathbb{H}}_{n_{4}}=-\pi
_{0}\left( \mathrm{R}\right) b_{n_{4}}^{\dagger }b_{n_{4}}\ ,  \notag \\
&&\widehat{\mathbb{H}}_{n_{3}}=\ ^{+}\mathcal{E}_{n_{3}}\
^{+}a_{n_{3}}^{\dag }(\mathrm{out})\ ^{+}a_{n_{3}}(\mathrm{out})-\ _{+}%
\mathcal{E}_{n_{3}}\ _{+}b_{n_{3}}^{\dag }(\mathrm{out})\ _{+}b_{n_{3}}(%
\mathrm{out})  \notag \\
&=&\ ^{-}\mathcal{E}_{n_{3}}\ ^{-}a_{n_{3}}^{\dag }(\mathrm{in})\
^{-}a_{n_{3}}(\mathrm{in})-\ _{-}\mathcal{E}_{n_{3}}\ _{-}b_{n_{3}}^{\dag }(%
\mathrm{in})\ _{-}b_{n_{3}}(\mathrm{in})\ ,  \notag \\
&&\widehat{\mathbb{H}}_{n_{5}}=-\ _{+}\mathcal{E}_{n_{5}}\
_{+}b_{n_{5}}^{\dag }(\mathrm{out})\ _{+}b_{n_{5}}(\mathrm{out})-\ ^{-}%
\mathcal{E}_{n_{5}}\ ^{-}b_{n_{5}}^{\dag }(\mathrm{out})\ ^{-}b_{n_{5}}(%
\mathrm{out})  \notag \\
&=&-\ _{-}\mathcal{E}_{n_{5}}\ _{-}b_{n_{5}}^{\dag }(\mathrm{in})\
_{-}b_{n_{5}}(\mathrm{in})-\ ^{+}\mathcal{E}_{n_{5}}\ ^{+}b_{n_{5}}^{\dag }(%
\mathrm{in})\ ^{+}b_{n_{5}}(\mathrm{in})\ .  \label{2.27}
\end{eqnarray}

We stress that the existence of two different representations for physical
observables in the ranges $\Omega _{1}$, $\Omega _{3}$, and $\Omega _{5}$
corresponds to the existence of two different complete sets of solutions in
these ranges.{\huge \ }According to eqs. (\ref{i18a}) and (\ref{i18b}), we
have%
\begin{equation*}
\ _{+}\mathcal{E}_{n_{3}}=\ _{-}\mathcal{E}_{n_{3}},\ \ _{-}\mathcal{E}%
_{n_{5}}+\ ^{+}\mathcal{E}_{n_{5}}=\ _{+}\mathcal{E}_{n_{5}}+\ ^{-}\mathcal{E%
}_{n_{5}}\ ,
\end{equation*}%
that is why the constant $H_{0}$ has the same value for any possible choice
of $\widehat{\mathbb{H}}_{n_{i}}$, $i=1,3,5$ in representation (\ref{2.27}).

The formal expression of\ the charge operator $\hat{Q}$ is%
\begin{equation}
\hat{Q}=\frac{q}{2}\int \left[ \hat{\Psi}^{\dagger }\left( X\right) ,\hat{%
\Psi}\left( X\right) \right] _{-}d\mathbf{r\ }.  \label{2.27a}
\end{equation}%
Its decomposition in the creation and annihilation operators introduced reads%
\begin{eqnarray}
&&\hat{Q}=\sum_{i=1}^{5}\sum_{n_{i}}\hat{Q}_{n_{i}}\ \mathbf{,}  \notag \\
&&\hat{Q}_{n_{1}}=-e\left[ \ _{+}a_{n_{1}}^{\dag }(\mathrm{in})\
_{+}a_{n_{1}}(\mathrm{in})+\ ^{-}a_{n_{1}}^{\dag }(\mathrm{in})\
^{-}a_{n_{1}}(\mathrm{in})\right]  \notag \\
&&\ =-e\left[ \ _{-}a_{n_{1}}^{\dag }(\mathrm{out})\ _{-}a_{n_{1}}(\mathrm{%
out})+\ ^{+}a_{n_{1}}^{\dag }(\mathrm{out})\ ^{+}a_{n_{1}}(\mathrm{out})%
\right] \ ,  \notag \\
&&\hat{Q}_{n_{2}}=-ea_{n_{2}}^{\dagger }a_{n_{2}},\;\;\hat{Q}%
_{n_{4}}=eb_{n_{4}}^{\dagger }b_{n_{4}}\ ,  \notag \\
&&\hat{Q}_{n_{3}}=-e\left[ \ ^{+}a_{n_{3}}^{\dag }(\mathrm{out})\
^{+}a_{n_{3}}(\mathrm{out})-\ _{+}b_{n_{3}}^{\dag }(\mathrm{out})\
_{+}b_{n_{3}}(\mathrm{out})\right]  \notag \\
&&\ =-e\left[ \ ^{-}a_{n_{3}}^{\dag }(\mathrm{in})\ ^{-}a_{n_{3}}(\mathrm{in}%
)-\ _{-}b_{n_{3}}^{\dag }(\mathrm{in})\ _{-}b_{n_{3}}(\mathrm{in})\right] ,
\notag \\
&&\hat{Q}_{n_{5}}=e\left[ \ _{+}b_{n_{5}}^{\dag }(\mathrm{out})\
_{+}b_{n_{5}}(\mathrm{out})+\ ^{-}b_{n_{5}}^{\dag }(\mathrm{out})\
^{-}b_{n_{5}}(\mathrm{out})\right]  \notag \\
&&\ =e\left[ \ _{-}b_{n_{5}}^{\dag }(\mathrm{in})\ _{-}b_{n_{5}}(\mathrm{in}%
)+\ ^{+}b_{n_{5}}^{\dag }(\mathrm{in})\ ^{+}b_{n_{5}}(\mathrm{in})\right] .
\label{2.28a}
\end{eqnarray}

We also will consider the energy flux of the Dirac field through the surface
$x=\mathrm{const}$. Its QFT operator has the form
\begin{equation}
\hat{F}\left( x\right) =\frac{1}{T}\int \hat{T}^{10}dtd\mathbf{r}_{\bot }\ ,
\label{2.25}
\end{equation}%
where $\hat{T}^{10}$ is the component of the operator of the energy momentum
tensor $\hat{T}^{\mu \nu },$ and the integral over $d\mathbf{r}_{\bot }$ is
defined in Eq. (\ref{IP}). For our purposes, it is enough to work with the
canonical energy momentum tensor
\begin{equation*}
T_{\mu \nu }=\frac{1}{2}\left\{ \bar{\psi}(x)\gamma _{\mu }P_{\nu }\psi (x)+%
\left[ P_{\nu }^{\ast }\bar{\psi}(x)\right] \gamma _{\mu }\psi (x)\right\} \
.
\end{equation*}%
Then the QFT operator $\hat{F}\left( x\right) $ reads
\begin{equation}
\hat{F}\left( x\right) =\frac{1}{T}\int \hat{\Psi}^{\dagger }\left( X\right)
\gamma ^{0}\gamma ^{1}\hat{H}^{\mathrm{kin}}\ \hat{\Psi}\left( X\right) dtd%
\mathbf{r}_{\bot }\ .  \label{cq9b}
\end{equation}

Another physical quantity useful for the further analysis is the electric
current of the Dirac field through the surface $x=\mathrm{const}$. The
corresponding QFT operator has the form%
\begin{equation}
\hat{J}=-\frac{e}{T}\int \hat{\Psi}^{\dag }\left( X\right) \gamma ^{0}\gamma
^{1}\hat{\Psi}\left( X\right) dtd\mathbf{r}_{\bot }\ .  \label{cq9}
\end{equation}

Inserting decompositions (\ref{2.20}), (\ref{2.23}), and (\ref{2.21}) in the
right-hand part of the quantities (\ref{cq9b}) and (\ref{cq9}), we can
obtain their representations in terms of the introduced creation and
annihilation operators. Then we can see that the right-hand part of (\ref%
{cq9b}) is diagonal with respect to the quantum numbers $n$ for $x\in S_{%
\mathrm{L}}$ and for $x\in S_{\mathrm{R}}$,{\Huge \ }whereas the right-hand
part of (\ref{cq9}) does not depend on $x$ and is diagonal with respect to
the quantum numbers $n$\ for any $x$.

\subsection{Partial and total vacuum states\label{SS4.3}}

Let us define two vacuum vectors $\left\vert 0,\mathrm{in}\right\rangle $
and $\left\vert 0,\mathrm{out}\right\rangle $, one of which is the\
zero-vector for all \textrm{in}-annihilation operators and the other is
zero-vector for all $\mathrm{out}$-annihilation operators. Besides, the both
vacua are zero-vectors for the annihilation operators $a_{n_{2}}$ and $%
b_{n_{4}},$ which is consistent with the anticommutation relations (\ref%
{2.24}). Thus, we have%
\begin{eqnarray}
&&\ _{+}a_{n_{1}}(\mathrm{in})\left\vert 0,\mathrm{in}\right\rangle =\
^{-}a_{n_{1}}(\mathrm{in})\left\vert 0,\mathrm{in}\right\rangle =0,  \notag
\\
&&\ _{-}b_{n_{5}}(\mathrm{in})\left\vert 0,\mathrm{in}\right\rangle =\
^{+}b_{n_{5}}(\mathrm{in})\left\vert 0,\mathrm{in}\right\rangle =0,  \notag
\\
&&\ ^{-}a_{n_{3}}(\mathrm{in})\left\vert 0,\mathrm{in}\right\rangle =\
_{-}b_{n_{3}}(\mathrm{in})\left\vert 0,\mathrm{in}\right\rangle =0,  \notag
\\
&&\ \ a_{n_{2}}\left\vert 0,\mathrm{in}\right\rangle =b_{n_{4}}\left\vert 0,%
\mathrm{in}\right\rangle =0,  \label{cq8a}
\end{eqnarray}

and%
\begin{eqnarray}
&&\ _{-}a_{n_{1}}(\mathrm{out})\left\vert 0,\mathrm{out}\right\rangle =\
^{+}a_{n_{1}}(\mathrm{out})\left\vert 0,\mathrm{out}\right\rangle =0,  \notag
\\
&&\ _{+}b_{n_{5}}(\mathrm{out})\left\vert 0,\mathrm{out}\right\rangle =\
^{-}b_{n_{5}}(\mathrm{out})\left\vert 0,\mathrm{out}\right\rangle =0,  \notag
\\
&&\ _{+}b_{n_{3}}(\mathrm{out})\left\vert 0,\mathrm{out}\right\rangle =\
^{+}a_{n_{3}}(\mathrm{out}\left\vert 0,\mathrm{out}\right\rangle =0,  \notag
\\
&&\ \ a_{n_{2}}\left\vert 0,\mathrm{out}\right\rangle =b_{n_{4}}\left\vert 0,%
\mathrm{out}\right\rangle =0\ .  \label{cq8b}
\end{eqnarray}

Then we postulate that the state space of the system under consideration is
the Fock space constructed, say, with the help of the vacuum $\left\vert 0,%
\mathrm{in}\right\rangle $ and the corresponding creation operators. One can
verify (see Sec. \ref{S5}) that this Fock space is unitary equivalent to
the\ other Fock space constructed with the help of the vacuum $\left\vert 0,%
\mathrm{out}\right\rangle $ and the corresponding creation operators.

In this case, one\ can see that vacuum mean values of the operator{\large \ }%
$\widehat{\mathbb{H}}^{\mathrm{kin}}${\large \ }(\ref{2.27}) and of the
charge operator{\large \ }$\hat{Q}${\large \ }are zero in the\ both Fock
spaces,{\large \ }%
\begin{equation*}
\langle 0,\mathrm{in}|\widehat{\mathbb{H}}^{\mathrm{kin}}\left\vert 0,%
\mathrm{in}\right\rangle =\langle 0,\mathrm{out}|\widehat{\mathbb{H}}^{%
\mathrm{kin}}\left\vert 0,\mathrm{out}\right\rangle =\langle 0,\mathrm{in}|%
\hat{Q}\left\vert 0,\mathrm{in}\right\rangle =\langle 0,\mathrm{out}|\hat{Q}%
\left\vert 0,\mathrm{out}\right\rangle =0.
\end{equation*}%
Thus, according to Eqs.~(\ref{i18a}), (\ref{i19a}), and (\ref{i19}), these
are uncharged states with minimal kinetic energy and the operator{\large \ }$%
\widehat{\mathbb{H}}^{\mathrm{kin}}${\large \ }is positively-defined in the
ranges{\large \ }$\Omega _{i}${\large , }$i=1,2,4,5${\large . }One can
verify that the introduced vacua have minimum energy with respect to any
uncharged quasistationary states in the range{\large \ }$\Omega _{3}$.
Indeed,{\large \ }in our construction{\large \ }it is assumed that electrons
and positrons with quantum numbers in this range being in one of
corresponding asymptotic regions occupy quasistationary states, i.e. they
are described by wave packets that maintain their forms sufficiently long in
these regions.\ Only such electron and positron wave packets have a physical
meaning. As we demonstrate in Secs. \ref{SS7.4} and \ref{Aloc} in the range%
{\large \ }$\Omega _{3}$, any electron wave packets that really have a
physical meaning can be localized only in the asymptotic region{\large \ }$%
S_{\mathrm{L}}$ (as in the range $\Omega _{2}$), whereas any positron wave
packets that really have a physical meaning can be localized only in the
asymptotic regions{\large \ }$S_{\mathrm{R}}$ (as in the range\ $\Omega _{4}$%
).\ Kinetic energies of these wave packets are formed for electrons by
contributions of{\large \ }$\pi _{0}\left( \mathrm{L}\right) ,${\large \ }%
for positrons by contribution{\large \ }$\left\vert \pi _{0}\left( \mathrm{R}%
\right) \right\vert ${\large \ }and are, therefore, always positive.

Because any annihilation operators with quantum numbers $n_{i}$
corresponding to different $i$ anticommute between themselves, we can
represent the introduced vacua as tensor products of the corresponding vacua
in the five ranges,%
\begin{equation}
\left\vert 0,\mathrm{in}\right\rangle =\sideset{}{^{\,\lower1mm\hbox{$%
\otimes$}}}\prod\limits_{i=1}^{5}\left\vert 0,\mathrm{in}\right\rangle
^{\left( i\right) }\ ,\ \ \left\vert 0,\mathrm{out}\right\rangle =%
\sideset{}{^{\,\lower1mm\hbox{$\otimes$}}}\prod\limits_{i=1}^{5}\left\vert 0,%
\mathrm{out}\right\rangle ^{\left( i\right) }\ ,  \label{2.29}
\end{equation}%
where the partial vacua $\left\vert 0,\mathrm{in}\right\rangle ^{\left(
i\right) }$ and $\left\vert 0,\mathrm{out}\right\rangle ^{\left( i\right) }$
obey relations (\ref{cq8a}) and (\ref{cq8b}) for any $n_{i}$ and $\zeta .$

It follows from relations (\ref{cq8a}) and (\ref{cq8b}) that%
\begin{equation}
\left\vert 0,\mathrm{in}\right\rangle ^{\left( i\right) }=\left\vert 0,%
\mathrm{out}\right\rangle ^{\left( i\right) },\ \ i=2,4.  \label{2.31}
\end{equation}

Let us rewrite relations (\ref{rel1}) for solutions with quantum numbers $%
n_{1}$ as follows
\begin{eqnarray}
&&_{+}\psi _{n}\left( X\right) =g\left( _{+}\left\vert ^{+}\right. \right)
^{-1}\left[ \eta _{\mathrm{L}}\;^{+}\psi _{n}\left( X\right) +\;_{-}\psi
_{n}\left( X\right) g\left( _{-}\left\vert ^{+}\right. \right) \right] ,
\notag \\
&&_{-}\psi _{n}\left( X\right) =g\left( _{-}\left\vert ^{-}\right. \right)
^{-1}\left[ -\eta _{\mathrm{L}}\;^{-}\psi _{n}\left( X\right) +\;_{+}\psi
_{n}\left( X\right) g\left( _{+}\left\vert ^{-}\right. \right) \right] ;
\notag \\
&&^{+}\psi _{n}\left( X\right) =g\left( ^{+}\left\vert _{+}\right. \right)
^{-1}\left[ \eta _{\mathrm{R}}\;_{+}\psi _{n}\left( X\right) +\;^{-}\psi
_{n}\left( X\right) g\left( ^{-}\left\vert _{+}\right. \right) \right] ,
\notag \\
&&^{-}\psi _{n}\left( X\right) =g\left( ^{-}\left\vert _{-}\right. \right)
^{-1}\left[ -\eta _{\mathrm{R}}\;_{-}\psi _{n}\left( X\right) +\;^{+}\psi
_{n}\left( X\right) g\left( ^{+}\left\vert _{-}\right. \right) \right] ,
\label{cq10}
\end{eqnarray}%
where pairs of solutions in the RHS of Eqs. (\ref{cq10}) are orthogonal due
to relations (\ref{i12}) at any fixed $t$. We recall that the relation%
\begin{equation}
\left\vert g\left( _{+}\left\vert ^{+}\right. \right) \right\vert
^{2}=\left\vert g\left( _{+}\left\vert ^{-}\right. \right) \right\vert ^{2}+1
\label{cq10a}
\end{equation}%
holds as a consequence of Eqs. (\ref{UR1}). Using these relations and two
possible representations (\ref{2.23}) for the operator $\hat{\Psi}_{1}\left(
X\right) ,$ one can find direct and inverse canonical transformations
between the initial and\emph{\ }final\emph{\ }pairs of annihilation
operators of electrons:
\begin{eqnarray}
&&\ ^{+}a_{n}(\mathrm{out})=\eta _{\mathrm{R}}g\left( _{+}\left\vert
^{+}\right. \right) ^{-1}\;_{+}a_{n}(\mathrm{in})+g\left( ^{-}\left\vert
_{-}\right. \right) ^{-1}g\left( ^{+}\left\vert _{-}\right. \right) \
^{-}a_{n}(\mathrm{in}),  \notag \\
&&_{-}a_{n}(\mathrm{out})=g\left( _{+}\left\vert ^{+}\right. \right)
^{-1}g\left( _{-}\left\vert ^{+}\right. \right) \;_{+}a_{n}(\mathrm{in}%
)-\eta _{\mathrm{L}}g\left( ^{-}\left\vert _{-}\right. \right) ^{-1}\
^{-}a_{n}(\mathrm{in});  \notag \\
&&\;_{+}a_{n}(\mathrm{in})=g\left( _{-}\left\vert ^{-}\right. \right)
^{-1}g\left( _{+}\left\vert ^{-}\right. \right) \;_{-}a_{n}(\mathrm{out}%
)+\eta _{\mathrm{L}}g\left( ^{+}\left\vert _{+}\right. \right) ^{-1}\
^{+}a_{n}(\mathrm{out}),  \notag \\
&&\ ^{-}a_{n}(\mathrm{in})=-\eta _{\mathrm{R}}g\left( _{-}\left\vert
^{-}\right. \right) ^{-1}\;_{-}a_{n}(\mathrm{out})+g\left( ^{+}\left\vert
_{+}\right. \right) ^{-1}g\left( ^{-}\left\vert _{+}\right. \right) \
^{+}a_{n}(\mathrm{out}).  \label{cq11}
\end{eqnarray}%
Canonical transformations between the\emph{\ }initial and final\emph{\ }%
pairs of creation operators of electrons can be derived from relations (\ref%
{cq11}).

In the same manner, using relations (\ref{rel1}) for solutions with quantum
numbers $n\in \Omega _{5}$, and two possible representations for the
operator $\hat{\Psi}_{5}\left( X\right) $ given by (\ref{2.23}), one can
find canonical transformations between the initial and final pairs of
positron creation operators. They can be obtained from relations (\ref{cq11}%
) by the substitution $_{+}a_{n_{1}}(\mathrm{in})\rightarrow
\;_{+}b_{n_{5}}^{\dag }(\mathrm{out}),$ $^{-}a_{n_{1}}(\mathrm{in}%
)\rightarrow \ ^{-}b_{n_{5}}^{\dag }(\mathrm{out}),$ $^{+}a_{n_{1}}(\mathrm{%
out})\rightarrow \ ^{+}b_{n_{5}}^{\dag }(\mathrm{in}),$ and $_{-}a_{n_{1}}(%
\mathrm{out})\rightarrow \;_{-}b_{n_{5}}^{\dag }(\mathrm{in})$.

Since the canonical transformations (\ref{cq11}) do not mix creation and
annihilation operators, we can choose%
\begin{equation}
\left\vert 0,\mathrm{in}\right\rangle ^{\left( i\right) }=\left\vert 0,%
\mathrm{out}\right\rangle ^{\left( i\right) },\ \ i=1,5.  \label{2.30}
\end{equation}

Together with their adjoint relations Eqs.~(\ref{cq11}) define an unitary
transformation $V_{\Omega _{1}}$ between \textrm{in}- and \textrm{out}%
-operators in the range $\Omega _{1}$,%
\begin{eqnarray*}
&&\left\{ \ _{+}a^{\dagger }(\mathrm{in}),\ ^{-}a^{\dag }(\mathrm{in}),\
_{+}a(\mathrm{in}),\ ^{-}a(\mathrm{in}),\right\} \\
&&\ =V_{\Omega _{1}}\left\{ \ ^{+}a^{\dagger }\left( \mathrm{out}\right) ,\
_{-}a^{\dag }\left( \mathrm{out}\right) ,\ ^{+}a\left( \mathrm{out}\right)
,\ _{-}a\left( \mathrm{out}\right) \right\} V_{\Omega _{1}}^{\dagger }\ .
\end{eqnarray*}%
The unitary operator $V_{\Omega _{1}}$ has the form%
\begin{eqnarray}
&&V_{\Omega _{1}}=\exp \left\{ \sum_{n\in \Omega _{1}}\ ^{+}a_{n}^{\dagger
}\left( \mathrm{out}\right) g\left( _{-}\left\vert ^{+}\right. \right)
^{-1}\ _{-}a_{n}\left( \mathrm{out}\right) \right\}  \notag \\
&&\times \exp \left\{ -\sum_{n\in \Omega _{1}}\ _{-}a_{n}^{\dag }\left(
\mathrm{out}\right) \ln \left[ g\left( ^{-}\left\vert _{-}\right. \right)
g\left( ^{-}\left\vert _{+}\right. \right) ^{-1}\right] \ _{-}a_{n}\left(
\mathrm{out}\right) \right\}  \notag \\
&&\times \exp \left\{ \sum_{n\in \Omega _{1}}\ ^{+}a_{n}^{\dagger }\left(
\mathrm{out}\right) \ln \left[ g\left( ^{+}\left\vert _{+}\right. \right)
g\left( ^{-}\left\vert _{+}\right. \right) ^{-1}\right] \ ^{+}a_{n}\left(
\mathrm{out}\right) \right\}  \notag \\
&&\times \exp \left\{ -\sum_{n\in \Omega _{1}}\ _{-}a_{n}^{\dag }\left(
\mathrm{out}\right) g\left( ^{-}\left\vert _{+}\right. \right)
^{-1}\;^{+}a_{n}\left( \mathrm{out}\right) \right\} .  \label{2.V1}
\end{eqnarray}

Similar results take place in the range $\Omega _{5}$:%
\begin{eqnarray}
&&\left\{ \ _{-}b^{\dagger }(\mathrm{in}),\ ^{+}b^{\dag }(\mathrm{in}),\
_{-}b(\mathrm{in}),\ ^{+}b(\mathrm{in})\right\}  \notag \\
&&\ =V_{\Omega _{5}}\left\{ \ ^{-}b^{\dagger }\left( \mathrm{out}\right) ,\
_{+}b^{\dag }\left( \mathrm{out}\right) ,\ ^{-}b\left( \mathrm{out}\right)
,\ _{+}b\left( \mathrm{out}\right) \right\} V_{\Omega _{5}}^{\dagger }\ ,
\notag \\
&&V_{\Omega _{5}}=\exp \left\{ -\sum_{n\in \Omega _{5}}\ _{+}b^{\dag }\left(
\mathrm{out}\right) g\left( _{+}\left\vert ^{-}\right. \right) ^{-1}\
^{-}b\left( \mathrm{out}\right) \right\}  \notag \\
&&\times \exp \left\{ -\sum_{n\in \Omega _{5}}\ ^{-}b^{\dagger }\left(
\mathrm{out}\right) \ln \left[ g\left( ^{-}\left\vert _{-}\right. \right)
g\left( ^{+}\left\vert _{-}\right. \right) ^{-1}\right] \ ^{-}b\left(
\mathrm{out}\right) \right\}  \notag \\
&&\times \exp \left\{ \sum_{n\in \Omega _{5}}\ _{+}b^{\dag }\left( \mathrm{%
out}\right) \ln \left[ g\left( ^{+}\left\vert _{+}\right. \right) g\left(
^{+}\left\vert _{-}\right. \right) ^{-1}\right] \ _{+}b\left( \mathrm{out}%
\right) \right\}  \notag \\
&&\times \exp \left\{ \sum_{n\in \Omega _{5}}\ ^{-}b^{\dagger }\left(
\mathrm{out}\right) g\left( ^{+}\left\vert _{-}\right. \right) ^{-1}\
_{+}b\left( \mathrm{out}\right) \right\} .  \label{2.V2}
\end{eqnarray}

We recall, that the\ similar result (\ref{2.31}) takes place\ in the regions
$\Omega _{i}$,\ $i=2,4.$ Both relations (\ref{2.31}) and (\ref{2.30}) mean
that the partial vacua $\left\vert 0,\mathrm{in}\right\rangle ^{\left(
i\right) },$ $i=1,2,4,5$ are stable under the action of the external field.
In what follows, we denote the tensor product of these partial vacua by $%
\left\vert 0\right\rangle $,%
\begin{equation}
\left\vert 0\right\rangle =\sideset{}{^{\,\lower1mm\hbox{$\otimes$}}}%
\prod\limits_{i=1,2,4,5}\left\vert 0,\mathrm{in}\right\rangle ^{\left(
i\right) }=\sideset{}{^{\,\lower1mm\hbox{$\otimes$}}}\prod%
\limits_{i=1,2,4,5}\left\vert 0,\mathrm{out}\right\rangle ^{\left( i\right)
}\ .  \label{2.34}
\end{equation}

Below, we are going to consider one-particle states $a_{n}^{\dag }\left(
\mathrm{in}/\mathrm{out}\right) \left\vert 0,\mathrm{in}/\mathrm{out}%
\right\rangle $ and $b_{n}^{\dag }\left( \mathrm{in}/\mathrm{out}\right)
\left\vert 0,\mathrm{in}/\mathrm{out}\right\rangle $ created by the
introduced creation operators from the vacuum.

The physical meaning of these states will be discussed separately for each
range of the quantum numbers $n$ .

\begin{figure}[tbp]
\centering\includegraphics[scale=.3]{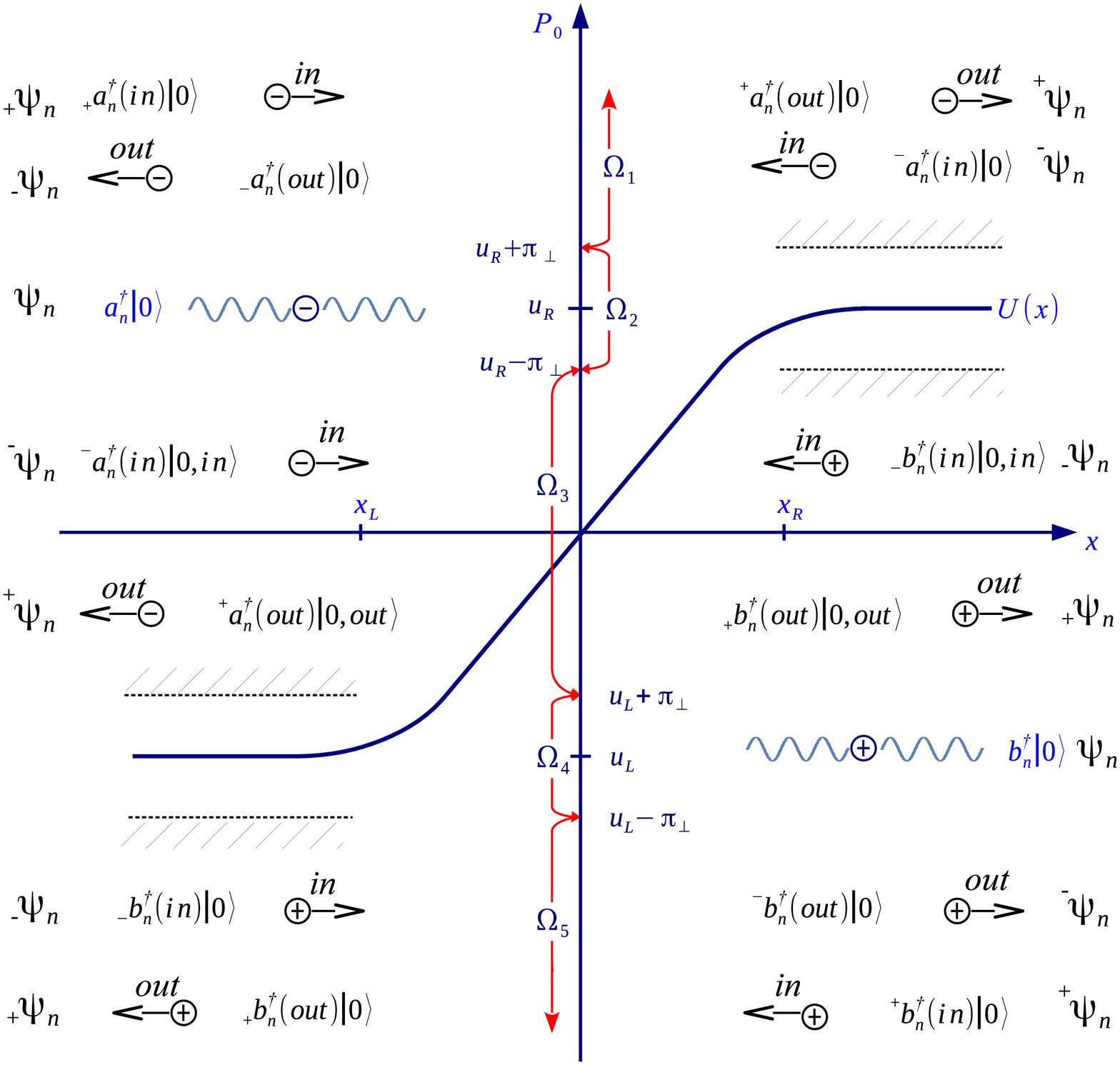}
\caption{\textrm{in} and \textrm{out}-particles near an $x$-potential step}
\label{3}
\end{figure}

\section{One-particle states in the ranges $\Omega _{1}$ and $\Omega _{5}$%
\label{S5}}

\subsection{General\label{SS5.1}}

We believe that according to the structure of the Dirac energy spectra in
the asymptotic regions $S_{\mathrm{L}}$ and $S_{\mathrm{R}}$, there exist
only one-electron states in the range $\Omega _{1},$ whereas in the range $%
\Omega _{5}$ there exist only one-positron states$.$

We remind that condition $\left( \ _{\zeta }\psi _{n},\ ^{-\zeta }\psi
_{n}\right) =0$ (\ref{i7}) for $n\in \Omega _{1}\cup \Omega _{5}$ means that
the sets of solution$s$ $\left\{ \ _{\zeta }\psi _{n}\left( X\right)
\right\} $ and $\left\{ \ ^{-\zeta }\psi _{n}\left( X\right) \right\} $ with
opposite $\zeta $ represent independent physical states.\emph{\ }

We\emph{\ }associate\emph{\ }the independent pair $\left\{ \ _{+}\psi
_{n_{1}}\left( X\right) ,\ ^{-}\psi _{n_{1}}\left( X\right) \right\} $ with
electron \textrm{in}-solutions and the independent pair $\left\{ \ _{-}\psi
_{n_{1}}\left( X\right) \right\} $ and $\left\{ \ ^{+}\psi _{n_{1}}\left(
X\right) \right\} $ with electron \textrm{out}-solutions. Correspondingly,
in the range $\Omega _{1},$ one-electron states are
\begin{equation}
\ _{+}a_{n_{1}}^{\dag }(\mathrm{in})\left\vert 0\right\rangle ,\ \ \
^{-}a_{n_{1}}^{\dag }(\mathrm{in})\left\vert 0\right\rangle ,\ \ \
_{-}a_{n_{1}}^{\dag }(\mathrm{out})\left\vert 0\right\rangle ,\ \ \
^{+}a_{n_{1}}^{\dag }(\mathrm{out})\left\vert 0\right\rangle .  \label{4.1a}
\end{equation}

We\emph{\ }associate\emph{\ }the independent pair $\left\{ \ _{-}\psi
_{n_{5}}\left( X\right) \right\} $ and $\left\{ \ ^{+}\psi _{n_{5}}\left(
X\right) \right\} $ with positron \textrm{in}-solutions and the independent
pair $\left\{ \ _{+}\psi _{n_{5}}\left( X\right) \right\} $ and $\left\{ \
^{-}\psi _{n_{5}}\left( X\right) \right\} $ with positron \textrm{out}%
-solutions. Correspondingly, in the range $\Omega _{5},$ one-positron states
are
\begin{equation}
\ _{+}b_{n_{5}}^{\dag }(\mathrm{out})\left\vert 0\right\rangle ,\ \ \
^{-}b_{n_{5}}^{\dag }(\mathrm{out})\left\vert 0\right\rangle ,\ \ \
_{-}b_{n_{5}}^{\dag }(\mathrm{in})\left\vert 0\right\rangle ,\ \ \
^{+}b_{n_{5}}^{\dag }(\mathrm{in})\left\vert 0\right\rangle .  \label{4.1b}
\end{equation}

\subsection{Interpretation of states in $\Omega _{1}$ and $\Omega _{5}$\label%
{SS5.3}}

To give an interpretation of states in {\large \ }$\Omega _{1}${\large \ and
}$\Omega _{5}$\ we have studied one-particle mean values of the charge, the
kinetic energy, the number of particles, the current, and the energy flux
through the surfaces{\large \ }$x=x_{\mathrm{L}}${\large \ }and{\large \ }$%
x=x_{\mathrm{R}}${\large . }The corresponding calculations are placed in the
Appendix \ref{SS5.2}{\large .} Based on results of such calculations, we can
finally conclude:

(1) The states (\ref{4.1a}) are states with the charge $-e.$ The states (\ref%
{4.1b}) are states with the charge $+e.$

(2) All these states have positive energies and therefore they can be
treated as physical particle states, namely states (\ref{4.1a}) represent
electrons, whereas states (\ref{4.1b}) represent positrons. Their currents (%
\ref{5.7}) and (\ref{5.8}) also\ confirm this interpretation.

(3) Mean energy fluxes (\ref{5.9}) and (\ref{5.10}) of states under
consideration through the surfaces $x=x_{\mathrm{L}}$ and $x=x_{\mathrm{R}}$
together with expressions for their currents allow us to believe that:

(a) electrons $\ _{+}a_{n_{1}}^{\dag }(\mathrm{in})\left\vert 0\right\rangle
$ and$\ ^{+}a_{n_{1}}^{\dag }(\mathrm{out})\left\vert 0\right\rangle \ $\
are moving to the right,

(b) electrons$\ \ _{-}a_{n_{1}}^{\dag }(\mathrm{out})\left\vert
0\right\rangle \ $and$\ \ ^{-}a_{n_{1}}^{\dag }(\mathrm{in})\left\vert
0\right\rangle $ \ are moving to the left;

(c) positrons $_{-}b_{n_{5}}^{\dag }(\mathrm{in})\left\vert 0\right\rangle \
^{-}b_{n_{5}}^{\dag }(\mathrm{out})\left\vert 0\right\rangle $\ are moving
to the right,

(d) positrons$\ \ _{+}b_{n_{5}}^{\dag }(\mathrm{out})\left\vert
0\right\rangle $ and$\ ^{+}b_{n_{5}}^{\dag }(\mathrm{in})\left\vert
0\right\rangle $\ \ are moving to the left.

Thus, we see that the asymptotic longitudinal physical momenta of electrons
are{\large \ }$p_{ph}^{\mathrm{L}}=p^{\mathrm{L}}${\large \ }and{\large \ }$%
p_{ph}^{\mathrm{R}}=p^{\mathrm{R}}$, whereas for positrons they are{\large \
}$p_{ph}^{\mathrm{L}}=-p^{\mathrm{L}}${\large \ }and{\large \ }$p_{ph}^{%
\mathrm{R}}=-p^{\mathrm{R}}${\large . }

(4) We classify electron states $\ _{+}a_{n_{1}}^{\dag }(\mathrm{in}%
)\left\vert 0\right\rangle $ and $\ ^{-}a_{n_{1}}^{\dag }(\mathrm{in}%
)\left\vert 0\right\rangle $ as \textrm{in}- states, because they are moving
to the step from the asymptotic regions $S_{\mathrm{L}}$ and $S_{\mathrm{R}%
}, $ respectively, with definite asymptotic behavior there. We classify
electron states $\ _{-}a_{n_{1}}^{\dag }(\mathrm{out})\left\vert
0\right\rangle \ $ and $\ ^{+}a_{n_{1}}^{\dag }(\mathrm{out})\left\vert
0\right\rangle $ as \textrm{out}- states because they are moving from the
step to the asymptotic regions $S_{\mathrm{L}}$ and $S_{\mathrm{R}}$,
respectively, having there definite asymptotics.

We classify positrons states $\ _{-}b_{n_{5}}^{\dag }(\mathrm{in})\left\vert
0\right\rangle \ $ and $\ ^{+}b_{n_{5}}^{\dag }(\mathrm{in})\left\vert
0\right\rangle \ $as \textrm{in}- states because they are moving to the step
from the asymptotic regions $S_{\mathrm{L}}$ and $S_{\mathrm{R}}$,
respectively, having their definite asymptotics. We classify positron states
$\ _{+}b_{n_{5}}^{\dag }(\mathrm{out})\left\vert 0\right\rangle \ $ and $\
^{-}b_{n_{5}}^{\dag }(\mathrm{out})\left\vert 0\right\rangle $ as \textrm{out%
}- states because they are moving from the step to the asymptotic regions $%
S_{\mathrm{L}}$ and $S_{\mathrm{R}}$, respectively, having their definite
asymptotics.

In Fig. 3 we show \textrm{in}- and \textrm{out}-electron states in the range
$\Omega _{1}$ and \textrm{in}- and \textrm{out}-positron states in the range
$\Omega _{5}.$ Here electrons are drawn by circles with the sign minus
inside and positrons with the sign plus inside. The associated arrows show
the energy flux directions given by Eqs. (\ref{5.9}) and (\ref{5.10}). Thus,
these arrows show the \ directions of motion.

To justify completely our interpretation of the \textrm{in}- and \textrm{out}%
-states, we first recall that it is impossible to refer\ (even in the
nonrelativistic quantum mechanics) to a\ direction of motion of plane waves,
which has no physical meaning, since they are not localized.{\Huge \ }The
scattering problem is formulated for particles that are represented by\emph{%
\ }wave packets localized in some space areas. What do we demand from such
localized packets? First of all, the localization areas have to belong to
one of the asymptotic regions $S_{\mathrm{L}}$ or $S_{\mathrm{R}}.$ Each of
the localized wave packets must be composed of states with asymptotic
physical momenta $p_{\mathrm{ph}}^{\mathrm{L}}$ or $p_{\mathrm{ph}}^{\mathrm{%
R}}$, respectively, that have the same directions and belong to one and the
same range $\Omega _{i}$. We call these packets quasilocalized, because of
not very rigid requirements for their localization. Such wave packets are
moving in the same direction as their constituent waves. In the scattering
problem under consideration, we consider four types of wave packets in each
range $\Omega _{1}$ and $\Omega _{5},$ two of them being\ quasilocalized in%
\textrm{\ }the asymptotic region $S_{\mathrm{L}}$ and two of them in the
asymptotic region $S_{\mathrm{R}}.$ All the packets quasilocalized in $S_{%
\mathrm{L}}$ are formed of solutions $\ _{\zeta }\psi _{n}\left( X\right) $,
whereas all the packets quasilocalized in $S_{\mathrm{R}}$ are formed of
solutions $\ ^{\zeta }\psi _{n}\left( X\right) $. Indeed, in the asymptotic
region $S_{\mathrm{L}}$ and $S_{\mathrm{R}}$ solutions $\ _{\zeta }\psi
_{n}\left( X\right) $ and $\ ^{\zeta }\psi _{n}\left( X\right) $,
respectively$,$ are reduced to waves with definite asymptotic physical
momenta $p_{ph}^{\mathrm{L}}$ and $p_{ph}^{\mathrm{R}}$, respectively. That
is why directions of motion of the wave packets in these regions are well
defined.

The electron wave packets in the range $\Omega _{1}$, composed of solutions $%
\ _{+}\psi _{n}\left( X\right) $ or $\ ^{-}\psi _{n}\left( X\right) $, are
moving to the region $S_{\mathrm{int}}$ (in the QM scattering theory they
are called incoming waves), whereas the electron wave packets composed of
solutions$\ _{-}\psi _{n}\left( X\right) $ or $\ ^{+}\psi _{n}\left(
X\right) $ are moving away from the region $S_{\mathrm{int}}$ (in the QM
scattering theory they are called outgoing waves). Thus, we believe{\huge \ }%
that the first type of the wave packets describe \textrm{in}-electron states
with asymptotic behavior formed before they meet the external field, and the
second type of wave packets describe \textrm{out}-electrons that have
asymptotic behavior observed after they have left the region where the
external field is present. That is the reason for our definitions of \textrm{%
in}- and \textrm{out}- creation and annihilation operators with quantum
numbers $n_{1}$ in the decomposition of the quantized Dirac field (\ref{2.23}%
). It is not difficult to give similar interpretation for positron wave
packets in the range $\Omega _{5}$.

According to these definitions, we introduce absolute $\tilde{R}$ and
relative $R$ amplitudes of an electron reflection, and absolute $\tilde{T}$\
and relative $T$ amplitudes of an electron transmission in the range $\Omega
_{1}$ as%
\begin{eqnarray}
\tilde{R}_{+,n_{1}} &=&c_{v}R_{+,n_{1}},\ R_{+,n_{1}}=\left\langle
0\left\vert \ _{-}a_{n_{1}}(\mathrm{out})\ _{+}a_{n_{1}}^{\dag }(\mathrm{in}%
)\right\vert 0\right\rangle ,  \notag \\
\tilde{T}_{+,n_{1}} &=&c_{v}T_{+,n_{1}},\ T_{+,n_{1}}=\left\langle
0\left\vert ^{+}a_{n_{1}}(\mathrm{out})\ _{+}a_{n_{1}}^{\dag }(\mathrm{in}%
)\right\vert 0\right\rangle ,  \notag \\
\tilde{R}_{-,n_{1}} &=&c_{v}R_{-,n_{1}},\ R_{-,n_{1}}=\left\langle
0\left\vert \ ^{+}a_{n_{1}}(\mathrm{out})\ ^{-}a_{n_{1}}^{\dag }(\mathrm{in}%
)\right\vert 0\right\rangle ,  \notag \\
\tilde{T}_{-,n_{1}} &=&c_{v}T_{-,n_{1}},\ T_{-,n_{1}}=\left\langle
0\left\vert _{-}a_{n_{1}}(\mathrm{out})\ ^{-}a_{n_{1}}^{\dag }(\mathrm{in}%
)\right\vert 0\right\rangle ,  \label{cq14}
\end{eqnarray}%
and similar quantities for a positron in the range $\Omega _{5}$ as
\begin{eqnarray}
\tilde{R}_{+,n_{5}} &=&c_{v}R_{+,n_{5}},\ R_{+,n_{5}}=\left\langle
0\left\vert \ ^{-}b_{n_{5}}(\mathrm{out})\ ^{+}b_{n_{5}}^{\dag }(\mathrm{in}%
)\right\vert 0\right\rangle ,  \notag \\
\tilde{T}_{+,n_{5}} &=&c_{v}T_{+,n_{5}},\ T_{+,n_{5}}=\left\langle
0\left\vert _{+}b_{n_{5}}(\mathrm{out})\ ^{+}b_{n_{5}}^{\dag }(\mathrm{in}%
)\right\vert 0\right\rangle ,  \notag \\
\tilde{R}_{-,n_{5}} &=&c_{v}R_{-,n_{5}},\ R_{-,n_{5}}=\left\langle
0\left\vert \ _{+}b_{n_{5}}(\mathrm{out})\ _{-}b_{n_{5}}^{\dag }(\mathrm{in}%
)\right\vert 0\right\rangle ,  \notag \\
\tilde{T}_{-,n_{5}} &=&c_{v}T_{-,n_{5}},\ T_{-,n_{5}}=\left\langle
0\left\vert ^{-}b_{n_{1}}(\mathrm{out})\ _{-}b_{n_{5}}^{\dag }(\mathrm{in}%
)\right\vert 0\right\rangle ,  \label{cq14b}
\end{eqnarray}%
where $R$ and $T$ are the corresponding relative amplitudes, and $%
c_{v}=\langle 0,\mathrm{out}|0,\mathrm{in}\rangle $, see Sec. \ref{SS7.5}.

Using canonical transformations (\ref{cq11}) one can calculate the relative
electron amplitudes,
\begin{eqnarray}
R_{+,n} &=&g\left( _{+}\left\vert ^{+}\right. \right) ^{-1}g\left(
_{-}\left\vert ^{+}\right. \right) ,\;\;T_{+,n}=\eta _{\mathrm{L}}g\left(
_{+}\left\vert ^{+}\right. \right) ^{-1},  \notag \\
R_{-,n} &=&g\left( ^{-}\left\vert _{-}\right. \right) ^{-1}g\left(
^{+}\left\vert _{-}\right. \right) ,\;\;T_{-,n}=-\eta _{\mathrm{R}}g\left(
^{-}\left\vert _{-}\right. \right) ^{-1}.  \label{cq15}
\end{eqnarray}%
Similar expressions can be obtained for the corresponding positron
amplitudes. They differ from Eqs. (\ref{cq15}) only by phases.

As it follows from Eqs. (\ref{UR2}) and (\ref{UR1}) \ the corresponding
probabilities satisfy the following relations%
\begin{equation}
\left\vert R_{+,n}\right\vert ^{2}=\left\vert R_{-,n}\right\vert
^{2},\;\;\left\vert T_{+,n}\right\vert ^{2}=\left\vert T_{-,n}\right\vert
^{2},\ \ \left\vert R_{\zeta ,n}\right\vert ^{2}+\left\vert T_{\zeta
,n}\right\vert ^{2}=1,\ n\in \Omega _{1},\Omega _{5}.  \label{cq16}
\end{equation}%
Equation (\ref{cq16}) is just the condition of the probability conservation,
written in terms of relative probabilities of reflection and transmission,
under the condition that in all other states with quantum numbers $m\neq n$\
partial vacua remain vacua.

Now we see that according to Eqs. (\ref{cq13}) the relative probabilities
coincide with the corresponding mean values,%
\begin{equation}
N_{\zeta ,n}^{(a)}\left( n,-\zeta \right) =\left\vert R_{\zeta
,n}\right\vert ^{2},\;\;N_{\zeta ,n}^{(a)}\left( n,\zeta \right) =\left\vert
T_{\zeta ,n}\right\vert ^{2},\ \ n\in \Omega _{1},\Omega _{5}.  \label{cq16a}
\end{equation}

This nontrivial result may be interpreted as QFT justification of rules of
time-independent potential scattering theory, see Ref. \cite{PS}, in the
ranges $\Omega _{1}$ and $\Omega _{5}$. To clarify this point of view, let
us consider one specific process in the range $\Omega _{1},$ namely, the
evolution of the \textrm{in}-state$\ _{+}a_{n_{1}}^{\dag }(\mathrm{in}%
)\left\vert 0\right\rangle .$ From the point of view of causal evolution
this state can be reflected, i.e., to pass to the \textrm{out}-state $\
_{-}a_{n_{1}}^{\dag }(\mathrm{out})\left\vert 0\right\rangle $ with the
probability $\left\vert R_{+,n}\right\vert ^{2}$ and can be transmitted,
i.e., to pass to the \textrm{out}-state$\ \ ^{+}a_{n_{1}}^{\dag }(\mathrm{out%
})\left\vert 0\right\rangle $ with the probability $\left\vert
T_{+,n}\right\vert ^{2}.$ Let us try to apply the potential scattering
theory to the same problem, using our QFT picture. Then, we have to
calculate two mean currents in our \textrm{in}-state, one $J_{R}$ related to
the \textrm{out}-particles $\ _{-}a_{n_{1}}^{\dag }(\mathrm{out})\left\vert
0\right\rangle $ and another one $J_{T}\ $related to the \textrm{out}%
-particles $\ ^{+}a_{n_{1}}^{\dag }(\mathrm{out})\left\vert 0\right\rangle .$
Both currents are proportional to the mean numbers of the corresponding
\textrm{out}-particles in our \textrm{in}-state and can be represented by
these numbers in the example under consideration. Then%
\begin{eqnarray*}
J_{R} &=&\left\langle 0\left\vert \ _{+}a_{n_{1}}(\mathrm{in})\left[ \
_{-}a_{n_{1}}^{\dag }(\mathrm{out})\ _{-}a_{n_{1}}(\mathrm{out})\right] \
_{+}a_{n_{1}}^{\dag }(\mathrm{in})\right\vert 0\right\rangle \\
&=&\left\vert g\left( _{+}\left\vert ^{+}\right. \right) \right\vert
^{-2}\left\vert g\left( _{-}\left\vert ^{+}\right. \right) \right\vert
^{2}=\left\vert R_{+,n}\right\vert ^{2}, \\
J_{T} &=&\left\langle 0\left\vert \ _{+}a_{n_{1}}(\mathrm{in})\left[ \
^{+}a_{n_{1}}^{\dag }(\mathrm{out})\ ^{+}a_{n_{1}}(\mathrm{out})\right] \
_{+}a_{n_{1}}^{\dag }(\mathrm{in})\right\vert 0\right\rangle \\
&=&\left\vert g\left( _{+}\left\vert ^{+}\right. \right) \right\vert
^{-2}=\left\vert T_{+,n}\right\vert ^{2}.
\end{eqnarray*}%
Thus, we see that in the range $\Omega _{1}$ realization of rules of the
potential scattering theory in the framework of QFT\ allows one to obtain
the correct result $J_{R}+J_{T}=1.$

\section{One-particle states in the ranges $\Omega _{2}$ and $\Omega _{4}$%
\label{S6}}

In the range $\Omega _{2}$ there exist only one-electron states $%
a_{n_{2}}^{\dag }\left\vert 0\right\rangle $, whereas in the range $\Omega
_{4}$ there exist only one-positron states $b_{n_{4}}^{\dag }\left\vert
0\right\rangle $,%
\begin{equation}
a_{n_{2}}^{\dag }\left\vert 0\right\rangle ,\ \ b_{n_{4}}^{\dag }\left\vert
0\right\rangle .  \label{6.1}
\end{equation}%
Below, we study their interpretations and properties.

Using Eqs. (\ref{c3}) we see that the renormalized QFT currents, given by
the operator $\widehat{\mathbb{J}}$ (\ref{5.7b}) is zero in the states under
consideration,%
\begin{equation}
J_{n_{2}}=\left\langle 0\left\vert a_{n_{2}}\widehat{\mathbb{J}}%
a_{n_{2}}^{\dag }\right\vert 0\right\rangle =J_{n_{4}}=\left\langle
0\left\vert b_{n_{4}}\widehat{\mathbb{J}}b_{n_{4}}^{\dag }\right\vert
0\right\rangle =0.  \label{currents}
\end{equation}

We interpret the QFT states (\ref{6.1}) as standing waves (stationary waves)
that present a result of interference between two waves traveling in
opposite directions, see Eqs. (\ref{2.46b}) and (\ref{2.46a}). In Fig. \ref%
{3}, we show these standing waves in the ranges $\Omega _{2}$ and $\Omega
_{4}.$ Here electron standing waves are drawn as circles with the minus
inside and positrons with the plus inside.

It should be stressed that the case of the ranges $\Omega _{2,4}$ can be
considered as a degenerate one with respect to the case of the ranges $%
\Omega _{1,5}.${\Huge \ }This case could formally by extracted from relation
(\ref{cq15}) by\ considering the limit $T_{n,+}=\eta _{\mathrm{L}}g\left(
_{+}\left\vert ^{+}\right. \right) ^{-1}\rightarrow 0$, which implies that $%
\left\vert g\left( _{+}\left\vert ^{+}\right. \right) \right\vert
^{2}=\left\vert T_{n}\right\vert ^{-2}\rightarrow \infty $ for the potential
step under consideration. Then it follows from the relation $\left\vert
R_{n}\right\vert ^{2}+\left\vert T_{n}\right\vert ^{2}=1$ that $\left\vert
R_{n}\right\vert ^{2}\rightarrow 1$, which corresponds to the almost total
reflection, when the currents of incoming and outgoing waves with a given $n$
almost cancel each other. According to the definition of the charge operator
(\ref{2.27a}), the space distributions of the charge are given by the
densities $\left\vert \ ^{\zeta }\psi _{n}\left( X\right) \right\vert ^{2}$
and $\left\vert \ _{\zeta }\psi _{n}\left( X\right) \right\vert ^{2}$. In
turn this means that linear charge density in the asymptotic regions $S_{%
\mathrm{L}}$ and $S_{\mathrm{R}}$ are given by the quantities $^{\zeta }%
\mathcal{R}_{\mathrm{L}/\mathrm{R}}/K^{\left( \mathrm{R}\right) }$ and $%
_{\zeta }\mathcal{R}_{\mathrm{L}/\mathrm{R}}/K^{\left( \mathrm{R}\right) }$
which were obtained in (\ref{i10}) and (\ref{i11}). We note that $\left\vert
g\left( _{+}\left\vert ^{-}\right. \right) \right\vert ^{2}\simeq \left\vert
g\left( _{+}\left\vert ^{+}\right. \right) \right\vert ^{2}$ in the case $%
\left\vert g\left( _{+}\left\vert ^{+}\right. \right) \right\vert ^{2}\gg 1$
such that distributions for the charge density in the ranges $\Omega _{1,5}$
become similar to the distributions in the ranges $\Omega _{2,4}$. It is
natural to expect that such a case is realized when quantum numbers $n_{1}$
are close to the upper bound of the range $\Omega _{2}$, and the quantum
numbers $n_{5}$ are close to the lower bound of the range $\Omega _{4}$.

Finally, in the ranges $\Omega _{2,4}$ only the total reflection takes
place. This process can be well{\Huge \ }described in the framework of
one-particle quantum mechanics, see \cite{LanLiQM}. Its rigorous description
in the framework of QFT as a time-dependent process leads to the same
results confirming heuristic one-particle quantum-mechanical interpretation.

One can, in principle, define \textrm{in}- and \textrm{out}-states related
to these opposite waves. It should be noted that, on the one hand, this is
not a trivial task, and, on the other hand, this problem is\ the scope of
our main goal, which is the consideration of particle creation processes.
The latter processes are specific for the $\Omega _{3}$ range, as it\ will
be clear in what follows. However, we represent below a brief discussion of
\textrm{in}- and \textrm{out}-states in the $\Omega _{2,4}$ ranges.

We believe that physical state vectors that correspond to localized \textrm{%
in}- and \textrm{out}-electrons or positrons are some wave packets composed
of formal solutions introduced in the ranges $\Omega _{2}$ and $\Omega _{4}$%
. In the region $S_{\mathrm{R}}$\ the constituent waves $\psi _{n_{2}}\left(
X\right) $\ have zero asymptotic values which implies that any wave packet
describing an electron is quasilocalized in the region $S_{\mathrm{L}}$. In
the region $S_{\mathrm{L}}$\ the constituent waves $\psi _{n_{4}}\left(
X\right) $\ have zero asymptotic values, which implies that any wave packet
describing a positron state$\ $is quasilocalized in the region $S_{\mathrm{R}%
}$.

\section{One-particle states in the range $\Omega _{3}$\label{S7}}

\subsection{General\label{SS7.1}}

First we recall (see Sec. \ref{SSS3.2.3}) that according to the structure of
the Dirac energy spectra in the asymptotic regions $S_{\mathrm{L}}$ and $S_{%
\mathrm{R}}$, there exist two sets $\left\{ \ _{\zeta }\psi _{n_{3}}\left(
X\right) \right\} $ and $\left\{ \ ^{\zeta }\psi _{n_{3}}\left( X\right)
\right\} $ of solutions in the range $\Omega _{3}$ that obey the
orthogonality relations $\left( \ _{\zeta }\psi _{n_{3}},\ ^{\zeta }\psi
_{n_{3}}\right) =0$, see (\ref{i9}). Thus, in this range\ we have two pairs $%
\left\{ \ _{-}\psi _{n_{3}}\left( X\right) ,\ ^{-}\psi _{n_{3}}\left(
X\right) \right\} \ $and$\ \left\{ \ _{+}\psi _{n_{3}}\left( X\right) ,\
^{+}\psi _{n_{3}}\left( X\right) \right\} $ of independent solutions. Each
pair forms a complete set of solutions in the range $\Omega _{3}$.

According to Eqs. (\ref{rel1}), there exist relations between solutions $%
\left\{ \ _{\zeta }\psi _{n_{3}}\left( X\right) \right\} $ and $\left\{ \
^{\zeta }\psi _{n_{3}}\left( X\right) \right\} $,%
\begin{eqnarray}
\ _{+}\psi _{n}\left( X\right) &=&g\left( _{+}\left\vert ^{-}\right. \right)
^{-1}\left[ \ _{-}\psi _{n}\left( X\right) g\left( _{-}\left\vert
^{-}\right. \right) +\ ^{-}\psi _{n}\left( X\right) \right] ,  \notag \\
\ _{-}\psi _{n}\left( X\right) &=&g\left( _{-}\left\vert ^{+}\right. \right)
^{-1}\left[ \ _{+}\psi _{n}\left( X\right) g\left( _{+}\left\vert
^{+}\right. \right) -\ ^{+}\psi _{n}\left( X\right) \right] ,  \notag \\
\ ^{+}\psi _{n}\left( X\right) &=&g\left( ^{+}\left\vert _{-}\right. \right)
^{-1}\left[ \ ^{-}\psi _{n}\left( X\right) g\left( ^{-}\left\vert
_{-}\right. \right) -\ _{-}\psi _{n}\left( X\right) \right] ,  \notag \\
\ ^{-}\psi _{n}\left( X\right) &=&g\left( ^{-}\left\vert _{+}\right. \right)
^{-1}\left[ \ ^{+}\psi _{n}\left( X\right) g\left( ^{+}\left\vert
_{+}\right. \right) +\ _{+}\psi _{n}\left( X\right) \right] .  \label{7.3}
\end{eqnarray}%
\textrm{\ }

As it follows from Eqs. (\ref{UR1}), in the range under consideration the
coefficients $g$ satisfy the following relation%
\begin{equation}
\left\vert g\left( _{+}\left\vert ^{-}\right. \right) \right\vert
^{2}=\left\vert g\left( _{+}\left\vert ^{+}\right. \right) \right\vert
^{2}+1,  \label{cq10b}
\end{equation}%
which differs from similar relation (\ref{cq10a}) in the range $\Omega _{1}$.

We\emph{\ }associate\emph{\ }the first independent pair $\left\{ \ _{-}\psi
_{n_{3}}\left( X\right) ,\ ^{-}\psi _{n_{3}}\left( X\right) \right\} $ with
\textrm{in}-solutions and the second independent pair $\left\{ \ _{+}\psi
_{n_{3}}\left( X\right) ,\ ^{+}\psi _{n_{3}}\left( X\right) \right\} $ with
\textrm{out}-solutions. Thus, solutions $\left\{ \ ^{-}\psi _{n_{3}}\left(
X\right) ,\ ^{+}\psi _{n_{3}}\left( X\right) \right\} \ $are associated with
\textrm{in-} and \textrm{out}-electron states, whereas solutions $\left\{ \
_{-}\psi _{n_{3}}\left( X\right) ,\ _{+}\psi _{n_{3}}\left( X\right)
\right\} $ are associated with \textrm{in-} and \textrm{out}-positron
states. Correspondingly, in the range $\Omega _{3}$ there exist four types
of one-particle QFT states,
\begin{equation}
\ ^{-}a_{n_{3}}^{\dag }(\mathrm{in})\left\vert 0,\mathrm{in}\right\rangle ,\
\ ^{+}a_{n_{3}}^{\dag }(\mathrm{out})\left\vert 0,\mathrm{out}\right\rangle
,\ \ _{-}b_{n_{3}}^{\dag }(\mathrm{in})\left\vert 0,\mathrm{in}\right\rangle
,\ \ _{+}b_{n_{3}}^{\dag }(\mathrm{out})\left\vert 0,\mathrm{out}%
\right\rangle \ .  \label{7.1}
\end{equation}%
Since no other ranges are considered in this section, the quantum numbers $%
n_{3}$ are sometimes denoted by $n$ for simplicity.

Using both alternative decompositions (\ref{2.23}) for $\Psi _{3}\left(
X\right) $ and relations (\ref{7.3}), we find the following linear canonical
transformations between the introduced \textrm{in}- and \textrm{out}-
creation and annihilation operators%
\begin{eqnarray}
\  &&^{+}a_{n}(\mathrm{out})=-g\left( _{-}\left\vert ^{+}\right. \right)
^{-1}\ _{-}b_{n}^{\dagger }(\mathrm{in})+g\left( ^{-}\left\vert _{+}\right.
\right) ^{-1}g\left( ^{+}\left\vert _{+}\right. \right) \ ^{-}a_{n}(\mathrm{%
in}),  \notag \\
&&\ _{+}b_{n}^{\dagger }(\mathrm{out})=g\left( _{-}\left\vert ^{+}\right.
\right) ^{-1}g\left( _{+}\left\vert ^{+}\right. \right) \ _{-}b_{n}^{\dagger
}(\mathrm{in})+g\left( ^{-}\left\vert _{+}\right. \right) ^{-1}\ ^{-}a_{n}(%
\mathrm{in}),  \notag \\
&&\ _{-}b_{n}^{\dagger }(\mathrm{in})=g\left( _{+}\left\vert ^{-}\right.
\right) ^{-1}g\left( _{-}\left\vert ^{-}\right. \right) \ _{+}b_{n}^{\dagger
}(\mathrm{out})-g\left( ^{+}\left\vert _{-}\right. \right) ^{-1}\ ^{+}a_{n}(%
\mathrm{out}),  \notag \\
&&\ ^{-}a_{n}(\mathrm{in})=g\left( _{+}\left\vert ^{-}\right. \right) ^{-1}\
_{+}b_{n}^{\dagger }(\mathrm{out})+g\left( ^{+}\left\vert _{-}\right.
\right) ^{-1}g\left( ^{-}\left\vert _{-}\right. \right) \ ^{+}a_{n}(\mathrm{%
out}).  \label{7.4}
\end{eqnarray}%
Because these transformations entangle annihilation and creation operators,
the partial vacua $\left\vert 0,\mathrm{in}\right\rangle ^{(3)}$ and $%
\left\vert 0,\mathrm{out}\right\rangle ^{(3)}$ are essentially different.
That is why, the total vacua $\left\vert 0,\mathrm{in}\right\rangle $ and $%
\left\vert 0,\mathrm{out}\right\rangle $ are different as well, see Eqs. (%
\ref{2.29}) and (\ref{2.34}).

\subsection{ Interpretation of states in $\Omega _{3}$\label{SS7.3}}

To give an interpretation of states in $\Omega _{3}$\ we have studied
one-particle mean values of the charge, the kinetic energy, the number of
particles, the current, and the energy flux through the surfaces{\large \ }$%
x=x_{\mathrm{L}}${\large \ }and{\large \ }$x=x_{\mathrm{R}}${\large . }The
corresponding calculations are placed in the Appendix \ref{SS7.2}. Based on
results of such calculations, we can finally conclude:

(1) The states
\begin{equation}
\ ^{-}a_{n_{3}}^{\dag }(\mathrm{in})\left\vert 0,\mathrm{in}\right\rangle ,\
\ ^{+}a_{n_{3}}^{\dag }(\mathrm{out})\left\vert 0,\mathrm{out}\right\rangle
\label{7.6b}
\end{equation}%
are states with the charge $-e.$ The states%
\begin{equation}
\ _{-}b_{n_{3}}^{\dag }(\mathrm{in})\left\vert 0,\mathrm{in}\right\rangle ,\
\ _{+}b_{n_{3}}^{\dag }(\mathrm{out})\left\vert 0,\mathrm{out}\right\rangle
\label{7.6c}
\end{equation}%
are states with the charge $+e.$ Equations (\ref{7.12}) confirm these
conclusions.

(2) Each of these states belongs to one of the physical wave packets that
have positive energies as is demonstrated in Sec. \ref{SS7.4}, and,
therefore, they can be treated as physical particle states, namely, states (%
\ref{7.6b}) represent electrons, whereas states (\ref{7.6c}) represent
positrons. Their currents (\ref{7.12}) also$\ $confirm this interpretation.

(3) Mean energy fluxes (\ref{cq20.2}) of states (\ref{7.1}) through the
surfaces $x=x_{\mathrm{L}}$ and $x=x_{\mathrm{R}}$ together with expressions
(\ref{7.12}) for their currents allow us to conclude that:

(a) electrons$\ \ ^{-}a_{n_{3}}^{\dag }(\mathrm{in})\left\vert 0,\mathrm{in}%
\right\rangle $ and positrons $\ _{+}b_{n_{3}}^{\dag }(\mathrm{out}%
)\left\vert 0,\mathrm{out}\right\rangle $\ are moving to the right;

(b) electrons $\ ^{+}a_{n_{3}}^{\dag }(\mathrm{out})\left\vert 0,\mathrm{out}%
\right\rangle $ and positrons $\ _{-}b_{n_{3}}^{\dag }(\mathrm{in}%
)\left\vert 0,\mathrm{in}\right\rangle $\ are traveling to the left.

We stress that in contrast to the ranges $\Omega _{1}$\ and $\Omega _{5}$,
in the range $\Omega _{3}$\ signs of the energy fluxes of the electrons are
determined by signs of the quantum number{\large \ }$p^{\mathrm{R}}${\large %
\ }whereas signs of the energy fluxes of the positrons are determined by
signs of the quantum number{\large \ }$p^{\mathrm{L}}${\large . }This
correlation follows from the fact that in the region $S_{\mathrm{R}}$ the
directions of the positron motion coincides with the directions of positron
current and in the region{\large \ }$S_{\mathrm{L}}${\large \ }the
directions of the electron motion is opposite to the directions of the
electron current.

(4) We classify electron states $\ ^{-}a_{n_{3}}^{\dag }(\mathrm{in}%
)\left\vert 0,\mathrm{in}\right\rangle $ as \textrm{in}- states, because
they can move to the step only from the asymptotic region $S_{\mathrm{L}}.$
The latter statement is based on our belief that the structure of the Dirac
spectrum in the region $S_{\mathrm{R}}$ forbids electrons to be present in
this region.

We classify electron states $\ ^{+}a_{n_{3}}^{\dag }(\mathrm{out})\left\vert
0,\mathrm{out}\right\rangle $ as \textrm{out}- states, because they can move
from the step only to the asymptotic regions $S_{\mathrm{L}}$. The latter
statement is based on our belief that the structure of the Dirac spectrum in
the region $S_{\mathrm{R}}$ forbids electrons to be present in this region.

We classify positrons states $\ _{-}b_{n_{3}}^{\dag }(\mathrm{in})\left\vert
0,\mathrm{in}\right\rangle \ $as \textrm{in}- states, because they can move
to the step only from the asymptotic region $S_{\mathrm{R}}$. The latter
statement is based on our belief that the structure of the Dirac spectrum in
the region $S_{\mathrm{L}}$ forbids positrons to be present in this region.

We classify positrons states $\ _{+}b_{n_{3}}^{\dag }(\mathrm{out}%
)\left\vert 0,\mathrm{out}\right\rangle \ $as \textrm{out}- states, because
they can move to the step only from the asymptotic region $S_{\mathrm{R}}$.
The latter statement is based on our belief that the structure of the Dirac
spectrum in the region $S_{\mathrm{L}}$ forbids positrons to be present in
this region.

In Fig.~\ref{3} we show \textrm{in}- and \textrm{out}-states in the range $%
\Omega _{3}.$

\subsection{Discussion of the localization properties\label{SS7.4}}

Taking into account Eqs. (\ref{UR2}), one can verify that all differential
mean numbers (\ref{7.5}) are equal. That is why, we introduce the unique
notation $N_{n}^{\mathrm{cr}}$ for all of them,
\begin{equation}
N_{n}^{\mathrm{cr}}=N_{n}^{b}\left( \mathrm{out}\right) =N_{n}^{a}\left(
\mathrm{out}\right) =N_{n}^{b}\left( \mathrm{in}\right) =N_{n}^{a}\left(
\mathrm{in}\right) =\left\vert g\left( _{+}\left\vert ^{-}\right. \right)
\right\vert ^{-2}.  \label{7.7}
\end{equation}%
For the fermions under consideration, it is natural that the quantity $%
N_{n}^{\mathrm{cr}}$ is always less or equal than one, $N_{n}^{\mathrm{cr}%
}\leq 1.$ We also note that the quantities $\pi _{0}\left( \mathrm{L}\right)
$ and $\left\vert \pi _{0}\left( \mathrm{R}\right) \right\vert $ achieve
their minimal values on the boundaries of the range $\Omega _{3}$, namely,%
\begin{equation}
\min \pi _{0}\left( \mathrm{L}\right) =\pi _{\bot },\ \min \left\vert \pi
_{0}\left( \mathrm{R}\right) \right\vert =\pi _{\bot }.  \label{7.8a}
\end{equation}

If $N_{n}^{\mathrm{cr}}$ tends to zero, $N_{n}^{\mathrm{cr}}\rightarrow 0$,
then $\left\vert g\left( _{+}\left\vert ^{-}\right. \right) \right\vert
^{2}\rightarrow \infty $ and, at the same time, $\left\vert g\left(
_{+}\left\vert ^{+}\right. \right) \right\vert ^{2}\rightarrow \infty $ in
accordance to relation (\ref{cq10b}).

The quantity $N_{n}^{\mathrm{cr}}$ can be calculated explicitly in the
so-called exactly solvable cases (see Introduction and section \ref{S9})).
However, we can derive some additional useful properties of this quantity in
the general case.

Taking into account the fact that no pair creation takes place in the
regions $\Omega _{2}$ and $\Omega _{4}$, we suppose that\emph{\ }pair
creation vanishes near the boundaries of the range $\Omega _{3}$,
\begin{eqnarray}
\left. N_{n_{3}}^{\mathrm{cr}}\right\vert _{n_{3}\rightarrow \Omega _{2}}
&\rightarrow &0\Longleftrightarrow \left. N_{n_{3}}^{\mathrm{cr}}\right\vert
_{\pi _{0}\left( \mathrm{R}\right) \rightarrow -\pi _{\bot }}\rightarrow 0,
\notag \\
\left. N_{n_{3}}^{\mathrm{cr}}\right\vert _{n_{3}\rightarrow \Omega _{4}}
&\rightarrow &0\Longleftrightarrow \left. N_{n_{3}}^{\mathrm{cr}}\right\vert
_{\pi _{0}\left( \mathrm{L}\right) \rightarrow \pi _{\bot }}\rightarrow 0.
\label{7.9}
\end{eqnarray}

Note that in this case kinetic energies of electron or positron plane waves
given by Eqs. (\ref{i18b}) tend to their values, $^{\zeta }\mathcal{E}%
_{n_{3}}\rightarrow \pi _{0}\left( \mathrm{L}\right) $ and $_{\zeta }%
\mathcal{E}_{n_{3}}\rightarrow \pi _{0}\left( \mathrm{R}\right) ,$ near the
boundaries of the range{\large \ }$\Omega _{3}$\textrm{. }Therefore,
conditions (\ref{7.9}) provide that%
\begin{equation}
^{\zeta }\mathcal{E}_{n_{3}}>0,\ \ _{\zeta }\mathcal{E}_{n_{3}}<0
\label{i20}
\end{equation}%
near the boundaries of the range $\Omega _{3}$\textrm{\ }and operator $%
\widehat{\mathbb{H}}^{\mathrm{kin}}$\ (\ref{2.27}) is positively-defined in
this part of the range $\Omega _{3}$\textrm{.\ }We believe that inequality (%
\ref{7.9}) takes place for all nonpathological\textrm{\ }$x$-electric steps.

As was already mentioned$,$ Eqs. (\ref{7.9}) imply that $\left\vert g\left(
_{+}\left\vert ^{-}\right. \right) \right\vert ^{2}\simeq \left\vert g\left(
_{+}\left\vert ^{+}\right. \right) \right\vert ^{2}\rightarrow \infty $.
Then it follows from Eqs. (\ref{i10}) and (\ref{i11}) that the electron
density $\left\vert ^{\zeta }\psi _{n}\left( X\right) \right\vert ^{2}$ is
concentrated in the region $S_{\mathrm{L}},$ whereas the positron density $%
\left\vert _{\zeta }\psi _{n}\left( X\right) \right\vert ^{2}$ \ is
concentrated in the region $S_{\mathrm{R}}$,%
\begin{eqnarray}
&&\left. \left\vert ^{\zeta }\psi _{n_{3}}\left( X\right) \right\vert
^{2}\right\vert _{\pi _{0}\left( \mathrm{R}\right) \rightarrow -\pi _{\bot
}}\rightarrow 0,\ \ x\in S_{\mathrm{R}}\ ,  \notag \\
&&\left. \left\vert _{\zeta }\psi _{n_{3}}\left( X\right) \right\vert
^{2}\right\vert _{\pi _{0}\left( \mathrm{L}\right) \rightarrow \pi _{\bot
}}\rightarrow 0,\ \ x\in S_{\mathrm{L}}\ ,  \label{7.10}
\end{eqnarray}%
which means that these densities are continuous near the boundaries of the
range $\Omega _{3}$. We see that conditions (\ref{7.9}) and (\ref{7.10}) are
equivalent. Thus we believe that conditions (\ref{7.10}) imply (\ref{7.9})
near the boundaries of the range $\Omega _{3}$.

For arbitrary $n\in \Omega _{3}$ some properties of $N_{n}^{\mathrm{cr}}$
can be established first in the case of weak external fields (but still
strong enough, $\mathbb{U}-2\pi _{\bot }>0$, to provide the existence of the
$\Omega _{3}$-range) using a semiclassical approximation. In the latter case
a really strong restriction has to be imposed, that $\ N_{n}^{\mathrm{cr}}$
is exponentially small, $N_{n}^{\mathrm{cr}}\ll 1.$ Then it is natural to
expect that for any finite $\mathbb{U}<\infty $\textrm{\ }and\textrm{\ }$%
m\neq 0$ the inequality $\left\vert g\left( _{+}\left\vert ^{-}\right.
\right) \right\vert ^{-2}<2\pi _{\bot }/\mathbb{U}$ holds, such that\textrm{%
\ }inequalities\textrm{\ }(\ref{i20})\textrm{\ }hold\ for arbitrary $n\in
\Omega _{3}$\textrm{. }The condition $N_{n}^{\mathrm{cr}}\ll 1$ and Eqs. (%
\ref{i10}) and (\ref{i11}) imply that for arbitrary $n\in \Omega _{3}$ the
electron density $\left\vert \ ^{\zeta }\psi _{n}\left( X\right) \right\vert
^{2}$ is concentrated in the region $S_{\mathrm{L}},$ whereas the positron
density $\left\vert \ _{\zeta }\psi _{n}\left( X\right) \right\vert ^{2}$ \
is concentrated in the region $S_{\mathrm{R}}$,
\begin{eqnarray}
&&\left\vert \ ^{\zeta }\psi _{n_{3}}\left( X\right) \right\vert
^{2}\rightarrow 0,\ \ x\in S_{\mathrm{R}}\ ,  \notag \\
&&\left\vert \ _{\zeta }\psi _{n_{3}}\left( X\right) \right\vert
^{2}\rightarrow 0,\ \ x\in S_{\mathrm{L}}\ .  \label{7.10b}
\end{eqnarray}

Thus, the wave functions $\ ^{\zeta }\psi _{n_{3}}\left( X\right) $ and $\
_{\zeta }\psi _{n_{3}}\left( X\right) $\textrm{\ }behave quite similarly to
the behavior of the corresponding functions with $n\in \Omega _{2},\Omega
_{4}$.

In the general case when the quantities $N_{n}^{\mathrm{cr}}$\ are not
small, it is natural to expect a\ similar behavior, namely: the region $S_{%
\mathrm{R}}$\ is not available for electrons, a the region $S_{\mathrm{L}}$
is not available for positrons. However, when the quantities $N_{n}^{\mathrm{%
cr}}$\ are not small, the latter property may hold only for the
corresponding wave packets, but not for the separate plane waves. That means
that Eq. (\ref{7.10b}) may not hold,\ that is, these plane waves may be
different from zero in the whole space. Within our context it is assumed \
that electrons and positrons in one of corresponding asymptotic regions may
occupy quasistationary states, i.e. they should be described by wave packets
that pertain their form sufficiently long in one of corresponding asymptotic
regions.\ In other words, only such electron and positron wave packets have
a physical meaning in the problem under consideration. This is what we shall
keep in mind when discussing the wave packets in what follows.\textrm{\ }We
can demonstrate that in the general case the electron wave packets that
really have a physical meaning can be localized only in the asymptotic
region $S_{\mathrm{L}}$, whereas the positron wave packets that really have
a physical meaning can be localized only in the asymptotic regions $S_{%
\mathrm{R}}$.\emph{\ }This is a consequence of a specific structure of plane
waves $\ ^{\zeta }\psi _{n_{3}}\left( X\right) $\ and $\ _{\zeta }\psi
_{n_{3}}\left( X\right) $\ in asymptotic regions $S_{\mathrm{L}}$\ and $S_{%
\mathrm{R}}$. Indeed, this structure is quite different from the structure
of plane waves in the ranges $\Omega _{1}$\ and $\Omega _{5}$\emph{. }As was
mentioned above, electron states with given quantum numbers $n_{3}$ are
states with a$\ $definite quantum number $p^{\mathrm{R}},$ whereas positron
states with given quantum numbers $n_{3}$ are states with a\ definite
quantum number $p^{\mathrm{L}}$. This fact together with relation (\ref{7.3})%
\textrm{\ }implies, for example, that a partial wave of an \textrm{in}%
-electron, $\ ^{-}\psi _{n_{3}}\left( X\right) ,$ in the region where the
electron can really be observed, i.e., in the region $S_{\mathrm{L}}$, is
always a superposition of two waves with opposite signs of the quantum
number $p^{\mathrm{L}}$, $\ _{+}\psi _{n_{3}}\left( X\right) $ and $\
_{-}\psi _{n_{3}}\left( X\right) $. In turn, this implies that in contrast
to the ranges $\Omega _{1}$ and $\Omega _{5}$, the sign of $p^{\mathrm{L}}$
is not related to the sign of the mean energy flux in the region $S_{\mathrm{%
L}}$. The same holds true for a partial wave of an $\mathrm{out}$-electron%
\emph{\ \ }$^{+}\psi _{n_{3}}\left( X\right) $\emph{. }Similarly, one can
see that partial waves of both \textrm{in}- positron $\ _{-}\psi
_{n_{3}}\left( X\right) $, and $\mathrm{out}$-positron $\ _{+}\psi
_{n_{3}}\left( X\right) $, in the region $S_{\mathrm{R}}$, are always
superpositions of two waves with quantum number $p^{\mathrm{R}}$ of opposite
signs\ and, therefore, signs of these quantum numbers are not connected to
the sign of the mean energy flux in the region $S_{\mathrm{R}}$\emph{. }%
However, as it was demonstrated above, these are states with well-defined
asymptotic energy flux, and therefore with the corresponding well-defined
asymptotic field momentum.\emph{\ }One can demonstrate that namely these
properties of the constituent plane waves are responsible for the fact that
stable electron wave packets can exist only in the region $S_{\mathrm{L}}$,
whereas stable positron wave packets can exist only in the region $S_{%
\mathrm{R}}$, see details in the Appendix \ref{Aloc}.

We stress that in the range $\Omega _{3}$, within each pair of independent
states with the same $n$, the$\ \mathrm{in}$- particles and $\mathrm{out}$-
particles always move in opposite directions.

Mean values (\ref{7.6}) and (\ref{7.12}) are typical for the total
reflection. Indeed, all the mean particle numbers do not change in the
course of the interaction with the step field, and the\ both electron
currents are equal in magnitude and have opposite directions, the same
holding for the both positron currents.{\Large \ }We have demonstrated that
electron densities in $\Omega _{3}$ have a behavior similar to that in\ the $%
\Omega _{2}$ range, vanishing in the $S_{\mathrm{R}}$ region$,$ while
positron densities in $\Omega _{3}$ behave similarly to the $\Omega _{4}$
range, vanishing in the $S_{\mathrm{L}}$ region.

Since the above-described properties of mean values (\ref{7.6}) and (\ref%
{7.12}) hold true in the whole range $\Omega _{3},$ we believe that all the
\textrm{in}-states in this range are subjected to the total reflection. Once
this is the case, the wave functions of the \textrm{in}-states and of\ the
\textrm{out}-states corresponding to them have to be concentrated in the
same regions on the left or on the right of the $x$-electric step, similar
to the behavior of the particles in the ranges $\Omega _{2}$ and $\Omega
_{4} $, respectively. Note that kinetic energies of these physical states
are sums of the positive partial energies $V_{\bot }M_{n}^{-1}\ ^{\zeta
}E_{n}^{\mathrm{L}}\ $for electrons and $-V_{\bot }M_{n}^{-1}\ _{\zeta
}E_{n}^{\mathrm{R}}$ for positrons, where the quantities $\ ^{\zeta }E_{n}^{%
\mathrm{L}}\ $\ and $\ _{\zeta }E_{n}^{\mathrm{R}}$ are given by Eqs.~(\ref%
{i17b}).

Some additional arguments in favor of the given interpretation are in order.

In the range $\Omega _{3}$ no\ particles exist\ that could maintain the
direction of their motion after the interaction with the external field.
This peculiarity allows one to classify these one-particle states as \textrm{%
in}- or \textrm{out}-states by using mean currents (\ref{7.12}). The
electric field under consideration accelerates positrons along the axis $x$
and electrons in the opposite direction, that is why in the range $\Omega
_{3}$ the current of \textrm{out}-electron states coincides with the
electric field direction, whereas the current of \textrm{in}-electron states
is opposite to the electric field direction. Thus, $\ ^{+}a_{n}^{\dagger }(%
\mathrm{out})\left\vert 0,\mathrm{out}\right\rangle $ and $\
_{+}b_{n_{3}}^{\dagger }(\mathrm{out})\left\vert 0,\mathrm{out}\right\rangle
$ are \textrm{out}- states of electrons and positrons, respectively, whereas
$^{-}a_{n_{3}}^{\dagger }($\textrm{in}$)\left\vert 0,\mathrm{in}%
\right\rangle $ and $_{-}b_{n_{3}}^{\dagger }($\textrm{in}$)\left\vert 0,%
\mathrm{in}\right\rangle $ are \textrm{in}$-$ states of electrons and
positrons, respectively. This also implies that $\left\vert 0,\mathrm{in}%
\right\rangle $ is the \textrm{in}$-$ vacuum and $\left\vert 0,\mathrm{out}%
\right\rangle $ is the \textrm{out}$-$ vacuum. Finally, $^{-}a_{n_{3}}^{%
\dagger }($\textrm{in}$)$, $^{-}a_{n_{3}}($\textrm{in}$)$, $%
_{-}b_{n_{3}}^{\dagger }($\textrm{in}$)$, $_{-}b_{n_{3}}($\textrm{in}$)$ are
creation and annihilation operators of \textrm{in}$-$ electrons and
positrons respectively, whereas $^{+}a_{n_{3}}^{\dagger }($\textrm{out}$)$, $%
^{+}a_{n_{3}}($\textrm{out}$)$, $_{+}b_{n_{3}}^{\dagger }($\textrm{out}$)$, $%
_{+}b_{n_{3}}($\textrm{out}$)$ are creation and annihilation operators of
\textrm{out}-electrons and \textrm{out}-positrons, respectively.

We come to the same conclusion considering the mean values (\ref{7.5}).
Thus, we see that the quantities $N_{n}^{a}\left( \mathrm{out}\right) $ and $%
N_{n}^{b}\left( \mathrm{out}\right) $ are differential mean numbers of
\textrm{out-}electrons and \textrm{out}-positrons$,$ respectively$,$ created
from the vacuum, since electric currents composed of the corresponding
states coincide with the direction of the electric field. In this case we
face electron and positron pairs outgoing from the state where no incoming
particles{\huge \ }were present. The mean numbers of created electrons are
equal to mean numbers of created positrons, in full agreement with the
charge conservation law.

Besides, it follows from Eq.~(\ref{7.5}) that $N_{n}^{b}\left( \mathrm{in}%
\right) $ and $N_{n}^{a}\left( \mathrm{in}\right) $\textrm{\ }are
differential mean numbers of electrons and positrons, respectively,
annihilated from the initial neutral state of an electron-positron pair. The%
{\huge \ }electric current corresponding to this initial pair is directed
opposite to the electric field, that is why after the annihilation its
electron and positron components are equally reduced by the quantity $e%
\mathcal{M}^{-1}N_{n}^{\mathrm{cr}}$.

Thus, we believe that the wave functions $\ ^{-}\psi _{n_{3}}\left( X\right)
$ and $\ _{-}\psi _{n_{3}}\left( X\right) $ describe initial or\emph{\ }%
incoming states of an electron and a positron, respectively, whereas the
wave functions $\ ^{+}\psi _{n_{3}}\left( X\right) $ and $\ _{+}\psi
_{n_{3}}\left( X\right) $ describe final or outgoing states of an electron
and a positron, respectively.

This causal identification coincides with the one proposed by Nikishov in
the frame of relativistic quantum mechanics in Refs. \cite{Nikis79,Nikis04}.
Note that it differs from another identification in the frame of
relativistic quantum mechanics given in Refs. \cite{HansRavn81} and
repeated, for example, in Refs. \cite{GMR85,DomCal99}. In the Sec. \ref%
{SS7.6} we discuss this problem in more detail.

\subsection{Reflection and creation of particles in $\Omega _{3}$\label%
{SS7.5}}

The total number $N_{n}^{\mathrm{cr}}$ of pairs\ created from the vacuum is
the sum over the range $\Omega _{3}$ of the differential mean numbers $%
N_{n}^{\mathrm{cr}}$,%
\begin{equation}
N=\sum_{n\in \Omega _{3}}N_{n}^{\mathrm{cr}}=\sum_{n\in \Omega
_{3}}\left\vert g\left( _{+}\left\vert ^{-}\right. \right) \right\vert ^{-2}.
\label{TN}
\end{equation}

Here we consider probability amplitudes of some simplest processes in the
range $\Omega _{3}$. First of all, this is the vacuum-to-vacuum transition
amplitude which coincides [due to Eq. (\ref{2.34})] with the total
vacuum-to-vacuum transition amplitude $c_{v}$ ,%
\begin{equation}
c_{v}^{\left( 3\right) }=\ ^{\left( 3\right) }\langle 0,\mathrm{out}|0,%
\mathrm{in}\rangle ^{(3)}\ =c_{v}=\langle 0,\mathrm{out}|0,\mathrm{in}%
\rangle .  \label{cq23}
\end{equation}%
Among other nonzero amplitudes we have to consider two relative scattering
amplitudes of electrons and positrons,%
\begin{eqnarray}
&&w\left( +|+\right) _{n^{\prime }n}=c_{v}^{-1}\langle 0,\mathrm{out}%
\left\vert \ ^{+}a_{n^{\prime }}\left( \mathrm{out}\right) \
^{-}a_{n}^{\dagger }(\mathrm{in})\right\vert 0,\mathrm{in}\rangle ,  \notag
\\
&&w\left( -|-\right) _{n^{\prime }n}=c_{v}^{-1}\langle 0,\mathrm{out}%
\left\vert \ _{+}b_{n^{\prime }}\left( \mathrm{out}\right) \
_{-}b_{n}^{\dagger }(\mathrm{in})\right\vert 0,\mathrm{in}\rangle \,,
\label{cq22a}
\end{eqnarray}%
and two relative amplitudes of a pair creation and a pair annihilation,%
\begin{eqnarray}
&&w\left( +-|0\right) _{n^{\prime }n}=c_{v}^{-1}\langle 0,\mathrm{out}%
\left\vert \ ^{+}a_{n^{\prime }}\left( \mathrm{out}\right) \ _{+}b_{n}\left(
\mathrm{out}\right) \right\vert 0,\mathrm{in}\rangle \,,  \notag \\
&&w\left( 0|-+\right) _{nn^{\prime }}=c_{v}^{-1}\langle 0,\mathrm{out}%
\left\vert \ _{-}b_{n}^{\dagger }(\mathrm{in})\ ^{-}a_{n^{\prime }}^{\dagger
}(\mathrm{in})\right\vert 0,\mathrm{in}\rangle \,.  \label{cq22}
\end{eqnarray}%
As can be derived from relations (\ref{7.4}), all the amplitudes (\ref{cq22a}%
) and (\ref{cq22}) are diagonal in the quantum numbers $n$ and can be
expressed in terms of the coefficients $g\left( ^{\zeta ^{\prime
}}\left\vert _{\zeta }\right. \right) $ as follows:%
\begin{eqnarray}
&&w\left( +|+\right) _{n^{\prime }n}=\delta _{n,n^{\prime }}w_{n}\left(
+|+\right) ,\ \ w_{n}\left( +|+\right) =g\left( ^{+}\left\vert _{-}\right.
\right) g\left( ^{-}\left\vert _{-}\right. \right) ^{-1}=g\left(
_{+}\left\vert ^{-}\right. \right) g\left( _{+}\left\vert ^{+}\right.
\right) ^{-1},  \notag \\
&&w\left( -|-\right) _{nn^{\prime }}=\delta _{n,n^{\prime }}w_{n}\left(
-|-\right) ,\ \ w_{n}\left( -|-\right) =g\left( ^{-}\left\vert _{+}\right.
\right) g\left( ^{-}\left\vert _{-}\right. \right) ^{-1}=g\left(
_{-}\left\vert ^{+}\right. \right) g\left( _{+}\left\vert ^{+}\right.
\right) ^{-1},  \notag \\
&&w\left( 0|-+\right) _{nn^{\prime }}=\delta _{n,n^{\prime }}w_{n}\left(
0|-+\right) ,\ \ w_{n}\left( 0|-+\right) =-g\left( ^{-}\left\vert
_{-}\right. \right) ^{-1},  \notag \\
&&w\left( +-|0\right) _{n^{\prime }n}=\delta _{n,n^{\prime }}w_{n}\left(
+-|0\right) ,\ \ w_{n}\left( +-|0\right) =g\left( _{+}\left\vert ^{+}\right.
\right) ^{-1}\ .  \label{cq25}
\end{eqnarray}

Recalling the physical meaning of the one-particle states (\ref{7.1}), we
conclude that in the range $\Omega _{3}$\ the total reflection is the only
possible form of particle scattering, with $w\left( +|+\right) _{n}$\ and $%
w\left( -|-\right) _{n}$ being relative probability amplitudes of a particle
reflection. The relative probability amplitude $w\left( +-|0\right) _{n}$\
describes creation of an electron-positron pair of \textrm{out}-particles
with given quantum numbers $n$, and the relative probability amplitude $%
w\left( 0|-+\right) _{n}$ describes annihilation of an electron-positron
pair of \textrm{in}-particles, each of them having quantum numbers $n$.

Unitary relations (\ref{UR}) and their consequences (\ref{UR2}) and \ (\ref%
{UR1}) imply the following connections for the introduced amplitudes $w$:%
\begin{eqnarray}
&&\left\vert w_{n}\left( +|+\right) \right\vert ^{2}=\left\vert w_{n}\left(
-|-\right) \right\vert ^{2},\;\left\vert w_{n}\left( +-|0\right) \right\vert
^{2}=\left\vert w_{n}\left( 0|-+\right) \right\vert ^{2},  \notag \\
&&\left\vert w_{n}\left( +|+\right) \right\vert ^{2}-\left\vert w_{n}\left(
+-|0\right) \right\vert ^{2}=1,\;\frac{w_{n}\left( -|-\right) ^{\ast }}{%
w_{n}\left( +|+\right) }=-\frac{w_{n}\left( +-|0\right) ^{\ast }}{%
w_{n}\left( 0|-+\right) }.  \label{cq26}
\end{eqnarray}

Referring to Eqs. (\ref{cq25}), relations (\ref{7.4}) can be rewritten as%
\begin{eqnarray}
\ ^{-}a_{n}\left( \mathrm{in}\right)  &=&w_{n}\left( +|+\right) ^{-1}\left[
\ ^{+}a_{n}\left( \mathrm{out}\right) +w_{n}\left( +-|0\right) \
_{+}b_{n}^{\dag }\left( \mathrm{out}\right) \right] ,  \notag \\
\ _{-}b_{n}\left( \mathrm{in}\right)  &=&w_{n}\left( -|-\right) ^{-1}\left[
\ _{+}b_{n}\left( \mathrm{out}\right) -w_{n}\left( +-|0\right) \
^{+}a_{n}^{\dag }\left( \mathrm{out}\right) \right] .  \label{cq27}
\end{eqnarray}%
Together with their adjoint relations they determine a unitary
transformation $V_{\Omega _{3}}$ between the \textrm{in}- and \textrm{out}%
-operators,%
\begin{equation}
\left\{ \ ^{-}a^{\dagger }(\mathrm{in}),\ ^{-}a(\mathrm{in}),\
_{-}b^{\dagger }(\mathrm{in}),\ _{-}b(\mathrm{in})\right\} =V_{\Omega
_{3}}\left\{ \ ^{+}a^{\dagger }\left( \mathrm{out}\right) ,\ ^{+}a\left(
\mathrm{out}\right) ,\ _{+}b^{\dagger }\left( \mathrm{out}\right) ,\
_{+}b\left( \mathrm{out}\right) \right\} V_{\Omega _{3}}^{\dagger }\ .
\notag
\end{equation}%
Since Eqs.~(\ref{cq26}) and (\ref{cq27}) coincide formally with the
corresponding equations for $t$-electric potential steps, the unitary
operator $V_{\Omega _{3}}$ can be taken from the works \cite{Gitman,GavGT06}%
. It has the form%
\begin{eqnarray}
&&V_{\Omega _{3}}=\exp \left\{ -\sum_{n\in \Omega _{3}}\ ^{+}a_{n}^{\dagger
}\left( \mathrm{out}\right) w_{n}\left( +-|0\right) \ _{+}b_{n}^{\dagger
}\left( \mathrm{out}\right) \right\}   \notag \\
&&\times \exp \left\{ -\sum_{n\in \Omega _{3}}\ _{+}b_{n}\left( \mathrm{out}%
\right) \ln w_{n}\left( -|-\right) \ _{+}b_{n}^{\dagger }\left( \mathrm{out}%
\right) \right\}   \notag \\
&&\times \exp \left\{ \sum_{n\in \Omega _{3}}\ ^{+}a_{n}^{\dagger }\left(
\mathrm{out}\right) \ln w_{n}\left( +|+\right) \ ^{+}a_{n}\left( \mathrm{out}%
\right) \right\}   \notag \\
&&\times \exp \left\{ -\sum_{n\in \Omega _{3}}\ _{+}b_{n}\left( \mathrm{out}%
\right) w_{n}\left( 0|-+\right) \ ^{+}a_{n}\left( \mathrm{out}\right)
\right\} .  \label{cq29}
\end{eqnarray}

At the same time, the operator $V_{\Omega _{3}}$ relates the \textrm{in}-
and \textrm{out}-vacua, $|0,\mathrm{in}\rangle =V_{\Omega _{3}}|0,\mathrm{out%
}\rangle $ and therefore it\ determines the vacuum-to-vacuum transition
amplitude $c_{v}$,%
\begin{equation}
c_{v}=\langle 0,\mathrm{out}|V_{\Omega _{3}}|0,\mathrm{out}\rangle
=\prod\limits_{n}w_{n}\left( -|-\right) ^{-1}\,.  \label{cq30}
\end{equation}

The probabilities of a particle reflection, a pair creation, and the
probability for a vacuum to remain a vacuum can be expressed via
differential mean numbers of created pairs $N_{n}^{\mathrm{cr}}$. By using
the relation $\left\vert w_{n}\left( -|-\right) \right\vert ^{2}=\left(
1-N_{n}^{\mathrm{cr}}\right) ^{-1}$, one finds
\begin{eqnarray}
&&P(+|+)_{n^{\prime },n}=|\langle 0,\mathrm{out}|\ ^{+}a_{n^{\prime }}(%
\mathrm{out})\ ^{-}a_{n}^{\dagger }(\mathrm{in})|0,\mathrm{in}\rangle
|^{2}=\delta _{n,n^{\prime }}\frac{1}{1-N_{n}^{\mathrm{cr}}}P_{v}\;,  \notag
\\
&&P(+-|0)_{n^{\prime },n}=|\langle 0,\mathrm{out}|\ ^{+}a_{n^{\prime }}(%
\mathrm{out})\ _{+}b_{n}(\mathrm{out})|0,\mathrm{in}\rangle |^{2}=\delta
_{n,n^{\prime }}\frac{N_{n}^{\mathrm{cr}}}{1-N_{n}^{\mathrm{cr}}}P_{v}\;,
\notag \\
&&P_{v}=|c_{v}|^{2}=\prod\limits_{n}p_{v}^{n},\;\;p_{v}^{n}=\left( 1-N_{n}^{%
\mathrm{cr}}\right) .  \label{cq31}
\end{eqnarray}

The probabilities for a positron scattering $P(-|-)_{n,n^{\prime }}$ and a
pair annihilation $P\left( 0|-+\right) _{n,n^{\prime }}$ coincide with the
expressions $P(+|+)$ and $P(+-|0)$, respectively.

We finish this section with some important remarks. Note that $p_{v}^{n}$
given by Eq. (\ref{cq31}) is the probability that the partial vacuum state
with given $n$ remains a vacuum. One can see that in the framework of
developed QFT quantization the total reflection of a particle off the $x$%
-electric potential step is described by the conditional probability of
reflection of a particle with given quantum numbers $n$, under the condition
that all other partial vacua remain vacua. Such a probability has the form $%
\left\vert w_{n}\left( +|+\right) \right\vert ^{2}p_{v}^{n}$ . Consequently,
$\left\vert w_{n}\left( +|+\right) \right\vert ^{2}p_{v}^{n}=1$. However,
the probability $P(+|+)_{n,n}$ is not equal to one as it follows from Eq. (%
\ref{cq31}). If all $N_{n}^{\mathrm{cr}}\ll 1$\ then%
\begin{equation}
1-P_{v}\approx N=\sum_{n\in \Omega _{3}}N_{n}^{\mathrm{cr}}.  \label{cq32}
\end{equation}%
The vacuum instability is not essential if $N_{n}^{\mathrm{cr}}\rightarrow 0$%
. Then $P_{v}\rightarrow 1$, $P(+|+)_{n,n}\rightarrow 1$ and $%
P(+-|0)_{n,n}\rightarrow N_{n}^{\mathrm{cr}}$.

It should be noted that semiclassical approximations can be used namely
under the latter condition. We recall that, by using the proper-time method,
Schwinger calculated the one-loop effective Lagrangian $L$ in electric field
and assumed that the probability $P_{v}$ $\ $that no actual pair-creation
has occurred in the history of the field during the time $T$ in the volume $V
$ can be presented as $P_{v}=\exp \{-VT2\mathrm{Im}L\}$ \cite{S51} (for a
subsequent development, see the review \cite{Dunn04}). Schwinger interpreted
$2\mathrm{{Im}L}$ as the probability, per time unit, and per volume unit, of
creating a pair by a constant electric field. This interpretation remains
approximately valid as long as the WKB calculation is applicable, that is, $%
VT2\mathrm{{Im}L\ll 1}$. Then the total probability of pair-creation reads
as $1-P_{v}\approx VT2\mathrm{{Im}L\,.}$ To calculate the differential
probabilities of pair-creation with quantum numbers $m$ (for instance,
momentum and spin polarization), one can represent the probability $P_{v}$
as an infinite product:
\begin{equation}
P_{v}=\prod_{m}e^{-2\mathrm{{Im}S_{m}}}\,,  \label{vst1}
\end{equation}%
where a certain discretization scheme is used, so that the effective action $%
S=VTL$ is written as $S=\sum_{m}S_{m}$. The above-said is possible only if $m
$ are selected as integrals of motion. Then, $e^{-2\mathrm{{Im}S_{m}}}$ is
the vacuum-persistence probability in a cell of the space of quantum numbers
$m$. Using the WKB approximation in the case $2\mathrm{{Im}S_{m}\ll 1,}$ one
obtains for the probability of a single pair-production with quantum numbers
$m$ and for the corresponding mean values $N_{m}^{\mathrm{cr}}$ of created
pairs the following relation:
\begin{equation}
N_{m}^{\mathrm{cr}}\approx P(+-|0)_{m,m}\approx 2\mathrm{{Im}S_{m}\,.}
\label{vst2}
\end{equation}

\subsection{Some comments on in- and out- states in the Klein zone\label%
{SS7.6}}

It should be noted that in the works \cite{Nikis79,Nikis70b} of Nikishov he
gave a consistent (and correct to our mind) resolution of the Klein's
paradox based on his original approach, which is a combination of elements
of second quantized theory and relativistic quantum mechanics. In
particular, treating the Klein step, he interpreted solutions $\ ^{\zeta
}\psi _{n}\left( X\right) $ and $\ _{\zeta }\psi _{n}\left( X\right) $ in
the way that correspond to our choice of \textrm{in-} and \textrm{out-}
states. After Nikishov's works there appear a work of Hansen and Ravndal
\cite{HansRavn81} where they tried to use second quantized quantum field
theory to describe quantum effects near $x$-electric potential steps.
However, their interpretation of the solutions $\ ^{\zeta }\psi _{n}\left(
X\right) $ and $\ _{\zeta }\psi _{n}\left( X\right) $ was different from
Nikishov's one. In terms of proposed by us quantization they interpreted our
\textrm{in}-states as \textrm{out}-states and our \textrm{out}-states as
\textrm{in}-states. Nikishov in his work \cite{Nikis04} has pointed out that
this is wrong interpretation. Our detailed analysis confirms his opinion.
Nevertheless the interpretation of Hansen and Ravndal was repeated in the
textbook \cite{GMR85}, in the review \cite{DomCal99}, and in some other
publications. To our mind Hansen and Ravndal came to their interpretation
treating processed in the Klein zone by a misleading analogy with the
one-particle scattering, which really takes place in the ranges $\Omega _{1}$
and $\Omega _{5}$. That is why they believed that in the range $\Omega _{3}$
signs of the quantum numbers $p^{\mathrm{L}}$ and $p^{\mathrm{R}}$ define
signs of particle asymptotic longitudinal kinetic momenta in the regions $S_{%
\mathrm{L}}$ and $S_{\mathrm{R}}$ respectively. If so, then one has to treat
solutions$\ \ _{\zeta }\psi _{n}$ as describing one-electron states and
solutions $\ ^{\zeta }\psi _{n}$ as describing one-positron states, such
that signs of the corresponding quantum numbers $p^{\mathrm{L}}$ and $p^{%
\mathrm{R}}$ label incoming and outgoing particles. However, our analysis of
states in the range $\Omega _{3}$, given in Sec. \ref{S7}, does not confirm
such a correlation between signs of the asymptotical physical longitudinal
energy flux and quantum numbers $p^{\mathrm{L}}$ and $p^{\mathrm{R}}$. We
have demonstrated that the wave functions $\ ^{-}\psi _{n_{3}}\left(
X\right) $ and $\ _{-}\psi _{n_{3}}\left( X\right) $ describe incoming
states of an electron and a positron, respectively, whereas the wave
functions $\ ^{+}\psi _{n_{3}}\left( X\right) $ and $\ _{+}\psi
_{n_{3}}\left( X\right) $ describe outgoing states of an electron and a
positron, respectively. That is why we believe that the
particle-antiparticle and causal identification of wave functions $\ ^{\zeta
}\psi _{n}\left( X\right) $ and $\ _{\zeta }\psi _{n}\left( X\right) $ given
in Ref. \cite{HansRavn81} is erroneous. Accepting this identification, one
makes a mistake in defining what are \textrm{in}- and \textrm{out}-vacua,
and calculates, e.g., the quantity $N_{n}^{b}\left( \mathrm{in}\right) $%
\emph{\ }(\ref{7.5})\emph{\ }instead of the mean number of created pairs $%
N_{n}^{a}\left( \mathrm{out}\right) $ with all the ensuing consequences. The
correct causal interpretation is extremely important in all problems where
one has to use the causal (\textrm{in-out}) propagator. Probably, the wrong
interpretation did not attract for a long time an attention since in
majority works devoted to the pair creation due to $x$-electric potential
steps, the only mean numbers $N_{n}^{b}\left( \mathrm{in}\right) $ (instead
of $N_{n}^{a}\left( \mathrm{out}\right) $) and functions of these numbers
were calculated.\emph{\ }However, in all the considered cases the quantities
$N_{n}^{b}\left( \mathrm{in}\right) $ and $N_{n}^{a}\left( \mathrm{out}%
\right) $ coincide numerically.

\section{Complete QED and massive particle propagators\label{S8}}

Finally, it should be noted that quantization of the Dirac field in the
presence of an $x$-electric potential steps developed in the present article
serves as the base for constructing the Furry picture in the complete QED,
which includes, apart from the matter (Dirac) field, the electromagnetic
field as well. Formally, this Furry picture can be formulated in full
analogy with the case of the $t$-electric potential steps considered in
detail in Refs. \cite{Gitman}. Such a theory allows one to consider all
quantum processes with charged particles moving in external background and
interacting with photons. Processes of zero order with respect to the
radiative interaction in such a theory do not include the interaction with
the photons, their treatment was already presented above in Secs. \ref{S4}-%
\ref{S7}.

In complete QED with an external background any possible process is
described by the following matrix element

\begin{equation*}
\langle 0,\mathrm{out}\left\vert a(\mathrm{out})\cdots b(\mathrm{out})\cdots
c\cdots S\left( \hat{\Psi},\hat{\Psi}^{\dagger },\hat{A}_{\mu }\right)
\cdots c^{\dagger }\cdots b^{\dagger }(\mathrm{in})\cdots a^{\dagger }(%
\mathrm{in})\right\vert 0,\mathrm{in}\rangle ,
\end{equation*}%
where $c$ and $c^{\dag }$ are creation and annihilation operators of photons
and $S\left( \hat{\Psi},\hat{\Psi}^{\dagger },\hat{A}_{\mu }\right) $ is the
$S$-matrix in the interaction representation.

One sees specific terms influenced by vacuum instability when the initial
and final states of charged particles belong to the range $\Omega _{3}$. In
this range, every operator $F$ can be expressed exclusively in terms of
\textrm{in}-annihilating operators and \textrm{out}-creation operators,
using relations (\ref{cq27}), and then divided into two parts,%
\begin{equation*}
F=F^{\left( -\right) }+F^{\left( +\right) },\ F^{\left( -\right) }\mid 0,%
\mathrm{in}\rangle =0,\ \ 0=\langle 0,\mathrm{out}\mid F^{\left( +\right) },
\end{equation*}%
Then one can introduce a generalized normal form $\mathcal{N}_{out-in}\left(
\ldots \right) $ in which all $F^{\left( -\right) }$ are placed to the right
from all $F^{\left( +\right) }.$ Wick's theorem holds with generalized
chronological couplings,%
\begin{eqnarray*}
&&\underbrace{FG}=FG-\mathcal{N}_{out-in}\left( FG\right) =\langle 0,\mathrm{%
out}\left\vert FG\right\vert 0,\mathrm{in}\rangle c_{v}^{-1},\ \  \\
&&\ \overbrace{F\left( X\right) G\left( Y\right) }=\hat{T}F\left( X\right)
G\left( Y\right) -\mathcal{N}_{out-in}\left[ F\left( X\right) G\left(
Y\right) \right] =\langle 0,\mathrm{out}\left\vert \hat{T}F\left( X\right)
G\left( Y\right) \right\vert 0,\mathrm{in}\rangle c_{v}^{-1},
\end{eqnarray*}%
where $\hat{T}$ denotes the chronological ordering operation, see \cite%
{Gitman}.

To reduce any functional of field operators to the generalized normal form
in the range $\Omega _{3}$, we have to represent the operator $\hat{\Psi}%
_{3}\left( X\right) $ (\ref{2.23}) with the help of Eqs.~(\ref{7.3}) and (%
\ref{cq25}) in the following form%
\begin{eqnarray}
\hat{\Psi}_{3}\left( X\right) &=&\sum_{n_{3}}\mathcal{M}_{n_{3}}^{-1/2}\left[
\ ^{-}a_{n_{3}}(\mathrm{in})w_{n_{3}}\left( +|+\right) \ ^{+}\psi
_{n_{3}}\left( X\right) \right.  \notag \\
&&+\left. \ _{+}b_{n_{3}}^{\dagger }(\mathrm{out})w_{n_{3}}\left( -|-\right)
\ _{-}\psi _{n_{3}}\left( X\right) \right] \ ,  \notag \\
\hat{\Psi}_{3}^{\dag }\left( X\right) &=&\sum_{n_{3}}\mathcal{M}%
_{n_{3}}^{-1/2}\left[ \ ^{+}a_{n_{3}}^{\dag }(\mathrm{out})w_{n_{3}}\left(
+|+\right) \ ^{-}\psi _{n_{3}}^{\dag }\left( X\right) \right.  \notag \\
&&+\left. \ _{-}b_{n_{3}}(\mathrm{in})w_{n_{3}}\left( -|-\right) \ _{+}\psi
_{n_{3}}^{\dag }\left( X\right) \right] \ .  \label{8.3}
\end{eqnarray}

Processes of higher orders are described by the Feynman diagrams with two
kinds of charged particle propagators in the external field under
consideration, namely, the so-called \textrm{in-out }propagator $%
S^{c}(X,X^{\prime }),$ which is just the causal Feynman propagator, and the
so-called \textrm{in-in} propagator $S_{\mathrm{in}}^{c}(X,X^{\prime })$,
\begin{eqnarray}
&&S^{c}(X,X^{\prime })=i\langle 0,\mathrm{out}|\hat{T}\hat{\Psi}\left(
X\right) \hat{\Psi}^{\dagger }\left( X^{\prime }\right) \gamma ^{0}|0,%
\mathrm{in}\rangle c_{v}^{-1}\,,  \notag \\
&&S_{\mathrm{in}}^{c}(X,X^{\prime })=i\langle 0,\mathrm{in}|\hat{T}\hat{\Psi}%
\left( X\right) \hat{\Psi}^{\dagger }\left( X^{\prime }\right) \gamma ^{0}|0,%
\mathrm{in}\rangle \,,  \label{8.1}
\end{eqnarray}%
where $\hat{T}$ in Eqs. (\ref{8.1}) denotes the chronological ordering
operation.

Using Eqs. (\ref{2.20}), (\ref{2.23}), (\ref{2.21}), and anticommutation
relations (\ref{2.24}), we find for the \textrm{in-in} propagator:%
\begin{align}
& S_{\mathrm{in}}^{c}(X,X^{\prime })=\theta (t-t^{\prime })S_{\mathrm{in}%
}^{-}(X,X^{\prime })-\theta (t^{\prime }-t)S_{\mathrm{in}}^{+}(X,X^{\prime
})\,,  \notag \\
& S_{\mathrm{in}}^{-}(X,X^{\prime })=i\sum_{j=1}^{2}G_{j}\left( X,X^{\prime
}\right) \gamma ^{0}+\tilde{S}_{\mathrm{in}}^{-}(X,X^{\prime }),  \notag \\
& S_{\mathrm{in}}^{+}(X,X^{\prime })=i\sum_{j=4}^{5}G_{j}\left( X,X^{\prime
}\right) \gamma ^{0}+\tilde{S}_{\mathrm{in}}^{+}(X,X^{\prime }),  \notag \\
& \tilde{S}_{\mathrm{in}}^{-}(X,X^{\prime })=i\sum_{n_{3}}\mathcal{M}%
_{n_{3}}^{-1}\ ^{-}\psi _{n_{3}}\left( X\right) \ ^{-}\bar{\psi}%
_{n_{3}}\left( X^{\prime }\right) ,  \notag \\
& \tilde{S}_{\mathrm{in}}^{+}(X,X^{\prime })=i\sum_{n_{3}}\mathcal{M}%
_{n_{3}}^{-1}\ _{-}\psi _{n_{3}}\left( X\right) \ _{-}\bar{\psi}%
_{n_{3}}\left( X^{\prime }\right) ,  \label{8.2}
\end{align}%
where the functions $G_{j}\left( X,X^{\prime }\right) $ are given by Eq.~(%
\ref{i14}).

Calculation of the \textrm{in-out }propagator can be done in a similar
manner. Then, taking into account Eq.~(\ref{cq27}), we obtain%
\begin{eqnarray}
S^{c}(X,X^{\prime }) &=&\theta (t-t^{\prime })\,S^{-}\left( x,x^{\prime
}\right) -\theta (t^{\prime }-t)\,S^{+}\left( x,x^{\prime }\right) \,,
\notag \\
S^{-}(X,X^{\prime }) &=&i\sum_{j=1}^{2}G_{j}\left( X,X^{\prime }\right)
\gamma ^{0}+\tilde{S}^{-}(X,X^{\prime }),  \notag \\
S^{+}(X,X^{\prime }) &=&i\sum_{j=4}^{5}G_{j}\left( X,X^{\prime }\right)
\gamma ^{0}+\tilde{S}^{+}(X,X^{\prime }),  \notag \\
\tilde{S}^{-}(X,X^{\prime }) &=&i\sum_{n_{3}}\mathcal{M}_{n_{3}}^{-1}\left[
\ ^{+}\psi _{n_{3}}\left( X\right) w_{n_{3}}\left( +|+\right) \ ^{-}\bar{\psi%
}_{n_{3}}\left( X^{\prime }\right) \right] ,  \notag \\
\tilde{S}^{+}(X,X^{\prime }) &=&i\sum_{n_{3}}\mathcal{M}_{n_{3}}^{-1}\left[
\ _{-}\psi _{n_{3}}\left( X\right) w_{n_{3}}\left( -|-\right) \ _{+}\bar{\psi%
}_{n_{3}}\left( X^{\prime }\right) \right] ,  \label{8.4}
\end{eqnarray}

Using relations (\ref{7.3}), we can represent the difference between the
both propagators as follows%
\begin{eqnarray}
&&\ \ S^{p}(X,X^{\prime })=S_{\mathrm{in}}^{c}(X,X^{\prime
})-S^{c}(X,X^{\prime })  \notag \\
&&\ =-i\sum_{n_{3}}\mathcal{M}_{n_{3}}^{-1}\left[ \ _{-}\psi _{n_{3}}\left(
X\right) w_{n_{3}}\left( 0|-+\right) \ ^{-}\bar{\psi}_{n_{3}}\left(
X^{\prime }\right) \right] .  \label{8.6}
\end{eqnarray}%
It is formed in the range $\Omega _{3}$ only and vanishes if there is no
pair creation.

\section{Sauter potential\label{S9}}

\subsection{Scattering, reflection, and pair creation on the Sauter
potential \label{SS9.1}}

Let us consider an $x$-electric potential step in the form of the Sauter
potential \cite{Sauter-pot}. In this case
\begin{eqnarray}
&&A_{0}(x)=-\alpha E\tanh \left( x/\alpha \right) ,\ \ \alpha >0,  \notag \\
&&E(x)=E\cosh ^{-2}\left( x/\alpha \right) ,\ U\left( x\right)
=-eA_{0}(x)=eE\alpha \tanh \left( x/\alpha \right) ,  \label{10.1}
\end{eqnarray}%
see Fig.~\ref{1}, and the asymptotic quantities introduced in Sec. \ref{S2}
are
\begin{equation*}
U_{\mathrm{R}}=-U_{\mathrm{L}}=U\left( +\infty \right) =eE\alpha ,\ \
\mathbb{U=\ }2eE\alpha ,\ \ \,\pi _{0}\left( \mathrm{L}\right)
=p_{0}+eE\alpha ,\ \ \ \ \,\pi _{0}\left( \mathrm{R}\right) =p_{0}-eE\alpha
\,.
\end{equation*}

Solutions (\ref{e7}) of Dirac equation (\ref{e1}) with special asymptotic
behavior at $x\rightarrow \pm \infty $ are expressed via the corresponding
solutions of Eq.~(\ref{e3}). The latter have the form%
\begin{eqnarray}
&&_{\;\zeta }\varphi _{n}\left( x\right) =\;_{\zeta }\mathcal{N}\exp \left(
i\zeta \left\vert p^{\mathrm{L}}\right\vert x\right) \left[ 1+e^{2x/\alpha }%
\right] ^{-i\left( \zeta \left\vert p^{\mathrm{L}}\right\vert +\left\vert p^{%
\mathrm{R}}\right\vert \right) \alpha /2}\ _{\zeta }u\left( x\right) ,
\notag \\
&&\ _{+}u\left( x\right) =F\left( a,b;c;\xi \right) ,\ _{-}u\left( x\right)
=F\left( a+1-c,b+1-c;2-c;\xi \right) ;  \notag \\
&&\ ^{\zeta }\varphi _{n}\left( x\right) =\;^{\zeta }\mathcal{N}\exp \left(
i\left\vert p^{\mathrm{L}}\right\vert x\right) \left[ 1+e^{2x/\alpha }\right]
^{i\left( \zeta \left\vert p^{\mathrm{R}}\right\vert -\left\vert p^{\mathrm{L%
}}\right\vert \right) \alpha /2}\ ^{\zeta }u\left( x\right) ,  \notag \\
&&\ ^{+}u\left( x\right) =F\left( c-a,c-b;c+1-a-b;1-\xi \right) ,  \notag \\
&&\ ^{-}u\left( x\right) =F\left( a,b;a+b+1-c;1-\xi \right) ,  \notag \\
&&a=\frac{i\alpha }{2}\left( \left\vert p^{\mathrm{L}}\right\vert
+\left\vert p^{\mathrm{R}}\right\vert \right) +\frac{1}{2}+\left( \frac{1}{4}%
-\left( eE\alpha ^{2}\right) ^{2}+i\chi eE\alpha ^{2}\right) ^{1/2},  \notag
\\
&&b=\frac{i\alpha }{2}\left( \left\vert p^{\mathrm{L}}\right\vert
+\left\vert p^{\mathrm{R}}\right\vert \right) +\frac{1}{2}-\left( \frac{1}{4}%
-\left( eE\alpha ^{2}\right) ^{2}+i\chi eE\alpha ^{2}\right) ^{1/2},  \notag
\\
&&c=1+i\alpha \left\vert p^{\mathrm{L}}\right\vert ,\ \ \xi =\frac{1}{2}%
\left( 1+\tanh \frac{x}{\alpha }\right) ,  \label{exs2}
\end{eqnarray}%
where $F\left( a,b;c;\xi \right) $ is the hypergeometric series of variable $%
\xi $ with the normalization $F\left( a,b;c;0\right) =1$, \cite{BatE53}. As
was already mentioned in Sec. \ref{S2}, the quantity $\chi $ can be chosen
to be either $\chi =+1$ or $\chi =-1$, and $\ _{\zeta }\mathcal{N}$ and $\
^{\zeta }\mathcal{N}$ are normalization factors given by Eq.~(\ref{e8b}).

A formal transition to the Bose case can be done by setting $\chi =0$\ in
Eqs. (\ref{exs2}). In this case $n=(p_{0},\mathbf{p}_{\bot })$, and $\
_{\zeta }\mathcal{N}$ and $\ ^{\zeta }\mathcal{N}$ are normalization factors
given by Eq.~(\ref{a3b}).

For fermions, using Kummer's relations and Eq.(\ref{c12}), one can find
coefficients $g\left( _{+}\left\vert ^{-}\right. \right) ^{\ast }$ to be%
\begin{equation}
g\left( _{+}\left\vert ^{-}\right. \right) ^{\ast }=-\eta _{\mathrm{R}}\frac{%
\ _{+}C\ \Gamma \left( c\right) \Gamma \left( c-a-b\right) }{\ ^{-}C\ \Gamma
\left( c-a\right) \Gamma \left( c-b\right) },  \label{exs3}
\end{equation}%
where $\ _{+}C$ and $\ ^{-}C$ are constants given by Eq.~(\ref{e8b}). Then%
\begin{equation}
\left\vert g\left( _{+}\left\vert ^{-}\right. \right) \right\vert ^{-2}=%
\frac{\sinh \left( \pi \alpha \left\vert p^{\mathrm{L}}\right\vert \right)
\sinh \left( \pi \alpha \left\vert p^{\mathrm{R}}\right\vert \right) }{%
\left\vert \sinh \left\{ \pi \alpha \left[ eE\alpha +\frac{1}{2}\left(
\left\vert p^{\mathrm{L}}\right\vert -\left\vert p^{\mathrm{R}}\right\vert
\right) \right] \right\} \sinh \left\{ \pi \alpha \left[ eE\alpha -\frac{1}{2%
}\left( \left\vert p^{\mathrm{L}}\right\vert -\left\vert p^{\mathrm{R}%
}\right\vert \right) \right] \right\} \right\vert }\ .  \label{exs4}
\end{equation}

In the similar manner we obtain coefficients $g\left( _{+}\left\vert
^{-}\right. \right) ^{\ast }$ for bosons,%
\begin{equation}
g\left( _{+}\left\vert ^{-}\right. \right) ^{\ast }=-\frac{\ _{+}C\ \Gamma
\left( c\right) \Gamma \left( c-a-b\right) }{\ ^{-}C\ \Gamma \left(
c-a\right) \Gamma \left( c-b\right) }\ ,  \label{exs3b}
\end{equation}%
where $\ _{+}C$ and $\ ^{-}C$ are given by Eqs.~(\ref{a3b}) and parameters $%
a,$ $b,$ and $c$ are given by Eq.~(\ref{exs2}) at $\chi =0$. Then%
\begin{equation}
\left\vert g\left( _{+}\left\vert ^{-}\right. \right) \right\vert ^{-2}=%
\frac{\sinh \left( \pi \alpha \left\vert p^{\mathrm{L}}\right\vert \right)
\sinh \left( \pi \alpha \left\vert p^{\mathrm{R}}\right\vert \right) }{\cosh
^{2}\left[ \pi \sqrt{\left( eE\alpha ^{2}\right) ^{2}-\frac{1}{4}}\right]
+\sinh ^{2}\left[ \frac{\pi \alpha }{2}\left( \left\vert p^{\mathrm{L}%
}\right\vert -\left\vert p^{\mathrm{R}}\right\vert \right) \right] }\ .
\label{exs4b}
\end{equation}

Relations (\ref{exs4}) and (\ref{exs4b}) for $\left\vert g\left(
_{+}\left\vert ^{-}\right. \right) \right\vert ^{-2}$ hold in the ranges $%
\Omega _{1}$, $\Omega _{5}$, and $\Omega _{3}$. However, their
interpretations in the range $\Omega _{3}$ and in the ranges $\Omega _{1}$, $%
\Omega _{5}$ are completely different.

Using Eqs. (\ref{cq15}) and (\ref{cq16}), we find reflection $\left\vert
R_{\zeta ,n}\right\vert ^{2}$ and transmission $\left\vert T_{\zeta
,n}\right\vert ^{2}$ probabilities for electrons in the range $\Omega _{1}$
and for the positrons in the range $\Omega _{5}$, that formally have the
same form in terms of the quantities $\left\vert g\left( _{+}\left\vert
^{-}\right. \right) \right\vert $,\textrm{\ }%
\begin{equation}
\left\vert R_{\zeta ,n}\right\vert ^{2}=1-\left\vert T_{\zeta ,n}\right\vert
^{2},\;\;\left\vert T_{\zeta ,n}\right\vert ^{2}=\left\vert g\left(
_{+}\left\vert ^{+}\right. \right) \right\vert ^{-2}=\left[ 1+\left\vert
g\left( _{+}\left\vert ^{-}\right. \right) \right\vert ^{2}\right] ^{-1}%
\text{.}  \label{g1}
\end{equation}%
However, the latter quantities are given by Eq.~(\ref{exs4}) for fermions
and by Eq. (\ref{exs4b}) for bosons. As was already noted in Sec. \ref{SS5.3}%
, Eqs. (\ref{g1}) imply that $\left\vert T_{\zeta ,n}\right\vert ^{2}\leq 1$.

In the range $\Omega _{3}$, the quantity $N_{n}^{\mathrm{cr}}$ has the form $%
N_{n}^{\mathrm{cr}}=\left\vert g\left( _{+}\left\vert ^{-}\right. \right)
\right\vert ^{-2}$ (\ref{7.7}), where $\left\vert g\left( _{+}\left\vert
^{-}\right. \right) \right\vert ^{-2}$ are given by Eq.~(\ref{exs4}) for
fermions and by Eq. (\ref{exs4b}) for bosons. As was already demonstrated,
in the general case, for fermions one has $N_{n}^{\mathrm{cr}}\leq 1$.

According to our interpretation presented in Sec. \ref{SS7.5}, it is known
that in the range $\Omega _{3}$, similar to the ranges $\Omega _{2}$ and $%
\Omega _{4}$,\ the \textrm{in}-electron or the \textrm{in}-positron are
subjected to the total reflection such that the corresponding transmission
probabilities vanish. In this case, $w_{n}\left( +|+\right) $\ and $%
w_{n}\left( -|-\right) $\ represent relative probability amplitudes of the
reflection. In spite of the fact that the transmission probabilities vanish,
the relative probabilities of the reflection are not equal to unit due to
the vacuum instability. The quantities $w_{n}\left( +-|0\right) $\ and $%
w_{n}\left( 0|-+\right) $\ are relative probability amplitudes of an
electron-positron pair creation and annihilation, respectively.\emph{\ }

Using Eqs. (\ref{cq25}) and (\ref{a42}), we can rewrite these relative
probabilities and the probability for a vacuum to remain a vacuum as follows:%
\begin{eqnarray}
&&\left\vert w_{n}\left( +-|0\right) \right\vert ^{2}=\left\vert g\left(
_{+}\left\vert ^{+}\right. \right) \right\vert ^{-2}=\left[ \left\vert
g\left( _{+}\left\vert ^{-}\right. \right) \right\vert ^{2}-\kappa \right]
^{-1}=N_{n}^{\mathrm{cr}}\left( 1-\kappa N_{n}^{\mathrm{cr}}\right) ^{-1},
\notag \\
&&\left[ p_{v}^{n}\right] ^{-\kappa }=\left\vert w_{n}\left( -|-\right)
\right\vert ^{2}=\left\vert g\left( _{+}\left\vert ^{-}\right. \right)
\right\vert ^{2}\left\vert g\left( _{+}\left\vert ^{+}\right. \right)
\right\vert ^{-2}=\left( 1-\kappa N_{n}^{\mathrm{cr}}\right) ^{-1},  \notag
\\
&&P_{v}=|c_{v}|^{2}=\prod\limits_{n\in \Omega
_{3}}p_{v}^{n}=\prod\limits_{n\in \Omega _{3}}\left( 1-\kappa N_{n}^{\mathrm{%
cr}}\right) ^{\kappa },\ \kappa =\left\{
\begin{array}{c}
+1,\ \mathrm{for\ fermions} \\
-1,\ \mathrm{for\ bosons}%
\end{array}%
\right. .  \label{10.2}
\end{eqnarray}

For the first time these formulas were obtained by Nikishov in the framework
of one-particle relativistic quantum mechanics in Refs. \cite%
{Nikis79,Nikis70b}.

Equations (\ref{exs4}) and (\ref{exs4b}) allow one to verify that for any $%
\pi _{\bot }\neq 0$ one of the following limits holds true:%
\begin{equation}
\left\vert g\left( _{+}\left\vert ^{-}\right. \right) \right\vert ^{-2}\sim
\left\vert \alpha p^{\mathrm{R}}\right\vert \rightarrow 0,\ \ \left\vert
g\left( _{+}\left\vert ^{-}\right. \right) \right\vert ^{-2}\sim \left\vert
\alpha p^{\mathrm{L}}\right\vert \rightarrow 0.  \label{exs5}
\end{equation}%
These limits imply the following properties of the coefficients $\left\vert
g\left( _{+}\left\vert ^{-}\right. \right) \right\vert $ in the case of the
Sauter step:

a) $\left\vert g\left( _{+}\left\vert ^{-}\right. \right) \right\vert
^{-2}\rightarrow 0$ in the range $\Omega _{1}$ if $n$ tends to the boundary
with the range $\Omega _{2}$ ($\left\vert p^{\mathrm{R}}\right\vert
\rightarrow 0$)

b) $\left\vert g\left( _{+}\left\vert ^{-}\right. \right) \right\vert
^{-2}\rightarrow 0$ in the range $\Omega _{5}$ if $n$ tends to the boundary
with the range $\Omega _{4}$ ($\left\vert p^{\mathrm{L}}\right\vert
\rightarrow 0$)

c) $\left\vert g\left( _{+}\left\vert ^{-}\right. \right) \right\vert
^{-2}\rightarrow 0$ in the range $\Omega _{3}$ if $n$ tends to the boundary
with the range $\Omega _{2}$ ($\left\vert p^{\mathrm{R}}\right\vert
\rightarrow 0$)

d) $\left\vert g\left( _{+}\left\vert ^{-}\right. \right) \right\vert
^{-2}\rightarrow 0$ in the range $\Omega _{3}$ if $n$ tends to the boundary
with the range $\Omega _{4}$ ($\left\vert p^{\mathrm{L}}\right\vert
\rightarrow 0$)

Namely these properties are essential for supporting the interpretation
proposed by us in Secs. \ref{S6} and \ref{SS7.4} .

\subsection{Integral quantities\label{SS9.2}}

Usually Sauter potential is used for imitating a slowly alternating electric
field or a small-gradient field. To this end the parameter $\alpha $ is
taken to be sufficiently large. Let us consider just this case, supposing
that%
\begin{equation}
eE\alpha ^{2}\gg 1.  \label{exs6}
\end{equation}

Let us consider the total number $N^{\mathrm{cr}}$ of pairs created from the
vacuum by Sauter potential with a large parameter $\alpha $. This quantity
can be calculated using Eq.~(\ref{TN}) with differential numbers $N_{n}^{%
\mathrm{cr}}$ given by Eq.~(\ref{exs9}) in the Appendix \ref{Auniform}. In
the case under consideration these numbers are the same for fermions and
bosons and do not depend on the spin polarization parameters $\sigma _{s}$.
Thus, for fermions, the probabilities and mean numbers summed over all $%
\sigma _{s}$ are $J_{(d)}=2^{\left[ \frac{d}{2}\right] -1}$ times greater
than the corresponding differential quantities. To get the total number $N^{%
\mathrm{cr}}$ of fermion pairs created in all possible states one has to sum
over the spin projections and then over the momenta $\mathbf{p}_{\bot }$\
and energy $p_{0}$. The latter sum can be easily transformed into an
integral as follows%
\begin{equation}
N^{\mathrm{cr}}=\sum_{\mathbf{p}_{\bot },p_{0}\in \Omega _{3}}\sum_{\sigma
}N_{n}^{\mathrm{cr}}=\frac{V_{\bot }TJ_{(d)}}{(2\pi )^{d-1}}\int_{\Omega
_{3}}dp_{0}d\mathbf{p}_{\bot }N_{n}^{\mathrm{cr}},  \label{exs30}
\end{equation}%
where $V_{\bot }$ is the spatial volume of the ($d-1)$- dimensional
hypersurface orthogonal to the electric field direction and $T$ is the time
duration of the electric field. The total number of boson pairs created in
all possible states follows from Eq. (\ref{exs30}) at $J_{(d)}=1$.

To calculate the integral in the right-hand side of Eq. (\ref{exs30}), we
can find a subrange $D\subset \Omega _{3}$ where this integral is collected.
It is demonstrated in Appendix \ref{Auniform} that the quantity $N_{n}^{%
\mathrm{cr}}$ is almost zero in some areas near the boundary of the range $%
\Omega _{3}$. Such areas are characterized by the conditions
\begin{equation*}
\pi \alpha \left\vert p^{\mathrm{R}}\right\vert <1\ \mathrm{or}\ \ \pi
\alpha \left\vert p^{\mathrm{L}}\right\vert <1.
\end{equation*}%
For areas that are closer to the center of the range $\Omega _{3},$ where
either $1\lesssim \pi \alpha \left\vert p^{\mathrm{R}}\right\vert \lesssim
\pi km\alpha $ or $1\lesssim \pi \alpha \left\vert p^{\mathrm{L}}\right\vert
\lesssim \pi km\alpha $, the quantity $N_{n}^{\mathrm{cr}}$\ satisfies Eqs.~
(\ref{exs17}). Therefore it is almost zero if%
\begin{equation}
k\ll \frac{\pi m\alpha }{2}.  \label{exs31}
\end{equation}

We assume that inequality (\ref{exs31}) holds true together with Eq. (\ref%
{exs12b}), found in the Appendix \ref{Auniform}. Therefore, the main
contribution to integral (\ref{exs30}) is due to the subrange $D\subset
\Omega _{3}$ that is defined by Eqs.~(\ref{exs11b}) and (\ref{exs21}) (see
the Appendix \ref{Auniform}). In this subrange the functions $N_{n}^{\mathrm{%
cr}}$ can be approximated by Eq. (\ref{exs22}), and the integral (\ref{exs30}%
) can be represented as%
\begin{equation}
N^{\mathrm{cr}}\approx \frac{V_{\bot }TJ_{(d)}}{(2\pi )^{d-1}}\int_{\alpha
\pi _{\bot }<K_{\bot }}d\mathbf{p}_{\bot }I_{p_{\bot }},\ \ I_{p_{\bot
}}=2\int_{0}^{eE\alpha -K/\alpha }dp_{0}e^{-\pi \tau },  \label{exs33}
\end{equation}%
where $\tau $\ is given by Eq. (\ref{exs22}). By using a variable $s$,
defined as $\tau =\lambda \left( s^{2}+1\right) ,$ one can represent the
quantity $I_{p_{\bot }}$ as follows%
\begin{equation}
I_{p_{\bot }}=2\int_{0}^{s_{\max }}e^{-\pi \lambda \left( s^{2}+1\right)
}f\left( p_{0}\left( s\right) \right) ds,  \label{exs34}
\end{equation}%
where the number $s_{\max }$ is defined by the relation $\tau _{\max
}=\lambda \left( s_{\max }^{2}+1\right) $ and $\tau _{\max }$ is given by
Eq.~(\ref{exs23.2}). Note that the expansion of $\tau $ in powers of $p_{0}$
has the form
\begin{equation}
\tau =\lambda +\frac{\lambda }{(eE\alpha )^{2}}p_{0}^{2}+\ldots
\label{exs35}
\end{equation}

The leading contribution to integral (\ref{exs34}) is formed at $%
s\rightarrow 0$, or equivalently as $p_{0}/eE\alpha \rightarrow 0$. Using
expansions
\begin{equation*}
p_{0}\left( s\right) =eE\alpha s\left( 1+c_{2}s^{2}+c_{4}s^{4}+\ldots
\right) \Longrightarrow f\left( p_{0}\left( s\right) \right) =eE\alpha
\left( 1+3c_{2}s^{2}+\ldots \right) ,
\end{equation*}%
where finite coefficients $c_{2}$, $c_{4}$, ..., can be found from Eq. (\ref%
{exs35}), we obtain the following{\large \ }asymptotic expressions for the
quantity $I_{p_{\bot }}$,%
\begin{equation}
I_{p_{\bot }}\approx 2eE\alpha \int_{0}^{s_{\max }}e^{-\pi \lambda \left(
s^{2}+1\right) }ds\approx 2eE\alpha \int_{0}^{\infty }e^{-\pi \lambda \left(
s^{2}+1\right) }ds.  \label{exs37}
\end{equation}%
Substituting it into integral (\ref{exs33}) and neglecting exponentially
small contribution from the integration over $\pi _{\bot }>K_{\bot }/\alpha $%
, we find
\begin{equation}
N^{\mathrm{cr}}\approx V_{\bot }Tn^{\mathrm{cr}},\;\;n^{\mathrm{cr}}=\frac{%
J_{(d)}2eE\alpha }{(2\pi )^{d-1}}\int_{0}^{\infty }ds\int d\mathbf{p}_{\bot
}e^{-\pi \lambda \left( s^{2}+1\right) }.  \label{exs38}
\end{equation}

It should be noted that the density $n^{\mathrm{cr}}$ (per the $d-1$ space
volume) of pairs created by $t$-electric potential step given by the
Sauter-type vector potential $A_{1}(x^{0})=\alpha E\tanh \left( t/\alpha
\right) $ with large $\alpha $ $\left( eE\alpha ^{2}\gg \max \left(
1,m^{2}/eE\right) \right) $ is given by the same integral (\ref{exs38}) as
was demonstrated for the first time in our work \cite{GavG96a}.

Finally, performing the integration over $\mathbf{p}_{\bot }$, we obtain
\begin{equation}
n^{\mathrm{cr}}=\frac{J_{(d)}\alpha \delta }{(2\pi )^{d-1}}\left( eE\right)
^{\frac{d}{2}}\exp \left\{ -\pi \frac{m^{2}}{eE}\right\} .  \label{exs39}
\end{equation}%
Here%
\begin{equation*}
\delta =\int_{0}^{\infty }dtt^{-1/2}(t+1)^{-\left( d-2\right) /2}\exp \left(
-t\pi \frac{m^{2}}{eE}\right) =\sqrt{\pi }\Psi \left( \frac{1}{2},-\frac{d-2%
}{2};\pi \frac{m^{2}}{eE}\right) ,
\end{equation*}%
where $\Psi \left( a,b;x\right) $ is the confluent hypergeometric function
\cite{BatE53}, and $J_{(d)}=1$ for bosons.

The vacuum-to-vacuum transition probability $P_{v}$ reads
\begin{eqnarray}
&&P_{v}=\exp \left( -\mu N^{\mathrm{cr}}\right) ,\;\;\mu =\sum_{j=0}^{\infty
}\frac{(-1)^{(1-\kappa )j/2}\epsilon _{j+1}}{(j+1)^{d/2}}\exp \left( -j\pi
\frac{m^{2}}{eE}\right) \;,  \notag \\
&&\epsilon _{j}=\delta ^{-1}\sqrt{\pi }\Psi \left( \frac{1}{2},-\frac{d-2}{2}%
;j\pi \frac{m^{2}}{eE}\right) .  \label{exs40}
\end{eqnarray}

If $eE/m^{2}<<1$, one can use the asymptotic expression for the $\Psi $%
-function \cite{BatE53},%
\begin{equation*}
\Psi \left( 1/2,-\left( d-2\right) /2;j\pi m^{2}/eE\right) =\left( eE/j\pi
m^{2}\right) ^{1/2}+O\left( \left[ eE/m^{2}\right] ^{3/2}\right) .
\end{equation*}%
Then $\delta \approx \sqrt{eE/m^{2}},\;\epsilon _{j}\approx j^{-\frac{1}{2}}$
and $\mu \approx 1$.

In $d=4,$ the formula (\ref{exs40}) reproduces a result obtained in Ref.
\cite{ChK09} for bosons, and a result obtained in Ref. \cite{ChK11} for
fermions.

\subsection{The Klein step\label{SS9.3}}

The Klein paradox was discovered by Klein \cite{Klein27} who calculated,
using the Dirac equation, reflection and transmission probabilities of
charged particles incident on a sufficiently high rectangular potential step
(Klein step) of the form%
\begin{equation}
qA_{0}\left( x\right) =\left\{
\begin{array}{l}
U_{L},\ x<0 \\
U_{R},\ x>0%
\end{array}%
\right. ,  \label{exs41}
\end{equation}%
where $U_{R}$ and $U_{L}$ are constants. According to calculations of Klein
and other authors, for certain energies and sufficient high magnitude $U=U_{%
\mathrm{R}}-U_{\mathrm{L}}$ of the Klein step, there are more reflected
fermions than incident. This is what many articles and books call the Klein
paradox. Let us study quantum processes near the Klein step, applying our
approach to the Sauter potential with $\alpha $ sufficiently small, $\alpha
\rightarrow 0$\emph{.}

The Sauter potential with constant asymptotic potentials, $U_{R}=-U_{L}=%
\mathbb{U}/2=eE\alpha $ and with small $\alpha ,$
\begin{equation}
\mathbb{U}\alpha \ll 1,  \label{exs42}
\end{equation}%
imitates the Klein step (\ref{exs41}) sufficiently well, and coincides with
the latter as $\alpha \rightarrow 0$. Thus, the Sauter potential can be
considered as the regularization of the Klein step.

In the ranges $\Omega _{1}$ and $\Omega _{5}$ the energy $\left\vert
p_{0}\right\vert $ is not restricted from the above, that is why in what
follows we consider only the subranges, where%
\begin{equation}
\max \left\{ \alpha \left\vert p^{\mathrm{L}}\right\vert ,\alpha \left\vert
p^{\mathrm{R}}\right\vert \right\} \ll 1.  \label{exs43}
\end{equation}%
Then in the leading-term approximation in\emph{\ }$\alpha $ it follows from
Eqs.~(\ref{exs4}) and (\ref{exs4b}) that%
\begin{equation}
\left\vert g\left( _{+}\left\vert ^{-}\right. \right) \right\vert
^{-2}\approx \frac{4k}{\left( 1-k\right) ^{2}},\ \ k=\left\{
\begin{array}{l}
k_{f}=k_{b}\frac{\pi _{0}\left( \mathrm{L}\right) +\pi _{\bot }}{\pi
_{0}\left( \mathrm{R}\right) +\pi _{\bot }},\ \mathrm{for}\ \mathrm{fermions}
\\
k_{b}=\frac{\left\vert p^{\mathrm{R}}\right\vert }{\left\vert p^{\mathrm{L}%
}\right\vert },\ \mathrm{for\ bosons}%
\end{array}%
\right. ,  \label{exs45}
\end{equation}%
\ where $k$ is called the kinematic factor.

Note that in the ranges $\Omega _{1}$ and $\Omega _{5}\ $we have that both $%
k_{b}$ and $k_{f}$ are positive and do not achieve the unit values, $%
k_{b}\neq 1,$ $k_{f}\neq 1$.

It should be noted that the quantity $\left\vert g\left( _{+}\left\vert
^{-}\right. \right) \right\vert ^{-2}$ was calculated in Refs. \cite%
{Klein27,Sauter31a,DomCal99,HansRavn81} only at $p_{\bot }=0$. Equation (\ref%
{exs45}) contains these results as a particular case.

In the ranges $\Omega _{1}$ and $\Omega _{5}$, coefficients $g$ satisfy the
same relations for bosons and fermions,%
\begin{equation}
\left\vert g\left( _{+}\left\vert ^{+}\right. \right) \right\vert
^{2}=\left\vert g\left( _{+}\left\vert ^{-}\right. \right) \right\vert
^{2}+1.  \label{exs47}
\end{equation}%
Therefore, reflection and transmission probabilities derived from Eqs~(\ref%
{exs45}) have the same forms, common for bosons and fermions%
\begin{eqnarray}
\left\vert T_{\zeta ,n}\right\vert ^{2} &=&\left\vert g\left( _{+}\left\vert
^{+}\right. \right) \right\vert ^{-2}=\frac{4k}{\left( 1+k\right) ^{2}},
\notag \\
\left\vert R_{\zeta ,n}\right\vert ^{2} &=&\left\vert g\left( _{+}\left\vert
^{-}\right. \right) \right\vert ^{2}\left\vert g\left( _{+}\left\vert
^{+}\right. \right) \right\vert ^{-2}=\frac{\left( 1-k\right) ^{2}}{\left(
1+k\right) ^{2}}.  \label{exs49}
\end{eqnarray}

To compare our exact results with results of the nonrelativistic
consideration obtained in any textbook for one dimensional quantum motion,
we set $p_{\bot }=0,$\ then $\pi _{\bot }=m,$ $\pi _{0}\left( \mathrm{L}%
\right) =p_{0}=m+E$, and $\pi _{0}\left( \mathrm{R}\right) =p_{0}-\mathbb{U}%
=m+E-\mathbb{U}$. In this case
\begin{equation*}
k_{f}=\mu k_{b},\ \mu =\frac{\pi _{0}\left( \mathrm{L}\right) +m}{\pi
_{0}\left( \mathrm{R}\right) +m}=\left[ 1-\mathbb{U}/\left( E+2m\right) %
\right] ^{-1}.
\end{equation*}%
For sufficiently small steps $\mathbb{U}\ll E+2m$, we have $\mu \approx 1+%
\mathbb{U}/\left( E+2m\right) $. In the nonrelativistic limit, when $E\ll m$%
, we obtain%
\begin{equation*}
k_{b}=k_{f}=k^{\mathrm{NR}}=\sqrt{\frac{E-\mathbb{U}}{E}},
\end{equation*}%
which can be identified with nonrelativistic results, e.g., see \cite%
{LanLiQM}. Relativistic corrections have different forms for bosons and
fermions,%
\begin{equation*}
k_{b}\approx k^{\mathrm{NR}}\left( 1-\frac{\mathbb{U}}{4m}\right)
,\;\;k_{f}\approx k^{\mathrm{NR}}\left( 1+\frac{\mathbb{U}}{4m}\right) .
\end{equation*}

Let us consider the range $\Omega _{3}$. Here quantum numbers $\mathbf{p}%
_{\bot }$ are restricted by the inequality$\ 2\pi _{\bot }\leq\mathbb{U}$
and for any of such $\pi _{\bot }$\ quantum numbers $p_{0}$ obey the strong
inequality (\ref{2.48}), see Fig. 3. In this range the quantity $\left\vert
g\left( _{+}\left\vert ^{-}\right. \right) \right\vert ^{-2}$ represents the
differential mean numbers of electron-positron pairs created from the
vacuum, $N_{n}^{\mathrm{cr}}=\left\vert g\left( _{+}\left\vert ^{-}\right.
\right) \right\vert ^{-2}$. In this range for any given $\pi _{\bot }$ the
absolute values of $\left\vert p^{\mathrm{R}}\right\vert $ and $\left\vert
p^{\mathrm{L}}\right\vert $ are restricted from above, see (\ref{d8}).
Therefore, condition (\ref{exs42}) implies Eq. (\ref{exs43}). Then it
follows from Eq.~(\ref{exs4}) that for fermions\ in the leading
approximation the following result holds true%
\begin{equation}
\left\vert g\left( _{+}\left\vert ^{-}\right. \right) \right\vert
^{-2}\approx \frac{4\left\vert p^{\mathrm{L}}\right\vert \left\vert p^{%
\mathrm{R}}\right\vert }{\mathbb{U}^{2}-\left( \left\vert p^{\mathrm{L}%
}\right\vert -\left\vert p^{\mathrm{R}}\right\vert \right) ^{2}}=\frac{%
4\left\vert k_{f}\right\vert }{\left( 1+\left\vert k_{f}\right\vert \right)
^{2}}.  \label{exs50}
\end{equation}

Note that expression (\ref{exs50}) differs from expression (\ref{exs45})
only by the sign of the kinematic factor $k_{f}$. This factor is positive in
$\Omega _{1}$ and $\Omega _{5}$, and it is negative in $\Omega _{3}$. In the
range $\Omega _{3}$, the difference $\left\vert p^{\mathrm{L}}\right\vert
-\left\vert p^{\mathrm{R}}\right\vert $ may be zero at $p_{0}=0$, which
corresponds to $k_{f}=-\left( \mathbb{U}+2\pi _{\bot }\right) /\left(
\mathbb{U}-2\pi _{\bot }\right) $. Namely in this case the quantity $%
\left\vert g\left( _{+}\left\vert ^{-}\right. \right) \right\vert ^{-2}$ has
a maximum at a given $\pi _{\bot }$,%
\begin{equation}
\max \left\vert g\left( _{+}\left\vert ^{-}\right. \right) \right\vert
^{-2}=1-\left( 2\pi _{\bot }/\mathbb{U}\right) ^{2}.  \label{exs50b}
\end{equation}

As it follows from Eq.~(\ref{exs4b}) for bosons in the range $\Omega _{3}$
and\ in the leading approximation, the quantity $\left\vert g\left(
_{+}\left\vert ^{-}\right. \right) \right\vert ^{-2}$reads%
\begin{equation}
\left\vert g\left( _{+}\left\vert ^{-}\right. \right) \right\vert
^{-2}\approx \frac{4\left\vert p^{\mathrm{L}}\right\vert \left\vert p^{%
\mathrm{R}}\right\vert }{\alpha ^{2}\mathbb{U}^{4}/2+\left( \left\vert p^{%
\mathrm{L}}\right\vert -\left\vert p^{\mathrm{R}}\right\vert \right) ^{2}}.
\label{exs51}
\end{equation}%
It has a maximum at $\left\vert p^{\mathrm{L}}\right\vert -\left\vert p^{%
\mathrm{R}}\right\vert =0$,%
\begin{equation}
\max \left\vert g\left( _{+}\left\vert ^{-}\right. \right) \right\vert ^{-2}=%
\frac{2}{\left( \alpha \mathbb{U}\right) ^{2}}\left[ 1-\left( \frac{2\pi
_{\bot }}{\mathbb{U}}\right) ^{2}\right] .  \label{51b}
\end{equation}

However, in contrast with the Fermi case the limit $\alpha \rightarrow 0$ in
(\ref{exs51}) is possible only when the difference $\left\vert p^{\mathrm{L}%
}\right\vert -\left\vert p^{\mathrm{R}}\right\vert $ is not very small,
namely when%
\begin{equation*}
\alpha ^{2}\mathbb{U}^{4}/2\ll \left( \left\vert p^{\mathrm{L}}\right\vert
-\left\vert p^{\mathrm{R}}\right\vert \right) ^{2}.
\end{equation*}%
Only under the latter condition one can neglect an $\alpha $-depending term
in Eq.~(\ref{exs51}) to obtain
\begin{equation}
\left\vert g\left( _{+}\left\vert ^{-}\right. \right) \right\vert
^{-2}\approx \frac{4k_{b}}{\left( 1-k_{b}\right) ^{2}},  \label{exs52}
\end{equation}%
which coincides with Eq. (\ref{exs45}).

The same results for $\left\vert g\left( _{+}\left\vert ^{-}\right. \right)
\right\vert ^{-2}$ in the forms (\ref{exs50}) and (\ref{exs52})\textrm{\ }at
$p_{\bot }=0$ were obtained in Refs. \cite%
{Klein27,Sauter31a,DomCal99,HansRavn81}.

In the range $\Omega _{3}$, relation (\ref{exs47}) still holds for bosons,
whereas for fermions we have
\begin{equation}
\left\vert g\left( _{+}\left\vert ^{+}\right. \right) \right\vert
^{2}=\left\vert g\left( _{+}\left\vert ^{-}\right. \right) \right\vert
^{2}-1.  \label{exs53}
\end{equation}

Relative probability amplitudes of the reflection and of electron-positron
pair creation follows from Eqs.~(\ref{10.2}), (\ref{exs50}) and (\ref{exs52}%
) to be%
\begin{eqnarray}
&&\left\vert w_{n}\left( +-|0\right) \right\vert ^{2}=\left\vert g\left(
_{+}\left\vert ^{+}\right. \right) \right\vert ^{-2}=\frac{4\left\vert
k\right\vert }{\left( 1+k\right) ^{2}},  \notag \\
&&\left\vert w_{n}\left( -|-\right) \right\vert ^{2}=\left\vert g\left(
_{+}\left\vert ^{-}\right. \right) \right\vert ^{2}\left\vert g\left(
_{+}\left\vert ^{+}\right. \right) \right\vert ^{-2}=\frac{\left( 1-k\right)
^{2}}{\left( 1+k\right) ^{2}}.  \label{exs54}
\end{eqnarray}

We see that expressions (\ref{exs54}) for $\left\vert w_{n}\left(
+-|0\right) \right\vert ^{2}$ and $\left\vert w_{n}\left( -|-\right)
\right\vert ^{2}$ are quite similar to the forms of transmission and
reflection probabilities{\Huge \ }given by Eqs.~(\ref{exs49}), respectively.
However, in case of fermions, the range of values of these functions is
quite different because $k_{f}<0$ in Eq.~(\ref{exs54}). This is natural,
since the interpretation of\emph{\ }these quantities in the range\textrm{\ }$%
\Omega _{3}$ differ essentially from their interpretation in the ranges $%
\Omega _{1}$\textrm{\ }and\textrm{\ }$\Omega _{5}$. This formal similarity
was the reason for the systematic misunderstanding in treating quantum
processes in the Klein zone.

\subsection{Klein paradox\label{SS9.4}}

We remind that the Klein paradox dating back to the works of Klein \cite%
{Klein27} and Sauter \cite{Sauter31a,Sauter-pot} (see Sommerfeld \cite%
{Somm60}, as well) is that when considering scattering of relativistic
electrons on a high step potential in the context of the Dirac equation one
comes to a strange result that there is more reflected electrons than
incoming. One of the initial resolutions of the paradox was reduced to the
impossibility that there is no possibility of establishing the appropriate
high-step potential. However, later Hund studied the paradox in connection
with pair production \cite{Hund40} and it seems that Feynman was the first
to point out that the paradox should disappear in a field-theoretical
treatment \cite{Feyn61}. A detailed historical review can be found in Refs.
\cite{DomCal99,HansRavn81}\emph{.} The absence of the Klein paradox in the
framework of appropriate field theoretical interpretation of solutions of
the Dirac and Klein-Gordon equations was first demonstrated by Nikishov in
Refs. \cite{Nikis79,Nikis70b}. In these works Nikishov used a reformulation
of the Green theorem to demonstrate an analogy between the propagation in
time $t$ and in the space coordinate $x$ and thus to identify solutions of
the Dirac equation that describe electrons and positrons. Nikishov had
tested his way of calculation using the special case of a constant and
uniform electric field. In this case, explicit solutions of the Dirac and
Klein-Gordon equations can be found in the constant electric field, which
can be described either by a vector potential with only one nonzero
component $A_{1}(t)=Et$, or by a scalar potential $A_{0}(x)=-Ex$ alone,
only, see details in \cite{Nikis04}. The first case can be treated as a
degenerated $t$-electric potential step and the second case can be treated
as a degenerated $x$-electric potential step. Comparison of exact solutions
in these cases allowed that author to confirm his interpretation for $x$-
potential steps referring to the well developed Feynman interpretation of
the $t$-electric potential step. His calculations give a clear qualitative
explanation of the physics involved in Klein scattering. However complete
consideration of the scattering on arbitrary $x$-electric potential steps in
the frame work of a consistent QFT consideration was not given.

Applying our general approach to particular cases that were studied by
Nikishov, we obtain the same results. In particular, his point of view that
the scattering theory that works within the ranges $\Omega _{1}$ and $\Omega
_{5}$ cannot be applied to the range $\Omega _{3}$ has a clear support in
our general approach.

In the ranges $\Omega _{1}$ and $\Omega _{5}$ transmission and reflection
probabilities for bosons and fermions are expressed via the coefficients $g$
as follows%
\begin{equation}
\left\vert T_{\zeta ,n}\right\vert ^{2}=\left\vert g\left( _{+}\left\vert
^{+}\right. \right) \right\vert ^{-2},\ \ \left\vert R_{\zeta ,n}\right\vert
^{2}=\left\vert g\left( _{+}\left\vert ^{-}\right. \right) \right\vert
^{2}\left\vert g\left( _{+}\left\vert ^{+}\right. \right) \right\vert ^{-2}.
\label{r1}
\end{equation}%
As follows from unitarity relations the sum of these probabilities satisfies
the relation of probability conservation,%
\begin{equation}
\left\vert R_{\zeta ,n}\right\vert ^{2}+\left\vert T_{\zeta ,n}\right\vert
^{2}=1.  \label{r3}
\end{equation}%
In particular, for the Klein step we obtain Eq.~(\ref{exs49})%
\begin{equation}
\left\vert g\left( _{+}\left\vert ^{+}\right. \right) \right\vert ^{-2}=%
\frac{4k}{\left( 1+k\right) ^{2}},\ \ \left\vert g\left( _{+}\left\vert
^{-}\right. \right) \right\vert ^{2}\left\vert g\left( _{+}\left\vert
^{+}\right. \right) \right\vert ^{-2}=\frac{\left( 1-k\right) ^{2}}{\left(
1+k\right) ^{2}},\ \ k=\left\{
\begin{array}{c}
k_{f},\ \mathrm{Fermi\ case} \\
k_{b},\ \mathrm{Bose\ case}%
\end{array}%
\right. .  \label{r2}
\end{equation}

In the range $\Omega _{3},$ we obtain for the Klein step\footnote{%
In Sec. \ref{SS9.3} we have demonstrated that for bosons Eq. (\ref{exs52})
holds true only in a part of $\Omega _{3}$, where the difference $\left\vert
p^{\mathrm{L}}\right\vert -\left\vert p^{\mathrm{R}}\right\vert $ is not
very small.}%
\begin{equation}
\left\vert g\left( _{+}\left\vert ^{+}\right. \right) \right\vert ^{-2}=%
\frac{4\left\vert k\right\vert }{\left( 1+k\right) ^{2}},\ \ \left\vert
g\left( _{+}\left\vert ^{-}\right. \right) \right\vert ^{2}\left\vert
g\left( _{+}\left\vert ^{+}\right. \right) \right\vert ^{-2}=\frac{\left(
1-k\right) ^{2}}{\left( 1+k\right) ^{2}},  \label{r4}
\end{equation}%
where for fermions one has $k=k_{f}<0,$ see Eq.~(\ref{exs50}) and Eq.~(\ref%
{exs52}).

If by analogy with the ranges $\Omega _{1}$ and $\Omega _{5}$ we believe
that $\left\vert g\left( _{+}\left\vert ^{+}\right. \right) \right\vert
^{-2} $ and $\left\vert g\left( _{+}\left\vert ^{-}\right. \right)
\right\vert ^{2}\left\vert g\left( _{+}\left\vert ^{+}\right. \right)
\right\vert ^{-2}$ are transmission and reflection probabilities,
respectively, then we have to accept that there will apparently be more
fermions reflected than coming in. This is the situation first considered by
Klein. Besides, in this case there will apparently be more fermions
transmitted than coming in and the following relation will hold true%
\begin{equation}
\left\vert g\left( _{+}\left\vert ^{-}\right. \right) \right\vert
^{2}\left\vert g\left( _{+}\left\vert ^{+}\right. \right) \right\vert
^{-2}-\left\vert g\left( _{+}\left\vert ^{+}\right. \right) \right\vert
^{-2}=1,  \label{r5}
\end{equation}%
which does not imply Eq. (\ref{r3}). These contradictions do not exist in
the framework of our approach with correct interpretation of the quantities (%
\ref{r4}). Indeed, as follows from Eq. (\ref{cq25}) the quantity $\left\vert
g\left( _{+}\left\vert ^{+}\right. \right) \right\vert ^{-2}$ is the
relative probability of the electron-positron pair creation,%
\begin{equation}
\left\vert g\left( _{+}\left\vert ^{+}\right. \right) \right\vert
^{-2}=\left\vert w_{n}\left( +-|0\right) \right\vert ^{2},  \label{r6}
\end{equation}%
and $\left\vert g\left( _{+}\left\vert ^{-}\right. \right) \right\vert
^{2}\left\vert g\left( _{+}\left\vert ^{+}\right. \right) \right\vert ^{-2}$
is the relative probability of the electron (positron) reflection,%
\begin{equation}
\left\vert g\left( _{+}\left\vert ^{-}\right. \right) \right\vert
^{2}\left\vert g\left( _{+}\left\vert ^{+}\right. \right) \right\vert
^{-2}=\left\vert w_{n}\left( -|-\right) \right\vert ^{2}=\left\vert
w_{n}\left( +|+\right) \right\vert ^{2},  \label{r7}
\end{equation}%
in the range $\Omega _{3}$. Besides, the $x$-electric potential step creates
pairs in the region $\Omega _{3},$ the differential numbers of such pairs
being $N_{n}^{\mathrm{cr}}=\left\vert g\left( _{+}\left\vert ^{-}\right.
\right) \right\vert ^{-2}$ (\ref{7.7}). In this situation the many-particle
nature of the problem is essential, and relations of probability
conservation are quite different in the range $\Omega _{3}$ and in the
ranges $\Omega _{1}$ and $\Omega _{5}$. Unitarity of canonical
transformation between the \textrm{in}- and \textrm{out}- creation and
annihilation operators was proved in the general case in Sec. \ref{SS7.5}.
Here it is enough to mention that for fermions the quantity%
\begin{equation}
p_{v}^{n}=\left\vert w_{n}\left( +|+\right) \right\vert ^{-2}=\left(
1-N_{n}^{cr}\right)  \label{r8}
\end{equation}%
is the probability that the partial vacuum state with a given $n$ remains a
vacuum. Due to the Pauli principle, if an initial state is vacuum, there are
only two possibilities in a cell of the space with given quantum number $n$%
\emph{,} namely, this partial vacuum remains a vacuum, or with the
probability $p_{v}^{n}\left\vert w_{n}\left( +-|0\right) \right\vert ^{2}$ a
pair with the quantum number $n$ will be created. Then Eq.~(\ref{r5}) is
just the condition of probability conservation,%
\begin{equation}
p_{v}^{n}+p_{v}^{n}\left\vert w_{n}\left( +-|0\right) \right\vert ^{2}=1.
\label{r9}
\end{equation}

It is obvious that many particle consideration is necessary for correct
interpretation of relation (\ref{r5}). The same is true when we consider
one-electron initial state with a given $n$. Here again due to the Pauli
principle, creation of a pair of fermions with the same quantum number is
impossible. Then the total reflection of the initial particle on the $x$%
-electric potential step is described by the conditional probability of
reflection under the condition that in all other cells of the space with
quantum numbers $m\neq n$\emph{\ } partial vacua remain vacua. Such a
probability has the form $\left\vert w_{n}\left( +|+\right) \right\vert
^{2}p_{v}^{n}$ . Consequently,%
\begin{equation}
\left\vert w_{n}\left( +|+\right) \right\vert ^{2}p_{v}^{n}=1.  \label{r10}
\end{equation}%
This result is consistent with calculations of mean occupation numbers given
in Sec. \ref{SS7.2}.

For bosons, in the range $\Omega _{3}$, the correct interpretation of
quantities $\left\vert g\left( _{+}\left\vert ^{+}\right. \right)
\right\vert ^{-2}$,

$\left\vert g\left( _{+}\left\vert ^{-}\right. \right) \right\vert
^{2}\left\vert g\left( _{+}\left\vert ^{+}\right. \right) \right\vert ^{-2}$
and $\left\vert g\left( _{+}\left\vert ^{-}\right. \right) \right\vert ^{-2}$
coincides with the one for fermions, they are relative probabilities of a
pair creation and electron/positron reflection respectively. However, for
bosons an additional relation holds,
\begin{equation}
\left\vert g\left( _{+}\left\vert ^{-}\right. \right) \right\vert
^{2}\left\vert g\left( _{+}\left\vert ^{+}\right. \right) \right\vert
^{-2}+\left\vert g\left( _{+}\left\vert ^{+}\right. \right) \right\vert
^{-2}=1.  \label{r11}
\end{equation}%
It formally can be understood as relation (\ref{r3}). However, here equation
(\ref{r11}) is a relation for relative probabilities, and does not admit an
interpretation in the framework of one-particle theory. Due to the vacuum
instability the probability conservation has to be considered in the
framework of many-particle theory, taking into account that in the bosons
case in a given state $n$ any number of pairs can be created. The unitarity
of canonical transformation between the \textrm{in}- and \textrm{out}-
operators was proven in Appendix \ref{Abosons}. Here it should be noted that
for boson the probability that the partial vacuum state with given $n$
remains a vacuum is%
\begin{equation}
p_{v}^{n}=\left\vert w_{n}\left( +|+\right) \right\vert ^{2}=\left(
1+N_{n}^{cr}\right) ^{-1}  \label{r12}
\end{equation}%
The conditional probability of a pair creation with a given quantum numbers $%
n$, under the condition that all other partial vacua with the quantum numbers%
\emph{\ }$m\neq n$ remain the vacua is the sum of probabilities of creation
for any number $l$ of pairs%
\begin{equation}
P(\mathrm{pairs}|0)_{n}=p_{v}^{n}\left[ \sum_{l=1}^{\infty }\left\vert
w_{n}\left( +-|0\right) \right\vert ^{2l}\right] .  \label{r13}
\end{equation}%
In this case Eq.(\ref{r11}) is the probability conservation law in the form
of a sum of probabilities of all possible events in a cell of the space of
quantum numbers $n$:%
\begin{equation}
P(\mathrm{pairs}|0)_{n}+p_{v}^{n}=1.  \label{r14}
\end{equation}

\section{Summary}

When quantizing charged fields (those of Dirac and Klein-Gordon) in the
presence of $x$-electric potential steps, we succeeded to describe quantum
theory of the systems under consideration in terms of adequate \textrm{in}-
and \textrm{out}-particles. These particles represent positive-energy
excitations above the corresponding \textrm{in}- and \textrm{out}-vacua and
have all natural physical properties inherent to such particles in various
examples of known QFT models. In fact, the idea of introducing such
particles is an advancement of the well-known Furry picture in QED with
external magnetic field \cite{Furry51}{\huge \ }and of the generalized Furry
picture in QED with{\huge \ }$t$-electric potential steps \cite{Gitman}. For
the class of external electromagnetic field, which we identify as the $x$%
-electric potential steps, we define special solutions of the relativistic
Dirac and Klein-Gordon wave equations that expand the corresponding
Heisenberg field operators in adequate \textrm{in}- and \textrm{out}%
-creation and annihilation operators related to the \textrm{in}- and \textrm{%
out}-particles. Solutions, which we have used in the implementation of\emph{%
\ }the quantization program, were chosen as stationary solutions with
special asymptotic behavior at the remote left and remote right sides\emph{\
}of the potential step (of course, this choice is not unique). These
solutions,{\LARGE \ }by their asymptotic behavior,{\LARGE \ } are labeled
also by a set of quantum numbers $n$ that include the total energy $p_{0},$
transverse momenta $\mathbf{p}_{\bot },$ and the spin polarization (the
latter in the case of the Dirac field). In the most general case of critical
steps with the potential difference $\mathbb{U}>2m,$ there exist five ranges
of quantum numbers $n$, where these solutions have similar forms, and
physical processes with the corresponding \textrm{in}- and \textrm{out}%
-particles have similar interpretation. A detailed consideration of various
physical processes with these particles had confirmed\emph{\ }justified
their definitions and had demonstrated that:

a) In the first range $p_{0}\geq U_{\mathrm{R}}+\pi _{\bot }$ ($\pi _{\bot }=%
\sqrt{\mathbf{p}_{\bot }^{2}+m^{2}}$)\textrm{\ }there exist only \textrm{in}%
- and \textrm{out}-electrons, whereas in the fifth range $p_{0}\leq U_{%
\mathrm{L}}-\pi _{\bot }$\ there exist only \textrm{in}- and \textrm{out}%
-positrons. In these ranges electrons and positrons are subjected to the
scattering and the reflection only. No particle creation in these ranges is
possible.

b) In the second range $U_{\mathrm{R}}-\pi _{\bot }<p_{0}<U_{\mathrm{R}}+\pi
_{\bot }$ if $2\pi _{\bot }\leq \mathbb{U},$ similar to the first range,
there exist only electrons that are subjected to the total\emph{\ }%
reflection. In the fourth range $U_{\mathrm{L}}-\pi _{\bot }<p_{0}<U_{%
\mathrm{L}}+\pi _{\bot }$ there exist only positrons that are also subjected
to the total\emph{\ }reflection.

c) In the third range $U_{\mathrm{L}}+\pi _{\bot }\leq p_{0}\leq U_{\mathrm{R%
}}-\pi _{\bot }$, which exists only for critical steps and for transversal
momenta that satisfy the inequality{\large \ }$2\pi _{\bot }\leq \mathbb{U}$%
, there exist \textrm{in}- and \textrm{out}-electrons that can be situated
only to the left of the step, and \textrm{in}- and \textrm{out}-positrons
that can be situated only to the right of the step. In the third range, all
the partial vacua are unstable, processes of pair creation are possible. The
pairs consist of \textrm{out}-electrons and \textrm{out}-positrons that
appear on the left and on the right of the step and move there{\large \ }to
the left and to the right, respectively. At the same time, the \textrm{in}%
-electrons that move\emph{\ }to the step from the left are subjected to the
total reflection. After\emph{\ }being reflected they move to the left of the
step already as \textrm{out}-electrons. Similarly, the \textrm{in}-positrons
that move\emph{\ } to the step from the right are subjected to the total
reflection.\ After being reflected they move to the right of the step
already as \textrm{out}-positrons.

We elaborated a technique that allows one to calculate all the above
described processes (zero-order processes) and also to calculate Feynman
diagrams that describe all the processes of interaction between the
introduced \textrm{in}- and \textrm{out}-particles and photons. These
diagrams have formally the usual form, but contain special propagators.
Constructions of these propagators in terms of introduced \textrm{in}- and
\textrm{out}-solutions are presented. It should be noted that calculations
in terms of these Feynman diagrams (as well as calculations of zero-order
processes) are nonperturbative, in such calculations interaction with
external field of $x$-electric potential steps are taken into account
exactly. Another interesting feature is worth noting: when considering
reflection and transmission of \textrm{in}-particles in the first and fifth
ranges the formalism of QFT allows one to calculate both the\emph{\ }%
probability amplitudes of transitions between \textrm{in}- and \textrm{out}%
-states and\emph{\ }the\emph{\ }mean currents of \textrm{out}-particles in
the \textrm{in}-states, testing in such a way the rules of one-particle
time-independent potential scattering theory and its applicability.

Finally, the developed theory is applied to exactly solvable cases of $x$%
-electric potential steps, namely, to the Sauter potential, and to the Klein
step. We present a consistent QFT treatment of processes, where a naive
one-particle consideration might lead to the Klein paradox. From this point
of view we comment various approaches known in the literature that use pure
one-particle consideration or its partial combination with elements of QFT.

\subparagraph{\protect\large Acknowledgement}

S.P.G. thanks FAPESP for support and University of S\~{a}o Paulo for the
hospitality. D.M.G. is grateful to the Brazilian foundations FAPESP and CNPq
for permanent support. The work of S.P.G. and D.M.G. is also partially
supported by the Tomsk State University Competitiveness Improvement Program.
The reported study of S.P.G. and D.M.G. was partially supported by RFBR,
research project No. 15-02-00293a.

\appendix

\section{Some details of scalar field quantization\label{Abosons}}

The quantization of scalar field in terms of adequate in- and out-particles
can be given along the same lines as the quantization for Dirac field.
Having results of Sec. \ref{S3},{\large \ }a quantum scalar field $\hat{\Psi}%
\left( X\right) ${\large \ }can be written as the sum $\hat{\Psi}\left(
X\right) =\sum_{i=1}^{5}\hat{\Psi}_{i}\left( X\right) $ of five operators,
each one $\hat{\Psi}_{i}\left( X\right) $ in the range $\Omega _{i}$. The
operators $\hat{\Psi}_{i}\left( X\right) $, $i=1,2,4,5$. have similar to the
fermionic case [see Eqs.~(\ref{2.23}) and (\ref{2.21})] decompositions in
terms of the creation and annihilation operators and only the operator $\hat{%
\Psi}_{3}\left( X\right) $,%
\begin{eqnarray}
&&\hat{\Psi}_{3}\left( X\right) =\sum_{n_{3}}\mathcal{M}_{n_{3}}^{-1/2}\left[
\ ^{+}a_{n_{3}}(\mathrm{in})\ ^{+}\psi _{n_{3}}\left( X\right) +\
_{+}b_{n_{3}}^{\dagger }(\mathrm{in})\ _{+}\psi _{n_{3}}\left( X\right) %
\right]   \notag \\
&&\ =\sum_{n_{3}}\mathcal{M}_{n_{3}}^{-1/2}\left[ \ ^{-}a_{n_{3}}(\mathrm{out%
})\ ^{-}\psi _{n_{3}}\left( X\right) +\ _{-}b_{n_{3}}^{\dagger }(\mathrm{out}%
)\ _{-}\psi _{n_{3}}\left( X\right) \right] ,  \label{a20}
\end{eqnarray}%
is distinct from the corresponding form in Eqs.~(\ref{2.23}) due to the
{\large \ }$\pm ${\large \ }signs in the right-hand side of Eq. (\ref{a14}).
In what follows, we consider some peculiarities of the quantization of the
scalar field in the range{\large \ }$\Omega _{3}${\large .}

Taking into account relations (\ref{a14}) and Eqs.~(\ref{a16}) - (\ref{a17b}%
), one can see that commutation relations (\ref{a19}) imply the commutation
rules for the introduced creation and annihilation \textrm{in}- and \textrm{%
out-}operators: all creation (annihilation) operators with different quantum
numbers $n$ commute between themselves; all the operators from different
ranges $\Omega _{i}$ commute between themselves, and satisfy the canonical
commutation relations. One defines two vacuum vectors $\left\vert 0,\mathrm{%
in}\right\rangle $ and $\left\vert 0,\mathrm{out}\right\rangle $, using
introduced annihilation operators. In the ranges $\Omega _{i}$,$\ i=1,2,4,5$
the corresponding equations have exactly the same form as in (\ref{cq8a})
and (\ref{cq8b}), whereas in the range $\Omega _{3}$ they are different
since here the \textrm{in}- and \textrm{out}-operators are different,
\begin{eqnarray}
&&\ ^{+}a_{n_{3}}(\mathrm{in})\left\vert 0,\mathrm{in}\right\rangle =\
_{+}b_{n_{3}}(\mathrm{in})\left\vert 0,\mathrm{in}\right\rangle =0,  \notag
\\
&&\ _{-}b_{n_{3}}(\mathrm{out})\left\vert 0,\mathrm{out}\right\rangle =\
^{-}a_{n_{3}}(\mathrm{out)}\left\vert 0,\mathrm{out}\right\rangle =0.
\label{a22}
\end{eqnarray}%
Partial and total vacuum states are related by Eqs.~(\ref{2.29}), (\ref{2.31}%
), (\ref{2.30}), and (\ref{2.34}). The introduced vacua have zero energy and
electric charge and all the excitations above the vacuum have positive
energies.

Canonical transformations between the \textrm{in-} and \textrm{out-}%
operators in the ranges \textrm{\ }$\Omega _{1}$ and $\Omega _{5}$ are
similar to the fermionic case. In particular, amplitudes of electron
reflection and transmission in the range $\Omega _{1}$ have the same form (%
\ref{cq15}) as in the fermionic case. The form of amplitudes of positron
reflection and transmission in the range $\Omega _{5}$ differs from Eqs. (%
\ref{cq15}) only by phases.

Justification of the presented choice of \textrm{in}- and \textrm{out}%
-operators can be done similarly to the fermionic case. In the range $\Omega
_{3}$ there appear some peculiarities, since here we have different
positions of $\pm $\ superscripts and subscripts in comparing to the
fermionic case, nevertheless the{\large \ }interpretation of sets$\ \left\{
\ ^{\zeta }\psi _{n}\left( X\right) \right\} $ as electron states and sets $%
\left\{ \ _{\zeta }\psi _{n}\left( X\right) \right\} $ as positron states is
the same. The canonical transformations between the \textrm{in}- and \textrm{%
out}-operators are%
\begin{eqnarray}
\  &&^{-}a_{n}(\mathrm{out})=g\left( _{+}\left\vert ^{-}\right. \right)
^{-1}\ _{+}b_{n}^{\dagger }(\mathrm{in})+g\left( ^{+}\left\vert _{-}\right.
\right) ^{-1}g\left( ^{-}\left\vert _{-}\right. \right) \ ^{+}a_{n}(\mathrm{%
in}),  \notag \\
&&\ _{-}b_{n}^{\dagger }(\mathrm{out})=g\left( _{+}\left\vert ^{-}\right.
\right) ^{-1}g\left( _{-}\left\vert ^{-}\right. \right) \ _{+}b_{n}^{\dagger
}(\mathrm{in})+g\left( ^{+}\left\vert _{-}\right. \right) ^{-1}\ ^{+}a_{n}(%
\mathrm{in}),  \notag \\
&&\ _{+}b_{n}^{\dagger }(\mathrm{in})=g\left( _{-}\left\vert ^{+}\right.
\right) ^{-1}g\left( _{+}\left\vert ^{+}\right. \right) \ _{-}b_{n}^{\dagger
}(\mathrm{out})-g\left( ^{-}\left\vert _{+}\right. \right) ^{-1}\ ^{-}a_{n}(%
\mathrm{out}),  \notag \\
&&\ ^{+}a_{n}(\mathrm{in})=-g\left( _{-}\left\vert ^{+}\right. \right)
^{-1}\ _{-}b_{n}^{\dagger }(\mathrm{out})+g\left( ^{-}\left\vert _{+}\right.
\right) ^{-1}g\left( ^{+}\left\vert _{+}\right. \right) \ ^{-}a_{n}(\mathrm{%
out}),\ \ n\in \Omega _{3}.  \label{a30}
\end{eqnarray}

Differential mean numbers of \textrm{out-}particles created from the \textrm{%
in-}vacuum are
\begin{eqnarray}
N_{n_{3}}^{a}\left( \mathrm{out}\right) &=&\left\langle 0,\mathrm{in}%
\left\vert \ ^{-}a_{n_{3}}^{\dagger }(\mathrm{out})\ ^{-}a_{n_{3}}(\mathrm{%
out})\right\vert 0,\mathrm{in}\right\rangle =\left\vert g\left(
_{+}\left\vert ^{-}\right. \right) \right\vert ^{-2},  \notag \\
N_{n_{3}}^{b}\left( \mathrm{out}\right) &=&\left\langle 0,\mathrm{in}%
\left\vert \ _{-}b_{n_{3}}^{\dagger }(\mathrm{out})\ _{-}b_{n_{3}}(\mathrm{%
out})\right\vert 0,\mathrm{in}\right\rangle =\left\vert g\left(
_{-}\left\vert ^{+}\right. \right) \right\vert ^{-2}.  \label{a31}
\end{eqnarray}%
It can be shown that they are equal and define differential mean number $%
N_{n_{3}}^{\mathrm{cr}}$ of created pairs similarly to the fermionic case
given by Eq.~(\ref{7.7}). In contrast to the case of fermions, the quantity$%
N_{n}^{\mathrm{cr}}$ is unbounded from above due to relation (\ref{UR1})
with formal setting $\eta _{\mathrm{L}}=\eta _{\mathrm{R}}=1$ for bosons. If
$N_{n}^{\mathrm{cr}}$ tends to zero, $N_{n}^{\mathrm{cr}}\rightarrow 0$,
then $\left\vert g\left( _{+}\left\vert ^{-}\right. \right) \right\vert
^{2}\rightarrow \infty $ and, at the same time, $\left\vert g\left(
_{+}\left\vert ^{+}\right. \right) \right\vert ^{2}\rightarrow \infty $
similar to the fermionic case. The total number $N$ of pairs\ created from
the vacuum is defined similarly to the fermionic case given by Eq.~(\ref{TN}%
).

One can see that due to relation (\ref{UR1}) with{\large \ }$\eta _{\mathrm{L%
}}=\eta _{\mathrm{R}}=1$ the differential mean numbers of \textrm{out-}%
particles in one-particle \textrm{in-}states are%
\begin{eqnarray}
\left\langle 0,\mathrm{in}\left\vert \ ^{+}a_{n}(\mathrm{in})\
^{-}a_{n}^{\dagger }(\mathrm{out})\ ^{-}a_{n}(\mathrm{out})\
^{+}a_{n}^{\dagger }(\mathrm{in})\right\vert 0,\mathrm{in}\right\rangle
&=&1+2N_{n}^{\mathrm{cr}},  \notag \\
\left\langle 0,\mathrm{in}\left\vert \ ^{+}a_{n}(\mathrm{in})\
_{-}b_{n}^{\dagger }(\mathrm{out})\ _{-}b_{n}(\mathrm{out})\
^{+}a_{n}^{\dagger }(\mathrm{in})\right\vert 0,\mathrm{in}\right\rangle
&=&2N_{n}^{\mathrm{cr}},  \notag \\
\left\langle 0,\mathrm{in}\left\vert \ _{+}b_{n}(\mathrm{in})\
_{-}b_{n}^{\dagger }(\mathrm{out})\ _{-}b_{n}(\mathrm{out})\
_{+}b_{n}^{\dagger }(\mathrm{in})\right\vert 0,\mathrm{in}\right\rangle
&=&1+2N_{n}^{\mathrm{cr}},  \notag \\
\left\langle 0,\mathrm{in}\left\vert \ _{+}b_{n}(\mathrm{in})\
^{-}a_{n}^{\dagger }(\mathrm{out})\ ^{-}a_{n}(\mathrm{out})\
_{+}b_{n}^{\dagger }(\mathrm{in})\right\vert 0,\mathrm{in}\right\rangle
&=&2N_{n}^{\mathrm{cr}},\ n\in \Omega _{3},  \label{a33b}
\end{eqnarray}%
and see that Eqs.(\ref{a33b}) are quite different from Eqs.(\ref{7.6})
obtained for fermions. This a consequence of the absence of the Pauli
principle. In this case $2N_{n}^{\mathrm{cr}}$ is the differential mean
number of scalar electrons (positrons) created by the external field. We see
that the presence of a particle at the initial state increases the mean
number of created bosons. It is known effect for bosons, e.g., see \cite%
{GavGT06}.

In the range $\Omega _{3}$, we consider relative scattering amplitudes of
scalar electrons and positrons,%
\begin{eqnarray}
&&w\left( +|+\right) _{n^{\prime }n}=c_{v}^{-1}\langle 0,\mathrm{out}%
\left\vert \ ^{-}a_{n^{\prime }}\left( \mathrm{out}\right) \
^{+}a_{n}^{\dagger }(\mathrm{in})\right\vert 0,\mathrm{in}\rangle ,  \notag
\\
&&w\left( -|-\right) _{n^{\prime }n}=c_{v}^{-1}\langle 0,\mathrm{out}%
\left\vert \ _{-}b_{n^{\prime }}\left( \mathrm{out}\right) \
_{+}b_{n}^{\dagger }(\mathrm{in})\right\vert 0,\mathrm{in}\rangle \,,
\label{a35}
\end{eqnarray}%
and relative amplitudes of a pair creation and a pair annihilation,%
\begin{eqnarray}
&&w\left( +-|0\right) _{n^{\prime }n}=c_{v}^{-1}\langle 0,\mathrm{out}%
\left\vert \ ^{-}a_{n^{\prime }}\left( \mathrm{out}\right) \ _{-}b_{n}\left(
\mathrm{out}\right) \right\vert 0,\mathrm{in}\rangle \,,  \notag \\
&&w\left( 0|-+\right) _{nn^{\prime }}=c_{v}^{-1}\langle 0,\mathrm{out}%
\left\vert \ _{+}b_{n}^{\dagger }(\mathrm{in})\ ^{+}a_{n^{\prime }}^{\dagger
}(\mathrm{in})\right\vert 0,\mathrm{in}\rangle \,,  \label{a36}
\end{eqnarray}%
where $c_{v}$ is the vacuum-to-vacuum transition amplitude for bosons
\begin{equation}
c_{v}^{\left( 3\right) }=\ ^{\left( 3\right) }\langle 0,\mathrm{out}|0,%
\mathrm{in}\rangle ^{(3)}\ =c_{v}=\langle 0,\mathrm{out}|0,\mathrm{in}%
\rangle .  \label{a34}
\end{equation}

As follows from relations (\ref{a30}), all the amplitudes (\ref{a35}) and (%
\ref{a36}) are diagonal in the quantum numbers $n$ and can be expressed in
terms of the coefficients $g\left( ^{\zeta ^{\prime }}\left\vert _{\zeta
}\right. \right) $ as follows:%
\begin{eqnarray}
&&w\left( +|+\right) _{n^{\prime }n}=\delta _{n,n^{\prime }}w_{n}\left(
+|+\right) ,\ \ w_{n}\left( +|+\right) =g\left( ^{-}\left\vert _{+}\right.
\right) g\left( ^{+}\left\vert _{+}\right. \right) ^{-1}=g\left(
_{-}\left\vert ^{+}\right. \right) g\left( _{-}\left\vert ^{-}\right.
\right) ^{-1},  \notag \\
&&w\left( -|-\right) _{nn^{\prime }}=\delta _{n,n^{\prime }}w_{n}\left(
-|-\right) ,\ \ w_{n}\left( -|-\right) =g\left( ^{+}\left\vert _{-}\right.
\right) g\left( ^{+}\left\vert _{+}\right. \right) ^{-1}=g\left(
_{+}\left\vert ^{-}\right. \right) g\left( _{-}\left\vert ^{-}\right.
\right) ^{-1},  \notag \\
&&w\left( 0|-+\right) _{nn^{\prime }}=\delta _{n,n^{\prime }}w_{n}\left(
0|-+\right) ,\ \ w_{n}\left( 0|-+\right) =-g\left( ^{+}\left\vert
_{+}\right. \right) ^{-1},  \notag \\
&&w\left( +-|0\right) _{n^{\prime }n}=\delta _{n,n^{\prime }}w_{n}\left(
+-|0\right) ,\ \ w_{n}\left( +-|0\right) =g\left( _{-}\left\vert ^{-}\right.
\right) ^{-1}\ ,\ \ n\in \Omega _{3}.  \label{a37}
\end{eqnarray}

In the range $\Omega _{3},$ similar to the case of fermions,\ the total
reflection is the only possible form of particle scattering, with $w\left(
+|+\right) _{n}$\ and $w\left( -|-\right) _{n}$ being relative probability
amplitudes of a particle reflection.

Unitary relations (\ref{UR}) and their consequences\ (\ref{UR2}) and (\ref%
{UR1}) with setting $\eta _{\mathrm{L}}=\eta _{\mathrm{R}}=1$ for bosons
imply the following connections for the introduced amplitudes $w$:%
\begin{eqnarray}
&&\left\vert w_{n}\left( +|+\right) \right\vert ^{2}=\left\vert w_{n}\left(
-|-\right) \right\vert ^{2},\;\left\vert w_{n}\left( +-|0\right) \right\vert
^{2}=\left\vert w_{n}\left( 0|-+\right) \right\vert ^{2},  \notag \\
&&\left\vert w_{n}\left( +|+\right) \right\vert ^{2}+\left\vert w_{n}\left(
+-|0\right) \right\vert ^{2}=1,\;\frac{w_{n}\left( -|-\right) ^{\ast }}{%
w_{n}\left( +|+\right) }=-\frac{w_{n}\left( +-|0\right) ^{\ast }}{%
w_{n}\left( 0|-+\right) }.  \label{a38}
\end{eqnarray}

Using Eqs. (\ref{a37}), two lower lines in relations (\ref{a30}) can be
rewritten as%
\begin{eqnarray}
\ ^{+}a_{n}\left( \mathrm{in}\right)  &=&w_{n}\left( +|+\right) ^{-1}\left[
\ ^{-}a_{n}\left( \mathrm{out}\right) -w_{n}\left( +-|0\right) \
_{-}b_{n}^{\dag }\left( \mathrm{out}\right) \right] ,  \notag \\
\ _{+}b_{n}\left( \mathrm{in}\right)  &=&w_{n}\left( -|-\right) ^{-1}\left[
\ _{-}b_{n}\left( \mathrm{out}\right) -w_{n}\left( +-|0\right) \
^{-}a_{n}^{\dag }\left( \mathrm{out}\right) \right] .  \label{a39}
\end{eqnarray}%
Together with their adjoint relations they define an unitary transformation $%
V_{\Omega _{3}}$ between the \textrm{in}- and \textrm{out}-operators,%
\begin{equation}
\left\{ \ ^{+}a^{\dagger }(\mathrm{in}),\ ^{+}a(\mathrm{in}),\
_{+}b^{\dagger }(\mathrm{in}),\ _{+}b(\mathrm{in})\right\} =V_{\Omega
_{3}}\left\{ \ ^{-}a^{\dagger }\left( \mathrm{out}\right) ,\ ^{-}a\left(
\mathrm{out}\right) ,\ _{-}b^{\dagger }\left( \mathrm{out}\right) ,\
_{-}b\left( \mathrm{out}\right) \right\} V_{\Omega _{3}}^{\dagger }\ .
\notag
\end{equation}%
Since Eqs~(\ref{a38}) and (\ref{a39}) formally coincide with the
corresponding equations for the general case of time-dependent external
field, the unitary operator $V_{\Omega _{3}}$ can be taken, for example,
from \cite{GavGT06} or book \cite{Gitman}. The operator $V_{\Omega _{3}}$
relates the \textrm{in}- and \textrm{out}-vacua, $|0,\mathrm{in}\rangle
=V_{\Omega _{3}}|0,\mathrm{out}\rangle $ and determines the vacuum-to-vacuum
transition amplitude,%
\begin{equation}
c_{v}=\langle 0,\mathrm{out}|V_{\Omega _{3}}|0,\mathrm{out}\rangle
=\prod\limits_{n}w_{n}\left( -|-\right) \,.  \label{a41}
\end{equation}

The probabilities of a particle reflection, a pair creation, and the
probability for a vacuum to remain a vacuum can be expressed via
differential mean numbers of created pairs $N_{n}^{\mathrm{cr}}$. By using
the relation $\left\vert w_{n}\left( -|-\right) \right\vert ^{2}=\left(
1+N_{n}^{\mathrm{cr}}\right) ^{-1}$, one finds
\begin{eqnarray}
&&P(+|+)_{n,n^{\prime }}=|\langle 0,\mathrm{out}|\ ^{+}a_{n}(\mathrm{out})\
^{-}a_{n^{\prime }}^{\dagger }(\mathrm{in})|0,\mathrm{in}\rangle
|^{2}=\delta _{n,n^{\prime }}\left( 1+N_{n}^{\mathrm{cr}}\right)
^{-1}P_{v}\;,  \notag \\
&&P(+-|0)_{n,n^{\prime }}=|\langle 0,\mathrm{out}|\ ^{+}a_{n}(\mathrm{out})\
_{+}b_{n^{\prime }}(\mathrm{out})|0,\mathrm{in}\rangle |^{2}=\delta
_{n,n^{\prime }}N_{n}^{\mathrm{cr}}\left( 1+N_{n}^{\mathrm{cr}}\right)
^{-1}P_{v}\;,  \notag \\
&&P_{v}=|c_{v}|^{2}=\prod\limits_{n\in \Omega
_{3}}p_{v}^{n},\;\;p_{v}^{n}=\left( 1+N_{n}^{\mathrm{cr}}\right) ^{-1}.
\label{a42}
\end{eqnarray}%
The probabilities for a positron scattering $P(-|-)$ and a pair annihilation
$P(0|-+)$ coincide with the expressions $P(+|+)$ and $P(+-|0)$, respectively.

Note that $p_{v}^{n}$ given by Eq. (\ref{a42}) is the probability that the
partial vacuum state with given $n$ remains a vacuum. If all $N_{n}^{\mathrm{%
cr}}\ll 1$ then in the leading approximation the relation $1-P_{v}\approx N$
is the same with the case of fermions. The vacuum instability is not
essential if $N_{n}^{\mathrm{cr}}\rightarrow 0$. Then like for fermions $%
P_{v}\rightarrow 1$, $P(+|+)_{n,n}\rightarrow 1$ and $P(+-|0)_{n,n}%
\rightarrow N_{n}^{\mathrm{cr}}$.

Processes of higher orders are described by the Feynman diagrams with two
kinds of charged scalar particle propagators in the external field under
consideration, namely, the so-called \textrm{in-out }propagator $\Delta
^{c}(X,X^{\prime }),$ which is just the causal Feynman propagator, and the
so-called \textrm{in-in} propagator $\Delta _{\mathrm{in}}^{c}(X,X^{\prime
}) $,%
\begin{eqnarray}
&&\Delta ^{c}(X,X^{\prime })=i\langle 0,\mathrm{out}|\hat{T}\hat{\Psi}\left(
X\right) \hat{\Psi}^{\dagger }\left( X^{\prime }\right) |0,\mathrm{in}%
\rangle c_{v}^{-1}\,,  \notag \\
&&\Delta _{\mathrm{in}}^{c}(X,X^{\prime })=i\langle 0,\mathrm{in}|\hat{T}%
\hat{\Psi}\left( X\right) \hat{\Psi}^{\dagger }\left( X^{\prime }\right) |0,%
\mathrm{in}\rangle \,.  \label{a43}
\end{eqnarray}%
We find the \textrm{in-in} propagator as%
\begin{align}
& \Delta _{\mathrm{in}}^{c}(X,X^{\prime })=\theta (t-t^{\prime })\Delta _{%
\mathrm{in}}^{-}(X,X^{\prime })-\theta (t^{\prime }-t)\Delta _{\mathrm{in}%
}^{+}(X,X^{\prime })\,,  \notag \\
& \Delta _{\mathrm{in}}^{-}(X,X^{\prime })=i\sum_{j=1}^{2}G_{j}\left(
X,X^{\prime }\right) +\tilde{\Delta}_{\mathrm{in}}^{-}(X,X^{\prime }),
\notag \\
& \Delta _{\mathrm{in}}^{+}(X,X^{\prime })=-i\sum_{j=4}^{5}G_{j}\left(
X,X^{\prime }\right) +\tilde{\Delta}_{\mathrm{in}}^{+}(X,X^{\prime }),
\notag \\
& \tilde{\Delta}_{\mathrm{in}}^{-}(X,X^{\prime })=i\sum_{n_{3}}\mathcal{M}%
_{n_{3}}^{-1}\ ^{+}\psi _{n_{3}}\left( X\right) \ ^{+}\psi _{n_{3}}^{\ast
}\left( X^{\prime }\right) ,  \notag \\
& \tilde{\Delta}_{\mathrm{in}}^{+}(X,X^{\prime })=-i\sum_{n_{3}}\mathcal{M}%
_{n_{3}}^{-1}\ _{+}\psi _{n_{3}}\left( X\right) \ _{+}\psi _{n_{3}}^{\ast
}\left( X^{\prime }\right) ,  \label{a44}
\end{align}%
where the functions $G_{j}\left( X,X^{\prime }\right) $ are given by Eq.~(%
\ref{i14}), and obtain for the \textrm{in-out }\ propagator that%
\begin{align}
& \Delta ^{c}(X,X^{\prime })=\theta (t-t^{\prime })\Delta ^{c}(X,X^{\prime
})-\theta (t^{\prime }-t)\Delta ^{+}(X,X^{\prime })\,,  \notag \\
& \Delta ^{-}(X,X^{\prime })=i\sum_{j=1}^{2}G_{j}\left( X,X^{\prime }\right)
+\tilde{\Delta}^{-}(X,X^{\prime }),  \notag \\
& \Delta ^{+}(X,X^{\prime })=-i\sum_{j=4}^{5}G_{j}\left( X,X^{\prime
}\right) +\tilde{\Delta}^{+}(X,X^{\prime }),  \notag \\
& \tilde{\Delta}^{-}(X,X^{\prime })=i\sum_{n_{3}}\mathcal{M}_{n_{3}}^{-1}%
\left[ \ ^{-}\psi _{n_{3}}\left( X\right) w_{n_{3}}\left( +|+\right) \
^{+}\psi _{n_{3}}^{\ast }\left( X^{\prime }\right) \right] ,  \notag \\
& \tilde{\Delta}^{+}(X,X^{\prime })=-i\sum_{n_{3}}\mathcal{M}_{n_{3}}^{-1}%
\left[ \ _{+}\psi _{n_{3}}\left( X\right) w_{n_{3}}\left( -|-\right) \
_{-}\psi _{n_{3}}^{\ast }\left( X^{\prime }\right) \right] .  \label{a45}
\end{align}

Using relations (\ref{rel1}) with{\large \ }$\eta _{\mathrm{L}}=\eta _{%
\mathrm{R}}=1$ for bosons, we can represent the difference between the both
propagators as follows%
\begin{eqnarray}
&&\ \ \Delta ^{p}(X,X^{\prime })=\Delta _{\mathrm{in}}^{c}(X,X^{\prime
})-\Delta ^{c}(X,X^{\prime })  \notag \\
&&\ =-i\sum_{n_{3}}\mathcal{M}_{n_{3}}^{-1}\left[ \ _{+}\psi _{n_{3}}\left(
X\right) w_{n_{3}}\left( 0|-+\right) \ ^{+}\psi _{n_{3}}^{\ast }\left(
X^{\prime }\right) \right] .  \label{a46}
\end{eqnarray}%
It is formed in the range $\Omega _{3}$ only and vanishes if there is no
pair creation.

\section{Orthogonality and normalization on $t$-constant hyperplane\label%
{t-const}}

Integrating in (\ref{t4}) over the coordinates $\mathbf{r}_{\bot }$ and
using the structure of constant spinors $v_{\sigma }$ that enter the states $%
\psi _{n}\left( X\right) $ and $\psi _{n^{\prime }}^{\prime }\left( X\right)
,$ we obtain:%
\begin{eqnarray}
&&\left( \psi _{n},\psi _{n^{\prime }}^{\prime }\right) =\delta _{\sigma
,\sigma ^{\prime }}\delta _{\mathbf{p}_{\bot },\mathbf{p}_{\bot }^{\prime
}}V_{\bot }\mathcal{R},\ \ \mathcal{R}=\int_{-K^{\left( \mathrm{L}\right)
}}^{K^{\left( \mathrm{R}\right) }}\Theta dx\ ,  \notag \\
&&\Theta =e^{i\left( p_{0}-p_{0}^{\prime }\right) t}\varphi _{n}^{\ast
}\left( x\right) \left[ p_{0}+p_{0}^{\prime }-2U\left( x\right) \right] %
\left[ p_{0}^{\prime }-U\left( x\right) +\chi i\partial _{x}\right] \varphi
_{n^{\prime }}^{\prime }\left( x\right) .  \label{i1}
\end{eqnarray}%
Then we represent the integral $\mathcal{R}$ as follows%
\begin{equation}
\mathcal{R}=\int_{-K^{\left( \mathrm{L}\right) }}^{x_{\mathrm{L}}}\Theta
dx+\int_{x_{\mathrm{L}}}^{x_{\mathrm{R}}}\Theta dx+\int_{x_{\mathrm{R}%
}}^{K^{\left( \mathrm{R}\right) }}\Theta dx\ .  \label{i2}
\end{equation}%
Due to our suppositions about the structure of the scalar potential $%
A_{0}\left( x\right) $ only the second terms in the right-hand side of Eq. (%
\ref{i2}) depends on the external field. At the same time, the smoothness of
the scalar potential allows us to believe that this integral is finite. The
first and the third terms are calculated as integrals over the areas where
the electric field is zero but the scalar potentials, as being constant,
differ from zero. Thus, functions $\varphi \left( x\right) $ and $\varphi
^{\prime }\left( x\right) $ entering the quantity $\Theta $ (\ref{i1}) are
different for the left and the right areas even for equal quantum numbers $n$
and $n^{\prime }$.

First, we evaluate the quantity $\mathcal{R}$ for coinciding quantum numbers
$n$ and $n^{\prime },$ and then we calculate the norm squared\textrm{\ }$%
\left( \psi _{n},\psi _{n}\right) $ of the introduced solutions for any $n.$
In this case Eq. (\ref{i2}) reads:%
\begin{eqnarray}
&&\left. \mathcal{R}\right\vert _{n=n^{\prime }}=\mathcal{R}_{\mathrm{L}}+%
\mathcal{R}_{\mathrm{int}}+\mathcal{R}_{\mathrm{R}},\ \   \notag \\
&&\ \mathcal{R}_{\mathrm{L}}=\int_{-K^{\left( \mathrm{L}\right) }}^{x_{%
\mathrm{L}}}\Theta _{\mathrm{L}}dx,\ \ \mathcal{R}_{\mathrm{int}}=\int_{x_{%
\mathrm{L}}}^{x_{\mathrm{R}}}\left. \Theta \right\vert _{n=n^{\prime
}}dx<\infty ,\ \ \mathcal{R}_{\mathrm{R}}=\int_{x_{\mathrm{R}}}^{K^{\left(
\mathrm{R}\right) }}\Theta _{\mathrm{R}}dx,  \notag \\
&&\Theta _{\mathrm{L}/\mathrm{R}}=\varphi _{n}^{\ast }\left( x\right) 2\pi
_{0}\left( \mathrm{L}/\mathrm{R}\right) \left[ \pi _{0}\left( \mathrm{L}/%
\mathrm{R}\right) +\chi i\partial _{x}\right] \varphi _{n}^{\prime }\left(
x\right) .  \label{i3}
\end{eqnarray}

If we consider only solutions from the sets $\left\{ \ _{\zeta }\psi
_{n}\left( X\right) \right\} $ and $\left\{ \ ^{\zeta }\psi _{n}\left(
X\right) \right\} ,$ then the first line in Eqs. (\ref{i3}) looks different
in the ranges $n\in \Omega _{1}\cup \Omega _{3}\cup \Omega _{5},$ $n\in
\Omega _{2},$ and $n\in \Omega _{4}.$ Namely,
\begin{eqnarray}
\left. \mathcal{R}\right\vert _{n=n^{\prime }} &=&\mathcal{R}_{\mathrm{L}}+%
\mathcal{R}_{\mathrm{R}}+O\left( 1\right) ,\ ,\ n\in \Omega _{1}\cup \Omega
_{3}\cup \Omega _{5},  \notag \\
\left. \mathcal{R}\right\vert _{n=n^{\prime }} &=&\mathcal{R}_{\mathrm{L}%
}+O\left( 1\right) ,\ \ n\in \Omega _{2},  \notag \\
\left. \mathcal{R}\right\vert _{n=n^{\prime }} &=&\mathcal{R}_{\mathrm{R}%
}+O\left( 1\right) ,\ \ n\in \Omega _{4},  \label{i3b}
\end{eqnarray}%
where the designation $O\left( 1\right) $ is used here and in what follows
for the terms that satisfy the relation%
\begin{equation*}
\lim_{K^{\left( \mathrm{L}/\mathrm{R}\right) }\rightarrow \infty }\frac{%
O\left( 1\right) }{K^{\left( \mathrm{L}/\mathrm{R}\right) }}=0.
\end{equation*}

Consider the quantities $\mathcal{R}_{\mathrm{L}/\mathrm{R}}$ (\ref{i3})
defined by the functions $_{\zeta }\varphi _{n}\left( x\right) $ and $%
^{\zeta }\varphi _{n}\left( x\right) $. In this case we attribute the
corresponding index $\zeta $ to these quantities as follows: $\mathcal{R}_{%
\mathrm{L}/\mathrm{R}}\rightarrow \ _{\zeta }\mathcal{R}_{\mathrm{L}/\mathrm{%
R}}$ or $\mathcal{R}_{\mathrm{L}/\mathrm{R}}\rightarrow \ ^{\zeta }\mathcal{R%
}_{\mathrm{L}/\mathrm{R}}.$ Using Eqs. (\ref{2.62a}), (\ref{2.6}), and (\ref%
{e8b}), we obtain%
\begin{equation}
\ _{\zeta }\mathcal{R}_{\mathrm{L}}=Y^{2}K^{\left( \mathrm{L}\right)
}\left\vert \frac{\pi _{0}\left( \mathrm{L}\right) }{p^{\mathrm{L}}}%
\right\vert +O\left( 1\right) ,\;\;^{\zeta }\mathcal{R}_{\mathrm{R}%
}=Y^{2}K^{\left( \mathrm{R}\right) }\left\vert \frac{\pi _{0}\left( \mathrm{R%
}\right) }{p^{\mathrm{R}}}\right\vert +O\left( 1\right) .  \label{i10}
\end{equation}

This result allows one to find the square norm of the states with $n\in
\Omega _{2}\cup \Omega _{4},$%
\begin{eqnarray}
&&\left( \psi _{n},\psi _{n}\right) =\mathcal{M}_{n},\;n\in \Omega _{2}\cup
\Omega _{4}\ ;  \notag \\
&&\mathcal{M}_{n_{2}}=2\frac{K^{\left( \mathrm{L}\right) }}{T}\left\vert
\frac{\pi _{0}\left( \mathrm{L}\right) }{p^{\mathrm{L}}}\right\vert +O\left(
1\right) ,\mathrm{\;}\mathcal{M}_{n_{4}}=2\frac{K^{\left( \mathrm{R}\right) }%
}{T}\left\vert \frac{\pi _{0}\left( \mathrm{R}\right) }{p^{\mathrm{R}}}%
\right\vert +O\left( 1\right) .  \label{i10a}
\end{eqnarray}

To calculate the quantities $_{\zeta }\mathcal{R}_{\mathrm{R}}$ and $^{\zeta
}\mathcal{R}_{\mathrm{L}}$ that correspond to functions $\varphi _{n}\left(
x\right) $ with$\ n\in \Omega _{1}\cup \Omega _{3}\cup \Omega _{5}$ we have
to use relations between the functions $_{\zeta }\varphi _{n}\left( x\right)
$ and $^{\;\zeta }\varphi _{n}\left( x\right) $. It follows from Eq. (\ref%
{i1}) that the matrix elements $\left( \psi _{n},\psi _{n^{\prime }}^{\prime
}\right) $ are diagonal in quantum numbers $\sigma .$ Using this fact, one
can easily see that relations (\ref{rel1}) remain valid under changing the
functions $\ _{\zeta }\psi _{n_{i}}\left( X\right) $ and $\ ^{\zeta ^{\prime
}}\psi _{n_{i}}\left( X\right) $ to $\ _{\zeta }\varphi _{n_{i}}\left(
x\right) $ and $\ ^{\zeta ^{\prime }}\varphi _{n_{i}}\left( x\right) $.
Using these relations, and taking into account eqs. (\ref{2.62a}), (\ref{2.6}%
), and (\ref{e8b}), we find
\begin{eqnarray}
_{\zeta }\mathcal{R}_{\mathrm{R}} &=&Y^{2}K^{\left( \mathrm{R}\right)
}\left\vert \frac{\pi _{0}\left( \mathrm{R}\right) }{p^{\mathrm{R}}}%
\right\vert \left[ \left\vert g\left( _{\zeta }\left\vert ^{+}\right.
\right) \right\vert ^{2}+\left\vert g\left( _{\zeta }\left\vert ^{-}\right.
\right) \right\vert ^{2}\right] +O\left( 1\right) ,  \notag \\
^{\zeta }\mathcal{R}_{\mathrm{L}} &=&Y^{2}K^{\left( \mathrm{L}\right)
}\left\vert \frac{\pi _{0}\left( \mathrm{L}\right) }{p^{\mathrm{L}}}%
\right\vert \left[ \left\vert g\left( _{+}\left\vert ^{\zeta }\right.
\right) \right\vert ^{2}+\left\vert g\left( _{-}\left\vert ^{\zeta }\right.
\right) \right\vert ^{2}\right] +O\left( 1\right) .  \label{i11}
\end{eqnarray}%
These results allow us to find square norms of states with $n\in \Omega
_{1}\cup \Omega _{3}\cup \Omega _{5}$. They are%
\begin{equation}
\left( _{\zeta }\psi _{n},_{\zeta }\psi _{n}\right) =\ _{\zeta }\mathcal{R}_{%
\mathrm{L}}+\ _{\zeta }\mathcal{R}_{\mathrm{R}},\;\;\left( \ ^{\zeta }\psi
_{n},\ ^{\zeta }\psi _{n}\right) =\ ^{\zeta }\mathcal{R}_{\mathrm{L}}+\
^{\zeta }\mathcal{R}_{\mathrm{R}}\ .  \label{i11b}
\end{equation}%
Note that these square norms are of the order of the large numbers $%
K^{\left( \mathrm{L}\right) }$\textrm{\ }and/or\textrm{\ }$K^{\left( \mathrm{%
R}\right) }$.

In the case $n\neq n^{\prime },$ we already know that $\left( \psi _{n},\psi
_{n^{\prime }}^{\prime }\right) \sim \delta _{\sigma ,\sigma ^{\prime
}}\delta _{\mathbf{p}_{\bot },\mathbf{p}_{\bot }^{\prime }}.$ Thus, it is
enough to study the quantity $\left. \mathcal{R}\right\vert _{\sigma =\sigma
^{\prime },\mathbf{p}_{\bot }=\mathbf{p}_{\bot }^{\prime },p_{0}\neq
p_{0}^{\prime }}$ in order\ to make up\ a conclusion about the complete
inner product (\ref{i1}) for $n\neq n^{\prime }$. Let us consider solutions $%
\psi _{n}\left( X\right) $ and $\psi _{n^{\prime }}^{\prime }\left( X\right)
$ with a given asymptotic behavior and for $\sigma =\sigma ^{\prime },%
\mathbf{p}_{\bot }=\mathbf{p}_{\bot }^{\prime },p_{0}\neq p_{0}^{\prime }$.
In this case{\Huge \ }$p_{0}\neq p_{0}^{\prime }$ implies $p^{\mathrm{L}%
}\neq p^{\mathrm{L}\prime }$\ and/or $p^{\mathrm{R}}\neq p^{\mathrm{R}\prime
}$.\emph{\ }That is why the quantities $\Theta $\ are oscillating functions
of $x$ in the both regions $S_{\mathrm{L}}$\emph{\ }and $S_{\mathrm{R}}$,
and the modulus of the quantity\emph{\ }$\left. \mathcal{R}\right\vert
_{\sigma =\sigma ^{\prime },\mathbf{p}_{\bot }=\mathbf{p}_{\bot }^{\prime
},p_{0}\neq p_{0}^{\prime }}$ is finite. Then for any $n,n^{\prime }\in
\Omega $ we have%
\begin{equation}
\left( \psi _{n},\psi _{n^{\prime }}^{\prime }\right) =O\left( 1\right) ,%
\mathrm{\;}n\neq n^{\prime }.  \label{i14b}
\end{equation}%
One can easily verify that the relation%
\begin{equation}
\left( \psi _{n},\hat{H}\psi _{n^{\prime }}^{\prime }\right) -\left( \hat{H}%
\psi _{n},\psi _{n^{\prime }}^{\prime }\right) =O\left( 1\right)
\label{i14c}
\end{equation}%
holds true for any stationary states. Thus, the Hamiltonian $\hat{H}$ is
Hermitian as $K^{\left( \mathrm{L}/\mathrm{R}\right) }\rightarrow \infty $.

In what follows, such matrix elements always appear divided by terms
proportional $K^{\left( \mathrm{L}/\mathrm{R}\right) }$ such that they can
be neglected in the limits $K^{\left( \mathrm{L}/\mathrm{R}\right)
}\rightarrow \infty $. Thus, we further assume that all the wave functions
(described in subsections \ref{SS3.1} and \ref{SS3.2}) having different
quantum numbers $n$ are orthogonal with respect to the introduced inner
product on the hyperplane $t=\mathrm{const}$,
\begin{equation}
\left( \psi _{n},\psi _{n^{\prime }}^{\prime }\right) =0,\mathrm{\;}n\neq
n^{\prime }.  \label{i4}
\end{equation}%
Then the complete orthonormality relations (\ref{i12}) for the ranges $%
\Omega _{2}$ and $\Omega _{4}$ follow from Eqs.~(\ref{i10a}) and (\ref{i4}).

Orthonormality relations for solutions with quantum numbers\emph{\ }$n\in
\Omega _{1}\cup \Omega _{3}\cup \Omega _{5}$ are considered below, they are
more complicated since such solutions have an additional quantum number $%
\zeta $.

There always exist two independent solutions with quantum numbers $n\in
\Omega _{1}\cup \Omega _{3}\cup \Omega _{5}$. In spite of the fact that
these solutions are obtained in the constant external field, we believe that
they represent asymptotic forms of some unknown solutions of the Dirac
equation with the external field that is switched on and off at $%
t\rightarrow \pm \infty $ and that effects of the switching on and off are
negligible. We believe that there exist orthogonal pairs of solutions
describing independent particle states at the initial and final time
instants, and [since the inner product (\ref{t4}) does not depend on $t$ in
the limits $K^{\left( \mathrm{L}/\mathrm{R}\right) }\rightarrow \infty $]
that \ such solutions remain orthogonal at arbitrary time instant. Below, we
are going to find out which solutions under consideration form such
orthogonal pairs.

Let us consider the inner products $\left( _{\zeta }\psi _{n},^{\zeta
^{\prime }}\psi _{n}\right) $, $n\in \Omega _{1}\cup \Omega _{3}\cup \Omega
_{5}$. They are written (see (\ref{i1}) and (\ref{i3})) in terms of the
quantities $\mathcal{R}_{\mathrm{L}/\mathrm{R}}\left( _{\zeta }\left\vert
^{\zeta ^{\prime }}\right. \right) $ as follows
\begin{eqnarray}
&&\mathcal{R}_{\mathrm{L}}\left( _{\zeta }\left\vert ^{\zeta ^{\prime
}}\right. \right) =\int_{-K^{\left( \mathrm{L}\right) }}^{x_{\mathrm{L}%
}}\Theta _{\mathrm{L}}\left( _{\zeta }\left\vert ^{\zeta ^{\prime }}\right.
\right) dx,\;\mathcal{R}_{\mathrm{R}}\left( _{\zeta }\left\vert ^{\zeta
^{\prime }}\right. \right) =\int_{x_{\mathrm{R}}}^{K^{\left( \mathrm{R}%
\right) }}\Theta _{\mathrm{R}}\left( _{\zeta }\left\vert ^{\zeta ^{\prime
}}\right. \right) dx,  \notag \\
&&\Theta _{\mathrm{L}/\mathrm{R}}\left( _{\zeta }\left\vert ^{\zeta ^{\prime
}}\right. \right) =\ _{\zeta }\varphi _{n}^{\ast }\left( x\right) 2\pi
_{0}\left( \mathrm{L}/\mathrm{R}\right) \left[ \pi _{0}\left( \mathrm{L}/%
\mathrm{R}\right) +\chi i\partial _{x}\right] \ ^{\zeta ^{\prime }}\varphi
_{n}\left( x\right) ,  \notag \\
&&n\in \Omega _{1}\cup \Omega _{3}\cup \Omega _{5}\ .  \label{i5a}
\end{eqnarray}%
As was mentioned before, relations (\ref{rel1}) remain valid if the
functions $\ _{\zeta }\psi _{n_{i}}\left( X\right) $ and $^{\;\zeta ^{\prime
}}\psi _{n_{i}}\left( X\right) $ are changed\ to the functions $_{\;\zeta
}\varphi _{n_{i}}\left( x\right) $ and $\ ^{\zeta ^{\prime }}\varphi
_{n_{i}}\left( x\right) $. Using this fact,\emph{\ }we express the functions
$\ ^{\zeta }\varphi _{n_{i}}\left( x\right) $ in terms of $\ _{\zeta
}\varphi _{n_{i}}\left( x\right) $ in $\mathcal{R}_{\mathrm{L}}\left(
_{\zeta }\left\vert ^{\zeta ^{\prime }}\right. \right) $, and the functions $%
\ _{\zeta }\varphi _{n_{i}}\left( x\right) $ in terms of $\ ^{\zeta }\varphi
_{n_{i}}\left( x\right) $ in $\mathcal{R}_{\mathrm{R}}\left( _{\zeta
}\left\vert ^{\zeta ^{\prime }}\right. \right) $. Then taking into account
Eqs. (\ref{e8b}), we obtain%
\begin{eqnarray}
\mathcal{R}_{\mathrm{L}}\left( _{\zeta }\left\vert ^{\zeta ^{\prime
}}\right. \right)  &=&\zeta \eta _{\mathrm{L}}Y^{2}K^{\left( \mathrm{L}%
\right) }\left\vert \frac{\pi _{0}\left( \mathrm{L}\right) }{p^{\mathrm{L}}}%
\right\vert g\left( _{\zeta }\left\vert ^{\zeta ^{\prime }}\right. \right)
+O\left( 1\right) ,  \notag \\
\mathcal{R}_{\mathrm{R}}\left( _{\zeta }\left\vert ^{\zeta ^{\prime
}}\right. \right)  &=&\zeta ^{\prime }\eta _{\mathrm{R}}Y^{2}K^{\left(
\mathrm{R}\right) }\left\vert \frac{\pi _{0}\left( \mathrm{R}\right) }{p^{%
\mathrm{R}}}\right\vert g\left( _{\zeta }\left\vert ^{\zeta ^{\prime
}}\right. \right) +O\left( 1\right) .  \label{i6}
\end{eqnarray}

Then we consider only the case $n\in \Omega _{1}\cup \Omega _{5}$. For $n\in
\Omega _{1}$, due to the inequalities $\pi _{0}\left( \mathrm{L}\right) >\pi
_{0}\left( \mathrm{R}\right) \geq \pi _{\bot }$, both sets of solutions
describe electrons. For $n\in \Omega _{5}\ ,$ due to the inequalities $\pi
_{0}\left( \mathrm{R}\right) <\pi _{0}\left( \mathrm{L}\right) \leq -\pi
_{\bot },$ both sets of solutions describe positrons. In both cases $\eta _{%
\mathrm{L}}=\eta _{\mathrm{R}}$. Then it follows from Eqs.\ (\ref{i3}) and (%
\ref{i6}) that
\begin{equation}
\left( _{\zeta }\psi _{n},^{-\zeta }\psi _{n}\right) =0,\ \ n\in \Omega
_{1}\cup \Omega _{5}\ ,  \label{i7}
\end{equation}%
if we assume that the quantities $K^{\left( \mathrm{L}/\mathrm{R}\right) }$
satisfy the following relation%
\begin{equation}
K^{\left( \mathrm{L}\right) }\left\vert \frac{\pi _{0}\left( \mathrm{L}%
\right) }{p^{\mathrm{L}}}\right\vert -K^{\left( \mathrm{R}\right)
}\left\vert \frac{\pi _{0}\left( \mathrm{R}\right) }{p^{\mathrm{R}}}%
\right\vert =O\left( 1\right) ,  \label{i8}
\end{equation}%
that was first proposed by Nikishov in Ref. \cite{Nikis04}. Condition (\ref%
{i7}) means that for $n\in \Omega _{1}\cup \Omega _{5}$\ solutions $\
_{\zeta }\psi _{n}\left( X\right) $ and $\ ^{-\zeta }\psi _{n}\left(
X\right) $, represent independent physical states.\emph{\ }The currents (\ref%
{c3}) of these independent physical states have opposite directions.\emph{\ }

Let us consider the range $n\in \Omega _{3}$. In this range, the
inequalities $\pi _{0}\left( \mathrm{L}\right) \geq \pi _{\bot }$\ and $\pi
_{0}\left( \mathrm{R}\right) \leq -\pi _{\bot }$ hold true and $\eta _{%
\mathrm{L}}=-\eta _{\mathrm{R}}=+1$. Then it follows from Eqs.\ (\ref{i3}), (%
\ref{i6}) and (\ref{i8}) that%
\begin{equation}
\left( _{\zeta }\psi _{n},^{\zeta }\psi _{n}\right) =0,\ \ n\in \Omega _{3}\
.  \label{i9}
\end{equation}%
Thus, for $n\in \Omega _{3}$\ solutions $\ _{\zeta }\psi _{n}\left( X\right)
$ and $\ ^{\zeta }\psi _{n}\left( X\right) $, represent independent physical
states. Formally, the difference between the\ two cases $n\in \Omega
_{1}\cup \Omega _{5}$ and $n\in \Omega _{3}$ is owing to the difference in
signs in the unitarity relations for these cases. The currents (\ref{c3}) of
these independent physical states have the same directions.\emph{\ }

Finally, using Eqs.~(\ref{UR}), (\ref{i11b}), and (\ref{i8}), we obtain the
orthonormality relations (\ref{i12}) for the ranges.$\Omega _{i}$, $i=1,3,5$.

\section{Some mean values\label{mean}}

\subsection{Mean values in $\Omega _{1}$ and $\Omega _{5}$\label{SS5.2}}

I. All mean values of the QFT charge operator $\hat{Q}$ given by Eq. (\ref%
{2.27a}) in states (\ref{4.1a}) are $-e,$ whereas all mean values of the QFT
charge operator $\hat{Q}$ in states (\ref{4.1b}) are $+e.$

II. Using Eqs. (\ref{2.27}) and (\ref{i19a}), we can verify that all the
kinetic energies of the states under consideration are positive. For the
electron states, we obtain:%
\begin{eqnarray}
&&\ \langle 0|\ _{+}a_{n_{1}}(\mathrm{in})\ \widehat{\mathbb{H}}^{\mathrm{kin%
}}\ _{+}a_{n_{1}}^{\dag }(\mathrm{in})|0\rangle =\ _{+}\mathcal{E}%
_{n_{1}}>0\ ,  \notag \\
&&\ \langle 0|\ ^{-}a_{n_{1}}(\mathrm{in})\widehat{\mathbb{H}}^{\mathrm{kin}%
}\ ^{-}a_{n_{1}}^{\dag }(\mathrm{in})|0\rangle =\ ^{-}\mathcal{E}_{n_{1}}>0\
,  \notag \\
&&\ \langle 0|\ _{-}a_{n_{1}}(\mathrm{out})\widehat{\mathbb{H}}^{\mathrm{kin}%
}\ _{-}a_{n_{1}}^{\dag }(\mathrm{out})|0\rangle =\ _{-}\mathcal{E}%
_{n_{1}}>0\ ,  \notag \\
&&\ \langle 0|\ ^{+}a_{n_{1}}(\mathrm{out})\ \widehat{\mathbb{H}}^{\mathrm{%
kin}}\ ^{+}a_{n_{1}}^{\dag }(\mathrm{out})|0\rangle =\ ^{+}\mathcal{E}%
_{n_{1}}>0\ ,  \label{5.1}
\end{eqnarray}%
whereas for the positron states, we have%
\begin{eqnarray}
&&\langle 0|\ _{-}b_{n_{5}}(\mathrm{in})\widehat{\mathbb{H}}^{\mathrm{kin}}\
_{-}b_{n_{5}}^{+}(\mathrm{in})|0\rangle =-\ _{-}\mathcal{E}_{n_{5}}>0\ ,
\notag \\
&&\langle 0|\ ^{+}b_{n_{5}}(\mathrm{in})\ \widehat{\mathbb{H}}^{\mathrm{kin}%
}\ ^{+}b_{n_{5}}^{+}(\mathrm{in})|0\rangle =-\ ^{+}\mathcal{E}_{n_{5}}>0\ ,
\notag \\
&&\langle 0|\ _{+}b_{n_{5}}(\mathrm{out})\ \widehat{\mathbb{H}}^{\mathrm{kin}%
}\ _{+}b_{n_{5}}^{+}(\mathrm{out})|0\rangle =-\ _{+}\mathcal{E}_{n_{1}}>0\ ,
\notag \\
&&\langle 0|\ ^{-}b_{n_{5}}(\mathrm{out})\ \widehat{\mathbb{H}}^{\mathrm{kin}%
}\ ^{-}b_{n_{5}}^{+}(\mathrm{out})|0\rangle =-\ ^{-}\mathcal{E}_{n_{5}}>0\ .
\label{4.3}
\end{eqnarray}

III. Let us calculate differential mean values of \textrm{out-}particles
with respect to different \textrm{in}-states (\ref{4.1a}) and (\ref{4.1b}).
To this end we have to find the corresponding mean values of the following
operators,%
\begin{eqnarray}
\hat{N}_{+,n_{1}}^{(a)} &=&\;^{+}a_{n_{1}}^{\dag }(\mathrm{out})\
^{+}a_{n_{1}}(\mathrm{out}),\ \ \hat{N}_{-,n_{1}}^{(a)}=\
_{-}a_{n_{1}}^{\dag }(\mathrm{out})\ _{-}a_{n_{1}}(\mathrm{out}),  \notag \\
\hat{N}_{-,n_{5}}^{(b)} &=&\;^{-}b_{n_{5}}^{\dag }(\mathrm{out})\
^{-}b_{n_{5}}(\mathrm{out}),\ \ \hat{N}_{+,n_{5}}^{(b)}=\
_{+}b_{n_{5}}^{\dag }(\mathrm{out})\ _{+}b_{n5}(\mathrm{out}).  \label{4.4}
\end{eqnarray}%
Technically it can be done by using canonical transformations (\ref{cq11})
between \textrm{in} and \textrm{out} operators, derived in Sec. \ref{SS4.3}.
Thus, we obtain\textrm{\ }%
\begin{eqnarray}
&&N_{\zeta ,n_{1}}^{(a)}\left( 0\right) =\left\langle 0\left\vert \hat{N}%
_{\zeta ,n_{1}}^{(a)}\right\vert 0\right\rangle =0,\ \ N_{\zeta
,n_{5}}^{(b)}\left( 0\right) =\left\langle 0\left\vert \hat{N}_{\zeta
,n_{5}}^{(b)}\right\vert 0\right\rangle =0,  \notag \\
&&N_{\zeta ,n_{1}}^{(a)}\left( n_{1},+\right) =\left\langle 0\left\vert \
_{+}a_{n_{1}}(\mathrm{in})\hat{N}_{\zeta ,n_{1}}^{(a)}\ _{+}a_{n_{1}}^{\dag
}(\mathrm{in})\right\vert 0\right\rangle =\left\{
\begin{array}{l}
\left\vert g\left( _{+}\left\vert ^{+}\right. \right) \right\vert ^{-2},\ \
\zeta =+ \\
\left\vert g\left( _{+}\left\vert ^{+}\right. \right) \right\vert
^{-2}\left\vert g\left( _{-}\left\vert ^{+}\right. \right) \right\vert
^{2},\ \ \zeta =-%
\end{array}%
\right. ,  \notag \\
&&N_{\zeta ,n_{1}}^{(a)}\left( n_{1},-\right) =\left\langle 0\left\vert \
^{-}a_{n_{1}}(\mathrm{in})\hat{N}_{\zeta ,n_{1}}^{(a)}\ ^{-}a_{n_{1}}^{\dag
}(\mathrm{in})\right\vert 0\right\rangle =\left\{
\begin{array}{l}
\left\vert g\left( _{+}\left\vert ^{+}\right. \right) \right\vert ^{-2},\ \
\zeta =- \\
\left\vert g\left( _{+}\left\vert ^{+}\right. \right) \right\vert
^{-2}\left\vert g\left( _{-}\left\vert ^{+}\right. \right) \right\vert
^{2},\ \ \zeta =+%
\end{array}%
\right. ,  \notag \\
&&N_{\zeta ,n_{5}}^{(b)}\left( n_{5},+\right) =\left\langle 0\left\vert \
^{+}b_{n_{5}}(\mathrm{in})\hat{N}_{\zeta ,n_{5}}^{(b)}\ ^{+}b_{n_{5}}^{\dag
}(\mathrm{in})\right\vert 0\right\rangle =\left\{
\begin{array}{l}
\left\vert g\left( _{+}\left\vert ^{+}\right. \right) \right\vert ^{-2},\ \
\zeta =+ \\
\left\vert g\left( _{+}\left\vert ^{+}\right. \right) \right\vert
^{-2}\left\vert g\left( _{-}\left\vert ^{+}\right. \right) \right\vert
^{2},\ \ \zeta =-%
\end{array}%
\right. ,  \notag \\
&&N_{\zeta ,n_{5}}^{(b)}\left( n_{5},-\right) =\left\langle 0\left\vert \
_{-}b_{n_{5}}(\mathrm{in})\hat{N}_{\zeta ,n_{5}}^{(b)}\ _{-}b_{n_{5}}^{\dag
}(\mathrm{in})\right\vert 0\right\rangle =\left\{
\begin{array}{l}
\left\vert g\left( _{+}\left\vert ^{+}\right. \right) \right\vert ^{-2},\ \
\zeta =- \\
\left\vert g\left( _{+}\left\vert ^{+}\right. \right) \right\vert
^{-2}\left\vert g\left( _{-}\left\vert ^{+}\right. \right) \right\vert
^{2},\ \ \zeta =+%
\end{array}%
\right. ,  \label{cq13}
\end{eqnarray}%
see (\ref{c12}) for the definition of the coefficients $g.$

Then it follows from relations (\ref{UR2}) and (\ref{UR1}) that%
\begin{eqnarray}
N_{+,n_{1}}^{(a)}\left( n_{1},+\right) +N_{-,n_{1}}^{(a)}\left(
n_{1},+\right) &=&N_{+,n_{1}}^{(a)}\left( n_{1},-\right)
+N_{-,n_{1}}^{(a)}\left( n_{1},-\right) =1;  \notag \\
N_{+,n_{5}}^{(b)}\left( n_{5},+\right) +N_{-,n_{5}}^{(b)}\left(
n_{5},+\right) &=&N_{+,n_{5}}^{(b)}\left( n_{5},-\right)
+N_{-,n_{5}}^{(b)}\left( n_{5},-\right) =1\ .  \label{5.6}
\end{eqnarray}%
Thus, the number of electrons with quantum numbers $n_{1}$ and positrons
with quantum numbers $n_{5}$ are conserved in the course of scattering off
the $x$-electric potential step.

III. Using the electric current operator $\hat{J}$ (\ref{cq9}), we construct
the corresponding renormalized operator $\widehat{\mathbb{J}}$. Its mean
value in the vacuum state is zero,%
\begin{equation}
\widehat{\mathbb{J}}=\hat{J}-\left\langle 0\left\vert \hat{J}\right\vert
0\right\rangle ,\ \ \left\langle 0\left\vert \widehat{\mathbb{J}}\right\vert
0\right\rangle =0.  \label{5.7b}
\end{equation}%
Then, using orthonormality condition (\ref{c3}), we calculate currents
created by one-electron states in the range $\Omega _{1}$:
\begin{eqnarray}
&&\left\langle 0\left\vert \ _{+}a_{n_{1}}(\mathrm{in})\widehat{\mathbb{J}}\
_{+}a_{n_{1}}^{\dag }(\mathrm{in})\right\vert 0\right\rangle =-e\left(
\mathcal{M}_{n_{1}}T\right) ^{-1}<0,  \notag \\
&&\left\langle 0\left\vert \ ^{-}a_{n_{1}}(\mathrm{in})\widehat{\mathbb{J}}\
^{-}a_{n_{1}}^{\dag }(\mathrm{in})\right\vert 0\right\rangle =e\left(
\mathcal{M}_{n_{1}}T\right) ^{-1}>0,  \notag \\
&&\left\langle 0\left\vert \ _{-}a_{n_{1}}(\mathrm{out})\widehat{\mathbb{J}}%
\ _{-}a_{n_{1}}^{\dag }(\mathrm{out})\right\vert 0\right\rangle =e\left(
\mathcal{M}_{n_{1}}T\right) ^{-1}>0\ ,  \notag \\
&&\left\langle 0\left\vert \ ^{+}a_{n_{1}}(\mathrm{out})\widehat{\mathbb{J}}%
\ ^{+}a_{n_{1}}^{\dag }(\mathrm{out})\right\vert 0\right\rangle =-e\left(
\mathcal{M}_{n_{1}}T\right) ^{-1}<0\ ,  \label{5.7}
\end{eqnarray}%
and currents created by one-positron states in the range $\Omega _{5}$:%
\begin{eqnarray}
&&\left\langle 0\left\vert \ _{-}b_{n_{5}}(\mathrm{in})\widehat{\mathbb{J}}\
_{-}b_{n_{5}}^{\dag }(\mathrm{in})\right\vert 0\right\rangle =e\left(
\mathcal{M}_{n_{5}}T\right) ^{-1}>0\ ,  \notag \\
&&\left\langle 0\left\vert \ ^{+}b_{n_{5}}(\mathrm{in})\widehat{\mathbb{J}}\
^{+}b_{n_{5}}^{\dag }(\mathrm{in})\right\vert 0\right\rangle =-e\left(
\mathcal{M}_{n_{5}}T\right) ^{-1}<0\ ,  \notag \\
&&\left\langle 0\left\vert \ \ _{+}b_{n_{5}}(\mathrm{out})\widehat{\mathbb{J}%
}\ _{+}b_{n_{5}}^{\dag }(\mathrm{out})\right\vert 0\right\rangle =-e\left(
\mathcal{M}_{n_{5}}T\right) ^{-1}<0,  \notag \\
&&\left\langle 0\left\vert \ ^{-}b_{n_{5}}(\mathrm{out})\widehat{\mathbb{J}}%
\ ^{-}b_{n_{5}}^{\dag }(\mathrm{out})\right\vert 0\right\rangle =e\left(
\mathcal{M}_{n_{5}}T\right) ^{-1}>0,  \label{5.8}
\end{eqnarray}

The quantities $\mathcal{M}_{n}$ are given by Eqs. (\ref{i13a}), and the
combination $\left( \mathcal{M}_{n}T\right) ^{-1}$ is the modulus of the
probability flux of a one-particle state through the hyperplane $x=\mathrm{%
const}$. One can see that signs of currents (\ref{5.7}) and (\ref{5.8}) are
always opposite to the signs of the asymptotic values $p^{\mathrm{R}}$ and $%
p^{\mathrm{L}},$ respectively. Thus, the one-particle quantum mechanical
interpretation of quantum numbers $p^{\mathrm{R}}$\ and $p^{\mathrm{L}}$ as
momenta holds true in the ranges $\Omega _{1}$ and $\Omega _{5}$.

IV. Using energy flux operator $\hat{F}\left( x\right) $ (\ref{cq9b}), we
construct the corresponding renormalized operator $\mathbb{\hat{F}}\left(
x\right) $. Its mean value in the vacuum state is zero,%
\begin{equation}
\mathbb{\hat{F}}\left( x\right) =\hat{F}\left( x\right) -\left\langle
0\left\vert \hat{F}\left( x\right) \right\vert 0\right\rangle ,\ \
\left\langle 0\left\vert \mathbb{\hat{F}}\left( x\right) \right\vert
0\right\rangle =0.  \label{5.8a}
\end{equation}%
Then, with the help of this operator, we\textrm{\ }calculate mean energy
fluxes created by one-particle states (\ref{4.1a}) and (\ref{4.1b}) through
the surfaces $x=x_{\mathrm{L}}$ and $x=x_{\mathrm{R}}$ . Using
orthonormality condition (\ref{c3}), we obtain for electrons in the range $%
\Omega _{1}$ :%
\begin{eqnarray}
&&\mathbb{F}_{n_{1},+}\left( \mathrm{in}\right) =\left\langle 0\left\vert \
_{+}a_{n_{1}}(\mathrm{in})\mathbb{\hat{F}}\left( x_{\mathrm{L}}\right) \
_{+}a_{n_{1}}^{\dag }(\mathrm{in})\right\vert 0\right\rangle =\left(
\mathcal{M}_{n_{1}}T\right) ^{-1}\pi _{0}\left( \mathrm{L}\right) >0,  \notag
\\
&&\mathbb{F}_{n_{1},+}\left( \mathrm{out}\right) =\left\langle 0\left\vert \
^{+}a_{n_{1}}(\mathrm{out})\mathbb{\hat{F}}\left( x_{\mathrm{R}}\right) \
^{+}a_{n_{1}}^{\dag }(\mathrm{out})\right\vert 0\right\rangle =\left(
\mathcal{M}T\right) ^{-1}\pi _{0}\left( \mathrm{R}\right) >0,  \notag \\
&&\mathbb{F}_{n_{1},-}\left( \mathrm{in}\right) =\left\langle 0\left\vert \
^{-}a_{n_{1}}(\mathrm{in})\mathbb{\hat{F}}\left( x_{\mathrm{R}}\right) \
^{-}a_{n_{1}}^{\dag }(\mathrm{in})\right\vert 0\right\rangle =-\left(
\mathcal{M}T\right) ^{-1}\pi _{0}\left( \mathrm{R}\right) <0,  \notag \\
&&\mathbb{F}_{n_{1},-}\left( \mathrm{out}\right) =\left\langle 0\left\vert \
_{-}a_{n_{1}}(\mathrm{out})\mathbb{\hat{F}}\left( x_{\mathrm{L}}\right) \
_{-}a_{n_{1}}^{\dag }(\mathrm{out})\right\vert 0\right\rangle =-\left(
\mathcal{M}T\right) ^{-1}\pi _{0}\left( \mathrm{L}\right) <0,  \label{5.9}
\end{eqnarray}%
and for positrons in the range $\Omega _{5}:$%
\begin{eqnarray}
&&\mathbb{F}_{n_{5},-}\left( \mathrm{in}\right) =\left\langle 0\left\vert \
_{-}b_{n_{5}}(\mathrm{in})\mathbb{\hat{F}}\left( x_{\mathrm{L}}\right) \
_{-}b_{n_{5}}^{\dag }(\mathrm{in})\right\vert 0\right\rangle =\left(
\mathcal{M}T\right) ^{-1}\left\vert \pi _{0}\left( \mathrm{L}\right)
\right\vert >0,  \notag \\
&&\mathbb{F}_{n_{5},-}\left( \mathrm{out}\right) =\left\langle 0\left\vert \
^{-}b_{n_{5}}(\mathrm{out})\mathbb{\hat{F}}\left( x_{\mathrm{R}}\right) \
^{-}b_{n_{5}}^{\dag }(\mathrm{out})\right\vert 0\right\rangle =\left(
\mathcal{M}T\right) ^{-1}\left\vert \pi _{0}\left( \mathrm{R}\right)
\right\vert >0,  \notag \\
&&\mathbb{F}_{n_{5},+}\left( \mathrm{in}\right) =\left\langle 0\left\vert \
^{+}b_{n_{5}}(\mathrm{in})\mathbb{\hat{F}}\left( x_{\mathrm{R}}\right) \
^{+}b_{n_{5}}^{\dag }(\mathrm{in})\right\vert 0\right\rangle =-\left(
\mathcal{M}T\right) ^{-1}\left\vert \pi _{0}\left( \mathrm{R}\right)
\right\vert <0,  \notag \\
&&\mathbb{F}_{n_{5},+}\left( \mathrm{out}\right) =\left\langle 0\left\vert \
_{+}b_{n_{5}}(\mathrm{out})\mathbb{\hat{F}}\left( x_{\mathrm{L}}\right) \
_{+}b_{n_{5}}^{\dag }(\mathrm{out})\right\vert 0\right\rangle =-\left(
\mathcal{M}T\right) ^{-1}\left\vert \pi _{0}\left( \mathrm{L}\right)
\right\vert <0.  \label{5.10}
\end{eqnarray}

We believe that the direction of the energy flux indicates the direction of
motion\ of the corresponding particle, which is shown on Fig.~\ref{3} by the
corresponding arrows.

This fact allows us to find longitudinal momenta of the waves in the
asymptotic regions $S_{\mathrm{L}}$ and $S_{\mathrm{R}}$ by integrating the
corresponding energy fluxes over $x.$ Such QFT quantities differ from the
quantum numbers $p^{\mathrm{L}}$ and $p^{\mathrm{R}}$ that are asymptotic
longitudinal momenta of the corresponding unit one-particle flux, for
simplicity, we call them QM longitudinal momenta of particles.

For example, QFT longitudinal momenta of electron states are%
\begin{eqnarray}
&&P_{n_{1},+}\left( \mathrm{in}\right) =\mathbb{F}_{n_{1},+}\left( \mathrm{in%
}\right) K^{\left( \mathrm{L}\right) }=\frac{1}{2}\left\vert p^{\mathrm{L}%
}\right\vert \left\vert g\left( _{+}\left\vert ^{+}\right. \right)
\right\vert ^{-2},  \notag \\
&&P_{n_{1},+}\left( \mathrm{out}\right) =\mathbb{F}_{n_{1},+}\left( \mathrm{%
out}\right) K^{\left( \mathrm{R}\right) }=\frac{1}{2}\left\vert p^{\mathrm{R}%
}\right\vert \left\vert g\left( _{+}\left\vert ^{+}\right. \right)
\right\vert ^{-2},  \notag \\
&&P_{n_{1},-}\left( \mathrm{out}\right) =\mathbb{F}_{n_{1},-}\left( \mathrm{%
out}\right) K^{\left( \mathrm{L}\right) }=-\frac{1}{2}\left\vert p^{\mathrm{L%
}}\right\vert \left\vert g\left( _{+}\left\vert ^{+}\right. \right)
\right\vert ^{-2},  \notag \\
&&P_{n_{1},-}\left( \mathrm{in}\right) =\mathbb{F}_{n_{1},-}\left( \mathrm{in%
}\right) K^{\left( \mathrm{R}\right) }=-\frac{1}{2}\left\vert p^{\mathrm{R}%
}\right\vert \left\vert g\left( _{+}\left\vert ^{+}\right. \right)
\right\vert ^{-2},  \label{cqf5}
\end{eqnarray}%
see Sec. \ref{SS3.3}. In the same manner, using Eq. (\ref{5.10}), we can
find longitudinal momenta of positron states.

These relations show that quantum numbers $\left\vert \pi _{0}\left( \mathrm{%
L}/\mathrm{R}\right) \right\vert $ can be interpreted as kinetic energies of
the unit particle flux, both for electrons and positrons in the asymptotic
regions $S_{\mathrm{L}}$ and $S_{\mathrm{R}},$ respectively. Then it follows
from Eqs. (\ref{cqf5}) that $\left\vert p^{\mathrm{L}}\right\vert $ and $%
\left\vert p^{\mathrm{R}}\right\vert $ are modulus of asymptotic
longitudinal momenta of the unit of the corresponding one-particle fluxes.
One can also see that the sign of the mean values $\mathbb{F}_{n,\zeta
}\left( \mathrm{in}/\mathrm{out}\right) $ in electron states is $\zeta $ and
in positron states is $-\zeta $. The directions of the energy fluxes
represent the directions of the motion of both electrons and positrons.
Thus, we see that the asymptotic longitudinal physical momenta of electrons
are $p_{ph}^{\mathrm{L}}=p^{\mathrm{L}}$ and $p_{ph}^{\mathrm{R}}=p^{\mathrm{%
R}}$, whereas for positrons they are $p_{ph}^{\mathrm{L}}=-p^{\mathrm{L}}$
and $p_{ph}^{\mathrm{R}}=-p^{\mathrm{R}}$. This matches with the standard
interpretation of quantum numbers of solutions of the Dirac equation.\textrm{%
\ }One can also see that the electric current of electrons is opposite to
the direction of their energy flux (and to their asymptotic longitudinal
physical momenta), whereas the electric current of positrons coincides with
the direction of their energy flux (and with the asymptotic longitudinal
physical momenta).

\subsection{Mean values in $\Omega _{3}$\label{SS7.2}}

I. By using relations (\ref{7.4}), we find differential mean numbers of
\textrm{out-}particles in the vacuum $\left\vert 0,\mathrm{in}\right\rangle $%
, and differential mean numbers of \textrm{in}-particles in the \textrm{out}%
-vacuum $\left\vert 0,\mathrm{out}\right\rangle $,%
\begin{eqnarray}
&&N_{n}^{a}\left( \mathrm{out}\right) =\left\langle 0,\mathrm{in}\left\vert
\ ^{+}a_{n}^{\dagger }(\mathrm{out})\ ^{+}a_{n}(\mathrm{out})\right\vert 0,%
\mathrm{in}\right\rangle =\left\vert g\left( _{-}\left\vert ^{+}\right.
\right) \right\vert ^{-2},  \notag \\
&&N_{n}^{b}\left( \mathrm{out}\right) =\left\langle 0,\mathrm{in}\left\vert
\ _{+}b_{n}^{\dagger }(\mathrm{out})\ _{+}b_{n}(\mathrm{out})\right\vert 0,%
\mathrm{in}\right\rangle =\left\vert g\left( _{+}\left\vert ^{-}\right.
\right) \right\vert ^{-2},  \notag \\
&&N_{n}^{a}\left( \mathrm{in}\right) =\left\langle 0,\mathrm{out}\left\vert
\ ^{-}a_{n}^{\dagger }(\mathrm{in})\ ^{-}a_{n}(\mathrm{in})\right\vert 0,%
\mathrm{out}\right\rangle =\left\vert g\left( _{+}\left\vert ^{-}\right.
\right) \right\vert ^{-2},  \notag \\
&&N_{n}^{b}\left( \mathrm{in}\right) =\left\langle 0,\mathrm{out}\left\vert
\ _{-}b_{n}^{\dagger }(\mathrm{in})\ _{-}b_{n}(\mathrm{in})\right\vert 0,%
\mathrm{out}\right\rangle =\left\vert g\left( _{-}\left\vert ^{+}\right.
\right) \right\vert ^{-2}.  \label{7.5}
\end{eqnarray}

II. By using\textrm{\ }relations (\ref{UR1}) we find differential mean
numbers of \textrm{out-}particles in one-particle \textrm{in-}states,%
\begin{eqnarray}
\left\langle 0,\mathrm{in}\left\vert \ ^{-}a_{n}(\mathrm{in})\
^{+}a_{n}^{\dagger }(\mathrm{out})\ ^{+}a_{n}(\mathrm{out})\
^{-}a_{n}^{\dagger }(\mathrm{in})\right\vert 0,\mathrm{in}\right\rangle &=&1,
\notag \\
\left\langle 0,\mathrm{in}\left\vert \ ^{-}a_{n}(\mathrm{in})\
_{+}b_{n}^{\dagger }(\mathrm{out})\ _{+}b_{n}(\mathrm{out})\
^{-}a_{n}^{\dagger }(\mathrm{in})\right\vert 0,\mathrm{in}\right\rangle &=&0,
\notag \\
\left\langle 0,\mathrm{in}\left\vert \ _{-}b_{n}(\mathrm{in})\
_{+}b_{n}^{\dagger }(\mathrm{out})\ _{+}b_{n}(\mathrm{out})\
_{-}b_{n}^{\dagger }(\mathrm{in})\right\vert 0,\mathrm{in}\right\rangle &=&1,
\notag \\
\left\langle 0,\mathrm{in}\left\vert \ _{-}b_{n}(\mathrm{in})\
^{+}a_{n}^{\dagger }(\mathrm{out})\ ^{+}a_{n}(\mathrm{out})\
_{-}b_{n}^{\dagger }(\mathrm{in})\right\vert 0,\mathrm{in}\right\rangle &=&0.
\label{7.6}
\end{eqnarray}%
Thus, the one-electron \textrm{in-}state contains only one \textrm{out-}%
electron and does not contain any \textrm{out}-positron, whereas the
one-positron \textrm{in-}state contains only one \textrm{out}-positron and
does not contain any \textrm{out}-electron, which is a consequence of the
Pauli principle.

III. Using the operator $\widehat{\mathbb{H}}^{\mathrm{kin}}$ (\ref{2.27}),
we calculate kinetic energies of all particles in the range $\Omega _{3},$%
{\large \ }
\begin{eqnarray}
&&\left\langle 0,\mathrm{in}\left\vert \ ^{-}a_{n}(\mathrm{in})\widehat{%
\mathbb{H}}^{\mathrm{kin}}\ ^{-}a_{n}^{\dagger }(\mathrm{in})\right\vert 0,%
\mathrm{in}\right\rangle =\ ^{-}\mathcal{E}_{n_{3}},  \notag \\
&&\left\langle 0,\mathrm{in}\left\vert \ _{-}b_{n}(\mathrm{in})\widehat{%
\mathbb{H}}^{\mathrm{kin}}\ _{-}b_{n}^{\dagger }(\mathrm{in})\right\vert 0,%
\mathrm{in}\right\rangle =-\ _{-}\mathcal{E}_{n_{3}},  \notag \\
&&\left\langle 0,\mathrm{out}\left\vert \ ^{+}a_{n}(\mathrm{out})\widehat{%
\mathbb{H}}^{\mathrm{kin}}\ ^{+}a_{n}^{\dagger }(\mathrm{out})\right\vert 0,%
\mathrm{out}\right\rangle =\ ^{+}\mathcal{E}_{n_{3}},  \notag \\
&&\left\langle 0,\mathrm{out}\left\vert \ _{+}b_{n}(\mathrm{out})\widehat{%
\mathbb{H}}^{\mathrm{kin}}\ _{+}b_{n}^{\dagger }(\mathrm{out})\right\vert 0,%
\mathrm{out}\right\rangle =-\ _{+}\mathcal{E}_{n_{3}}.  \label{7.2}
\end{eqnarray}%
One can verify that, with\ the account taken of (\ref{i19}), the
combinations $\left( ^{\zeta }\mathcal{E}_{n_{3}}-\ _{\zeta }\mathcal{E}%
_{n_{3}}\right) $ are positive (as we demonstrate in Sec. \ref{SS7.4},
kinetic energies of physical wave packets of electrons and positrons are
also positive).

IV. Let us introduce renormalized (with respect to the corresponding vacua)
\textrm{in-} and \textrm{out-}operators of the electric current flowing\
through the surface $x=\mathrm{const}$,%
\begin{equation}
\widehat{\mathbb{J}}\left( \mathrm{in}\right) =\hat{J}-\left\langle 0,%
\mathrm{in}\left\vert \hat{J}\right\vert 0,\mathrm{in}\right\rangle ,\;\;%
\widehat{\mathbb{J}}\left( \mathrm{out}\right) =\hat{J}-\left\langle 0,%
\mathrm{out}\left\vert \hat{J}\right\vert 0,\mathrm{out}\right\rangle ,
\label{7.11}
\end{equation}%
where the operator $\hat{J}$ is given by Eq. (\ref{cq9}). With the help of
Eq. (\ref{c3}) we find differential mean values of these operators in all
one-particle states (\ref{7.1}),
\begin{eqnarray}
&&J_{n}^{a}\left( \mathrm{in}\right) =\left\langle 0,\mathrm{in}\left\vert \
^{-}a_{n}(\mathrm{in})\widehat{\mathbb{J}}\left( \mathrm{in}\right) \
^{-}a_{n}^{\dagger }(\mathrm{in})\right\vert 0,\mathrm{in}\right\rangle
=-e\left( \mathcal{M}_{n}T\right) ^{-1},  \notag \\
&&J_{n}^{b}\left( \mathrm{in}\right) =\left\langle 0,\mathrm{in}\left\vert \
_{-}b_{n}(\mathrm{in})\widehat{\mathbb{J}}\left( \mathrm{in}\right) \
_{-}b_{n}^{\dagger }(\mathrm{in})\right\vert 0,\mathrm{in}\right\rangle
=-e\left( \mathcal{M}_{n}T\right) ^{-1},  \notag \\
&&J_{n}^{a}\left( \mathrm{out}\right) =\left\langle 0,\mathrm{out}\left\vert
\ ^{+}a_{n}(\mathrm{out})\widehat{\mathbb{J}}\left( \mathrm{out}\right) \
^{+}a_{n}^{\dagger }(\mathrm{out})\right\vert 0,\mathrm{out}\right\rangle
=e\left( \mathcal{M}_{n}T\right) ^{-1},  \notag \\
&&J_{n}^{b}\left( \mathrm{out}\right) =\left\langle 0,\mathrm{out}\left\vert
\ _{+}b_{n}(\mathrm{out})\widehat{\mathbb{J}}\left( \mathrm{out}\right) \
_{+}b_{n}^{\dagger }(\mathrm{out})\right\vert 0,\mathrm{out}\right\rangle
=e\left( \mathcal{M}_{n}T\right) ^{-1}.  \label{7.12}
\end{eqnarray}%
One can see that the mean currents $J_{n}^{a}\left( \mathrm{out}\right) $
and $J_{n}^{b}\left( \mathrm{out}\right) $ are positive and have the same
direction as the applied external electric field, whereas the mean currents $%
J_{n}^{a}\left( \mathrm{in}\right) $ and $J_{n}^{b}\left( \mathrm{in}\right)
$ are negative and have the opposite direction to the applied external
electric field. Both \textrm{in-} and \textrm{out}-electron states (\ref{7.1}%
) are states with definite quantum numbers $p^{\mathrm{R}}$, whereas both
\textrm{in-} and \textrm{out}-positron states (\ref{7.1}) are states with
definite quantum numbers $p^{\mathrm{L}}$. Therefore, signs of both currents
$J_{n}^{a}\left( \mathrm{in}\right) $ and $J_{n}^{a}\left( \mathrm{out}%
\right) $ coincide with the sign of $p^{\mathrm{R}}$, whereas signs of both
currents $J_{n}^{b}\left( \mathrm{in}\right) $ and $J_{n}^{b}\left( \mathrm{%
out}\right) $ coincide with the sign of $p^{\mathrm{L}}$.

V. Using energy flux operator $\hat{F}\left( x\right) $ (\ref{cq9b}), we
construct the corresponding renormalized operators $\mathbb{\hat{F}}\left( x|%
\mathrm{in}\right) $ and $\mathbb{\hat{F}}\left( x|\mathrm{out}\right) $,%
\begin{equation}
\mathbb{\hat{F}}\left( x|\mathrm{in}\right) =\hat{F}\left( x\right)
-\left\langle 0,\mathrm{in}\left\vert \hat{F}\left( x\right) \right\vert 0,%
\mathrm{in}\right\rangle ,\ \ \mathbb{\hat{F}}\left( x|\mathrm{out}\right) =%
\hat{F}\left( x\right) -\left\langle 0,\mathrm{out}\left\vert \mathbb{\hat{F}%
}\left( x\right) \right\vert 0,\mathrm{out}\right\rangle .  \label{7.13}
\end{equation}%
With the help of these operators, we calculate mean energy fluxes through
the surfaces $x=x_{\mathrm{L}}$ and $x=x_{\mathrm{R}}$. Since electron wave
packets are localized in the region $S_{\mathrm{L}}$, and positron wave
packets are localized in the region $S_{\mathrm{R}}$, the mean energy flux
of electron partial waves is to be defined through the surface $x=x_{\mathrm{%
L}}$, and of positron partial waves through the surface $x=x_{\mathrm{R}}$.
These mean values are expressed via energy fluxes of the Dirac field through
the surfaces $x=x_{\mathrm{L}}$ and $x=x_{\mathrm{R}},$ respectively. The
latter fluxes can be calculated using Eqs. (\ref{c3}),%
\begin{eqnarray}
&&F_{n}^{a}\left( \mathrm{in}\right) =\left\langle 0,\mathrm{in}\left\vert \
^{-}a_{n}(\mathrm{in})\mathbb{\hat{F}}\left( x_{\mathrm{L}},\mathrm{in}%
\right) \ ^{-}a_{n}^{\dag }(\mathrm{in})\right\vert 0,\mathrm{in}%
\right\rangle =\left( \mathcal{M}_{n}T\right) ^{-1}\pi _{0}\left( \mathrm{L}%
\right) ,  \notag \\
&&F_{n}^{a}\left( \mathrm{out}\right) =\left\langle 0,\mathrm{out}\left\vert
\ ^{+}a_{n}(\mathrm{out})\mathbb{\hat{F}}\left( x_{\mathrm{L}},\mathrm{out}%
\right) \ ^{+}a_{n}^{\dag }(\mathrm{out})\right\vert 0,\mathrm{out}%
\right\rangle =-\left( \mathcal{M}_{n}T\right) ^{-1}\pi _{0}\left( \mathrm{L}%
\right) ,  \notag \\
&&F_{n}^{b}\left( \mathrm{in}\right) =\left\langle 0,\mathrm{in}\left\vert \
_{-}b_{n}(\mathrm{in})\mathbb{\hat{F}}\left( x_{\mathrm{R}},\mathrm{in}%
\right) \ _{-}b_{n}^{\dag }(\mathrm{in})\right\vert 0,\mathrm{in}%
\right\rangle =-\left( \mathcal{M}_{n}T\right) ^{-1}\left\vert \pi
_{0}\left( \mathrm{R}\right) \right\vert ,  \notag \\
&&F_{n}^{b}\left( \mathrm{out}\right) =\left\langle 0,\mathrm{out}\left\vert
\ _{+}b_{n}(\mathrm{out})\mathbb{\hat{F}}\left( x_{\mathrm{R}},\mathrm{out}%
\right) \ _{+}b_{n}^{\dag }(\mathrm{out})\right\vert 0,\mathrm{out}%
\right\rangle =\left( \mathcal{M}_{n}T\right) ^{-1}\left\vert \pi _{0}\left(
\mathrm{R}\right) \right\vert .  \label{cq20.2}
\end{eqnarray}

To find the longitudinal momenta of particles in the asymptotic regions $S_{%
\mathrm{L}}$ and $S_{\mathrm{R}}$, we have to integrate these energy fluxes
over $x$. Thus we obtain:%
\begin{eqnarray}
P_{n}^{a}\left( \mathrm{in}/\mathrm{out}\right) &=&F_{n}^{a}\left( \mathrm{in%
}/\mathrm{out}\right) K^{\left( \mathrm{L}\right) }=\pm \frac{1}{2}%
\left\vert p^{\mathrm{L}}\right\vert \left\vert g\left( _{+}\left\vert
^{-}\right. \right) \right\vert ^{-2},  \notag \\
P_{n}^{b}\left( \mathrm{in}/\mathrm{out}\right) &=&F_{n}^{b}\left( \mathrm{in%
}/\mathrm{out}\right) K^{\left( \mathrm{R}\right) }=\mp \frac{1}{2}%
\left\vert p^{\mathrm{R}}\right\vert \left\vert g\left( _{+}\left\vert
^{-}\right. \right) \right\vert ^{-2}.  \label{cq20.3}
\end{eqnarray}

We stress that in contrast to the ranges $\Omega _{1}$ and $\Omega _{5}$,
signs of the quantities related to the electrons in eqs. (\ref{cq20.2}) and (%
\ref{cq20.3}) are determined by the signs of the quantum number $p^{\mathrm{R%
}}$, but not by the signs of the quantum number $p^{\mathrm{L}}$, whereas
signs of the quantities related to the positrons are determined by the signs
of the quantum number $p^{\mathrm{L}}$, but not by the signs of the quantum
number $p^{\mathrm{R}}$,%
\begin{equation}
\mathrm{sgn}\left( P_{n}^{a}\left( \mathrm{in}/\mathrm{out}\right) \right) =-%
\mathrm{sgn}\left( p^{\mathrm{R}}\right) ,\;\mathrm{sgn}\left(
P_{n}^{b}\left( \mathrm{in}/\mathrm{out}\right) \right) =\mathrm{sgn}\left(
p^{\mathrm{L}}\right) .  \label{cq20.3a}
\end{equation}

Relations (\ref{cq20.3a}) indicate a direct correlation between directions
of $\mathrm{in}$- and $\mathrm{out}$- energy fluxes and directions of the
corresponding currents, which is:
\begin{equation}
\mathrm{sgn}p^{\mathrm{R}}=\mathrm{sgn}J_{n}^{b}\left( \mathrm{out}\right) =%
\mathrm{sgn}J_{n}^{a}\left( \mathrm{out}\right) ,\;\mathrm{sgn}p^{\mathrm{L}%
}=\mathrm{sgn}J_{n}^{b}\left( \mathrm{in}\right) =\mathrm{sgn}%
J_{n}^{a}\left( \mathrm{in}\right) .  \label{7.6a}
\end{equation}

One can see from these relations that in the range $\Omega _{3}$, as well as
in the ranges $\Omega _{1}$ and $\Omega _{5},$ the quantum numbers $%
\left\vert \pi _{0}\left( \mathrm{L}/\mathrm{R}\right) \right\vert $ are
kinetic energies per unit of one-particle flux in the corresponding
asymptotic regions. Direction of the energy flux indicates the direction of
the particle motion. This direction coincides with the direction of positron
currents situated in the region $S_{\mathrm{R}}$ and is opposite to the
direction of the electron currents situated in the region $S_{\mathrm{L}}$.

\section{Total reflection in the Klein zone\label{Aloc}}

To consider the total reflection in the range $\Omega _{3}$ in the general
case of arbitrary strong electric field when $N_{n}^{\mathrm{cr}}\rightarrow
1$ for some $n$ $\in \Omega _{3}^{str}\subset \Omega _{3}$ it is necessary
to associate particles with special wave packets that are stable enough. We
consider the Dirac case. The case of scalar field can be given along the
same lines.

There are \textrm{in} and \textrm{out} electron and positron states each of
them associated with the corresponding wave packet. Electron \textrm{in}%
-states are composed from the plane waves $\ ^{-}\psi _{n_{3}}\left(
X\right) ,$ whereas electron \textrm{out}-states are composed from the plane
waves $\ ^{+}\psi _{n_{3}}\left( X\right) .$ Positron \textrm{in}-states are
composed from the plane waves $\ _{-}\psi _{n_{3}}\left( X\right) ,$ whereas
positron \textrm{out}-states are composed from the plane waves $\ _{+}\psi
_{n_{3}}\left( X\right) $. It is natural to suppose that all these wave
packets are localized at some time instants in some areas that have a finite
length on the axis $x.$ Let $x=x_{\mathrm{F}}$ be a center of such an area
at some time instant. We represent such wave packets for electrons and
positrons as follows%
\begin{eqnarray}
\ ^{\zeta }\psi _{x_{\mathrm{F}}}\left( X\right) &=&\ ^{\zeta }N_{x_{\mathrm{%
F}}}\sum_{n\in \Omega _{3}}\ ^{\zeta }c_{n}^{\left( x_{\mathrm{F}}\right) }%
\mathcal{M}_{n}^{-1/2}\ ^{\zeta }\psi _{n}\left( X\right) \ -\ \mathrm{\
in/out\ }(\zeta =-/+)\ \mathrm{-\ electron\ states,}  \notag \\
\ _{\zeta }\psi _{x_{F}}\left( X\right) &=&\ _{\zeta }N_{x_{F}}\sum_{n\in
\Omega _{3}}\ _{\zeta }c_{n}^{\left( x_{\mathrm{F}}\right) }\mathcal{M}%
_{n}^{-1/2}\ _{\zeta }\psi _{n}\left( X\right) \ -\ \mathrm{in/out\ }(\zeta
=-/+)\ \mathrm{-positron\ states,}  \label{d3}
\end{eqnarray}%
where $\ ^{\zeta }c_{n}^{\left( x_{\mathrm{F}}\right) }$ and $\ _{\zeta
}c_{n}^{\left( x_{\mathrm{F}}\right) }$ are some coefficients and $\ ^{\zeta
}N_{x_{\mathrm{F}}}$ and $\ _{\zeta }N_{x_{F}}$ are normalization factors,%
\begin{equation*}
\left\vert ^{\zeta }N_{x_{\mathrm{F}}}\right\vert ^{-2}=\sum_{n\in \Omega
_{3}}\left\vert ^{\zeta }c_{n}^{\left( x_{\mathrm{F}}\right) }\right\vert
^{2},\;\left\vert _{\zeta }N_{x_{F}}\right\vert ^{-2}=\sum_{n\in \Omega
_{3}}\left\vert _{\zeta }c_{n}^{\left( x_{\mathrm{F}}\right) }\right\vert
^{2}.
\end{equation*}

We are interested in the cases when particle wave packets are localized in
the asymptotic region $S_{\mathrm{L}}$ far enough from the asymptotic region
$S_{\mathrm{R}}$, which means $x_{\mathrm{F}}<x_{\mathrm{L}}^{\mathrm{F}}<x_{%
\mathrm{L}}$, or in the asymptotic region $S_{\mathrm{R}}$\ far enough from
the asymptotic region $S_{\mathrm{L}}$, which means $x_{\mathrm{F}}>x_{%
\mathrm{R}}^{\mathrm{F}}>x_{\mathrm{R}}$. We assume that minimal extension $%
\Delta _{\mathrm{F}}$ of the whole wave packet along the axis $x$, is much
less than the distance between the points $x_{\mathrm{L}}^{\mathrm{F}}$\ and
$x_{\mathrm{R}}$, or $x_{\mathrm{R}}^{\mathrm{F}}$\ and $x_{\mathrm{L}}$, $%
\Delta _{\mathrm{F}}\ll x_{\mathrm{R}}^{\mathrm{F}}-x_{\mathrm{L}}$, $x_{%
\mathrm{R}}-x_{\mathrm{L}}^{\mathrm{F}}$.{\large \ } Similar to the
discussion related to Eqs.~(\ref{i3}) and (\ref{i3b}), one can study square
norms of the introduced wave packets $\left( \ ^{\zeta }\psi _{x_{\mathrm{F}%
}},\ ^{\zeta }\psi _{x_{\mathrm{F}}}\right) $ and $\left( \ _{\zeta }\psi
_{x_{\mathrm{F}}},\ _{\zeta }\psi _{x_{\mathrm{F}}}\right) .$ One can
separate contributions to these square norms from the asymptotic regions $S_{%
\mathrm{L}}$ and $S_{\mathrm{R}}$, as follows%
\begin{eqnarray}
\left( ^{\zeta }\psi _{x_{\mathrm{F}}},\;^{\zeta }\psi _{x_{\mathrm{F}%
}}\right) &=&\left( ^{\zeta }\psi _{x_{\mathrm{F}}},\;^{\zeta }\psi _{x_{%
\mathrm{F}}}\right) _{\mathrm{L}}+\left( ^{\zeta }\psi _{x_{\mathrm{F}%
}},\;^{\zeta }\psi _{x_{\mathrm{F}}}\right) _{\mathrm{R}}+O\left( 1\right) ,
\notag \\
\left( _{\zeta }\psi _{x_{\mathrm{F}}},\;_{\zeta }\psi _{x_{\mathrm{F}%
}}\right) &=&\left( _{\zeta }\psi _{x_{\mathrm{F}}},\;_{\zeta }\psi _{x_{%
\mathrm{F}}}\right) _{\mathrm{L}}+\left( _{\zeta }\psi _{x_{\mathrm{F}%
}},\;_{\zeta }\psi _{x_{\mathrm{F}}}\right) _{\mathrm{R}}+O\left( 1\right) ,
\label{d4c}
\end{eqnarray}%
where
\begin{eqnarray}
&&\left( ^{\zeta }\psi _{x_{\mathrm{F}}},\;^{\zeta }\psi _{x_{\mathrm{F}%
}}\right) _{\mathrm{L}}=\int_{V_{\bot }}d\mathbf{r}_{\bot }\int_{-K^{\left(
\mathrm{L}\right) }}^{x_{\mathrm{L}}}\;^{\zeta }\psi _{x_{\mathrm{F}}}^{\dag
}\left( X\right) \;^{\zeta }\psi _{x_{\mathrm{F}}}\left( X\right) dx,  \notag
\\
&&\left( ^{\zeta }\psi _{x_{\mathrm{F}}},\;^{\zeta }\psi _{x_{\mathrm{F}%
}}\right) _{\mathrm{R}}=\int_{V_{\bot }}d\mathbf{r}_{\bot }\int_{x_{\mathrm{R%
}}}^{K^{\left( \mathrm{R}\right) }}\;^{\zeta }\psi _{x_{\mathrm{F}}}^{\dag
}\left( X\right) \;^{\zeta }\psi _{x_{\mathrm{F}}}\left( X\right) dx;  \notag
\\
&&\left( _{\zeta }\psi _{x_{\mathrm{F}}},\;_{\zeta }\psi _{x_{\mathrm{F}%
}}\right) _{\mathrm{L}}=\int_{V_{\bot }}d\mathbf{r}_{\bot }\int_{-K^{\left(
\mathrm{L}\right) }}^{x_{\mathrm{L}}}\;_{\zeta }\psi _{x_{\mathrm{F}}}^{\dag
}\left( X\right) \;_{\zeta }\psi _{x_{\mathrm{F}}}\left( X\right) dx,  \notag
\\
&&\left( _{\zeta }\psi _{x_{\mathrm{F}}},\;_{\zeta }\psi _{x_{\mathrm{F}%
}}\right) _{\mathrm{R}}=\int_{V_{\bot }}d\mathbf{r}_{\bot }\int_{x_{\mathrm{R%
}}}^{K^{\left( \mathrm{R}\right) }}\;_{\zeta }\psi _{x_{\mathrm{F}}}^{\dag
}\left( X\right) \;_{\zeta }\psi _{x_{\mathrm{F}}}\left( X\right) dx.
\label{d4a}
\end{eqnarray}%
Let us study integrals (\ref{d4a}) following the procedure described in
Appendix \ref{t-const}. Taking into account the mutual decompositions of the
plane waves (\ref{rel1}) and using spin and coordinate factorization of
Dirac spinors given by Eq.~(\ref{e7}), one can represent integrals (\ref{d4a}%
) in the following forms%
\begin{eqnarray}
\left( ^{\zeta }\psi _{x_{\mathrm{F}}},\;^{\zeta }\psi _{x_{\mathrm{F}%
}}\right) _{\mathrm{L}} &=&\left\vert \;^{\zeta }N_{x_{\mathrm{F}%
}}\right\vert ^{2}\int_{V_{\bot }}d\mathbf{r}_{\bot }\int_{-K^{\left(
\mathrm{L}\right) }}^{x_{\mathrm{L}}}\left\vert \;^{\zeta }\varphi _{x_{%
\mathrm{F}}}^{\mathrm{L}}\left( X\right) \right\vert ^{2}dx,  \notag \\
\left( ^{\zeta }\psi _{x_{\mathrm{F}}},\;^{\zeta }\psi _{x_{\mathrm{F}%
}}\right) _{\mathrm{R}} &=&\left\vert \;^{\zeta }N_{x_{\mathrm{F}%
}}\right\vert ^{2}\int_{V_{\bot }}d\mathbf{r}_{\bot }\int_{x_{\mathrm{R}%
}}^{K^{\left( \mathrm{R}\right) }}\left\vert \;^{\zeta }\varphi _{x_{\mathrm{%
F}}}^{\mathrm{R}}\left( X\right) \right\vert ^{2}dx;  \notag \\
\left( _{\zeta }\psi _{x_{\mathrm{F}}},\;_{\zeta }\psi _{x_{\mathrm{F}%
}}\right) _{\mathrm{L}} &=&\left\vert \;_{\zeta }N_{x_{F}}\right\vert
^{2}\int_{V_{\bot }}d\mathbf{r}_{\bot }\int_{-K^{\left( \mathrm{L}\right)
}}^{x_{\mathrm{L}}}\left\vert \;_{\zeta }\varphi _{x_{\mathrm{F}}}^{\mathrm{L%
}}\left( X\right) \right\vert ^{2}dx,  \notag \\
\left( _{\zeta }\psi _{x_{\mathrm{F}}},\;_{\zeta }\psi _{x_{\mathrm{F}%
}}\right) _{\mathrm{R}} &=&\left\vert \;_{\zeta }N_{x_{F}}\right\vert
^{2}\int_{V_{\bot }}d\mathbf{r}_{\bot }\int_{x_{\mathrm{R}}}^{K^{\left(
\mathrm{R}\right) }}\left\vert \;_{\zeta }\varphi _{x_{\mathrm{F}}}^{\mathrm{%
R}}\left( X\right) \right\vert ^{2}dx,  \label{d5}
\end{eqnarray}%
where%
\begin{eqnarray}
^{\zeta }\varphi _{x_{\mathrm{F}}}^{\mathrm{R}}\left( X\right) &=&\sum_{n\in
\Omega _{3}}\frac{\ ^{\zeta }c_{n}^{\left( x_{\mathrm{F}}\right) }\exp
\left( -ip_{0}t+ip^{\mathrm{R}}\ x+i\mathbf{p}_{\bot }\mathbf{r}_{\bot
}\right) }{\sqrt{2}\left\vert g\left( _{+}\left\vert ^{-}\right. \right)
\right\vert },  \notag \\
^{\zeta }\varphi _{x_{\mathrm{F}}}^{\mathrm{L}}\left( X\right) &=&\sum_{n\in
\Omega _{3}}\left[ g\left( _{+}\left\vert ^{\zeta }\right. \right)
e^{i\left\vert p^{\mathrm{L}}\right\vert x}-g\left( _{-}\left\vert ^{\zeta
}\right. \right) e^{-i\left\vert p^{\mathrm{L}}\right\vert x}\right] \frac{\
^{\zeta }c_{n}^{\left( x_{\mathrm{F}}\right) }\exp \left( -ip_{0}t+i\mathbf{p%
}_{\bot }\mathbf{r}_{\bot }\right) }{\sqrt{2}\left\vert g\left(
_{+}\left\vert ^{-}\right. \right) \right\vert };  \notag \\
_{\zeta }\varphi _{x_{\mathrm{F}}}^{\mathrm{R}}\left( X\right) &=&\sum_{n\in
\Omega _{3}}\left[ g\left( ^{+}\left\vert _{\zeta }\right. \right)
e^{i\left\vert p^{\mathrm{R}}\right\vert x}-g\left( ^{-}\left\vert _{\zeta
}\right. \right) e^{-i\left\vert p^{\mathrm{R}}\right\vert x}\right] \frac{\
_{\zeta }c_{n}^{\left( x_{\mathrm{F}}\right) }\exp \left( -ip_{0}t+i\mathbf{p%
}_{\bot }\mathbf{r}_{\bot }\right) }{\sqrt{2}\left\vert g\left(
_{+}\left\vert ^{-}\right. \right) \right\vert },  \notag \\
_{\zeta }\varphi _{x_{\mathrm{F}}}^{\mathrm{L}}\left( X\right) &=&\sum_{n\in
\Omega _{3}}\frac{\ _{\zeta }c_{n}^{\left( x_{\mathrm{F}}\right) }\exp
\left( -ip_{0}t+ip^{\mathrm{L}}x+i\mathbf{p}_{\bot }\mathbf{r}_{\bot
}\right) }{\sqrt{2}\left\vert g\left( _{+}\left\vert ^{-}\right. \right)
\right\vert }.  \label{d6}
\end{eqnarray}

Absolute values of the asymptotic momenta $\left\vert p^{\mathrm{L}%
}\right\vert $ and $\left\vert p^{\mathrm{R}}\right\vert \ $are determined
by the quantum numbers $p_{0}$ and $\mathbf{p}_{\bot },$ see Eqs. (\ref%
{2.62b}) and (\ref{2.6a}). This fact imposes certain correlations between
both quantities. In particular, one can see from Eq. (\ref{2.61b}) that $%
d\left\vert p^{\mathrm{L}}\right\vert /d\left\vert p^{\mathrm{R}}\right\vert
<0,$ and at any given $\mathbf{p}_{\bot }$ these quantities are restricted
inside the range $\Omega _{3}$,%
\begin{equation}
0\leq \left\vert p^{\mathrm{R/L}}\right\vert \leq p^{\mathrm{\max }},\;\;p^{%
\mathrm{\max }}=\sqrt{\mathbb{U}\left( \mathbb{U}-2\pi _{\bot }\right) }.
\label{d8}
\end{equation}

As an example, let us consider an electron wave packet with a given spin
polarization $\sigma $ and transversal momentum $\mathbf{p}_{\bot }$,
\begin{equation}
\ ^{\zeta }\psi _{x_{\mathrm{F}}}\left( X\right) =\frac{\ ^{\zeta }N_{x_{%
\mathrm{F}}}T}{2\pi }\int_{p_{0}\in \Omega _{3}}\mathcal{M}_{n}^{-1/2}\
^{\zeta }c_{n}^{\left( x_{\mathrm{F}}\right) }\ ^{\zeta }\psi _{n}\left(
X\right) dp_{0},  \label{d9}
\end{equation}%
where the integration over $p_{0}$ is fulfilled for a given fixed $\mathbf{p}%
_{\bot }$ and the corresponding asymptotic scalar functions (\ref{d6}) are%
\begin{eqnarray}
\ ^{\zeta }\varphi _{x_{\mathrm{F}}}^{\mathrm{L}}\left( X\right)  &=&\frac{T%
}{2\pi }\int_{p_{0}\in \Omega _{3}}\frac{\ ^{\zeta }c_{n}^{\left( x_{\mathrm{%
F}}\right) }\exp \left( -ip_{0}t+i\mathbf{p}_{\bot }\mathbf{r}_{\bot
}\right) }{\sqrt{2}\left\vert g\left( _{+}\left\vert ^{-}\right. \right)
\right\vert }\left[ g\left( _{+}\left\vert ^{\zeta }\right. \right)
e^{i\left\vert p^{\mathrm{L}}\right\vert x}-g\left( _{-}\left\vert ^{\zeta
}\right. \right) e^{-i\left\vert p^{\mathrm{L}}\right\vert x}\right] dp_{0},
\notag \\
^{\zeta }\varphi _{x_{\mathrm{F}}}^{\mathrm{R}}\left( X\right)  &=&\frac{T}{%
2\pi }\int_{p_{0}\in \Omega _{3}}\frac{\ ^{\zeta }c_{n}^{\left( x_{\mathrm{F}%
}\right) }\exp \left( -ip_{0}t+ip^{\mathrm{R}}x+i\mathbf{p}_{\bot }\mathbf{r}%
_{\bot }\right) }{\sqrt{2}\left\vert g\left( _{+}\left\vert ^{-}\right.
\right) \right\vert }dp_{0}.\   \label{d10}
\end{eqnarray}%
Using Eqs.~(\ref{2.59}) and (\ref{2.62b}), we express $p_{0}$ via $p^{%
\mathrm{R}}$ as
\begin{equation*}
p_{0}=U_{\mathrm{R}}-\sqrt{\left( p^{\mathrm{R}}\right) ^{2}+\pi _{\bot }^{2}%
}.
\end{equation*}%
Then we denote the mean value of $p^{\mathrm{R}}$ in a wave packet in $S_{%
\mathrm{R}}$ as $\bar{p}^{\mathrm{R}}$ , and the mean value of $|p^{\mathrm{L%
}}|$ of the same packet in $S_{\mathrm{L}}$ as $\left\vert \bar{p}^{\mathrm{L%
}}\right\vert $. Afterwards we chose coefficients $\ ^{\zeta }c_{n}^{\left(
x_{\mathrm{F}}\right) }$ as follows%
\begin{equation}
\frac{\ ^{\zeta }c_{n}^{\left( x_{\mathrm{F}}\right) }}{\left\vert g\left(
_{+}\left\vert ^{-}\right. \right) \right\vert }dp_{0}=-d\left\vert p^{%
\mathrm{R}}\right\vert \frac{1}{\Delta _{\mathrm{F}}}\int_{-\Delta _{\mathrm{%
F}}/2}^{+\Delta _{\mathrm{F}}/2}d\delta x_{\mathrm{F}}\frac{\exp \left[
-ip_{0}\zeta T/2-i\zeta \left\vert p^{\mathrm{R}}\right\vert \ \left( x_{%
\mathrm{F}}+\delta x_{\mathrm{F}}\right) \right] }{\left\vert \bar{g}\left(
_{+}\left\vert ^{-}\right. \right) \right\vert },  \label{d12}
\end{equation}%
where the quantity $\left\vert \bar{g}\left( _{+}\left\vert ^{-}\right.
\right) \right\vert =\left\vert g\left( _{+}\left\vert ^{-}\right. \right)
\right\vert _{p^{\mathrm{R}}=\bar{p}^{\mathrm{R}}}$ does not depend on $%
\left\vert p^{\mathrm{R}}\right\vert $. This allows one to represent wave
packets (\ref{d10}) as integrals over $\left\vert p^{\mathrm{R}}\right\vert $%
,%
\begin{eqnarray}
&&\ ^{\zeta }\varphi _{x_{\mathrm{F}}}^{\mathrm{R}}\left( X\right) =\frac{\
^{\zeta }D}{\Delta _{\mathrm{F}}}\int_{-\Delta _{\mathrm{F}}/2}^{+\Delta _{%
\mathrm{F}}/2}d\delta x_{\mathrm{F}}\int_{0}^{p^{\mathrm{\max }}}d\left\vert
p^{\mathrm{R}}\right\vert   \notag \\
&&\times \exp \left[ i\sqrt{\left( p^{\mathrm{R}}\right) ^{2}+\pi _{\bot
}^{2}}\left( t+\zeta T/2\right) +i\zeta \left\vert p^{\mathrm{R}}\right\vert
\left( x-x_{\mathrm{F}}-\delta x_{\mathrm{F}}\right) \right] ,  \notag \\
&&\ ^{\zeta }\varphi _{x_{\mathrm{F}}}^{\mathrm{L}}\left( X\right) =\frac{\
^{\zeta }D}{\Delta _{\mathrm{F}}}\int_{-\Delta _{\mathrm{F}}/2}^{+\Delta _{%
\mathrm{F}}/2}d\delta x_{\mathrm{F}}\int_{0}^{p^{\mathrm{\max }}}d\left\vert
p^{\mathrm{R}}\right\vert   \notag \\
&&\times \exp \left[ i\sqrt{\left( p^{\mathrm{R}}\right) ^{2}+\pi _{\bot
}^{2}}\left( t+\zeta T/2\right) -i\zeta \left\vert p^{\mathrm{R}}\right\vert
\left( x_{\mathrm{F}}+\delta x_{\mathrm{F}}\right) \right]   \notag \\
&&\times \left[ g\left( _{+}\left\vert ^{\zeta }\right. \right)
e^{i\left\vert p^{\mathrm{L}}\right\vert \ x}-\ g\left( _{-}\left\vert
^{\zeta }\right. \right) e^{-i\left\vert p^{\mathrm{L}}\right\vert \ x}%
\right] ,\ \   \label{d13} \\
&&\ ^{\zeta }D=\frac{T\exp \left[ -iU_{\mathrm{R}}\left( t+\zeta T/2\right)
+i\mathbf{p}_{\bot }\mathbf{r}_{\bot }\right] }{2\sqrt{2}\pi \left\vert \bar{%
g}\left( _{+}\left\vert ^{-}\right. \right) \right\vert }.  \notag
\end{eqnarray}

The case $p^{\mathrm{\max }}\rightarrow 0$ where $\left\vert p^{\mathrm{\max
}}\left( x_{\mathrm{R}}-x_{\mathrm{F}}-\delta x_{\mathrm{F}}\right)
\right\vert \lesssim 1$, takes place for relatively weak fields, or near the
border between $\Omega _{3}$ and $\Omega _{2}$, and, as was already said
above, is characterized by big values of the quantity $\left\vert g\left(
_{+}\left\vert ^{-}\right. \right) \right\vert \sim \left\vert g\left(
_{+}\left\vert ^{+}\right. \right) \right\vert \rightarrow \infty $. In this
case, it follows from (\ref{d13}) that the asymptotic densities $\left\vert
\ ^{\zeta }\varphi _{x_{\mathrm{F}}}^{\mathrm{R}}\left( X\right) \right\vert
^{2}$ tend to zero, i.e., electron wave packets do not penetrate in the
asymptotic region $S_{\mathrm{R}}$, whereas the absolute values of the
coefficients in front of the incoming and outgoing plane waves in the
expression for $\ ^{\zeta }\varphi _{x_{\mathrm{F}}}^{\mathrm{L}}\left(
X\right) $ are equal.

In the case when $p^{\mathrm{\max }}$ is not small i.e., $p^{\mathrm{\max }%
}\Delta _{\mathrm{F}}\gg 1$, we consider first the situation when $x_{%
\mathrm{F}}\in $ $S_{\mathrm{L}}$, $x_{\mathrm{F}}<x_{\mathrm{L}}^{\mathrm{F}%
}$, and $x_{\mathrm{F}}+\delta x_{\mathrm{F}}<x_{\mathrm{L}}^{\mathrm{F}}$
for all $\delta x_{\mathrm{F}}$. In our general setting of the problem we
have $-T/2<t<T/2$ that is why $\zeta \left( t+\zeta T/2\right) >0$ and
therefore in the region $S_{\mathrm{R}}$ where $x>x_{\mathrm{R}}>0$ the
exponent index in the expression (\ref{d13}) for $\ ^{\zeta }\varphi _{x_{%
\mathrm{F}}}^{\mathrm{R}}\left( X\right) $ is not zero. Moreover, at any
time instant, high-frequency oscillations in the latter expression lead to
vanishing the asymptotic densities $\left\vert \ ^{\zeta }\varphi _{x_{%
\mathrm{F}}}^{\mathrm{R}}\left( X\right) \right\vert ^{2}\rightarrow 0$. It
is easy to see that situation is quite different in the asymptotic region $%
S_{\mathrm{L}}$, where $x<x_{\mathrm{L}}^{\mathrm{F}}<x_{\mathrm{L}}<0$.
Here $\ ^{\zeta }\varphi _{x_{\mathrm{F}}}^{\mathrm{L}}\left( X\right) $ is
a superposition of two types of plane waves with opposite signs of the
quantum number $p^{\mathrm{L}}$. That is why there always exists such an
area on the axis $x$ where the exponent index in the expression (\ref{d13})
for $\ ^{\zeta }\varphi _{x_{\mathrm{F}}}^{\mathrm{L}}\left( X\right) $ is
zero. In particular, when $\left\vert t+\zeta T/2\right\vert \sim 0$, there
always exists an $x$ such that $\left\vert p^{\mathrm{R}}\right\vert \left(
x_{\mathrm{F}}+\delta x_{\mathrm{F}}\right) -\left\vert p^{\mathrm{L}%
}\right\vert \ x=0$ for any $\zeta $.

This corresponds to incoming wave packets at $t\rightarrow -T/2$\ for $\zeta
=-$\ and outgoing wave packets at $t\rightarrow T/2$\ for $\zeta =+$. Note
that mean currents of these electron and positron wave packets are zero
unlike the mean currents of constituent plane waves. To understand such a
distinct behavior, it useful to recall that mean currents are defined by the
inner product (\ref{IP}), which is, in particular, the average value over
the period of time from $-T/2$\ to $T/2$,\ where $T$\ is the time dimension
of a large space-time box. Thus, a certain direction of a wave packet in a
given time instant matches with the zero average current of this wave packet.

Let us suppose now that $x_{\mathrm{F}}\in $ $S_{\mathrm{R}}$, $x_{\mathrm{F}%
}>x_{\mathrm{R}}^{\mathrm{F}}$, and $x_{\mathrm{F}}+\delta x_{\mathrm{F}}>x_{%
\mathrm{R}}^{\mathrm{F}}$ for all $\delta x_{\mathrm{F}}$. In such a case
there exist a coordinate $x\in S_{\mathrm{R}}$ and $x>x_{\mathrm{R}}^{%
\mathrm{F}}>x_{\mathrm{R}}>0$, such that the exponent index in the
expression (\ref{d13}) for $\ ^{\zeta }\varphi _{x_{\mathrm{F}}}^{\mathrm{R}%
}\left( X\right) $ is zero. However, since $\ ^{\zeta }\varphi _{x_{\mathrm{F%
}}}^{\mathrm{L}}\left( X\right) $ is a superposition of two types of\emph{\ }%
plane waves with opposite signs of the quantum number $p^{\mathrm{L}}$, in
the asymptotic region $S_{\mathrm{L}}$ there always exists such an area on
the axis $x$ where the exponent index in the expression (\ref{d13}) for $\
^{\zeta }\varphi _{x_{\mathrm{F}}}^{\mathrm{L}}\left( X\right) $ is also
zero. This means that such wave packets cannot represent an electron. This
result holds true for any electron wave packets. Indeed, in our reasonings
we have used only the general structure (\ref{d6}) of functions $\ ^{\zeta
}\varphi _{x_{\mathrm{F}}}^{\mathrm{R}}\left( X\right) $ which consist of
only one type of plane waves with the same sign of the quantum number $p^{%
\mathrm{R}},$ whereas the functions $\ ^{\zeta }\varphi _{x_{\mathrm{F}}}^{%
\mathrm{L}}\left( X\right) $ represent superpositions of the two types of
plane waves with opposite signs of the quantum number $p^{\mathrm{L}}$.
Namely this is the reason why electron packets cannot be localized only in $%
S_{\mathrm{R}}$ and cannot represent stable states describing electrons. We
see that in the framework of our consideration there are no electrons in the
region\emph{\ }$S_{\mathrm{R}}$ with quantum numbers from the range $\Omega
_{3}$.

It is not difficult to give similar example for positron wave packets and
prove that they can be localized only in one asymptotic region, namely in $%
S_{\mathrm{R}}$.

Thus, in the range $\Omega _{3}$\ there exists the same localization of
electrons as in the range $\Omega _{2}$ and positron localization as in the
range $\Omega _{4}$. That is why in contrast to the ranges $\Omega _{1}$%
\emph{\ }and\emph{\ }$\Omega _{5}$, any initial and final wave packets in
the range $\Omega _{3}$ may come in and go out only to the same asymptotic
region, which corresponds to the total reflection both for electrons and
positrons.\emph{\ }

\section{Differential mean number in slowly alternating field\label{Auniform}%
}

The absolute values of $\left\vert p^{\mathrm{R}}\right\vert $ and $%
\left\vert p^{\mathrm{L}}\right\vert $ are related by Eq.~(\ref{2.61b}). In
the range $\Omega _{3}$ these relations imply Eq.~(\ref{d8}) and%
\begin{equation}
0\leq \left\vert \left\vert p^{\mathrm{L}}\right\vert -\left\vert p^{\mathrm{%
R}}\right\vert \right\vert \leq p^{\mathrm{\max }}  \label{g8}
\end{equation}%
Therefore, for big $\alpha $ that satisfies Eq.~(\ref{2.61b}), we obtain%
\begin{equation}
\pi \alpha \left[ eE\alpha -\frac{1}{2}\left\vert \left\vert p^{\mathrm{L}%
}\right\vert -\left\vert p^{\mathrm{R}}\right\vert \right\vert \right] \gg 1.
\label{g9}
\end{equation}%
As a consequence of (\ref{g9}) the quantities $N_{n}^{\mathrm{cr}}$ given by
Eq. (\ref{exs4}) for fermions, and by Eq. (\ref{exs4b}) for bosons have
approximately the same form%
\begin{equation}
N_{n}^{\mathrm{cr}}=\left\vert g\left( _{+}\left\vert ^{-}\right. \right)
\right\vert ^{-2}\approx 4\sinh \left( \pi \alpha \left\vert p^{\mathrm{L}%
}\right\vert \right) \sinh \left( \pi \alpha \left\vert p^{\mathrm{R}%
}\right\vert \right) \exp \left( -2\pi eE\alpha ^{2}\right) .  \label{exs9}
\end{equation}

Then it follows from Eq. (\ref{exs9}) that if the range $\Omega _{3}$ is
small enough
\begin{equation}
eE\alpha -\pi _{\bot }\rightarrow 0\Longrightarrow \pi \alpha p^{\mathrm{%
\max }}\ll 1.  \label{exs10}
\end{equation}%
then the the quantities $N_{n}^{\mathrm{cr}}$\ are exponentially small.

Let us consider the opposite case of big ranges $\Omega _{3}$ when%
\begin{equation}
\pi \alpha p^{\mathrm{\max }}\gg 1  \label{exs11}
\end{equation}%
and the quantities $N_{n}^{\mathrm{cr}}$ are not small. Such ranges do exist
if%
\begin{equation}
eE\alpha \gg m  \label{exs12}
\end{equation}%
and
\begin{equation}
\alpha \pi _{\bot }<K_{\bot },  \label{exs11b}
\end{equation}%
where $K_{\bot }$ is a given arbitrary number, $m\alpha \ll K_{\bot }\ll
eE\alpha ^{2}$.

In this case, we consider first finite subranges adjoining the range $\Omega
_{3}$ from inside. In such subranges%
\begin{eqnarray}
N_{n}^{\mathrm{cr}} &\simeq &2\sinh \left( \pi \alpha \left\vert p^{\mathrm{R%
}}\right\vert \right) \exp \left[ -\pi \alpha \sqrt{\left( p^{\mathrm{R}%
}\right) ^{2}+\pi _{\bot }^{2}}\right] \ \mathrm{if}\ \ \,\pi \alpha
\left\vert p^{\mathrm{R}}\right\vert <km\alpha ,  \notag \\
N_{n}^{\mathrm{cr}} &\simeq &2\sinh \left( \pi \alpha \left\vert p^{\mathrm{L%
}}\right\vert \right) \exp \left[ -\pi \alpha \sqrt{\left( p^{\mathrm{L}%
}\right) ^{2}+\pi _{\bot }^{2}}\right] \ \mathrm{if}\ \ \,\pi \alpha
\left\vert p^{\mathrm{L}}\right\vert <km\alpha ,  \label{exs13a}
\end{eqnarray}%
where $k\gtrsim 1$ is a given arbitrary number, obeying the inequality%
\begin{equation}
km\alpha \ll eE\alpha ^{2}.  \label{exs12b}
\end{equation}

Near the borders of the range $\Omega _{3},$ we have%
\begin{eqnarray}
N_{n}^{\mathrm{cr}} &\approx &2\sinh \left( \pi \alpha \left\vert p^{\mathrm{%
R}}\right\vert \right) e^{-\pi \alpha \pi _{\bot }}\ \mathrm{if}\ \ \,\pi
\alpha \left\vert p^{\mathrm{R}}\right\vert <K^{0},  \notag \\
N_{n}^{\mathrm{cr}} &\approx &2\sinh \left( \pi \alpha \left\vert p^{\mathrm{%
L}}\right\vert \right) e^{-\pi \alpha \pi _{\bot }}\ \mathrm{if}\ \ \,\pi
\alpha \left\vert p^{\mathrm{L}}\right\vert <K^{0},  \label{exs13b}
\end{eqnarray}%
where $K^{0}<1$ is an arbitrary number. The numbers $N_{n}^{\mathrm{cr}}$
given by Eq.~(\ref{exs13b}) are exponentially small, $N_{n}^{\mathrm{cr}%
}\lesssim 2e^{-\pi m\alpha }.$

For border areas situated more close to the center of the range $\Omega _{3}$%
, where $1\lesssim \pi \alpha \left\vert p^{\mathrm{R,L}}\right\vert $, we
have%
\begin{eqnarray}
N_{n}^{\mathrm{cr}} &\approx &\exp \left[ -\pi \alpha \left[ \sqrt{\left( p^{%
\mathrm{R}}\right) ^{2}+\pi _{\bot }^{2}}-\left\vert p^{\mathrm{R}%
}\right\vert \right] \right] \ \mathrm{if}\ \ \,1\lesssim \pi \alpha
\left\vert p^{\mathrm{R}}\right\vert \lesssim \pi km\alpha ,  \notag \\
N_{n}^{\mathrm{cr}} &\approx &\exp \left[ -\pi \alpha \left[ \sqrt{\left( p^{%
\mathrm{L}}\right) ^{2}+\pi _{\bot }^{2}}-\left\vert p^{\mathrm{L}%
}\right\vert \right] \right] \ \mathrm{if}\ \ \,1\lesssim \pi \alpha
\left\vert p^{\mathrm{L}}\right\vert \lesssim \pi km\alpha .  \label{exs16}
\end{eqnarray}%
The numbers $N_{n}^{\mathrm{cr}}$ are growing as $n$ recedes from the
borders of the range $\Omega _{3}$ and for any fixed $\pi _{\bot }$ achieves
his maximum when $\pi \alpha \left\vert p^{\mathrm{R}}\right\vert
\rightarrow \pi km\alpha $, or $\alpha \left\vert p^{\mathrm{L}}\right\vert
\rightarrow \pi km\alpha $.{\Huge \ }In turn, this maximum value grows as $%
\pi _{\bot }\rightarrow m$. Thus, in the subranges under consideration, we
can estimate the quantities $N_{n}^{\mathrm{cr}}$ from above as
\begin{equation}
N_{n}^{\mathrm{cr}}<\exp \left( -\frac{\pi m\alpha }{2k}\right) .
\label{exs17}
\end{equation}%
This quantity is exponentially small\textrm{\ }if for any given $k$ the
ratio $\frac{\pi m\alpha }{2k}>1$ is big enough.

The main contribution to the particle creation is due to the inner part of
the range $\Omega _{3}$. This range is defined by the following inequalities%
\begin{equation}
\pi \alpha \left\vert p^{\mathrm{R}}\right\vert >\pi km\alpha ,\ \ \,\pi
\alpha \left\vert p^{\mathrm{R}}\right\vert >\pi km\alpha .  \label{exs20}
\end{equation}%
They correspond to the following restrictions on the energy $p_{0}:$
\begin{equation}
\alpha \left\vert p_{0}\right\vert <eE\alpha ^{2}-K,\ \ \,K=\alpha \sqrt{%
\left( km\right) ^{2}+\pi _{\bot }^{2}}.  \label{exs21}
\end{equation}

Inequalities (\ref{exs11b}) and (\ref{exs12b}) imply that $K\ll eE\alpha
^{2} $. In such a case, we can approximate the numbers (\ref{exs9}) as%
\begin{equation}
N_{n}^{\mathrm{cr}}\approx N_{p_{0},\mathbf{p}_{\bot }}^{\mathrm{as}%
}=e^{-\pi \tau },\ \,\tau =\alpha \left( 2eE\alpha -\left\vert p^{\mathrm{R}%
}\right\vert -\left\vert p^{\mathrm{L}}\right\vert \right)  \label{exs22}
\end{equation}

The function $\tau $ is minimal at $p_{0}=0$,%
\begin{equation}
\min \tau =\left. \tau \right\vert _{p_{0}=0}=\lambda =\frac{\pi _{\bot }^{2}%
}{eE},  \label{exs23}
\end{equation}%
it grows monotonically as $\left\vert p_{0}\right\vert $ grows, and takes
its maximum value%
\begin{equation}
\tau _{\max }=\left. \tau \right\vert _{\left\vert p_{0}\right\vert
=eE\alpha -K/\alpha }\approx K-km\alpha +\lambda /4.  \label{exs23.2}
\end{equation}%
on the boundary of the $\Omega _{3}$ range. In the wide range of energies
where $\alpha \left\vert p_{0}\right\vert \ll eE\alpha ^{2}$, the numbers $%
N_{n}^{\mathrm{cr}}$ do not depend practically on the parameter $\alpha $
and have the form of the differential number of created particles in an
uniform electric field \cite{Nikis79,Nikis70b},%
\begin{equation}
N_{n}^{\mathrm{cr}}\approx e^{-\pi \lambda }.  \label{exs23b}
\end{equation}


\begin{thebibliography}{99}
\bibitem{Nikis70a} A.I. Nikishov, \textit{Pair production by a constant
electric field}, Zh. Eksp. Teor. Fiz. \textbf{57}, 1210 (1969) [Transl. Sov.
Phys. JETP \textbf{30}, 660 (1970)].

\bibitem{Nikis79} A.I. Nikishov, \textit{Problems of intense external field
in quantum electrodynamics}, in \emph{Quantum Electrodynamics of Phenomena
in Intense Fields}, Proc. P.N. Lebedev Phys. Inst. (Nauka, Moscow, 1979),
Vol. 111, p. 153.

\bibitem{GMR85} W. Greiner, B. M\"{u}ller, and J. Rafelski, \emph{Quantum
Electrodynamics of Strong Fields} (Springer-Verlag, Berlin, 1985).

\bibitem{ruffini} R. Ruffini, G. Vereshchagin, and S. Xue, \textit{%
Electron-positron pairs in physics and astrophysics: from heavy nuclei to
black holes}, Phys. Rep. \textbf{487}, 1 (2010).

\bibitem{Gitman} D.M. Gitman, \textit{Processes of arbitrary order in
quantum electrodynamics with a pair-creating external field}, J. Phys. A
\textbf{10,} 2007 (1977); E.S. Fradkin and D.M. Gitman, \textit{Furry
picture for quantum electrodynamics with pair-creating external field},
Fortschr. Phys. \textbf{29,} 381 (1981); E.S. Fradkin, D.M. Gitman, and S.M.
Shvartsman, \emph{Quantum Electrodynamics with Unstable Vacuum}
(Springer-Verlag, Berlin, 1991).

\bibitem{Klein27} O. Klein, \textit{Die Reflexion von Elektronen einem
Potentialsprung nach der relativistischen Dynamik von Dirac}, Z. Phys.
\textbf{53}, 157 (1929); \textit{Elelrtrodynamik und Wellenmechanik vom
Standpunkt des Korrespondenzprinzips}, Z. Phys. \textbf{41}, 407 (1927).

\bibitem{Sauter31a} F. Sauter, \textit{\"{U}ber das Verhalten eines
Elektrons im homogenen elektrischen Feld nach der relativistischen Theorie
Diracs},\emph{\ }Z. Phys. \textbf{69,} 742 (1931).

\bibitem{Sauter-pot} F. Sauter, \textit{Zum "Klenschen Paradoxon"}\emph{,}
Z. Phys. \textbf{73}, 547 ( 1932).

\bibitem{Hol98} B. R. Holstein, \textit{Klein's paradox}, Am. J. Phys.
\textbf{66}, 507 (1998).

\bibitem{KrekSuGr04} P. Krekora, Q. Su, and R. Grobe, \textit{Klein Paradox
in Spatial and Temporal Resolution}, Phys. Rev. Lett. \textbf{92}, 040406
(2004).

\bibitem{Gerr+etal11} R. Gerritsma, B. P. Lanyon, G. Kirchmair, F. Z\"{a}%
hringer, C. Hempel, J. Casanova, J. J. Garc\'{\i}a-Ripoll, E. Solano, R.
Blatt, and C. F. Roos, \textit{Quantum Simulation of the Klein Paradox with
Trapped Ions}, Phys. Rev. Lett. \textbf{106}, 060503 (2011).

\bibitem{DomCal99} N. Dombey and A. Calogeracos, \textit{Seventy years of
the Klein paradox}, Phys. Rep. \textbf{315, }41 (1999); \textit{History and
Physics of the Klein Paradox}, Contemp. Phys. \textbf{40,} 313 (1999)
[arXiv:quant-ph/9905076].

\bibitem{Nikis70b} A.I. Nikishov, \textit{Barrier scattering in field
theory: removal of Klein paradox},\emph{\ }Nucl. Phys. \textbf{B21, }346
(1970).

\bibitem{HansRavn81} A. Hansen and F. Ravndal, \textit{Klein's Paradox and
Its Resolution}, Phys. Scr. \textbf{23}, 1036 (1981).

\bibitem{Damour77} T. Damour, \textit{Klein paradox and vacuum polarization}%
, in \emph{Proceedings of the} \emph{First Marcel Grossmann Meeting on
General Relativity,} edited by R. Ruffini (North-Holland, Amsterdam, 1977),
p. 459.

\bibitem{WongW88} R-Ch. Wang and Ch-Y. Wong, \textit{Finite-size effect in
the Schwinger particle-production mechanism}, Phys. Rev. D \textbf{38}, 348%
\textit{\ (1988).}

\bibitem{Nikis04} \textit{A.I. Nikishov,\ Scattering and pair production by
a potential barrier}, Yad. Fiz. \textbf{67, }1503 (2004) [Transl. Phys.
Atom. Nucl. \textbf{67, }1478 (2004); arXiv:hep-th/0304174]; \textit{On the
Theory of Scalar Pair Production by a Potential Barrier}, Problems of Atomic
Science and Technology (Kharkov, Ukraine, 2001), p. 103
[arXiv:hep-th/0111137].

\bibitem{RQM93} J. D. Bjorken and S. D. Drell, \emph{Relativistic Quantum
Mechanics} (McGraw--Hill, New York, 1964); F. Gross,\emph{\ Relativistic
Quantum Mechanics and Field Theory} (Wiley, New York, 1993); B. R. Holstein,
\emph{Topics in Advanced Quantum Mechanics} (Addison--Wesley, Reading, MA,
1992).

\bibitem{NKM99} H. Nitta, T. Kudo, and H. Minowa, \textit{Motion of a wave
packet in the Klein paradox}, Am. J. Phys. \textbf{67}, 966 (1999).

\bibitem{CHM08} L. C. B. Crispino, A. Higuchi, and G. E. A. Matsas, \textit{%
The Unruh effect and its applications}, Rev. Mod. Phys. \textbf{80}, 787
(2008).

\bibitem{AndMot14} P. R. Anderson and E. Mottola, \textit{On the Instability
of Global de Sitter Space to Particle Creation}, Phys. Rev. D \textbf{89},
104038 (2014) [arXiv:1310.0030].

\bibitem{Dun09} G. V. Dunne,\emph{\ }\textit{New strong-field QED effects at
ELI: Nonperturbative vacuum pair production}, Eur. Phys. J. D \textbf{55},
327 (2009); A. Di Piazza, C. M\"{u}ller, K. Z. Hatsagortsyan, and C. H.
Keitel, \textit{Extremely high-intensity laser interactions with fundamental
quantum systems}, Rev. Mod. Phys. \textbf{84}, 1177 (2012); G. Mourou and T.
Tajima, \textit{Summary of the IZEST science and aspiration}, Eur. Phys. J.
Special Topics \textbf{223}, 979 (2014); G. V. Dunne, \textit{Extreme
quantum field theory and particle physics with IZEST}, Eur. Phys. J. Special
Topics \textbf{223}, 1055 (2014); B. M. Hegelich, G. Mourou, and J.
Rafelski, \emph{P}\textit{robing the quantum vacuum with ultra intense laser
pulses}, Eur. Phys. J. Special Topics \textbf{223}, 1093 (2014).

\bibitem{NJBN12} P. D. Nation, J. R. Johansson, M. P. Blencowe, and F. Nori,
\textit{Stimulating uncertainty: Amplifying the quantum vacuum with
superconducting circuits}, Rev. Mod. Phys. \textbf{84}, 1 (2012)
[arXiv:1103.0835].

\bibitem{castroneto} A. H. Castro Neto, F. Guinea, N. M. R. Peres, K. S.
Novoselov, and A. K. Geim, \textit{The electronic properties of graphene},
Rev. Mod. Phys. \textbf{81,} 109 (2009); N. M. R. Peres, \textit{The
transport properties of graphene: An introduction}, Rev. Mod. Phys. \textbf{%
82,} 2673 (2010); M.A.H. Vozmediano, M.I. Katsnelson, and F. Guinea, \textit{%
Gauge fields in graphene}, Phys. Rep. \textbf{496}, 109 (2010).

\bibitem{dassarma} D. Das Sarma, S. Adam, E. H. Hwang, and E. Rossi, \textit{%
Electronic transport in two-dimensional graphene}, Rev. Mod. Phys. \textbf{%
83,} 407 (2011) [arXiv:1007.2849].

\bibitem{top-insul11} M.Z. Hasan and C.L. Kane, \emph{T}\textit{opological
insulators}, Rev. Mod. Phys. \textbf{82,} 3045 (2010); X.-L. Qi and S.-C.
Zhang, \textit{Topological insulators and superconductors}, Rev. Mod. Phys.
\textbf{83,} 1057 (2011).

\bibitem{VafVish14} O. Vafek and A. Vishwanath, \textit{Dirac Fermions in
Solids - from High Tc cuprates and Graphene to Topological Insulators and
Weyl Semimetals}, Annu. Rev. Condens. Matter Phys. \textbf{5}, 83 (2014)
[arXiv:1306.2272].

\bibitem{Vandecasteele10} N. Vandecasteele, A. Barreiro, M. Lazzeri, A.
Bachtold, and F. Mauri, \textit{Current-voltage characteristics of graphene
devices: Interplay between Zener-Klein tunneling and defects}, Phys. Rev. B
\textbf{82}, 045416 (2010) [arXiv:1003.2072].

\bibitem{GavGitY12} S. P. Gavrilov, D. M. Gitman, and N. Yokomizo, \textit{%
Dirac fermions in strong electric field and quantum transport in graphene},
Phys. Rev. D \textbf{86}, 125022 (2012).

\bibitem{KatsNovG06} M. I. Katsnelson, K. S. Novoselov, and A. K. Geim,
\textit{Chiral tunneling and the Klein paradox in graphene}, Nat. Phys.
\textbf{2}, 620 (2006); V.V. Cheianov and V.I. Fal'ko, \textit{Selective
transmission of Dirac electrons and ballistic magnetoresistance of n-p
junctions in graphene}, Phys. Rev. B \textbf{74}, 041403(R) (2006).

\bibitem{Klein-tunn-exp09} N. Stander, B. Huard, and D. Goldhaber-Gordon,
\textit{Evidence for Klein tunneling in graphene p-n junctions}, Phys. Rev.
Lett. \textbf{102}, 026807 (2009); A. F. Young and Ph. Kim, \textit{Quantum
interference and Klein tunnelling in graphene heterojunctions}, Nat. Phys.
\textbf{5}, 222 (2009).

\bibitem{been08} C. W. J. Beenakker, \textit{Andreev reflection and Klein
tunneling in graphene}, Rev. Mod. Phys. \textbf{80}, 1337 (2008).

\bibitem{Log+etal15} R. Logemann, K. J. A. Reijnders, T. Tudorovskiy, M. I.
Katsnelson, and S. Yuan, \textit{Modeling Klein tunneling and caustics of
electron waves in graphene}, Phys. Rev. B \textbf{91}, 045420 (2015)
[arXiv:1409.1277].

\bibitem{allor} D. Allor, T. D. Cohen, and D. A. McGady, \textit{The
Schwinger mechanism and graphene}, Phys. Rev. D \textbf{78}, 096009 (2008).

\bibitem{quench15} L. Bucciantini, S. Sotiriadis, T. Macr\`{\i}, \textit{%
Probing Klein tunneling through quantum quenches}, J. Phys. A \textbf{49},
025002 (2016) [arXiv:1503.05142].

\bibitem{KimPage06} S. P. Kim and D. N. Page, \textit{Schwinger pair
production via instantons in strong electric fields}, Phys. Rev. D \textbf{65%
}, 105002 (2002); \textit{Schwinger pair production in electric and magnetic
fields}, Phys. Rev. D \textbf{73}, 065020 (2006); \textit{Improved
Approximations for Fermion Pair Production in Inhomogeneous Electric Fields}%
, Phys. Rev. D \textbf{75}, 045013 (2007).

\bibitem{KimLeeY10} S. P. Kim, H. K. Lee, and Y. Yoon, \textit{Effective
action of QED in electric field backgrounds. II. Spatially localized fields}%
, Phys. Rev. D \textbf{82}, 025015 (2010).

\bibitem{GieK05} H. Gies and K. Klingmuller, \textit{Pair production in
inhomogeneous field}, Phys. Rev. D \textbf{72},\textbf{\ }065001 (2005)
[hep-th/0505099].

\bibitem{DunS05} G.V. Dunne and C. Schubert, \textit{Worldline instantons
and pair production in inhomogeneous fields}, Phys. Rev. D \textbf{72},
105004 (2005) [hep-th/0507174]; G.V. Dunne, H. Gies, C. Schubert, Q. Wang,
\textit{Worldline Instantons II: The Fluctuation Prefactor}, Phys.Rev. D
\textbf{73},\textbf{\ }065028 (2006) [hep-th/0602176]; D.D. Dietrich and
G.V. Dunne, \textit{Gutzwiller's Trace Formula and Vacuum Pair Production},
J. Phys. A: Math. Theor. \textbf{40},\textbf{\ }F825 (2007)
[arXiv:0706.4006]; R. Sch\"{u}tzhold, H. Gies, G. Dunne, \textit{Dynamically
assisted Schwinger mechanism}, Phys. Rev. Lett. \textbf{101}, 130404 (2008)
[arXiv:0807.0754].

\bibitem{DunW06} G.V. Dunne and Q. Wang, \textit{Multidimensional worldline
instantons}, Phys. Rev. D \textbf{\ 74}, 065015 (2006).

\bibitem{ChK09} A. Chervyakov and H. Kleinert, \textit{Exact pair production
rate for a smooth potential step}, Phys. Rev. D \textbf{80}, 065010 (2009).

\bibitem{ChK11} A. Chervyakov and H. Kleinert, \textit{On Electron-Positron
Pair Production by a Spatially Nonuniform Electric Field}, arXiv:1112.4120.

\bibitem{AkhmP15} E.T. Akhmedov and F.K. Popov, \textit{A few more comments
on secularly growing loop corrections in strong electric fields}, J. High
Energy Phys. 09 (2015) 085 [arXiv:1412.1554].

\bibitem{Schwe61} S. Schweber, \emph{An Introduction to Relativistic Quantum
Field Theory} (Harper \& Row, New York, 1961).

\bibitem{BraWe35} R. Brauer and H. Weyl, \textit{Spinors in}\emph{\ }$n$%
\textit{-dimensions}, Am. J. Math. \textbf{57, }425 (1935).

\bibitem{GitTy90} D.M. Gitman and I.V. Tyutin, \emph{Quantization of Fields
with Constraints} (Berlin--Heidelberg, Springer-Verlag, 1990).

\bibitem{PS} V. De Alfaro and T. Regge,\ \emph{Potential Scattering }%
(Interscience Publishers, New York, 1965).

\bibitem{LanLiQM} L.D. Landau and E.M. Lifshitz, \emph{Quantum Mechanics:
Non-Relativistic Theory} (Pergamon Press, Oxford, 1977).

\bibitem{GavGT06} S.P. Gavrilov, D.M. Gitman, and J.L. Tomazelli, \textit{%
Density matrix of a quantum field in a particle-creating background}, Nucl.
Phys. \textbf{B795, }645 (2008).

\bibitem{S51} J. Schwinger, \textit{On Gauge Invariance and Vacuum
Polarization}, Phys. Rev. \textbf{82}, 664 (1951).

\bibitem{Dunn04} G.V. Dunne, \textit{Heisenberg-Euler Effective Lagrangians:
Basics and extensions}, in I. Kogan Memorial Volume, \emph{From fields to
strings: Circumnavigating theoretical physics}, edited by M Shifman, A.
Vainshtein, and J. Wheater (World Scientific, Singapore, 2005)
[arXiv:hep-th/0406216].

\bibitem{BatE53} \emph{Higher Transcendental Functions} (Bateman Manuscript
Project), edited by A. Erdelyi \emph{et al.} (McGraw-Hill, New York, 1953),
Vols. 1 and 2.

\bibitem{GavG96a} S.P. Gavrilov and D.M. Gitman, \textit{Vacuum instability
in external fields}, Phys. Rev. D \textbf{53}, 7162 (1996).

\bibitem{Somm60} A. Sommerfeld, \emph{Atombau and Spektrallinien} (Friedr.
Vieweg \& Sohn, Braunschweig, 1960), Vol. 11.

\bibitem{Hund40} F. Hund, \textit{Materieerzeugung im anschaulichen und im
gequantelten Wellenbild der Materie}, Z. Phys. \textbf{117}, 1 (1941).

\bibitem{Feyn61} R. P. Feynman, \emph{Quantum Electrodynamics} (W. A.
Benjamin, New York, 1961).

\bibitem{Furry51} W.H. Furry,\emph{\ }\textit{On Bound States and Scattering
in Positron Theory}, Phys. Rev. \textbf{81},115 (1951).
\end{thebibliography}
\end{document}